\documentclass[rmp,twocolumn]{revtex4}

\usepackage{amsmath}
\usepackage{graphicx}
\usepackage{amssymb}
\usepackage{amsthm, amscd}
\usepackage{amsfonts}
\usepackage{cancel}
\usepackage{times}
\usepackage{pstricks}
\usepackage{wasysym}
\usepackage{ulem}
\usepackage[dvips]{epsfig}

\newcommand{\C}{\mathbb{C}}

\newtheorem{thm}{Theorem}[section]

\theoremstyle{remark}

\theoremstyle{definition}
\newtheorem{defn}[thm]{Definition}

\newcounter{mysubsection}
\def\mysubsection#1{\vspace*{3pt}
 \addtocounter{mysubsection}{1}{\small
{\bf{\textsf{(\alph{mysubsection}) #1:}}}}}
\def\mylabel#1{\newcounter{#1}{\setcounter{#1}{\value{mysubsection}}}}
\def\fib{\tau}
\def\pfepsilon{\psi}
\def\pfnotepsilon{\epsilon}

\begin{document}

%%%%%%%%%%%%
\title{Non-Abelian Anyons and Topological Quantum Computation}

\author{Chetan Nayak$^{1,2}$, Steven H. Simon$^3$,
Ady Stern$^4$, Michael Freedman$^1$, Sankar Das Sarma$^{5}$, 
}
\affiliation{
$^1$Microsoft Station Q, University of
California, Santa Barbara, CA 93108\\
$^2$Department of Physics and Astronomy, University of
California, Los Angeles, CA 90095-1547\\
$^3$Alcatel-Lucent, Bell Labs, 600 Mountain Avenue,
Murray Hill, New Jersey 07974\\
$^4$Department of Condensed Matter Physics, Weizmann Institute of Science,
Rehovot 76100, Israel\\
$^5$Department of Physics, University of Maryland,
College Park, MD 20742}

\begin{abstract}
Topological quantum computation has recently emerged as one of the
most exciting approaches to constructing a
fault-tolerant quantum computer. The proposal relies on the
existence of topological states of matter whose quasiparticle
excitations are neither bosons nor fermions, but are particles
known as {\it Non-Abelian anyons}, meaning that they obey
{\it non-Abelian braiding statistics}. Quantum information
is stored in states with multiple quasiparticles, which have a topological
degeneracy. The unitary gate operations
which are necessary for quantum computation are carried out
by braiding quasiparticles, and then measuring the multi-quasiparticle
states. The fault-tolerance of a topological quantum computer
arises from the non-local encoding of the states of the quasiparticles,
which makes them immune to errors caused by local perturbations.
To date, the only such
topological states thought to have been found in nature are
fractional quantum Hall states, most prominently the $\nu=5/2$ state,
although several other prospective candidates have been proposed
in systems as disparate as ultra-cold atoms in optical lattices
and thin film superconductors. In this review article,
we describe current research in this field,
focusing on the general theoretical concepts of non-Abelian
statistics as it relates to topological quantum computation, on
understanding non-Abelian quantum Hall states, on proposed
experiments to detect non-Abelian anyons, and on proposed
architectures for a topological quantum computer. We address
both the mathematical underpinnings of topological quantum computation
and the physics of the subject using the $\nu=5/2$ fractional
quantum Hall state as the archetype of a
non-Abelian topological state enabling fault-tolerant quantum computation.
\end{abstract}

\maketitle
%%%%%%%%%%%%

\tableofcontents

\section{Introduction}

In recent years, physicists' understanding of
the quantum properties of matter has undergone a major revolution
precipitated by surprising experimental discoveries and profound
theoretical revelations. Landmarks include the discoveries of the fractional
quantum Hall effect and high-temperature superconductivity
and the advent of topological quantum field theories.
At the same time, new potential applications for quantum
matter burst on the scene, punctuated by the discoveries of
Shor's factorization algorithm and quantum error correction protocols.
Remarkably, there has
been a convergence between these developments.
Nowhere is this more dramatic than in topological
quantum computation, which seeks to exploit the emergent
properties of many-particle systems to encode and manipulate
quantum information in a manner which is resistant to error.

It is rare for a new scientific paradigm, with its attendant concepts
and mathematical formalism, to develop in parallel
with potential applications, with all of their detailed technical issues.
However, the physics of topological phases of matter is not only evolving
alongside topological quantum computation but is even informed by it.
Therefore, this review must necessarily be rather sweeping in scope,
simply to introduce the concepts of non-Abelian anyons
and topological quantum computation, their
inter-connections, and how they may be realized in physical systems,
particularly in several fractional quantum Hall states. (For a
popular account, see \onlinecite{Collins06}; for a slightly more technical one,
see \onlinecite{DasSarma06b}.) This exposition will take us
on a tour extending from knot theory and topological quantum field theory to
conformal field theory and the quantum Hall effect to quantum
computation and all the way to the physics of gallium arsenide devices.

The body of this paper is composed of three parts, Sections
\ref{part1}, \ref{part2}, and \ref{part3}. Section
\ref{part1} is rather general, avoids technical details, and aims
to introduce concepts at a qualitative level. Section \ref{part1} should be
of interest, and should be accessible, to all readers.  In Section \ref{part2}
we describe the theory of topological phases in more detail.
In Section \ref{part3}, we describe how a topological
phase can be used as a platform for fault-tolerant quantum computation.
The second and third parts are probably of more
interest to theorists, experienced researchers, and those who hope
to conduct research in this field.

Section \ref{sec:Non-Abelian_quantum_statistics} begins by
discussing the concept of braiding statistics in $2+1$-dimensions.
We define the idea of a non-Abelian anyon, a particle exhibiting
non-Abelian braiding statistics. Section \ref{sec:Berry-phase}
discusses how non-Abelian anyons can arise in a many-particle system.
We then review the basic ideas
of quantum computation, and the problems of errors and decoherence
in section \ref{sec:Quantum_Computation}. Those familiar with
quantum computation may be able to skip much of this section. We
explain in section \ref{sec:Topological_Quantum_Computation}
how non-Abelian statistics naturally leads to the
idea of topological quantum computation, and explain why it is a
good approach to error-free quantum computation.  In section
\ref{sec:Non-Abelian_quantum_Hall_states}, we briefly describe the
non-Abelian quantum Hall systems which are the most likely arena for
observing non-Abelian anyons (and, hence, for producing a
topological quantum computer).  Section \ref{sec:quantumHallreview}
gives a very basic review of quantum Hall physics.
Experts in quantum Hall physics may be able to
skip much of this section.   Section \ref{sec:Non-AbelianQHE}
introduces non-Abelian quantum Hall states. 
This section also
explains the importance (and summarizes the results) of numerical
work in this field for determining which quantum Hall states are (or
might be) non-Abelian.  Section \ref{sec:interference} describes
some of the proposed interference experiments which may be able to
distinguish Abelian from non-Abelian quantum Hall states.
Section \ref{sec:FQHE-qc} shows how qubits and elementary gates
can be realized in a quantum Hall device.
Section \ref{sec:PhysicalSystems} discusses some of the engineering
issues associated with the physical systems where quantum Hall
physics is observed.  In section \ref{sec:Othersystems} we discuss
some of the other, non-quantum-Hall systems where it has been
proposed that non-Abelian anyons (and hence topological quantum
computation) might occur.

Sections \ref{part2} and \ref{part3} are still written to be
accessible to the broadest possible audiences, but they should be
expected to be somewhat harder going than Section \ref{part1}.
Section \ref{part2} introduces the theory of topological phases in
detail. Topological quantum computation can only become a reality if
some physical system `condenses' into a non-Abelian topological
phase. In Section \ref{part2}, we describe the universal low-energy,
long-distance physics of such phases. We also discuss how they can
be experimentally detected in the quantum Hall regime, and when they
might occur in other physical systems. Our focus is on a sequence of
universality classes of non-Abelian topological phases, associated
with SU(2)$_k$ Chern-Simons theory which we describe in section
\ref{sec:CS-theory}. The first interesting member of this sequence,
$k=2$, is realized in chiral p-wave superconductors and in the
leading theoretical model for the $\nu=5/2$ fractional quantum Hall
state. Section \ref{sec:pwave} shows how this universality class can
be understood with conventional BCS theory. In section
\ref{sec:Jones}, we describe how the topological properties of the
entire sequence of universality classes (of which $k=2$ is a special
case) can be understood using Witten's celebrated connection between
Chern-Simons theory and the Jones polynomial of knot theory. In
section \ref{sec:FQHE}, we describe an alternate formalism for
understanding the topological properties of Chern-Simons theory,
namely through conformal field theory. The discussion revolves
around the application of this formalism to fractional quantum Hall
states and explains how non-Abelian quantum Hall wavefunctions can
be constructed with conformal field theory. Appendix
\ref{section:CFT} gives a highly-condensed introduction to conformal
field theory. In Section \ref{sec:edge}, we discuss the gapless edge
excitations which necessarily accompany chiral (i.e. parity, $P$ and
time-reversal $T$-violating) topological phases. These excitations
are useful for interferometry experiments, as we discuss in Section
\ref{sec:experiments2}. Finally, in Section \ref{sec:P-T-Invariant},
we discuss topological phases which do not violate parity and
time-reversal symmetries. These phases emerge in models of
electrons, spins, or bosons on lattices which could describe
transition metal oxides, Josephson junction arrays, or ultra-cold
atoms in optical lattices.

In Section \ref{part3}, we discuss
how quasiparticles in topological phases can be used for
quantum computation. We first discuss the case of SU(2)$_2$,
which is the leading candidate for the $\nu=5/2$ fractional quantum
Hall state. We show in Section \ref{sec:5/2-qubits} how qubits and
gates can be manipulated in a gated GaAs device supporting
this quantum Hall state. We discuss
why quasiparticle braiding alone is not sufficient for universal
quantum computation and how this limitation of the $\nu=5/2$
state can be circumvented.
Section \ref{sec:fibonacci} discusses in detail
how topological computations can be performed
in the simplest non-Abelian theory that is capable of universal topological quantum computation, the so-called ``Fibonacci-Anyon" theory.
In \ref{sec:universal-tqc}, we show that the SU(2)$_k$
theories support universal topological quantum
computation for all integers $k$ except $k=1,2,4$.
In \ref{sec:errors}, we discuss the physical processes which
will cause errors in a topological quantum computer.

Finally, we briefly conclude in section \ref{sec:conclusion}.
We discuss questions for the immediate future, primarily
centered on the $\nu=5/2$ and $\nu=12/5$ fractional quantum
Hall states. We also discuss a broader set of question relating
to non-Abelian topological phases and fault-tolerant quantum computation.

\section{Basic Concepts}
\label{part1}

\subsection{Non-Abelian Anyons}
\label{sec:non-Abelian-Anyons}

\subsubsection{Non-Abelian Braiding Statistics}
\label{sec:Non-Abelian_quantum_statistics}

Quantum statistics is one of the basic pillars of the quantum mechanical view of
the world. It is the property which distinguishes fermions from
bosons: the wave function that describes a system of many identical particles should
%both be a solution of the system's Schroedinger equation {\it and}}
satisfy the proper symmetry under the interchange of any two
particles. In $3$ spatial dimension and one time dimension ($3+1$ D)
there are only two possible symmetries --- the wave function of
bosons is symmetric under exchange while that of fermions is
anti-symmetric.
One cannot overemphasize, of course, the importance
of the symmetry of the wavefunction, which is the root of
the Pauli principle, superfluidity, the metallic state,
Bose-Einstein condensation, and a long list of other phenomena.

The limitation to one of two possible types of quantum symmetry originates
from the observation that a process in which two particles are
adiabatically interchanged twice is equivalent to a process in which
one of the particles is adiabatically taken around the other. Since,
in three dimensions, wrapping one particle all the way around another
is topologically equivalent to a process in which none of the
particles move at all, the wave function
should be left unchanged by two such interchanges of particles.
The only two possibilities are for the wavefunction to change by
a $\pm$ sign under a single interchange, corresponding to
the cases of bosons and fermions, respectively.

We can recast this in path integral language.
Suppose we consider all possible
trajectories in $3+1$ dimensions
which take $N$ particles from initial positions
$R_1$, $R_2$, $\ldots$, $R_N$ at time $t_i$ to final positions
$R_1$, $R_2$, $\ldots$, $R_N$ at time $t_f$.
If the particles are distinguishable, then there are no topologically
non-trivial trajectories, i.e. all trajectories can be continuously
deformed into the trajectory in which the particles do not move
at all (straight lines in the time direction). If the particles are
indistinguishable, then the different trajectories fall into
topological classes corresponding to the elements of the
permutation group $S_N$, with each element of the group
specifying how the initial positions are permuted to
obtain the final positions.
To define the quantum evolution
of such a system, we must specify how the permutation
group acts on the states of the system. Fermions and bosons correspond
to the only two one-dimensional irreducible
representations of the permutation group of $N$ identical
particles.\footnote{Higher dimensional representations of the permutation group,
known as `parastatistics', can always be decomposed into fermions
or bosons with an additional quantum number attached to each particle
\cite{Doplicher71a,Doplicher71b}.}

%Two and one spatial dimensions are qualitatively different from
%three in that respect. In (non-compact) one dimensional systems an
%adiabatic exchange of two particles is impossible without the
%particles passing through one another. Thus, when particles interact
%through a hard core interaction, bosons become equivalent to
%fermions.

%shs removed discussion of 1D.  I think it was a bit distracting
%-- particularly so early in the paper when we
% are just trying to introduce the topic.

%Two dimensional systems are the most intriguing ones in this
%context, due to the following topological consideration. In three
%dimensions a particle loop enclosing another particle may be
%deformed continuously (See Fig. \ref{fig:unwrap}) into a point
%without any singularity (i.e., without going through the  enclosed
%particle). In contrast, in a two dimensional system such a deformation
%necessarily crosses the enclosed particle.

Two-dimensional systems are qualitatively different from three
(and higher dimensions) in this respect.
A particle loop that encircles another particle in
two dimensions cannot be deformed to a point without cutting through
the other particle. Consequently, the notion of a winding of one
particle around another in two dimensions is well-defined. Then,
when two particles are interchanged twice in a clockwise manner, their
trajectory involves a non-trivial winding, and the
system does not necessarily come back to the same state. This
topological difference between two and three dimensions, first
realized by \onlinecite{Leinaas77} and by
\onlinecite{Wilczek82a}, leads to a profound difference in the
possible quantum mechanical properties, at least as a matter of
principle, for quantum systems when particles are confined to $2+1$
D (see also \onlinecite{Goldin81} and \onlinecite{Wu84}).
(As an aside, we mention that in $1+1$ D,
quantum statistics is not well-defined since particle interchange is
impossible without one particle going through another, and bosons
with hard-core repulsion are equivalent to fermions.)

Suppose that we have two identical particles in two dimensions. Then when
one particle is exchanged in a counter-clockwise manner
with the other, the wavefunction can change by an arbitrary phase,
\begin{equation}
\label{eqn:exchange-phase}
\psi\left({\bf r_1},{\bf r_2}\right) \rightarrow e^{i\theta} \psi\left({\bf r_1},{\bf r_2}\right)
\end{equation}
The phase need not be merely a $\pm$ sign because
a second counter-clockwise exchange
need not lead back to the initial state but can result in
a non-trivial phase:
\begin{equation}
\label{eqn:two-exchange-phase}
\psi\left({\bf r_1},{\bf r_2}\right) \rightarrow e^{2i\theta} \psi\left({\bf r_1},{\bf r_2}
\right)
\end{equation}
The special cases $\theta=0,\pi$ correspond to
bosons and fermions, respectively. Particles with
other values of the `statistical angle' $\theta$ are
called {\it anyons} \cite{Wilczek90}. We will often refer to such particles
as anyons with statistics $\theta$.

\begin{figure} \ifx\JPicScale\undefined\def\JPicScale{.3}\fi
\psset{unit=.4 mm} \vspace*{30pt}\hspace*{10pt} \psset{unit=0.012mm}
\psset{yunit=0.012mm} \psset{xunit=-0.016mm} \hspace*{0pt}
\psset{linewidth=50,dotsep=1,hatchwidth=0.3,hatchsep=1.5,shadowsize=1,dimen=middle}
%\psset{dotsize=0.7 2.5,dotscale=1 1,fillcolor=black}
%\psset{arrowsize=12,arrowlength=1,arrowinset=0.25,tbarsize=0.75,bracketlength=0.15,rbracketlength=0.15}
\begin{pspicture}(0,7600)(400,8400)
\psline[linestyle=dashed,arrows=->](1400,7600)(1400,8800)
\rput[angle=90](1400,7400){\large time}
\psline(0.00,7600.00)(-0.04,8000.00)
\psline(400.00,7600.00)(399.96,8000.04)
\psbezier(-0.04,8000.00)(-0.04,8200.00)(399.96,8200.02)(399.96,8400.00)
\pspolygon[linewidth=0pt,fillcolor=white,fillstyle=solid,linecolor=white](199.96,8033.35)(33.29,8200.00)(199.96,8366.67)(366.63,8200.02)
\psbezier(399.96,8000.04)(399.96,8200.02)(-0.04,8200.00)(-0.04,8400.00)
\psline[arrows=->](-0.04,8400.00)(-0.04,8800.00)
\psline[arrows=->](399.96,8400.00)(399.96,8800.00)
\psline(0.00,7600.00)(-0.04,8000.00)
\psline[arrows=->](-380,7600)(-380,8800)
\end{pspicture}
\hspace*{95pt}  \psset{unit=0.005 mm} \psset{unit=0.012mm}
\psset{yunit=0.012mm} \psset{xunit=-0.016mm}
%\psset{linewidth=0.3,dotsep=1,hatchwidth=0.3,hatchsep=1.5,shadowsize=1,dimen=middle}
%\psset{dotsize=0.7 2.5,dotscale=1 1,fillcolor=black}
%\psset{arrowsize=12,arrowlength=1,arrowinset=0.25,tbarsize=0.75,bracketlength=0.15,rbracketlength=0.15}
\begin{pspicture}
\psline(0.00,7600.00)(-0.04,8000.00)
\psline(400.00,7600.00)(399.96,8000.04)
\psbezier(-0.04,8000.00)(-0.04,8200.00)(399.96,8200.02)(399.96,8400.00)
\pspolygon[linewidth=0pt,fillcolor=white,fillstyle=solid,linecolor=white](199.96,8033.35)(33.29,8200.00)(199.96,8366.67)(366.63,8200.02)
\psbezier(399.96,8000.04)(399.96,8200.02)(-0.04,8200.00)(-0.04,8400.00)
\psline[arrows=->](-0.04,8400.00)(-0.04,8800.00)
\psline[arrows=->](399.96,8400.00)(399.96,8800.00)
\psline(0.00,7600.00)(-0.04,8000.00)
\psline[arrows=->](750,7600)(750,8800) \rput(2700,8150){\large
$\sigma_1$} \rput(-175,8150){\large $\sigma_2$}
\end{pspicture}
\psset{yunit=0.010mm}
\\ \vspace*{40pt} \hspace*{15pt}
\begin{pspicture}(0,7600)(400,8400)
\psline(0.00,7600.00)(-0.04,8000.00)
\psline(400.00,7600.00)(399.96,8000.04)
\psbezier(-0.04,8000.00)(-0.04,8200.00)(399.96,8200.02)(399.96,8400.00)
\pspolygon[linewidth=0pt,fillcolor=white,fillstyle=solid,linecolor=white](199.96,8033.35)(33.29,8200.00)(199.96,8366.67)(366.63,8200.02)
\psbezier(399.96,8000.04)(399.96,8200.02)(-0.04,8200.00)(-0.04,8400.00)
\psline[arrows=->](-0.04,8400.00)(-0.04,8800.00)
\psline[arrows=->](399.96,8400.00)(399.96,8800.00)
\psline(0.00,7600.00)(-0.04,8000.00)
\psline[arrows=->](-380,7600)(-380,8800)
\end{pspicture}
\hspace*{90pt}
\begin{pspicture}
\psline(0.00,7600.00)(-0.04,8000.00)
\psline(400.00,7600.00)(399.96,8000.04)
\psbezier(-0.04,8000.00)(-0.04,8200.00)(399.96,8200.02)(399.96,8400.00)
\pspolygon[linewidth=0pt,fillcolor=white,fillstyle=solid,linecolor=white](199.96,8033.35)(33.29,8200.00)(199.96,8366.67)(366.63,8200.02)
\psbezier(399.96,8000.04)(399.96,8200.02)(-0.04,8200.00)(-0.04,8400.00)
\psline[arrows=->](-0.04,8400.00)(-0.04,8800.00)
\psline[arrows=->](399.96,8400.00)(399.96,8800.00)
\psline(0.00,7600.00)(-0.04,8000.00)
\psline[arrows=->](750,7600)(750,8800)
\end{pspicture}
\\ \vspace*{10pt} \hspace*{35pt}
\begin{pspicture}
\psline(-20,7600.00)(-20,8000.00)
\psline(370.00,7600.00)(370.,8000.04)
\psbezier(-0.04,8000.00)(-0.04,8200.00)(370,8200.02)(370,8400.00)
\pspolygon[linewidth=0pt,fillcolor=white,fillstyle=solid,linecolor=white](199.96,8033.35)(33.29,8200.00)(199.96,8366.67)(366.63,8200.02)
\psbezier(370,8000.04)(370,8200.02)(-0.04,8200.00)(-0.04,8400.00)
\psline[arrows=->](-0.04,8400.00)(-0.04,8800.00)
\psline[arrows=->](370,8400.00)(370,8800.00)
\psline(0.00,7600.00)(-0.04,8000.00)
\psline[arrows=->](760,7600)(760,8800) \rput(-500,8900){\large
$\neq$}
\end{pspicture}
\hspace*{38pt}
\begin{pspicture}(0,7600)(400,8400)
\psline(40,7600.00)(40,8000.00)
\psline(400.00,7600.00)(399.96,8000.04)
\psbezier(40,8000.00)(40,8200.00)(399.96,8200.02)(399.96,8400.00)
\pspolygon[linewidth=0pt,fillcolor=white,fillstyle=solid,linecolor=white](199.96,8033.35)(33.29,8200.00)(199.96,8366.67)(366.63,8200.02)
\psbezier(399.96,8000.04)(399.96,8200.02)(40,8200.00)(40,8400.00)
\psline[arrows=->](40,8400.00)(40,8800.00)
\psline[arrows=->](399.96,8400.00)(399.96,8800.00)
\psline(30,7600.00)(30,8000.00)
\psline[arrows=->](-360,7600)(-360,8800)
\end{pspicture}
\psset{yunit=0.010mm}
\\ \vspace*{40pt} \hspace*{15pt}
\begin{pspicture}(0,7600)(400,8400)
\psline(0.00,7600.00)(-0.04,8000.00)
\psline(400.00,7600.00)(399.96,8000.04)
\psbezier(-0.04,8000.00)(-0.04,8200.00)(399.96,8200.02)(399.96,8400.00)
\pspolygon[linewidth=0pt,fillcolor=white,fillstyle=solid,linecolor=white](199.96,8033.35)(33.29,8200.00)(199.96,8366.67)(366.63,8200.02)
\psbezier(399.96,8000.04)(399.96,8200.02)(-0.04,8200.00)(-0.04,8400.00)
\psline[arrows=->](-0.04,8400.00)(-0.04,8800.00)
\psline[arrows=->](399.96,8400.00)(399.96,8800.00)
\psline(0.00,7600.00)(-0.04,8000.00)
\psline[arrows=->](-380,7600)(-380,8800)
\end{pspicture}
\hspace*{90pt}
\begin{pspicture}
\psline(0.00,7600.00)(-0.04,8000.00)
\psline(400.00,7600.00)(399.96,8000.04)
\psbezier(-0.04,8000.00)(-0.04,8200.00)(399.96,8200.02)(399.96,8400.00)
\pspolygon[linewidth=0pt,fillcolor=white,fillstyle=solid,linecolor=white](199.96,8033.35)(33.29,8200.00)(199.96,8366.67)(366.63,8200.02)
\psbezier(399.96,8000.04)(399.96,8200.02)(-0.04,8200.00)(-0.04,8400.00)
\psline[arrows=->](-0.04,8400.00)(-0.04,8800.00)
\psline[arrows=->](399.96,8400.00)(399.96,8800.00)
\psline(0.00,7600.00)(-0.04,8000.00)
\psline[arrows=->](750,7600)(750,8800)
\end{pspicture}
\\ \vspace*{10pt} \hspace*{35pt}
\begin{pspicture}
\psline(-20,7600.00)(-20,8000.00)
\psline(370.00,7600.00)(370.,8000.04)
\psbezier(-0.04,8000.00)(-0.04,8200.00)(370,8200.02)(370,8400.00)
\pspolygon[linewidth=0pt,fillcolor=white,fillstyle=solid,linecolor=white](199.96,8033.35)(33.29,8200.00)(199.96,8366.67)(366.63,8200.02)
\psbezier(370,8000.04)(370,8200.02)(-0.04,8200.00)(-0.04,8400.00)
\psline[arrows=->](-0.04,8400.00)(-0.04,8800.00)
\psline[arrows=->](370,8400.00)(370,8800.00)
\psline(0.00,7600.00)(-0.04,8000.00)
\psline[arrows=->](760,7600)(760,8800) \rput(-500,8000){\large $=$}
\end{pspicture}
\hspace*{38pt}
\begin{pspicture}(0,7600)(400,8400)
\psline(40,7600.00)(40,8000.00)
\psline(400.00,7600.00)(399.96,8000.04)
\psbezier(40,8000.00)(40,8200.00)(399.96,8200.02)(399.96,8400.00)
\pspolygon[linewidth=0pt,fillcolor=white,fillstyle=solid,linecolor=white](199.96,8033.35)(33.29,8200.00)(199.96,8366.67)(366.63,8200.02)
\psbezier(399.96,8000.04)(399.96,8200.02)(40,8200.00)(40,8400.00)
\psline[arrows=->](40,8400.00)(40,8800.00)
\psline[arrows=->](399.96,8400.00)(399.96,8800.00)
\psline(30,7600.00)(30,8000.00)
\psline[arrows=->](-360,7600)(-360,8800)
\end{pspicture}
\\ \vspace*{10pt} \hspace*{15pt}
\begin{pspicture}(0,7600)(400,8400)
\psline(0.00,7600.00)(-0.04,8000.00)
\psline(400.00,7600.00)(399.96,8000.04)
\psbezier(-0.04,8000.00)(-0.04,8200.00)(399.96,8200.02)(399.96,8400.00)
\pspolygon[linewidth=0pt,fillcolor=white,fillstyle=solid,linecolor=white](199.96,8033.35)(33.29,8200.00)(199.96,8366.67)(366.63,8200.02)
\psbezier(399.96,8000.04)(399.96,8200.02)(-0.04,8200.00)(-0.04,8400.00)
\psline[arrows=->](-0.04,8400.00)(-0.04,8800.00)
\psline[arrows=->](399.96,8400.00)(399.96,8800.00)
\psline(0.00,7600.00)(-0.04,8000.00)
\psline[arrows=->](-380,7600)(-380,8800)
\end{pspicture}
\hspace*{90pt}
\begin{pspicture}
\psline(0.00,7600.00)(-0.04,8000.00)
\psline(400.00,7600.00)(399.96,8000.04)
\psbezier(-0.04,8000.00)(-0.04,8200.00)(399.96,8200.02)(399.96,8400.00)
\pspolygon[linewidth=0pt,fillcolor=white,fillstyle=solid,linecolor=white](199.96,8033.35)(33.29,8200.00)(199.96,8366.67)(366.63,8200.02)
\psbezier(399.96,8000.04)(399.96,8200.02)(-0.04,8200.00)(-0.04,8400.00)
\psline[arrows=->](-0.04,8400.00)(-0.04,8800.00)
\psline[arrows=->](399.96,8400.00)(399.96,8800.00)
\psline(0.00,7600.00)(-0.04,8000.00)
\psline[arrows=->](750,7600)(750,8800)
\end{pspicture}
\caption{{\bf Top:} The two elementary braid operations $\sigma_1$
and  $\sigma_2$ on three particles. {\bf Middle:} Here we show
$\sigma_2 \sigma_1 \neq \sigma_1 \sigma_2$, hence the braid group is
Non-Abelian.    {\bf Bottom:}  The braid relation  (Eq.
\ref{eq:braidrelation1}) $\sigma_i \sigma_{i+1} \sigma_i =
\sigma_{i+1} \sigma_i \sigma_{i+1}$.}
  \label{fig:braidexample}
\end{figure}
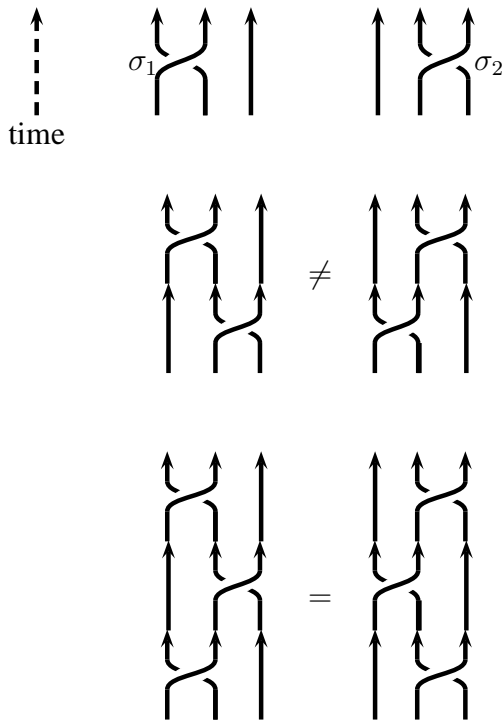

Let us now consider the general case of $N$ particles,
where a more complex structure arises.
The topological classes of trajectories which
take these particles from initial positions
$R_1$, $R_2$, $\ldots$, $R_N$ at time $t_i$ to final positions
$R_1$, $R_2$, $\ldots$, $R_N$ at time $t_f$ are in
one-to-one correspondence with the elements of the braid
group ${\cal B}_N$.
An element of the braid group can be visualized by thinking
of trajectories of particles as world-lines (or strands)
in 2+1 dimensional space-time originating at
initial positions and terminating at final positions,
as shown in Figure~\ref{fig:braidexample}.
The time direction will be represented vertically on the page,
with the initial time at the bottom and the final time
at the top.  An element of the $N$-particle braid group is
an equivalence class of such trajectories up to
smooth deformations. To represent an
element of a class, we will
draw the trajectories on paper with the initial and final
points ordered along lines at the initial and final times. When
drawing the trajectories, we must be careful to distinguish
when one strand passes over or under another, corresponding
to a clockwise or counter-clockwise exchange.
We also require that any intermediate time slice must
intersect $N$ strands. Strands cannot `double back',
which would amount to particle creation/annihilation
at intermediate stages. We do not allow this because we
assume that the particle number is known.
(We will consider particle creation/annihilation later
in this paper when we discuss field theories of anyons
and, from a mathematical perspective, when we discuss
the idea of a ``category'' in section \ref{part3} below.)
Then, the multiplication of two elements of the braid group
is simply the successive execution of the corresponding trajectories,
i.e. the vertical stacking of the two drawings.
(As may be seen from the figure, the order in which they are
multiplied is important because the group is non-Abelian,
meaning that multiplication is not commutative.)

The braid group can be represented algebraically in terms of
generators $\sigma_i$, with $1 \leq i \leq N-1$.
We choose an arbitrary ordering of the particles
$1,2,\ldots,N$.\footnote{Choosing a different ordering would
amount to a relabeling of the elements of the braid group,
as given by conjugation by the braid which transforms one ordering
into the other.}
 $\sigma_i$ is a counter-clockwise exchange of the $i^\text{th}$ and
$(i+1)^\text{th}$ particles. $\sigma_i^{-1}$ is, therefore, a
clockwise exchange of the $i^\text{th}$ and $(i+1)^\text{th}$
particles. The $\sigma_i$s satisfy the defining relations (see
Fig. \ref{fig:braidexample}),
\begin{eqnarray}
{\sigma_i} {\sigma_j} &=& {\sigma_j} {\sigma_i} \hskip 0.5 cm
\mbox{for } |i-j|\geq 2\cr {\sigma_i} \sigma_{i+1} {\sigma_i} &=&
\sigma_{i+1} {\sigma_i}\, \sigma_{i+1} \hskip 0.5 cm \mbox{for }
1\leq i \leq n-1 \label{eq:braidrelation1}
\end{eqnarray}
The only difference from the permutation group $S_N$ is that
${\sigma_i^2}~\neq~1$, but this makes an enormous difference.
While the permutation group is finite, the number
of elements in the group $\left|{S_N}\right|=N!$,
the braid group is infinite, even for just two particles.
Furthermore, there are non-trivial topological classes of
trajectories even when the particles are distinguishable,
e.g. in the two-particle case those trajectories in which
one particle winds around the other an integer number of times.
These topological classes correspond to the elements
of the `pure' braid group, which is the subgroup of the braid group
containing only elements which bring each particle back to
its own initial position, not the initial position of one of the other particles.
The richness of the braid group is the key fact enabling quantum
computation through quasiparticle braiding.

To define the quantum evolution of a system, we must now
specify how the braid group acts on the states of the system.
The simplest possibilities are one-dimensional representations
of the braid group. In these cases, the wavefunction acquires
a phase $\theta$ when one particle is taken around another,
analogous to Eqs. \ref{eqn:exchange-phase}, \ref{eqn:two-exchange-phase}.
The special cases $\theta=0,\pi$ are bosons and fermions, respectively,
while particles with other values of $\theta$ are {\it anyons} \cite{Wilczek90}.
These are straightforward many-particle generalizations of
the two-particle case considered above.
An arbitrary element of the braid group is represented by
the factor $e^{im\theta}$ where $m$ is the total number of times that
one particle winds around another in a counter-clockwise manner
(minus the number of times that a particle winds around another
in a clockwise manner). These representations are Abelian since
the order of braiding operations in unimportant. However, they
can still have a quite rich structure since there can be $n_s$
different particle species with parameters $\theta_{ab}$,
where $a,b=1,2,\ldots,{n_s}$, specifying the phases resulting
from braiding a particle of type $a$ around a particle of type $b$.
Since distinguishable particles can braid
non-trivially, i.e. $\theta_{ab}$ can be non-zero for $a\neq b$
as well as for $a=b$, anyonic `statistics' is, perhaps,
better understood as a kind of topological
interaction between particles.

We now turn to non-Abelian braiding statistics, which are
associated with higher-dimensional representations of
the braid group. Higher-dimensional representations
can occur when there is a degenerate set of $g$ states with
particles at fixed positions $R_1$, $R_2$, $\ldots$, $R_n$.
Let us define an orthonormal basis $\psi_\alpha$, $\alpha=1,2,\ldots,g$
of these degenerate states.
Then an element of the braid group -- say $\sigma_1$, which exchanges
particles 1 and 2 -- is represented by
a $g\times g$ unitary matrix $\rho({\sigma_1})$ acting on these states.
\begin{equation}
{\psi_\alpha} \rightarrow {\left[\rho({\sigma_1})\right]_{\alpha\beta}}\,{\psi_\beta}
\end{equation}
On the other hand, exchanging particles 2 and 3 leads to:
\begin{equation}
{\psi_\alpha} \rightarrow {\left[\rho({\sigma_2})\right]_{\alpha\beta}}\,{\psi_\beta}
\end{equation}
Both $\rho({\sigma_1})$ and $\rho({\sigma_2})$
are $g\times g$ dimensional unitary matrices, which
define unitary transformation within the subspace of degenerate
ground states. If $\rho({\sigma_1})$ and $\rho({\sigma_1})$ do not commute, ${\left[\rho({\sigma_1})\right]_{\alpha\beta}}
{\left[\rho({\sigma_2})\right]_{\beta\gamma}}\neq
{\left[\rho({\sigma_2})\right]_{\alpha\beta}}
{\left[\rho({\sigma_1})\right]_{\beta\gamma}}$, the particles obey
{\it non-Abelian braiding statistics}.  Unless they commute for any
interchange of particles, in which case the particles' braiding statistics is
Abelian, braiding quasiparticles will cause non-trivial rotations
within the degenerate many-quasiparticle Hilbert space.
Furthermore, it will essentially be true at low energies
that the {\it only} way to make non-trivial unitary operations
on this degenerate space is by braiding quasiparticles around each other.
This statement is equivalent to a
statement that no local perturbation can have nonzero matrix
elements within this degenerate space.

A system with anyonic particles must generally have multiple
types of anyons. For instance, in a system with Abelian
anyons with statistics $\theta$, a bound state of two such particles
has statistics $4\theta$. Even if no such stable bound state exists,
we may wish to bring two anyons close together while all
other particles are much further away. Then the two anyons can
be approximated as a single particle whose quantum numbers are obtained
by combining the quantum numbers, including the topological
quantum numbers, of the two particles. As a result, a complete
description of the system must also include these `higher' particle species.
For instance, if there are $\theta=\pi/m$ anyons in system, then there are also
$\theta=4\pi/m,9\pi/m,\ldots,{(m-1)^2}\pi/m$. Since the statistics parameter
is only well-defined up to $2\pi$, $\theta={(m-1)^2}\pi/m=-\pi/m$ for $m$ even
and $\pi-\pi/m$ for $m$ odd. The formation of a different type of
anyon by bringing together two anyons is called {\it fusion}.
When a statistics $\pi/m$ particle is fused with a statistics $-\pi/m$
particle, the result has statistics $\theta=0$. It is convenient to
call this the `trivial' particle. As far as topological properties are concerned,
such a boson is just as good as the absence of any particle,
so the `trivial' particle is also sometimes simply called the `vacuum'.
We will often denote the trivial particle by ${\bf 1}$.

With Abelian anyons which are made by forming successively
larger composites of $\pi/m$ particles, the {\it fusion rule} is:
$\frac{{n^2}\pi}{m}\times\frac{{k^2}\pi}{m}=\frac{{(n+k)^2}\pi}{m}$.
(We will use $a\times b$ to denote $a$ fused with
$b$.)
However, for non-Abelian anyons, the situation is more complicated.
As with ordinary quantum numbers,
there might not be a unique way of combining topological quantum numbers
(e.g. two spin-$1/2$ particles could combine to
form either a spin-$0$ or a spin-$1$ particle).
The different possibilities are called the different {\it fusion channels}.
This is usually denoted by
\begin{equation}
{\phi_a}\times{\phi_b}={\sum_c}N^c_{ab}{\phi_c}
\end{equation}
which represents the fact that when
a particle of species $a$ fuses with one of species $b$,
the result can be a particle of species $c$ if $N^c_{ab}\neq 0$.
For Abelian anyons, the fusion multiplicities
$N^c_{ab}=1$ for only one value of $c$ and $N^{c'}_{ab}=0$ for all $c'\neq c$.
For particles of type $k$ with statistics ${\theta_k}=\pi{k^2}/m$,
i.e. $N^{k''}_{kk'}=\delta_{k+k',k''}$.
For non-Abelian anyons, there is at least one $a$, $b$ such that
there are multiple fusion channels $c$ with $N^c_{ab}\neq 0$.
In the examples which we will be considering in this paper,
$N^c_{ab}= 0$ or $1$, but there are theories for which
$N^c_{ab}>1$ for some $a,b,c$. In this case, $a$ and $b$ can
fuse to form $c$ in $N^c_{ab}>1$ different distinct ways.
We will use $\bar{a}$ to denote the antiparticle of particle species $a$.
When $a$ and $\bar{a}$ fuse, they can always fuse to $1$ in precisely
one way, i.e. $N^1_{a\bar{a}}=1$; in the non-Abelian case,
they may or may not be able to fuse to other particle types as well.

The different fusion channels are one way of accounting for the different
degenerate multi-particle states. Let us see how this works
in one simple model of non-Abelian anyons which we
discuss in more detail in section \ref{part2}.
As we discuss in section \ref{part2}, this model is associated with
`Ising anyons' (which are so-named for reasons which will become
clear in sections \ref{sec:FQHE} and \ref{sec:edge}),
SU(2)$_2$, and chiral $p$-superconductors. There
are slight differences between these three theories, relating to
Abelian phases, but these are unimportant for the present discussion.
This model has three different types of anyons,
which can be variously called $1,\sigma,\psi$ or $0,\frac{1}{2},1$.
(Unfortunately, the notation is a little confusing because the trivial
particle is called `{\bf 1}' in the first model but `0` in the second,
however, we will avoid confusion by using bold-faced ${\bf 1}$
to denote the trivial particle.)
The fusion rules for such anyons are
\begin{eqnarray}
\sigma\times\sigma &=& {\bf 1}+\psi,\:\:\:\:
\sigma\times\psi=\sigma,\:\:\:\: \psi\times\psi=
{\bf 1},\cr {\bf 1}\times x &=& x
\:\:\mbox{ for }\:\:x={\bf 1},\sigma,\psi
\label{eq:isingfusion1}
\end{eqnarray}
(Translating these rules into
the notation of SU(2)$_2$, we see that
these fusion rules are very similar to the decomposition rules for tensor
products of irreducible SU(2) representations, but differ in the important
respect that $1$ is the maximum spin so that $\frac{1}{2}\times\frac{1}{2}=0+1$,
as in the SU(2) case, but $\frac{1}{2}\times 1=\frac{1}{2}$ and
$1\times 1 = 0$.) Note that there are two different fusion channels
for two $\sigma$s. As a result, if there are four $\sigma$s which
fuse together to give ${\bf 1}$, there is a two-dimensional space
of such states. If we divided the four $\sigma$s into two pairs,
by grouping particles $1, 2$ and $3,4$, then
a basis for the two-dimensional space is given by the state
in which $1,3$ fuse to ${\bf 1}$ or $1,3$ fuse to $\psi$
($2,4$ must fuse to the same particle type as $1,3$ do
in order that all four particles fuse to ${\bf 1}$).
We can call these states $\Psi_1$ and $\Psi_\psi$;
they are a basis for the four-quasiparticle Hilbert space
with total topological charge ${\bf 1}$.
(Similarly, if they all fused to give $\psi$, there would be another
two-dimensional degenerate space; one basis is given by the state in
which the first pair fuses to ${\bf 1}$ while the second fuses to $\psi$
and the state in which the opposite occurs.)

Of course, our division of the four $\sigma$s into two pairs
was arbitrary. We could have divided them differently, say, into
the pairs $1,3$ and $2,4$. We would thereby
obtain two different basis states, ${\tilde \Psi}_1$ and
${\tilde \Psi}_\psi$,
in which both pairs fuse to ${\bf 1}$ or to $\psi$, respectively.
This is just a different basis in the same two-dimensional
space. The matrix parametrizing this basis change (see also
Appendix \ref{section:CFT}) is called
the $F$-matrix: ${{\tilde \Psi}_a} = F_{ab}{\Psi_b}$, where $a,b=1,\psi$.
There should really be $6$ indices on $F$
if we include indices to specify the $4$ particle types:
$\left[F^{ijk}_{l}\right]_{ab}$, but we have dropped these
other indices since $i=j=k=l=\sigma$ in our case.
The $F$-matrices are sometimes called $6j$ symbols since
they are analogous to the corresponding quantities for SU(2)
representations. Recall that in SU(2), there are multiple states
in which spins ${{\bf j}_1}$, ${{\bf j}_2}$, ${{\bf j}_3}$
couple to form a total spin ${\bf J}$.
For instance, ${{\bf j}_1}$ and ${{\bf j}_2}$ can add to
form ${\bf j}_{12}$, which can then
add with ${\bf j}_3$ to give ${\bf J}$. The eigenstates of
$\left({\bf j}_{12}\right)^2$ form a basis of the different states with fixed ${{\bf j}_1}$,
${{\bf j}_2}$, ${{\bf j}_3}$, and ${\bf J}$. Alternatively, ${{\bf j}_2}$
and ${{\bf j}_3}$ can add to form ${\bf j}_{23}$, which can then add
with ${\bf j}_1$ to give ${\bf J}$. The eigenstates of $\left({\bf j}_{23}\right)^2$
form a different basis. The $6j$ symbol gives the basis change
between the two. The $F$-matrix of a system of anyons plays
the same role when particles of topological charges $i,j,k$ fuse to total topological
charge $l$. If $i$ and $j$ fuse to $a$, which then fuses with $k$
to give topological charge $l$, the different allowed $a$ define a basis.
If $j$ and $k$ fuse to $b$ and then fuse with $i$ to give topological charge $l$,
this defines another basis, and the $F$-matrix is the unitary transformation
between the two bases. States with more than 4 quasiparticles
can be understood by successively fusing additional particles,
in a manner described in Section \ref{sec:CS-theory}. The $F$-matrix
can be applied to any set of 4 consecutively fused particles.

The different states in this degenerate multi-anyon state space transform
into each other under braiding. However, two particles cannot
change their fusion channel simply by braiding with each other
since their total topological charge can be measured along a
far distant loop enclosing the two particles.
They must braid with a third particle in order to change
their fusion channel. Consequently, when two particles fuse
in a particular channel (rather than a linear superposition of channels),
the effect of taking one particle around
the other is just multiplication by a phase. This phase resulting from
a counter-clockwise exchange of particles of types $a$ and $b$ which
fuse to a particle of type $c$ is called $R^{ab}_c$.
In the Ising anyon case, as we will derive in section \ref{part2}
and Appendix \ref{section:CFT}.1,
${R^{\sigma\sigma}_1}=e^{-\pi i/8}$, ${R^{\sigma\sigma}_\psi}=e^{3\pi i/8}$,
${R^{\psi\psi}_1}=-1$, ${R^{\sigma\psi}_\sigma}=i$. For an example
of how this works,
suppose that we create a pair of $\sigma$ quasiparticles
out of the vacuum. They will necessarily fuse to ${\bf 1}$.
If we take one around another, the state will change by a phase $e^{-\pi i/8}$.
If we take a third $\sigma$ quasiparticle and take it around one,
but not both, of the first two, then the first two will now fuse to $\psi$,
as we will show in Sec. \ref{part2}. If we now take one of
the first two around the other, the state will change by a
phase $e^{3\pi i/8}$.

In order to fully specify the braiding statistics
of a system of anyons, it is necessary to specify (1) the particle species,
(2) the fusion rules $N_{ab}^c$, (3) the $F$-matrices, and (4) the $R$-matrices.
In section \ref{part3}, we will introduce the other sets of parameters,
namely the topological spins $\Theta_a$ and the $S$-matrix,
which, together with the parameters 1-4 above fully characterize
the topological properties of a system of anyons.
Some readers may be familiar with the incarnation
of these mathematical structures in conformal field theory (CFT),
where they occur for reasons which we explain
in section \ref{sec:FQHE}; we briefly review these properties
in the CFT context in Appendix A.

Quasiparticles obeying non-Abelian braiding
statistics or, simply non-Abelian anyons, were first considered
in the context of conformal field theory by \onlinecite{Moore88,Moore89}
and in the context of Chern-Simons theory by \onlinecite{Witten89}.
They were discussed in the context of discrete gauge
theories and linked to the representation theory of {\it quantum groups}
by \onlinecite{Bais80,Bais92,Bais93a,Bais93b}.
They were discussed in a more general context
by \onlinecite{Fredenhagen89} and \onlinecite{Frohlich90}.
The properties of non-Abelian quasiparticles make them appealing
for use in a quantum computer. But before discussing this, we will briefly
review how they could occur in nature and then
the basic ideas behind quantum computation.

\subsubsection{Emergent Anyons}
\label{sec:Berry-phase}

The preceding considerations show that exotic braiding
statistics is a theoretical possibility in $2+1$-D, but they do not
tell us when and where they might occur in nature.
Electrons, protons, atoms, and photons, are all
either fermions or bosons even when they are confined
to move in a two-dimensional plane. However, if
a system of many electrons (or bosons, atoms, etc.)
confined to a two-dimensional plane has excitations which are
localized disturbances of its quantum-mechanical ground state, known as
{\it quasiparticles}, then these quasiparticles can be anyons. When a system
has anyonic quasiparticle excitations above its ground state,
it is in a {\it topological phase of matter}. (A more precise definition
of a topological phase of matter will be given in Section \ref{part2}.)

Let us see how anyons might arise as an emergent property
of a many-particle system. For the sake of concreteness,
consider the ground state of a $2+1$ dimensional system of
of electrons, whose coordinates are $(r_1, \ldots, r_n)$. We assume that the
ground state is separated from the excited states by an energy gap
(i.e, it is incompressible), as is the situation in fractional
quantum Hall states in 2D electron systems.
The lowest energy electrically-charged
excitations are known as quasiparticles or quasiholes, depending
on the sign of their electric charge. (The term ``quasiparticle"
is also sometimes used in a generic sense to mean both quasiparticle
and quasihole as in the previous paragraph).  These quasiparticles
are local disturbances to the wavefunction of the electrons
corresponding to a quantized amount of total charge.
%If the quasiparticle's charge is screened, local
%measurements may discover the quasiparticle's existence only when
%carried out close to the quasiparticle's position.
%These quasiparticles are very significant for the low energy
%physics of the system, since
%As the lowest energy excitations, at temperatures well below the
%energy gap, these quasiparticles should be the only ``active"
%degrees of freedom the system possesses.

We now introduce into the system's Hamiltonian a scalar potential
composed of many local ``traps", each sufficient to capture exactly
one quasiparticle.  These traps may be created by impurities, by
very small gates, or by the potential created by tips of scanning
microscopes.  The quasiparticle's charge screens the potential
introduced by the trap and the ``quasiparticle-tip" combination
cannot be observed by local measurements from far away.
Let us denote the positions of these traps to be
$(R_1, \ldots, R_k)$, and assume that these positions are well spaced from each
other compared to the microscopic length scales.
A state with quasiparticles at these positions can be viewed
as an excited state of the Hamiltonian of the system without
the trap potential or, alternatively, as the ground state in the presence
of the trap potential. When we refer to the ground state(s)
of the system, we will often be referring to multi-quasiparticle
states in the latter context. The quasiparticles'
coordinates $(R_1, \ldots, R_k)$ are parameters both in the
Hamiltonian and in the resulting ground state wavefunction for the
electrons.

We are concerned here with the effect of taking these quasiparticles
around each other. We imagine making the
quasiparticles coordinates ${\bf R} = (R_1, \ldots, R_k)$
adiabatically time-dependent. In particular, we consider a
trajectory in which the final configuration of quasiparticles is
just a permutation of the initial configuration (i.e. at the end,
the positions of the quasiparticles are identical to the intial
positions, but some quasiparticles may have interchanged positions with
others.) If the ground state wave function is single-valued with
respect to $(R_1,..,R_k)$, {\it and if there is only one ground
state for any given set of $R_i$'s}, then the final ground state to
which the system returns to after the winding is identical to the
initial one, up to a phase. Part of this phase is simply the dynamical
phase which depends on the energy of the quasiparticle state and
the length of time for the process. In the adiabatic limit, it is
$\int dt E(\vec{R}(t))$. There is also a a geometric phase
which does not depend on how long the process takes.
This Berry phase is \cite{Berry84},
\begin{equation}
\label{eqn:Berry-phase}
\alpha= i\oint d {\bf R} \cdot\langle \psi ({\bf R})|\nabla_{\vec
R}|\psi({\bf R}) \rangle
\end{equation}
where $|\psi({\bf R})\rangle$ is the ground state with the
quasiparticles at positions ${\bf R}$, and where the integral is
taken along the trajectory ${\bf R}(t)$. It is manifestly dependent
only on the trajectory taken by the particles and not on
how long it takes to move along this trajectory.

The phase $\alpha$ has a piece that depends on the geometry of the path traversed
(typically proportional to the area enclosed by all of the loops),
and a piece $\theta$ that depends only on the topology of the
loops created.  If $\theta\neq 0$, then the quasiparticles
excitations of the system are anyons. In particular,
if we consider the case where only two quasiparticles are
interchanged clockwise (without wrapping around any
other quasiparticles), $\theta$ is the
statistical angle of the quasiparticles.

There were two key conditions to our above discussion of the Berry
phase. The single valuedness of the wave function is a technical
issue. The non-degeneracy of the ground state, however, is an
important physical condition. In fact, most of this paper deals with
the situation in which this condition does not hold.
We will generally be considering
systems in which, once the positions $(R_1,..,R_k)$ of the
quasiparticles are fixed, there remain multiple degenerate ground
states (i.e. ground states in the presence of a potential which
captures quasiparticles at positions $(R_1,..,R_k)$),
which are distinguished by a set of internal quantum
numbers. For reasons that will become clear later, we will refer to
these quantum numbers as ``topological".

When the ground state is degenerate, the effect of a closed
trajectory of the $R_i$'s is not necessarily {\it just} a phase
factor. The system starts and ends in ground states, but the initial
and final ground states may be different members of this degenerate
space.  The constraint imposed by adiabaticity in this case is that
the adiabatic evolution of the state of the system is confined to
the subspace of ground states. Thus, it may be expressed as a
unitary transformation within this subspace. The inner product
in (\ref{eqn:Berry-phase}) must be generalized to a matrix
of such inner products:
\begin{equation}
\label{eqn:non-Abelian-Berry1}
{\bf m}_{ab} =
\langle {\psi_a} ({\bf R})|\vec{ \nabla}_{\bf R}|{\psi_b}({\bf R}) \rangle
\end{equation}
where $|{\psi_a}({\bf R}) \rangle$, $a=1,2,\ldots,g$ are the
$g$ degenerate ground states. Since these matrices at different
points $\vec{R}$ do not commute, we must path-order the integral
in order to compute the transformation rule for the state,
${\psi_a} \rightarrow {M_{ab}}\,{\psi_b}$ where
\begin{multline}
\label{eqn:non-Abelian-Berry2}
{M_{ab}} = {\cal P}\exp\left(i\oint d{\bf R}\cdot{\bf m}\right)\\
= {\sum_{n=0}^\infty} {i^n}{\int_0^{2\pi}}\!\!{ds_1}
  {\int_0^{s_1}}\!\!{ds_2}\ldots
  {\int_0^{s_{n-1}}}\!\!{ds_n}\Bigl[
  \dot{\bf R}({s_1})\cdot{{\bf m}_{a{a_1}}}
  \left({\bf R}({s_1})\right)
  \ldots\,\\
  \dot{\bf R}({s_n})\cdot{{\bf m}_{{a_n}b}}
  \left({\bf R}({s_n})\right)\Bigr]
\end{multline}
Where ${\bf R}(s)$, $s\in [0,2\pi]$ is the closed trajectory
of the particles and the path-ordering symbol ${\cal P}$ is defined by the
second equality. Again, the matrix $M_{ab}$
may be the product of topological and non-topological parts. In
a system in which quasiparticles obey non-Abelian braiding
statistics, the non-topological part will be Abelian, that
is, proportional to the unit matrix. Only the topological part
will be non-Abelian.

The requirements for quasiparticles to follow non-Abelian
statistics are then, first, that the $N$-quasiparticle ground state
is degenerate. In general, the degeneracy will not be exact,
but it should vanish exponentially as the quasiparticle separations
are increased. Second, that adiabatic interchange of
quasiparticles applies a unitary transformation on the ground
state, whose non-Abelian part is determined only by the topology of
the braid, while its non-topological part is Abelian. If the particles
are not infinitely far apart, and the degeneracy is only
approximate, then the adiabatic interchange must be done
faster than the inverse of the energy splitting 
\cite{Thouless91} between
states in the nearly-degenerate subspace (but, of course,
still much slower than the energy gap between this
subspace and the excited states).
Third, the only way to make unitary operations
on the degenerate ground state space, so long
as the particles are kept far apart, is by braiding.
The simplest (albeit uninteresting)
example of degenerate ground states may arise if each
of the quasiparticles carried a spin $1/2$ with a vanishing
$g$--factor. If that were the case, the system would satisfy the
first requirement. Spin orbit coupling may conceivably lead to the
second requirement being satisfied. Satisfying the third one,
however, is much harder, and requires the subtle structure that we
describe below.

%arise due, e.g., to many degenerate ground state of the system's nuclei. This
%
%
%
%{\bf As mentioned above,
%degeneracy of the ground state in the presence of quasiparticles,
%can be achieved naturally if the quasiparticles carry some
%additional so-called "topological" quantum number.   The existence
%of the right type of non-Abelian braiding statistics is much less
%trivial to achieve. (Indeed, even the existence of fractional
%statistics in the case where there is a unique ground state is
%rather exotic!) As an example, we might returning to our toy
%example from above of quasiparticles with spin 1/2.   While such
%a model does indeed have a ground state degeneracy, it is clearly
%not true that the only way to make rotations in the degenerate
%space is by braiding (for example photons might be able to flip
%over the spin).  }

The degeneracy of $N$-quasiparticle ground states is conditioned on the
quasiparticles being well separated from one another.  When
quasiparticles are allowed to approach one another too closely, the
degeneracy is lifted. In other words, when non-Abelian anyonic
quasiparticles are close together, their different fusion channels
are split in energy. This dependence is analogous to the way the energy
of a system of spins depends on their internal quantum numbers
when the spins are close together and their coupling becomes
significant. The splitting between different fusion channels is
a means for a measurement of the internal quantum state,
a measurement that is of importance in the context of quantum computation.

\subsection{Topological Quantum Computation}
\label{sec:Top-Quantum-Comp}

\subsubsection{Basics of Quantum Computation}
\label{sec:Quantum_Computation}

As the components of computers become smaller and smaller, we are
approaching the limit in which quantum effects become important. One
might ask whether this is a problem or an opportunity. The founders
of the field of quantum computation (\onlinecite{Manin80},
\onlinecite{Feynman82,Feynman86}, \onlinecite{Deutsch85},
and most dramatically, \onlinecite{Shor94}) answered in favor of the latter.
They showed that a computer which operates coherently on quantum states has
potentially much greater power than a classical
computer \cite{Nielsen00}.

The problem which Feynman had in mind for a quantum computer was the
simulation of a quantum system \cite{Feynman82}. He showed that certain
many-body quantum Hamiltonians could be simulated {\it exponentially
faster} on a quantum computer than they could be on a classical
computer. This is an extremely important potential application of a
quantum computer since it would enable us to understand the
properties of complex materials, e.g. solve high-temperature
superconductivity. Digital simulations of large scale quantum many-body
Hamiltonians are essentially hopeless on classical computers
because of the exponentially-large size of the Hilbert space.
A quantum computer, using the physical resource of an exponentially-large
Hilbert space, may also enable progress in the solution of
lattice gauge theory and quantum chromodynamics, thus shedding
light on strongly-interacting nuclear forces.

In 1994 Peter Shor found an application of a quantum
computer which generated widespread interest not just inside but
also outside of the physics community \cite{Shor94}. He invented an
algorithm by which a quantum computer could find the prime factors
of an $m$ digit number in a length of time $\sim m^2 \log m \log \log m$.
This is much faster than the fastest known algorithm for a classical
computer, which takes $\sim \exp(m^{1/3})$ time. Since many
encryption schemes depend on the difficulty of finding the solution
to problems similar to finding the prime
factors of a large number, there is an obvious application of a
quantum computer which is of great basic and applied interest.

The computation model set forth by these pioneers of
quantum computing (and refined in \onlinecite{DiVincenzo00}),
is based on three steps: initialization, unitary
evolution and measurement. We assume that we have a system at our
disposal with Hilbert space ${\cal H}$. We further assume that we
can initialize the system in some known state $|{\psi_0}\rangle$.
We unitarily evolve the system until it is in some final state
$U(t) |{\psi_0}\rangle$. This evolution will occur according to
some Hamiltonian $H(t)$ such that $dU/dt = iH(t) \,U(t)/\hbar$. We
require that we have enough control over this Hamiltonian so that
$U(t)$ can be made to be any unitary transformation that we
desire. Finally, we need to measure the state of the system at the
end of this evolution. Such a process is called {\it quantum
computation} \cite{Nielsen00}. The Hamiltonian $H(t)$ is the
software program to be run. The initial state is the input to the
calculation, and the final measurement is the
output.

The need for versatility, i.e., for one computer to efficiently
solve many different problems, requires the construction of the
computer out of smaller pieces that can be manipulated and
reconfigured individually. Typically the fundamental piece is taken
to be a quantum two state system known as a  ``qubit" which is the
quantum analog of a bit. (Of course, one could equally well
take general ``dits", for which the fundamental unit is some
$d$-state system with $d$ not too large). While a classical bit, i.e., a classical
two-state system, can be either ``zero" or ``one" at any given time,
a qubit can be in one of the infinitely many
superpositions $a|0\rangle + b|1\rangle$. For $n$ qubits, the state
becomes a vector in a $2^n$--dimensional Hilbert space, in which the
different qubits are generally entangled with one another.

%{\bf It should be now clear why Feynman's simulation problem is so
%appropriate for a quantum computer.   A quantum system of $N$
%spins has a $2^N$-dimensional Hilbert space, so solving it would
%entail diagonalizing a ${2^N}\times{2^N}$ matrix. However, this is
%the kind of problem for which a quantum computer is  perfect since
%a quantum system of $M>N$ quantum bits (`qubits') has a
%$2^M$-dimensional Hilbert space, which is large enough to simulate
%the spin system. }

The quantum phenomenon of superposition allows a system to
traverse many trajectories in parallel, and determine its state by
their coherent sum. In some sense this coherent sum amounts to a
massive quantum parallelism. It should not, however, be confused
with classical parallel computing, where many computers are run in
parallel, and no coherent sum takes place.

%So given the potential power of quantum computation, why hasn't
%everyone run out and built a quantum computer in the 10 years
%since Shor's paper? There is a basic obstacle, namely the
%occurrence of errors. In the more colorful language of A. Peres:
%``Quantum phenomena do not occur in a Hilbert space. They occur in
%a laboratory.'' Of course, errors can occur even in classical
%computers, but they can be surmounted by keeping multiple copies
%of information and checking against these copies.  Indeed, the greatest source of
%error for most computers is human error, which can be alleviated
%by frequently backing up one's hard disk.

The biggest obstacle to building a practical quantum computer is
posed by errors, which would invariably happen during any
computation, quantum or classical. For any computation to be
successful one must devise practical schemes for error correction
which can be effectively implemented (and which must be sufficiently
fault-tolerant).
%Although quantum
%error correction theorems \cite{Shor95} exist, establishing the
%possibility of quantum error correction, actual quantum error
%correction is, in general, extremely difficult.
Errors are typically corrected in classical computers through redundancies,
i.e., by keeping multiple copies of information and checking against
these copies.

With a quantum computer, however, the situation is more complex. If
we measure a quantum state during an intermediate stage of a
calculation to see if an error has occurred, we collapse the wave
function and thus destroy quantum superpositions and ruin the
calculation.  Furthermore, errors need not be merely a discrete flip
of $|0\rangle$ to $|1\rangle$, but can be continuous: the state
$a|0\rangle + b|1\rangle$ may drift, due to an error, to the state
$\rightarrow a|0\rangle + be^{i\theta}|1\rangle$ with arbitrary
$\theta$.

Remarkably, in spite of these difficulties, error correction is
possible for quantum computers
\cite{Shor95,Steane96a,Calderbank96,Gottesman98,Preskill04}.
One can represent information redundantly so that errors can be identified
without measuring the information. For instance, if we use three
spins to represent each qubit, $|0\rangle \rightarrow |000\rangle$,
$|1\rangle \rightarrow |111\rangle$, and the spin-flip rate is low,
then we can identify errors by checking whether all three spins are
the same (here, we represent an up spin by $0$ and a down spin by
$1$). Suppose that our spins are in in the state
$\alpha|000\rangle+\beta |111\rangle$. If the first spin has flipped
erroneously, then our spins are in the state
$\alpha|100\rangle+\beta  |011\rangle$. We can detect this error by
checking whether the first spin is the same as the other two; this
does not require us to measure the state of the qubit. (``We measure
the errors, rather than the information.'' \cite{Preskill04}) If the first
spin is different from the other two, then we just need to flip it.
We repeat this process with the second and third spins. So long as
we can be sure that two spins have not erroneously flipped (i.e. so
long as the basic spin-flip rate is low), this procedure will
correct spin-flip errors. A more elaborate encoding is necessary in
order to correct phase errors, but the key observation is that a
phase error in the $\sigma_z$ basis is a bit flip error in the
$\sigma_x$ basis.

However, the error correction process may itself be a little noisy.
More errors could then occur during error correction, and the whole
procedure will fail unless the basic error rate is very small.
Estimates of the threshold error rate above which error correction
is impossible depend on the particular error correction scheme, but
fall in the range $10^{-4}-10^{-6}$ (see, e.g.
\onlinecite{Aharonov97,Knill98}). This means that we must be able to
perform $10^{4}-10^{6}$ operations perfectly before an error occurs.
This is an extremely stringent constraint and it is presently
unclear if local qubit-based quantum computation can ever be made
fault-tolerant through quantum error correction protocols.

Random errors are caused by the interaction between the quantum
computer and the environment. As a result of this interaction, the
quantum computer, which is initially in a pure superposition state,
becomes entangled with its environment. This can cause errors as
follows. Suppose that the quantum computer is in the state
$|0\rangle$ and the environment is in the state $|{E_0}\rangle$ so
that their combined state is $|0\rangle |{E_0}\rangle$. The
interaction between the computer and the environment could cause
this state to evolve to $\alpha |0\rangle |{E_0}\rangle +
\beta|1\rangle |{E_1}\rangle$, where $|{E_1}\rangle$ is another
state of the environment (not necessarily orthogonal to
$|{E_0}\rangle$). The computer undergoes a transition to the state
$|1\rangle$ with probability $|\beta|^2$. Furthermore, the computer
and the environment are now entangled, so the reduced
density matrix for the computer alone describes a mixed state, e.g.
$\rho=\text{diag}(|\alpha|^2,|\beta|^2)$ if $\langle {E_0} |
{E_1}\rangle = 0$. Since we cannot measure the state of the
environment accurately, information is lost, as reflected in the
evolution of the density matrix of the computer from a pure state to
a mixed one. In other words, the environment has caused {\it
decoherence}. Decoherence can destroy quantum information even if
the state of the computer does not undergo a transition. Although
whether or not a transition occurs is basis-dependent (a bit flip in
the $\sigma_z$ basis is a phase flip in the $\sigma_x$ basis), it is
a useful distinction because many systems have a preferred basis,
for instance the ground state $|0\rangle$ and excited state
$|1\rangle$ of an ion in a trap. Suppose the state $|0\rangle$
evolves as above, but with $\alpha=1$, $\beta=0$ so that no
transition occurs, while the state $|1\rangle|{E_0}\rangle$ evolves
to $|1\rangle|{E'_1}\rangle$ with $\langle {E'_1} | {E_1}\rangle =
0$. Then an initial pure state $\left(a|0\rangle + b
|1\rangle\right)|{E_0}\rangle$ evolves to a mixed state with density
matrix $\rho=\text{diag}(|a|^2,|b|^2)$. The correlations in which
our quantum information resides is now transferred to correlation
between the quantum computer and the environment. The quantum state
of a system invariably loses coherence in this way over a
characteristic time scale $T_\text{coh}$. It was universally assumed
until the advent of quantum error correction \cite{Shor95,Steane96a}
that quantum computation is intrinsically impossible since
decoherence-induced quantum errors simply cannot be corrected in any
real physical system. However, when error-correcting codes are used,
the entanglement is transferred from the quantum computer to
ancillary qubits so that the quantum information remains pure while
the entropy is in the ancillary qubits.

Of course, even if the coupling to the environment were completely eliminated,
so that there were no random errors, there could still be
systematic errors. These are unitary errors which occur while we process
quantum information. For instance, we may wish to rotate
a qubit by $90$ degrees but might inadvertently rotate it
by 90.01 degrees.

From a practical standpoint, it is often useful to divide errors
into two categories: (i) errors that occur when a qubit is
being processed (i.e., when computations are being performed on that
qubit) and (ii) errors that occur when a qubit is simply storing
quantum information and is not being processed (i.e., when it is
acting as a quantum memory). From a fundamental standpoint, this is
a bit of a false dichotomy, since one can think of quantum
information storage (or quantum memory) as being a computer that
applies the identity operation over and over to the qubit (i.e.,
leaves it unchanged). Nonetheless, the problems faced in the two
categories might be quite different. For quantum information
processing, unitary errors, such as rotating a qubit
by 90.01 degrees instead of 90, are an issue
of how precisely one can manipulate the system.  On the other hand,
when a qubit is simply storing information, one is likely to be more
concerned about errors caused by interactions with the environment.
This is instead an issue of how well isolated one can make
the system. As we will see below, a topological quantum computer
is protected from problems in both of these categories.

\subsubsection{Fault-Tolerance from Non-Abelian Anyons}
\label{sec:Topological_Quantum_Computation}

Topological quantum computation is a scheme for using a system whose
excitations satisfy non-Abelian braiding statistics to perform quantum
computation in a way that is naturally immune to errors.
The Hilbert space $\cal H$ used for quantum computation
is the subspace of the total Hilbert space of the system comprised
of the degenerate ground states with a fixed number of quasiparticles
at fixed positions. Operations within this subspace are
carried out by braiding quasiparticles. As we discussed above, the
subspace of degenerate ground states is separated from the rest of
the spectrum by an energy gap. Hence, if the temperature is much
lower than the gap and the system is weakly perturbed using
frequencies much smaller than the gap, the system evolves only
within the ground state subspace. Furthermore, that evolution is
severely constrained, since it is essentially the case (with
exceptions which we will discuss) that {\it the only
way the system can undergo a non-trivial unitary evolution - that
is, an evolution that takes it from one ground state to another - is
by having its quasiparticles braided}. The reason for this
exceptional stability is that any local perturbation (such as the
electron-phonon interaction and the hyperfine electron-nuclear
interaction, two major causes for decoherence in non-topological
solid state spin-based quantum computers \cite{Witzel06})
has no nontrivial matrix elements within the ground state subspace.
Thus, the system is rather immune from decoherence \cite{Kitaev97}.
Unitary errors are also unlikely
since the unitary transformations associated with braiding
quasiparticles are sensitive only to the topology of the quasiparticle
trajectories, and not to their geometry or dynamics.

A model in which non-Abelian quasiparticles are utilized for
quantum computation starts with the construction of qubits. In sharp
contrast to most realizations of a quantum computer, a qubit here is
a non-local entity, being comprised of several well-separated
quasiparticles, with the two states of the qubit being two
different values for the internal quantum numbers of this set of
quasiparticles.  In the simplest non-Abelian quantum Hall state, which
has Landau-level filling factor $\nu=5/2$, two quasiparticles can be put
together to form a qubit (see Sections \ref{sec:FQHE-qc} and
\ref{sec:5/2-qubits}).   Unfortunately, as we
will discuss below in Sections \ref{sec:5/2-qubits}
and \ref{sec:universal-tqc}, this system
turns out to be incapable of universal topological quantum computation
using only braiding operations; some unprotected operations
are necessary in order to perform universal quantum computation.
The simplest system that is capable of universal topological quantum
computation is discussed in Section \ref{sec:fibonacci}, and
utilizes three quasiparticles to form one qubit.

%The first conceptual step for the construction of a model for a topological
%quantum computer is the representation of a qubit. To that end,
%We will now outline in more detail how a quantum computation is
%carried out. Before we begin, it is useful to think about how we
%might represent a qubit in a non-Abelian system.  As we discussed
%above, the degeneracy of the ground state can be thought of as a
%set of quantum numbers associated with the quasiparticles.  We
%will generally be able to group the quasiparticles in clusters
%(not bringing them too close together, but just drawing an
%imaginary circle around the cluster  -- a notation that we will
%use below in section ***) and identify one of the topological
%quantum numbers of the cluster to represent a qubit (There may be
%other quantum numbers of the cluster that we will not be concerned
%with).  For example, we may put two quasiparticles in an
%imaginary circle and (for many topological systems) we may
%discover that the overall quantum number of these two
%quasiparticles has two possible values.  Thus we can declare
%these two quasiparticles to represent a qubit. Indeed, we will
%discuss a system with such a two-quasiparticle qubit below in
%section ***. However, in the simplest non-Abelian system capable of
%universal quantum computation (which we will discuss in section
%*** below) we will actually use three quasiparticles to represent a qubit.

As mentioned above, to perform a quantum
computation, one must be able to initialize the state of qubits at
the beginning, perform arbitrary controlled unitary operations on
the state, and then measure the state of qubits at the end.  We
now address each of these in turn.

Initialization may be performed by preparing the quasiparticles
in a specific way.   For example, if a
quasiparticle-anti-quasiparticle pair is created by ``pulling" it
apart from the vacuum (e.g. pair creation from the vacuum by an
electric field), the pair will begin in an initial state with the
pair necessarily having conjugate quantum numbers (i.e., the
``total" quantum number of the pair remains the same as that of the
vacuum). This gives us a known initial state to start with. It is
also possible to use measurement and unitary evolution (both to be
discussed below) as an initialization scheme
--- if one can measure the quantum numbers of some quasiparticles,
one can then perform a controlled unitary operation to put them
into any desired initial state.

Once the system is initialized, controlled unitary operations are then
performed by physically dragging quasiparticles around one another
in some specified way. When quasiparticles belonging to different
qubits braid, the state of the qubits changes. Since the resulting unitary evolution
depends only on the topology of the braid that is
formed and not on the details of how it is done, it is insensitive
to wiggles in the path, resulting, e.g., from the quasiparticles
being scattered by phonons or photons. Determining which
braid corresponds to which computation is a complicated but
eminently solvable task, which will be discussed in more depth in
section \ref{sec:fibonacci}.3.

Once the unitary evolution is completed, there are two ways to
measure the state of the qubits. The first relies on the
fact that the degeneracy of multi-quasiparticle states is
split when quasiparticles are brought close together (within
some microscopic length scale).  When two quasiparticles
are brought close together, for instance, a measurement of this energy
(or a measurement of the force between two
quasiparticles) measures the the topological
charge of the pair. A second way to measure the topological
charge of a group of
quasiparticles is by carrying out an Aharanov-Bohm type
interference experiment. We take a ``beam" of test quasiparticles, send it
through a beamsplitter, send one partial wave to the right of the
group to be measured and another partial wave to the left of the
group and then re-interfere the two waves
(see Figure \ref{fig:Fabry-Perot} and the surrounding discussion).
Since the two different beams make
different braids around the test group, they will experience
different unitary evolution depending on the topological quantum
numbers of the test group. Thus, the re-interference of these two
beams will reflect the topological quantum number of the group of
quasiparticles enclosed.

This concludes a rough description of the way a topological quantum
computation is to be performed. While the unitary transformation
associated with a braid depends only on the topology of the braid,
one may be concerned that errors could occur if one does not return
the quasiparticles to precisely the correct position at the end of
the braiding. This apparent problem, however, is evaded by the
nature of the computations, which correspond to closed world lines
that have no loose ends: when the computation involves creation and
annihilation of a quasiparticle quasi-hole pair, the world-line is
a closed curve in space-time. If the measurement occurs by
bringing two particles together to measure their quantum charge, it
does not matter where precisely they are brought together.
Alternatively, when the measurement involves an interference
experiment, the interfering particle must close a loop. In other
words, a computation corresponds to a set of {\it links} rather than
open braids, and the initialization and measurement techniques {\it
necessarily} involve bringing quasiparticles together in some way,
closing up the trajectories and making the full process from
initialization to measurement completely topological.

Due to its special characteristics, then, topological quantum
computation intrinsically guarantees fault-tolerance, at the level
of ``hardware", without ``software"-based error
correction schemes that are so essential for non-topological quantum
computers. This immunity to errors results from the stability of the
ground state subspace with respect to external local perturbations.
In non-topological quantum computers, the qubits are local, and the
operations on them are local, leading to a sensitivity to errors
induced by local perturbations. In a topological quantum computer
the qubits are non-local, and the operations --- quasiparticle
braiding --- are non-local, leading to an immunity to local
perturbations.

Such immunity to local perturbation gives topolgical quantum
memories exceptional protection from errors due to the interaction
with the environment. However, it
is crucial to note that topological quantum computers are also
exceptionally immune to unitary errors due to imprecise gate
operation. Unlike other types of quantum computers, the
operations that can be performed on a topological quantum computer
(braids) naturally take a discrete set of values. As discussed
above, when one makes a 90 degree rotation of a spin-based qubit,
for example, it is possible that one will mistakenly rotate by 90.01
degrees thus introducing a small error. In contrast, braids are
discrete: either a particle is taken around another, or it is not.
There is no way to make a small error by having slight imprecision
in the way the quasiparticles are moved. (Taking a particle only part
of the way around another particle rather than all of the way
does not introduce errors so long as the topological class of the
link formed by the particle trajectories -- as described above
-- is unchanged.)

Given the exceptional stability of the ground states, and their
insensitivity to local perturbations that do not involve excitations
to excited states, one may ask then which physical processes do
cause errors in such a topological quantum computer.
Due to the topological stability of the unitary transformations
associated with braids, the only error processes that we must be
concerned about are processes that might cause us to form the wrong
link, and hence the wrong computation. Certainly, one must keep
careful track of the positions of {\it all} of the quasiparticles
in the system during the computation and assure that one makes the
correct braid  to do the correct computation. This includes not just
the ``intended" quasiparticles which we need to manipulate for our
quantum computation, but also any ``unintended'' quasiparticle which
might be lurking in our system without our knowledge.   Two possible
sources of these unintended quasiparticles are thermally excited
quasiparticle-quasihole pairs, and randomly localized
quasiparticles trapped by disorder (e.g. impurities, surface roughness, etc.).
In a typical thermal
fluctuation, for example,  a quasiparticle-quasihole pair is
thermally created from the vacuum, braids with existing intended
quasiparticles, and then gets annihilated.   Typically,
such a pair has opposite electrical charges, so its
constituents will be attracted back to each other and annihilate.
However, entropy or temperature may lead the quasiparticle
and quasihole to split fully
apart and wander relatively freely through part of the system before
coming back together and annihilating. This type of process may
change the state of the qubits encoded in the intended
quasiparticles, and hence disrupt the computation. Fortunately, as
we will see in Section \ref{sec:fibonacci} below there is a whole class of such
processes that do not in fact cause error. This
class includes all of the most likely such thermal processes to
occur: including when a pair is created, encircles a single already
existing quasiparticle and then re-annihilates, or when a pair is
created and one of the pair annihilates an already existing
quasiparticle. For errors to be caused, the excited pair must braid
at least two intended quasiparticles. Nonetheless, the possibility
of thermally-excited quasiparticles wandering through the system
creating unintended braids and thereby causing error is a serious
one. For this reason, topological quantum computation must be
performed at temperatures well below the energy gap for
quasiparticle-quasihole creation so that these errors will be
exponentially suppressed.

Similarly, localized quasiparticles that are induced by disorder
(e.g. randomly-distributed impurities, surface roughness, etc.)
are another serious obstacle to overcome, since they enlarge the
dimension of the subspace of degenerate ground states in a way that
is hard to control. In particular, these unaccounted-for
quasiparticles may couple by tunneling to their intended
counterparts, thereby introducing dynamics to what is supposed to be
a topology-controlled system, and possibly ruining the quantum
computation. We further note that, in quantum Hall systems (as we will discuss in the next section), slight deviations in density or magentic field will also create unintented quasiparticles that must be carefully avoided.

Finally, we also note that while non-Abelian quasiparticles are
natural candidates for the realization of topological qubits, not
every system where quasiparticles satisfy non-Abelian statistics is
suitable for quantum computation. For this suitability it is
essential that the set of unitary transformations induced by
braiding quasiparticles is rich enough to allow for all operations
needed for computation. The necessary and sufficient conditions
for universal topological quantum computation are discussed in
Section \ref{sec:universal-tqc}.

\subsection{Non-Abelian Quantum Hall States}
\label{sec:Non-Abelian_quantum_Hall_states}

A necessary condition for topological quantum computation using
non-Abelian anyons is the existence of a physical system where
non-Abelian anyons can be found, manipulated (e.g. braided), and
conveniently read out. Several theoretical models and proposals for
systems having these properties have been introduced in recent years
\cite{Freedman05a,Levin05a,Fendley05,Kitaev06a}, and in section
\ref{sec:Othersystems} below we will mention some of these
possibilities briefly.   Despite the theoretical work in these
directions, the only real physical system where there is even
indirect experimental evidence that non-Abelian anyons exist are
quantum Hall systems in two-dimensional (2D) electron gases (2DEGs)
in high magnetic fields. Consequently, we will devote a considerable
part of our discussion to putative non-Abelian quantum Hall systems
which are also of great interest in their own right.

\subsubsection{Rapid Review of Quantum Hall Physics}
\label{sec:quantumHallreview}

A comprehensive review of the quantum Hall effect is well beyond the
scope of this article and can be found in the
literature \cite{Prange90,DasSarma97}. This effect, realized for two
dimensional electronic systems in a strong magnetic field, is
characterized by a gap between the ground state and the excited
states (incompressibility); a vanishing longitudinal resistivity
$\rho_{xx}=0$, which implies a dissipationless flow of current; and the
quantization of the Hall resistivity precisely to values of
$\rho_{xy}=\frac{1}{\nu}\frac{h}{e^2}$, with $\nu$ being an integer
(the integer quantum Hall effect), or a fraction (the fractional
quantum Hall effect).   These values of the two resistivities imply
a vanishing longitudinal conductivity $\sigma_{xx}=0$ and a
quantized Hall conductivity $\sigma_{xy}=\nu\frac{e^2}{h}$.

To understand the quantized Hall effect, we begin by
ignoring electron-electron Coulomb interactions, then
the energy eigenstates of the single-electron
Hamiltonian in a magnetic field,
${H_0}=\frac{1}{2m}\left({{\bf p}_i}-\frac{e}{c}{\bf A}({{\bf x}_i})\right)^2$
break up into an equally-spaced set of degenerate levels
called Landau levels. In symmetric gauge,
${\bf A}({\bf x})=\frac{1}{2}{\bf B}\times{\bf x}$,
a basis of single particle wavefunctions in the lowest Landau level (LLL) is
given by $\varphi_m(z)=z^m \exp(-|z|^2/(4{\ell_0}^2))$,
where $z= x + i y$. If the electrons are confined to a disk
of area $A$ pierced by magnetic flux $B\cdot A$,
then there are ${N_\Phi}=BA/{\Phi_0}=BAe/hc$
states in the lowest Landau level (and in each higher Landau level),
where $B$ is the magnetic field; $h,c$, and $e$ are, respectively,
Planck's constant, the speed of light, and the electron
charge; and ${\Phi_0}=hc/e$ is the flux quantum.
In the absence of disorder, these single-particle states are
all precisely degenerate. When the chemical potential lies
between the $\nu^\text{th}$ and $(\nu+1)^\text{th}$ Landau levels,
the Hall conductance takes the
quantized value $\sigma_{xy}=\nu\,\frac{e^2}{h}$ while
$\sigma_{xx}=0$. The two-dimensional electron density, $n$,
is related to $\nu$ via the formula $n = \nu e B / (h c )$.
In the presence of a periodic potential and/or disorder (e.g. impurities),
the Landau levels broaden into bands. However, except at the center
of a band, all states are localized when disorder is present (see
\onlinecite{Prange90,DasSarma97} and refs. therein).
When the chemical potential lies in the region of localized states between
the centers of the $\nu^\text{th}$ and $(\nu+1)^\text{th}$ Landau bands,
the Hall conductance again takes the quantized value
$\sigma_{xy}=\nu\,\frac{e^2}{h}$ while
$\sigma_{xx}=0$. The density will be near but not necessarily equal
to $\nu e B / (h c )$. This is known as the Integer quantum
Hall effect (since $\nu$ is an integer).

The neglect of Coulomb interactions is justified
when an integer number of Landau levels is filled,
so long as the energy splitting between Landau levels,
$\hbar {\omega_c}=\frac{\hbar e B}{mc}$ is much larger
than the scale of the Coulomb energy, $\frac{e^2}{\ell_0}$,
where ${\ell_0}=\sqrt{hc/eB}$ is the magnetic length.
When the electron density is such that a Landau level
is only partially filled, Coulomb interactions may be important.

In the absence of disorder, a partially-filled Landau level has
a very highly degenerate set of multi-particle states.
This degeneracy is broken by electron-electron interactions.
For instance, when the number of electrons is $N={N_\Phi}/3$,
i.e. $\nu=1/3$, the ground state is non-degenerate and there is a gap to
all excitations. When the electrons
interact through Coulomb repulsion, the Laughlin state
\begin{equation}
\label{eqn:Laughlin-first-time}
\Psi = \prod_{i>j} \left({z_i}-{z_j}\right)^3 \,e^{-{\sum_i}{|{z_i}|^2}/4{\ell_0}^2}
\end{equation}
is an approximation to the ground state (and is
the exact ground state for a repulsive ultra-short-ranged model interaction,
see for instance the article by Haldane in \onlinecite{Prange90}).
Such ground states survive even in the presence of disorder
if it is sufficiently weak compared to the gap to excited
states. More delicate states with smaller excitation gaps are, therefore,
only seen in extremely clean devices, as described in subsection
\ref{sec:PhysicalSystems}. However, some disorder is necessary
to pin the charged quasiparticle excitations which are created if
the density or magnetic field are slightly varied. When these excitations
are localized, they do not contribute to the Hall conductance and
a plateau is observed.

Quasiparticle excitations above fractional quantum Hall ground states,
such as the $\nu=1/3$ Laughlin state (\ref{eqn:Laughlin-first-time}), are emergent
anyons in the sense described in section \ref{sec:Berry-phase}.
An explicit calculation of the Berry phase, along the lines of
Eq. \ref{eqn:Berry-phase} shows that quasiparticle excitations
above the $\nu=1/k$ Laughlin states have charge $e/k$ and
statistical angle $\theta=\pi/k$ \cite{Arovas84}. The charge
is obtained from the non-topological part of the Berry phase
which is proportional to the flux enclosed by a particle's trajectory
times the quasiparticle charge.
This is in agreement with a general argument that
such quasiparticles must have fractional charge \cite{Laughlin83}.
The result for the statistics of the quasiparticles follows from
the topological part of the Berry phase; it is in agreement
with strong theoretical arguments which suggest that fractionally
charged excitations are necessarily Abelian anyons (see
\onlinecite{Wilczek90} and refs. therein).
Definitive experimental evidence for the existence of fractionally
charged excitations at $\nu=1/3$ has been
accumulating in the last few years \cite{Goldman95,Picciotto97,Saminadayar97}.
The observation of fractional statistics is much more subtle. First
steps in that direction have been recently reported \cite{Camino05} but
are still debated \cite{Rosenow07a,Godfrey07}.

The Laughlin states, with $\nu=1/k$, are the best understood
fractional quantum Hall states, both theoretically and experimentally.
To explain more complicated observed fractions, with $\nu$ not
of the form $\nu=1/k$,
Haldane and Halperin \cite{Haldane83,Halperin84,Prange90} used a
hierarchical construction in which quasiparticles of a principle
$\nu=1/k$ state can then themselves condense into a quantized state.
In this way, quantized Hall states can be constructed for any
odd-denominator fraction $\nu$ -- but only for odd-denominator
fractions. These states all have quasiparticles
with fractional charge and Abelian fractional statistics. Later, it was noticed by
Jain \cite{Jain89,Heinonen98} that the most prominent
fractional quantum Hall states are of
the form $\nu=p/(2p + 1)$, which can be explained
by noting that a system of electrons in a high magnetic
field can be approximated by a system of auxiliary fermions,
called `composite fermions' , in a lower magnetic field.
If the the electrons are at $\nu=p/(2p + 1)$, then
the lower magnetic field seen by the `composite fermions'
is such that they fill an integer number of Landau
levels $\nu' = p$. (See \onlinecite{Halperin93,Lopez91}
for a field-theoretic implementations.)
Since the latter state has a gap, one can hope that
the approximation is valid. The composite fermion picture
of fractional quantum Hall states has proven to be qualitatively and
semi-quantitatively correct in the LLL \cite{Murthy03}.

Systems with filling fraction $\nu >1$, can be mapped to
$\nu'\leq 1$ by keeping the fractional part of $\nu$ and using an
appropriately modified Coulomb interaction to account
for the difference between cyclotron orbits in the LLL and
those in higher Landau levels \cite{Prange90}. This involves the
assumption that the inter-Landau level coupling is negligibly small.
We note that this may not be a particularly good assumption for higher
Landau levels, where the composite fermion picture less successful.

Our confidence in the picture described above for the
$\nu=1/k$ Laughlin states and the hierarchy of
odd-denominator states which descend from them
derives largely from numerical studies.
Experimentally, most of what is known about quantum Hall states
comes from transport experiments --- measurements of the conductance
(or resistance) tensor.   While such measurements make it reasonably
clear when a quantum Hall plateau exists at a given filling
fraction, the nature of the plateau (i.e., the details of the low-energy
theory) is extremely hard to discern.   Because of this
difficulty, numerical studies of small systems
(exact diagonalizations and Monte Carlo) have played
a very prominent role in providing further insight.
Indeed, even Laughlin's original work \cite{Laughlin83} on the
$\nu=1/3$ state relied heavily on accompanying numerical work.
The approach taken was the following. One assumed that
the splitting between Landau levels is the largest energy
in the problem. The Hamiltonian is projected into the lowest
Landau level, where, for a finite number of electrons and a fixed
magnetic flux, the Hilbert space is finite-dimensional.
Typically, the system is given periodic boundary conditions (i.e. is on a torus)
or else is placed on a sphere; occasionally, one works on the disk,
e.g. to study edge excitations. 
The Hamiltonian is then a finite-sized matrix which
can be diagonalized by a computer so long as the number of
electrons is not too large. Originally, Laughlin examined only 3
electrons, but modern computers can handle sometimes as many as 18
electrons. The resulting ground state wavefunction can be compared
to a proposed trial wavefunction.
Throughout the history of the field, this approach
has proven to be extremely powerful in identifying the nature of
experimentally-observed quantum Hall states when the system in
question is deep within a quantum Hall phase, so that the associated
correlation length is short and the basic physics is already apparent
in small systems.

There are several serious challenges in using such numerical work to
interpret experiments.  First of all, there is always the challenge of
extrapolating finite-size results to the thermodynamic limit.
Secondly, simple overlaps between a proposed
trial state and an exact ground state may not be sufficiently
informative. For example, it is possible that an exact ground state
will be adiabatically connected to a particular trial state, i.e.,
the two wavefunctions represent the same phase of matter, but the
overlaps may not be very high. For this reason, it is necessary
to also examine quantum numbers and symmetries of the ground state,
as well as the response of the ground state to various perturbations,
particularly the response to changes in boundary conditions and
in the flux.

Another difficulty is the choice of Hamiltonian to diagonalize.
One may think that the Hamiltonian for a quantum Hall
system is just that of 2D electrons in a magnetic field interacting
via Coulomb forces. However, the small but finite width (perpendicular to
the plane of the system) of the quantum well
slightly alters the effective interaction between electrons.
Similarly, screening (from any nearby conductors, or from
inter-Landau-level virtual excitations), in-plane magnetic fields,
and even various types of disorder may alter the Hamiltonian in
subtle ways. To make matters worse,
one may not even know all the physical parameters (dimensions,
doping levels, detailed chemical composition, etc.) of any
particular experimental system very accurately.
Finally, Landau-level mixing is not small because
the energy splitting between Landau levels is not much larger
than the other energies in the problem. Thus, it is not even
clear that it is correct to truncate the Hilbert space to
the finite-dimensional Hilbert space of a single Landau level.

In the case of very robust states, such as the $\nu=1/3$ state,
these subtle effects are unimportant; the ground state is
essentially the same irrespective of these small deviations
from the idealized Hamiltonian. However, in the case
of weaker states, such as those observed
between $\nu=2$ and $\nu=4$ (some of which we will discuss
below), it appears that very
small changes in the Hamiltonian can indeed greatly affect
the resulting ground state. Therefore,
a very valuable approach has been to guess a likely
Hamiltonian, and search a space of ``nearby"  Hamiltonians, slightly
varying the parameters of the Hamiltonian, to map out the phase
diagram of the system. These phase diagrams
suggest the exciting technological
possibility that detailed numerics will allow us to engineer samples
with just the right small perturbations so as display certain
quantum Hall states more clearly \cite{Manfra07,Peterson07}.

\subsubsection{Possible Non-Abelian States}
\label{sec:Non-AbelianQHE}

The observation of a quantum Hall state with an even denominator
filling fraction \cite{Willett87}, the $\nu=5/2$ state,
was the first indication that not all fractional quantum
Hall states fit the above hierarchy (or equivalently composite
fermion) picture.
Independently, it was recognized \onlinecite{Fubini91a,Fubini91b,Moore91}
that conformal field theory gives a way to write
a variety of trial wavefunctions for quantum Hall states,
as we describe in Section \ref{sec:FQHE} below.
Using this approach, the so-called Moore-Read Pfaffian wavefunction
was constructed \cite{Moore91}:
\begin{equation}
\label{eqn:MR-Pfaffian1}
 \Psi_{\rm Pf}
   = \text{Pf}\!\left(\frac{1}{z_i - z_j}\right)\, \mbox{$\prod_{i<j}$}(z_i - z_j)^{m}
   e^{-{\sum_i}{|{z_i}|^2}/4{\ell_0}^2}
\end{equation}
The Pfaffian is the square root of the determinant of an anti-symmetric
matrix or, equivalently, the antisymmetrized sum over pairs:
\begin{equation}
{\rm Pf}\!\left( \frac{1}{z_j - z_k }\right) =
{\cal A}\left(\frac{1}{{z_1}-{z_2}}\frac{1}{{z_3}-{z_4}}\ldots\right)
\end{equation}
For $m$ even, this is an even-denominator quantum Hall state
in the lowest Landau level. \onlinecite{Moore91}
suggested that its quasiparticle
excitations would exhibit non-Abelian statistics \cite{Moore91}.
This wavefunction is the exact ground state of a $3$-body
repulsive interaction; as we discuss below, it is also an
approximate ground state for more realistic interactions.
This wavefunction is a representative of a universality class which
has remarkable properties which we discuss in detail
in this paper. In particular, the quasiparticle excitations
above this state realize the second scenario discussed in
Eqs. \ref{eqn:non-Abelian-Berry1}, \ref{eqn:non-Abelian-Berry2} in
section \ref{sec:Berry-phase}. There are
$2^{n-1}$ states with $2n$ quasiholes at fixed positions,
thereby establishing the degeneracy of multi-quasiparticle states
which is required for non-Abelian statistics \cite{Nayak96c}.
Furthermore, these quasihole wavefunctions can also
be related to conformal field theory
(as we discuss in section \ref{sec:FQHE}), from which it
can be deduced that the $2^{n-1}$-dimensional vector space of states
can be understood as the spinor representation
of SO(2n); braiding particles $i$ and $j$ has the
action of a $\pi/2$ rotation in the $i-j$ plane in $\mathbb{R}^{2n}$
\cite{Nayak96c}. In short, these quasiparticles are
essentially Ising anyons (with the difference being an additional
Abelian component to their statistics).
Although these properties were uncovered
using specific wavefunctions which are eigenstates of
the $3$-body interaction for which the Pfaffian wavefunction
is the exact ground state, they are representative of an entire universality
class. The effective field theory for this universality class
is $SU(2)$ Chern-Simons theory at level $k=2$
together with an additional Abelian Chern-Simons term
\cite{Fradkin98,Fradkin01}.
Chern-Simons theory is the archetypal topological
quantum field theory (TQFT), and we discuss it extensively
in section \ref{part2}. As we describe, Chern-Simons theory
is related to the Jones polynomial of knot theory \cite{Witten89};
consequently, the current
through an interferometer in such a non-Abelian
quantum Hall state would give a direct measure of the Jones
polynomial for the link produced by the quasiparticle trajectories
\cite{Fradkin98}!

One interesting feature of the 
Pfaffian wavefunction is that it is the quantum Hall analog
of a $p+ip$ superconductor: the antisymmetrized product
over pairs is the real-space form of the BCS wavefunction
\cite{Greiter92}. \onlinecite{Read00} showed that
the same topological properties mentioned above
are realized by a $p+ip$-wave superconductor, thereby cementing
the identification between such a paired state
and the Moore-Read state. \onlinecite{Ivanov01}
computed the braiding matrices by this approach
(see also \onlinecite{Stern04,Stone06}). Consequently,
we will often be able to discuss $p+ip$-wave superconductors
and superfluids in parallel with the $\nu=5/2$ quantum Hall state, although
the experimental probes are significantly different.

As we discuss below, all of these theoretical developments garnered greater
interest when numerical work \cite{Morf98,Rezayi00}
showed that the ground state of systems of up
to 18 electrons in the $N=1$ Landau level at filling fraction
1/2 is in the universality class of the
Moore-Read state. These results revived the conjecture that
the lowest Landau level ($N=0$) of both spins is filled and inert and the
electrons in the $N=1$ Landau level form the
analog of the Pfaffian state \cite{Greiter92}.
Consequently, it is the leading
candidate for the experimentally-observed
$\nu=5/2$ state.

\onlinecite{Read99} constructed a series of
non-Abelian quantum Hall states at filling fraction
$\nu= N + k/(Mk  + 2)$ with $M$ odd, which generalize the
Moore-Read state in a way which
we discuss in section \ref{part2}. These states are referred
to as the Read-Rezayi $\mathbb{Z}_k$ parafermion states for reasons
discussed in section \ref{sec:FQHE}.
Recently, a quantum Hall state was observed experimentally
with $\nu=12/5$ \cite{Xia04}. It is suspected (see below)
that the $\nu=12/5$ state may be (the particle hole conjugate of)
the $\mathbb{Z}_3$ Read-Rezayi state, although
it is also possible that 12/5 belongs to the conventional Abelian hierarchy
as the $2/5$ state does. Such an option is not possible
at $\nu=5/2$ as a result of the even denominator.

In summary, it is well-established that if the observed $\nu=5/2$
state is in the same universality class as the Moore-Read Pfaffian
state, then its quasiparticle excitations are non-Abelian anyons. Similarly,
if the $\nu=12/5$ state is in the universality class of the
$\mathbb{Z}_3$ Read-Rezayi state, its quasiparticles are
non-Abelian anyons. There is no
direct experimental evidence that the $\nu=5/2$ is in this
particular universality class, but there is evidence from
numerics, as we further discuss below. There is even less
evidence in the case of the $\nu=12/5$ state. In subsections
\ref{sec:interference} and \ref{sec:FQHE-qc},
we will discuss proposed experiments which could directly verify
the non-Abelian character of the $\nu=5/2$ state and will briefly mention
their extension to the $\nu=12/5$ case. Both of these states,
as well as others (e.g. \onlinecite{Ardonne99,Simon07a}),
were constructed on the basis of very deep connections
between conformal field theory, knot theory, and
low-dimensional topology \cite{Witten89}. Using
methods from these different branches of theoretical physics
and mathematics, we will explain the structure of the non-Abelian
statistics of the $\nu=5/2$ and $12/5$ states within the context
of a large class of non-Abelian topological states.
We will see in section \ref{sec:Jones} that
this circle of ideas enables us to use the theory of knots to
understand experiments on non-Abelian anyons.

In the paragraphs below, we will discuss numerical results for
$\nu=5/2,12/5,$ and other candidates in greater detail.

\setcounter{mysubsection}{0} \mysubsection{5/2 State} The $\nu=5/2$ fractional quantum Hall state is a useful case history for
how numerics can elucidate experiments.
This incompressible state is easily destroyed by the application of an in-plane magnetic field \cite{Eisenstein90}. At first it was assumed that this
implied that the 5/2 state is spin-unpolarized or partially
polarized since the in-plane magnetic field presumably couples
only to the electron spin.  Careful
finite-size numerical work changed this perception, leading to our
current belief that the 5/2 FQH state is actually in the universality
class of the spin-polarized Moore-Read Pfaffian state.

In rather pivotal work \cite{Morf98}, it was shown that spin-polarized states at
$\nu=5/2$ have lower energy than spin-unpolarized states.
Furthermore, it was shown that varying the Hamiltonian
slightly caused a phase transition between a gapped
phase that has high overlap with the Moore-Read wavefunction and a
compressible phase. The proposal put forth was that the most
important effect of the in-plane field was not on the electron spins,
but rather was to slightly alter the shape of the electron wavefunction
perpendicular to the sample which, in turn, slightly alters the
effective electron-electron interaction, pushing the system over a
phase boundary and destroying the gapped state. Further experimental
work showed that the effect of the in-plane magnetic field is to
drive the system across a phase transition from a gapped
quantum Hall phase into an anisotropic compressible
phase \cite{Pan99a,Lilly99b}. Further numerical
work \cite{Rezayi00} then mapped out a full phase diagram
showing the transition between gapped and compressible phases
and showing further that the experimental systems lie exceedingly close to the phase boundary. The correspondence between numerics and experiment
has been made more quantitative by comparisons between the
energy gap obtained from numerics and the one measured in
experiments \cite{Morf02,Morf03}.  
Very recently, this case has been further strengthened
by the application of the density-matrix renormalization
group method (DMRG) to this problem \cite{Feiguin07b}.

One issue worth considering is possible competitors to
the Moore-Read Pfaffian state. Experiments have already told
us that there is a fractional quantum Hall state at $\nu=5/2$. Therefore,
our job is to determine which of the possible states is realized there.
Serious alternatives to the Moore-Read Pfaffian state fall into two
categories. On the one hand, there is the possibility that the ground
state at $\nu=5/2$ is not fully spin-polarized. If it were completely unpolarized,
the so-called $(3,3,1)$ state \cite{Halperin83,DasSarma97} would be a possibility.
However, Morf's numerics \cite{Morf98} and a recent variational
Monte Carlo study \cite{Dimov07} indicate that an unpolarized
state is higher in energy than a fully-polarized state. This can be understood
as a consequence of a tendency towards spontaneous ferromagnetism;
however, a partially-polarized alternative (which may be either Abelian
or non-Abelian) to the Pfaffian is not ruled out \cite{Dimov07}.
Secondly, even if the ground state at $\nu=5/2$ is fully spin-polarized,
the Pfaffian is not the only possibility. It was very recently
noticed that the Pfaffian state
is not symmetric under a particle-hole transformation
of a single Landau level (which, in this case, is the $N=1$ Landau level,
with the $N=0$ Landau level filled and assumed inert),
even though this is an exact symmetry of the Hamiltonian
in the limit that the energy splitting between Landau levels
is infinity. Therefore, there is a distinct state, dubbed the
anti-Pfaffian \cite{LeeSS07,Levin07}, which is an equally
good state in this limit. Quasiparticles in this state are also
essentially Ising anyons, but they differ from Pfaffian
quasiparticles by Abelian statistical phases. In experiments,
Landau-level mixing is not small, so one or the other state
is lower in energy. On a finite torus, the symmetric combination
of the Pfaffian and the anti-Pfaffian will be lower in energy,
but as the thermodynamic limit
is approached, the anti-symmetric combination will become
equal in energy. This is a possible factor which complicates
the extrapolation of numerics to the thermodynamic limit.
On a finite sphere, particle-hole symmetry is not exact;
it relates a system with $2N-3$ flux quanta with a system
with $2N+1$ flux quanta. Thus, the anti-Pfaffian would
not be apparent unless one looked at a different value of
the flux.
To summarize, the only known alternatives to the Pfaffian state --
partially-polarized states and the anti-Pfaffian -- have not really been
tested by numerics, either because the spin-polarization was assumed to be
0\% or 100\% \cite{Morf98} or because Landau-level mixing was neglected.

With this caveat in mind,
it is instructive to compare the evidence placing the
$\nu=5/2$ FQH state in the Moore-Read Pfaffian universality class
with the evidence placing the $\nu=1/3$ FQH state in the corresponding
Laughlin universality class. In the latter case, there have been
several spectacular experiments \cite{Goldman95,Picciotto97,Saminadayar97}
which have observed quasiparticles with electrical charge $e/3$,
in agreement with the prediction of the Laughlin universality class.
In the case of the $\nu=5/2$ FQH state, we do not yet have
the corresponding measurements of the quasiparticle charge,
which should be $e/4$. However, the observation of charge $e/3$,
while consistent with the Laughlin universality class, does not uniquely fix
the observed state in this class (see, for example, \onlinecite{Simon07a,Wojs01}.
Thus, much of our confidence derives
from the amazing (99\% or better) overlap between the ground state
obtained from exact diagonalization for a finite size 2D system
with up to 14 electrons and the Laughlin wavefunction.
In the case of the $\nu=5/2$ FQH state, the corresponding overlap
(for 18 electrons on the sphere) between the $\nu=5/2$ ground state and the
Moore-Read Pfaffian state is reasonably impressive ($\sim$ 80\%).
This can be improved by modifying the wavefunction
at short distances without leaving the Pfaffian phase \cite{Moller07}.
However, on the torus, as we mentioned above,
the symmetric combination of the Pfaffian and the anti-Pfaffian
is a better candidate wavefunction in a finite-size system
than the Pfaffian itself (or the anti-Pfaffian). Indeed,
the symmetric combination of the Pfaffian and the anti-Pfaffian
has an overlap of 97\% for 14 electrons \cite{Rezayi00}.

To summarize, the overlap is somewhat smaller
in the $5/2$ case than in the $1/3$ case when particle-hole
symmetry is not accounted for, but only slightly smaller
when it is. This is an indication that Landau-level mixing -- which will
favor either the Pfaffian or the anti-Pfaffian -- is an
important effect at $\nu=5/2$, unlike at $\nu=1/3$.
Moreover, Landau-level mixing is likely to be large
because the 5/2 FQH state is typically realized at relatively low
magnetic fields, making the Landau level separation energy relatively small.

Given that potentially large effects have been neglected,
it is not too surprising that the gap obtained by
extrapolating numerical results for finite-size systems \cite{Morf02,Morf03}
is substantially larger than the experimentally-measured activation gap.
Also, the corresponding excitation gap obtained from numerics
for the $\nu=1/3$ state is much larger
than the measured activation gap. The discrepancy between the theoretical excitation gap and the measured activation gap is a generic problem of all FQH states, and may be related to poorly understood disorder effects and
Landau-level mixing.

Finally, it is important to mention that several very recent (2006-07) numerical works in the literature have raised some questions about the
identification of the observed $5/2$ FQH state with the Moore-Read
Pfaffian \cite{Toke06,Toke07,Wojs06b}. Considering the absence of a viable alternative
(apart from the anti-Pfaffian and partially-polarized states, which
were not considered by these authors)
it seems unlikely that these doubts will continue to persist,
as more thorough numerical work indicates
\cite{Moller07,Rezayi07,Peterson07}.

\mysubsection{12/5 State} While our current understanding of the 5/2
state is relatively good, the situation for the experimentally observed
12/5 state is more murky, although the possibilities are even more
exciting, at least from the perspective of topological quantum
computation. One (relatively dull) possibility is that the $12/5$ state
is essentially the same as the observed $\nu=2/5$ state,
which is Abelian.
However, Read and Rezayi, in their initial work on non-Abelian
generalizations of the Moore-Read state \cite{Read99} proposed that
the 12/5 state might be (the particle-hole conjugate of) their
$\mathbb{Z}_3$ parafermion (or $SU(2)$ level 3) state.  This is
quite an exciting possibility because, unlike the non-Abelian
Moore-Read state at 5/2, the $\mathbb{Z}_3$ parafermion state would
have braiding statistics that allow universal topological quantum
computation.

The initial numerics by Read and Rezayi \cite{Read99} indicated that
the 12/5 state is very close to a phase transition between the
Abelian hierarchy state and the non-Abelian parafermion state. More
recent work by the same authors \cite{Rezayi06} has mapped out a
detailed phase diagram showing precisely for what range of
parameters a system should be in the non-Abelian phase.  It was
found that the non-Abelian phase is not very ``far" from the results
that would be expected from most real experimental systems.   This
again suggests that (if the system is not already in the non-Abelian
phase), we may be able to engineer slight changes in an experimental
sample that would push the system over the phase boundary into the
non-Abelian phase.

Experimentally, very little is actually known about the 12/5 state.
Indeed, a well quantized plateau has only ever been seen in a single
published \cite{Xia04} experiment.  Furthermore, there is no
experimental information about spin polarization (the non-Abelian
phase should be polarized whereas the Abelian phase could be either
polarized or unpolarized), and it is not at all clear why the 12/5
state has been seen, but its particle-hole conjugate, the 13/5
state, has not (in the limit of infinite Landau level separation, these
two states will be identical in energy).
 Nonetheless, despite the substantial uncertainties, there is a great
deal of excitement about the possibility that this state will
provide a route to topological quantum computation.

\mysubsection{Other Quantum Hall States}  The most strongly
observed fractional quantum Hall states are the composite fermion
states $\nu=p/(2p + 1)$, or are simple generalizations of them. There is
little debate that these states are likely to be Abelian.  However,
there are a number of observed exotic states whose origin is not
currently agreed upon.  An optimist may look at any state of unknown
origin and suggest that it is a non-Abelian state.  Indeed,
non-Abelian proposals (published and unpublished) have been made for
a great variety states of uncertain origin
including \cite{Scarola02,Wojs06a,Simon07a,Simon07b,Jolicoeur07}
3/8, 4/11, 8/3, and 7/3.
Of course, other more conventional Abelian proposals have been made for
each of these states too \cite{Lopez04,Chang04,Goerbig04,Wojs02,Wojs04}. For each of
these states, there is a great deal of research left to be done,
both theoretical and experimental, before any sort of definitive
conclusion is reached.

In this context, it is worthwhile to mention another class of
quantum Hall systems where non-Abelian anyons could exist,
namely bilayer or multilayer $2D$ systems
\cite{Greiter91,He91,He93,DasSarma97}. More work is necessary
in investigating the possibility of non-Abelian multilayer quantum Hall states.

%Finite size exact diagonalization calculations carried out on the
%5/2 FQHE state by Morf and coworkers and by Haldane and Rezayi
%showed the following important features: (1) The exact numerical
%finite-size wavefunction (for 10-18 electrons) seems to have a high
%$(~90\%)$ overlap with the analytic (and completely spin-polarized)
%Moore-Read Pfaffian wavefunction for realistic 2D system parameters;
%and, (2) the numerical incompressible 5/2 state is strongly
%suppressed (and its overlap with the Pfaffian non-Abelian
%wavefunction decreases drastically) due to the orbital coupling of
%the in-plane magnetic field to the electron confirming wavefunction
%in the direction normal to the 2D plane; (3) there are competing
%compressible ground states (the so-called striped phase states),
%which are very close in energy to the incompressible FQH state, near
%the 5/2 filling, and the application of the in-plane field (or other
%small perturbations in the system Hamiltonian) tends to stabilize
%these compressible states at 5/2 filling over the incompressible
%FQHE 5/2 state; (4) the fully spin-polarized 5/2 incompressible
%state has lower energy than spin-unpolarized states. The current
%conclusion, based on these compelling numerical exact
%diagonalization studies, is that the observed 5/2 FQHE state is
%indeed the spin-polarized Moore-Read Pfaffian state, but it is a
%fragile weak-pairing state with a small excitation gap which could
%easily be destroyed by small external perturbations (e.g. in-plane
%magnetic field, finite thickness of the 2D layer).

\subsubsection{Interference Experiments}
\label{sec:interference}

While numerics give useful insight about the topological
nature of observed quantum Hall states, experimental
measurements will ultimately play the decisive role.
So far, rather little has been directly measured experimentally
about the topological nature of the $\nu=5/2$ state and even less is known
about other putative non-Abelian quantum Hall states such as $\nu=12/5$.
In particular, there is no direct experimental evidence for the
non-Abelian nature of the quasiparticles. The existence of a
degenerate, or almost degenerate, subspace of ground states leads to
a zero-temperature entropy and heat capacity, but those are very
hard to measure experimentally. Furthermore, this degeneracy is just
one requirement for non-Abelian statistics to take place.
How then does one demonstrate experimentally that fractional quantum
Hall states, particularly the $\nu=5/2$ state, are indeed
non-Abelian?

The fundamental quasiparticles (i.e. the ones with the
smallest electrical charge) of the Moore-Read Pfaffian
state have charge $e/4$ \cite{Moore91,Greiter92}.
The fractional charge does not uniquely identify the
state -- the Abelian $(3,3,1)$ state has the same
quasiparticle charge --
but a different value of the minimal quasiparticle charge
at $\nu=5/2$ would certainly rule out the
Pfaffian state. Hence, the first important measurement
is the quasiparticle charge, which was done more than 10 years
ago in the case of the $\nu=1/3$ state \cite{Goldman95,Picciotto97,Saminadayar97}.

\begin{figure}[t!]
\centerline{\includegraphics[width=3.5in]{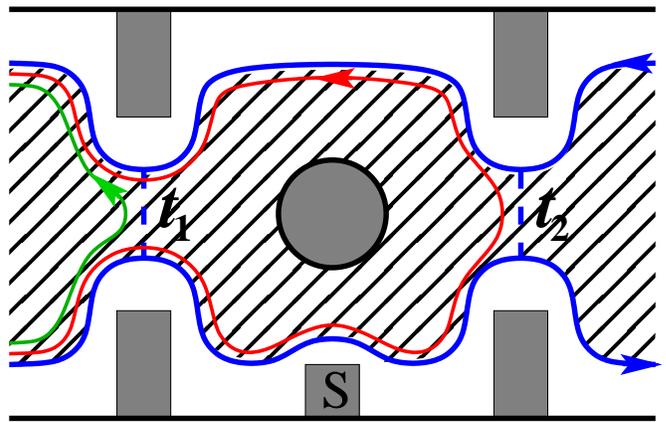}}
\caption{A quantum Hall analog of a Fabry-Perot interferometer.
Quasiparticles can tunnel from one edge to the other at either of
two point contacts. To lowest order in the tunneling amplitudes, the
backscattering probability, and hence the conductance, is determined
by the interference between these two processes. The area in the
cell can be varied by means of a side gate $S$ in order to observe
an interference pattern.}
\label{fig:Fabry-Perot}
\end{figure}

If the quasiparticle charge is shown to be $e/4$, then further
experiments which probe the braiding statistics of the charge $e/4$
quasiparticles will be necessary to pin down the topological
structure of the state. One way to do this is to use a mesoscopic
interference device. Consider a Fabry-Perot interferometer, as
depicted in Fig. (\ref{fig:Fabry-Perot}). A Hall bar lying parallel
to the $x$--axis is put in a field such that it is at filling
fraction $\nu=5/2$. It is perturbed by two constrictions, as shown
in the figure. The two constrictions introduce two amplitudes for
inter-edge tunnelling, $t_{1,2}$. To lowest order in $t_{1,2}$, the
four-terminal longitudinal conductance of the Hall bar, is:
\begin{equation}
G_L \propto |{t_1}|^2+|{t_2}|^2+2\mbox{Re}\!\left\{t_1^* t_2
e^{i\phi}\right\} \label{bsprob}
\end{equation}
For an integer Landau filling, the relative phase $\phi$ may be
varied either by a variation of the magnetic field or by a variation
of the area of the ``cell" defined by the two edges and the two
constrictions, since that phase is $2\pi\Phi/\Phi_0$, with $\Phi=BA$
being the flux enclosed in the cell, $A$ the area of the cell, and
$\Phi_0$ the flux quantum. Thus, when the area of the cell is varied
by means of a side gate (labeled $S$ in the figure),
the back-scattered current should oscillate.

For fractional quantum Hall states, the situation is
different \cite{Chamon97}. In an approximation in which the electronic
density is determined by the requirement of charge neutrality, a
variation of the area of the cell varies the flux it encloses and
keeps its bulk Landau filling unaltered. In contrast, a variation of
the magnetic field changes the filling fraction in the bulk, and
consequently introduces quasiparticles in the bulk. Since the
statistics of the quasiparticles is fractional, they contribute to the phase
$\phi$. The back-scattering probability is then determined not only
by the two constrictions and the area of the cell they define, but
also by the number of localized quasiparticles that the cell
encloses. By varying the voltage applied to an anti-dot
in the cell (the grey circle in Fig. \ref{fig:Fabry-Perot}), we can
independently vary the number of quasiparticles in the cell.
Again, however, as the area of the cell is varied, the
back-scattered current oscillates.

For non-Abelian quantum Hall states, the situation is more
interesting
\cite{Fradkin98,DasSarma05,Stern06,Bonderson06a,Bonderson06b,Chung06}.
Consider the case of the Moore-Read Pfaffian state. For clarity, we
assume that there are localized $e/4$ quasiparticles only within the
cell (either at the anti-dot or elsewhere in the cell).
If the current in Fig. (\ref{fig:Fabry-Perot}) comes from the
left, the portion of the current that is back-reflected from the
left constriction does not encircle any of these quasiparticles, and
thus does not interact with them. The part of the current that is
back-scattered from the right constriction, on the other hand, does
encircle the cell, and therefore applies a unitary transformation on
the subspace of degenerate ground states. The final state of the
ground state subspace that is coupled to the left back--scattered
wave, $|\xi_0\rangle$, is then different from the state coupled to
the right partial wave, ${\hat U}|\xi_0\rangle$. Here ${\hat U}$ is
the unitary transformation that results from the encircling of the
cell by the wave scattered from the right constriction. The
interference term in the four-terminal longitudinal conductance, the
final term in Eq. \ref{bsprob}, is then multiplied by the matrix
element $\langle \xi_0|{\hat U}|\xi_0\rangle$:
\begin{equation}
G_L \propto |t_1|^2+|t_1|^2+2\mbox{Re}\!\left\{t_1^*t_2 e^{i\phi}
\bigl\langle \xi_0 \bigl|{\hat U}\bigr|\xi_0\bigr\rangle\right\}
\label{bsprob-non-Abelian}
\end{equation}
In section \ref{part2}, we explain how $\langle
\xi_0|{\hat U}|\xi_0\rangle$ can be calculated by several
different methods. Here we just give a brief description
of the result.

For the Moore-Read Pfaffian state, which is believed to
be realized at $\nu=5/2$, the expectation value  $\langle\xi_0|{\hat
U}|\xi_0\rangle$ depends first and foremost on the parity of the
number of $e/4$ quasiparticles localized in the cell. When that number is
odd, the resulting expectation value is zero. When that number is
even, the expectation value is non-zero and may assume one of two
possible values, that differ by a minus sign. As a consequence, when
the number of localized quasiparticles is odd, {\it no interference
pattern is seen}, and the back-scattered current does not oscillate
with small variations of the area of the cell. When that number is
even, the back-scattered current oscillates as a function of the
area of the cell.

A way to understand this striking result is to observe that the
localized quasiparticles in the cell can be viewed as being created
in pairs from the vacuum. Let us suppose that we want to have $N$
quasiparticles in the cell. If $N$ is odd, then we can create
$(N+1)/2$ pairs and take one of the resulting quasiparticles outside
of the cell, where it is localized. Fusing all $N+1$ of these
particles gives the trivial particle since they were created from
the vacuum. Now consider what happens when a current-carrying
quasiparticle tunnels at one of the two point contacts. If it
tunnels at the second one, it braids around the $N$ quasiparticles
in the cell (but not the $N+1^\text{th}$, which is outside the
cell). This changes the fusion channel of the $N+1$ localized
quasiparticles. In the language introduced in subsection
\ref{sec:Non-Abelian_quantum_statistics}, each $e/4$ quasiparticle
is a $\sigma$ particle. An odd number $N$ of them can only fuse to
$\sigma$; fused now with the $N+1^\text{th}$, they can either give
${\bf 1}$ or $\psi$. Current-carrying quasiparticles, when they
braid with the $N$ in the cell, toggle the system between these two
possibilities. Since the state of the localized quasiparticles has
been changed, such a process cannot interfere with a process in
which the current-carrying quasiparticle tunnels at the first
junction and does not encircle any of the localized quasiparticles.
Therefore, the localized quasiparticles `measure' which trajectory
the current-carrying quasiparticles
take\cite{Overbosch01,Bonderson07}. If $N$ is even, then we
can create $(N+2)/2$ pairs and take two of the resulting
quasiparticles outside of the cell. If the $N$ quasiparticles in the
cell all fuse to the trivial particle, then this is not necessary,
we can just create $N/2$ pairs. However, if they fuse to a neutral
fermion $\psi$, then we will need a pair outside the cell which also
fuses to $\psi$ so that the total fuses to ${\bf 1}$, as it must for
pair creation from the vacuum. A current-carrying quasiparticle
picks up a phase depending on whether the $N$ quasiparticles in the
cell fuse to ${\bf 1}$ or $\psi$.

%Since the number of localized quasiparticles depends on the
%magnetic field, in a two dimensional parameter plane in which the
%axes are the area of the cell and the magnetic field, there will be
%stripes within which interference is observed, interweaved with
%stripes in which no interference is observed.

The Fabry-Perot interferometer depicted in Fig. \ref{fig:Fabry-Perot}
allows also for the interference of waves that are back-reflected
several times. For an integer filling factor, in the limit of strong
back-scattering at the constrictions, the sinusoidal dependence of
the Hall bar's conductance on the area of the cell gives way to a
resonance-like dependence: the conductance is zero unless a Coulomb
peak develops. For the $\nu=5/2$ state, again, the parity of the
number of localized quasiparticles matters: when it is odd, the
Coulomb blockade peaks are equally spaced. When it is even, the
spacing between the peaks alternate between two values \cite{Stern06}.

The Moore-Read Pfaffian state, which is possibly realized at
$\nu=5/2$, is the simplest of the non-Abelian states. The other
states are more complex, but also richer. The geometry of the
Fabry-Perot interferometer may be analyzed for these states as well.
In general, for all non-Abelian states the conductance of the Hall
bar depends on the internal state of the quasiparticles localized
between the constrictions -- i.e. the quasiparticle to which they
fuse. However, only for the Moore-Read Pfaffian state is
the effect quite so dramatic. For example, for the the $\mathbb{Z}_3$
parafermion state which may be realized at $\nu=12/5$, when the
number of localized quasiparticles is larger than three, the fusion
channel of the quasiparticles determines whether the interference is
fully visible or suppressed by a factor of $-\varphi^{-2}$
(with $\varphi$ being the golden ratio  $(\sqrt{5}+1)/2$)
\cite{Bonderson06b,Chung06}. The
number of quasiparticles, on the other hand, affects only the phase
of the interference pattern. Similar to the case of $\nu=5/2$ here
too the position of Coulomb blockade peaks on the two parameter
plane of area and magnetic field reflects the non-Abelian nature of
the quasiparticles \cite{Ilan07}.

%The non-abelian nature of the quasi-particles in Read-Rezayi
%states should manifest itself also in the spacing between
%Coulomb blockade peaks of a quantum dot, similar to the way
%in which it would happen in the $\nu=5/2$ state.
%For a $Z_k$ Read-Rezayi state, a periodic structure of $k$ Coulomb
%blockade peaks is formed as a function of the area of the dot, with each group of $k$ %peaks breaking into two sub-groups of equally spaced peaks.
%The number of peaks in each subgroup is determined by the
%number of quasi-particles localized in the bulk \cite{Ilan07}.

\subsubsection{A Fractional Quantum Hall Quantum Computer}
\label{sec:FQHE-qc}

\begin{figure}[t!]
\centerline{\includegraphics[width=3.5in]{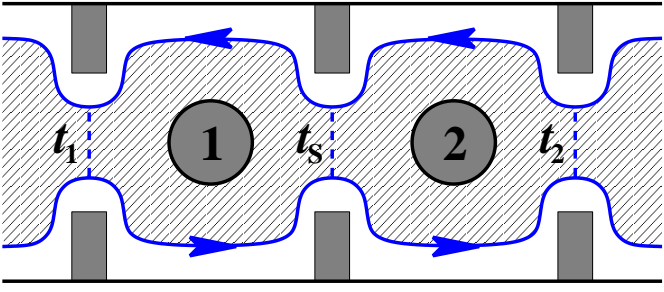}}
\caption{If a third constriction is added between the other
two, the cell is broken into two halves. We suppose that there
is one quasiparticle (or any odd number) in each half.
These two quasiparticles (labeled $1$ and $2$)
form a qubit which can be read
by measuring the conductance of the interferometer
if there is no backscattering at the middle constriction.
When a single quasiparticle tunnels from one edge to the other
at the middle constriction, a $\sigma_x$ or NOT
gate is applied to the qubit.}
\label{fig:Hall-qubit}
\end{figure}

We now describe how the
constricted Hall bar may be utilized as a quantum bit \cite{DasSarma05}. To that end,
an even number of $e/4$ quasiparticles should be trapped in the cell
between the constrictions, and a new, tunable, constriction should
be added between the other two so that the cell is broken
into two cells with an odd number of quasiparticles in each
(See Fig. (\ref{fig:Hall-qubit})). One way to tune the number of
quasiparticles in each half is to have two antidots in
the Hall bar. By tuning the voltage on the antidots, we
can change the number of quasiholes on each.
Let us assume that we thereby fix the number of
quasiparticles in each half of the cell to be odd.
For concreteness, let us take
this odd number to be one (i.e. let us assume that
we are in the idealized situation in which there are no quasiparticles
in the bulk, and one quasihole on each antidot).
These two quasiholes then form a
two-level system, i.e. a qubit. This two-level system can be understood in
several ways, which we discuss in detail in section \ref{part2}.
In brief, the two states correspond
to whether the two $\sigma$s fuse to ${\bf 1}$ or $\psi$ or,
in the language of chiral $p$-wave superconductivity,
the presence or absence of a neutral
(`Majorana') fermion; or, equivalently, as the fusion of two quasiparticles
carrying the spin-$1/2$ representation of an SU(2)
gauge symmetry in the spin-$0$ or spin-$1$ channels.

The interference between the $t_1$ and $t_2$ processes depends on
the state of the two-level system, so the qubit can be read by a
measurement of the four-terminal longitudinal conductance
\begin{equation}
\label{eq:read-out-sigma} G_L  \propto |t_1|^2+|t_2|^2 \pm2
\mbox{Re}\!\left\{t_1^*t_2 e^{i\phi}\right\}
\end{equation}
where the $\pm$ comes from the dependence of
$\langle\xi_0|{\hat U}|\xi_0\rangle$ on the state
of the qubit, as we discuss in section \ref{part2}.

The purpose of the middle constriction is to allow us to manipulate
the qubit. The state may be flipped, i.e. a $\sigma_x$ or NOT gate
can be applied, by the passage of a single quasiparticle from one
edge to the other, provided that its trajectory passes in between
the two localized quasiparticles. This is a simple example of how
braiding causes non-trivial transformations of multi-quasiparticle
states of non-Abelian quasiparticles, which we discuss in more
detail in section \ref{part2}. If we measure the four-terminal
longitudinal conductance $G_L$ before and after applying this NOT
gate, we will observe different values according to
(\ref{eq:read-out-sigma}).

For this operation to be a NOT gate, it is important that just a
single quasiparticle (or any odd number) tunnel from one edge to the
other across the middle constriction. In order to regulate the
number of quasiparticles which pass across the constriction, it may
be useful to have a small anti-dot in the middle of the constriction
with a large charging energy so that only a single quasiparticle can
pass through at a time. If we do not have good control over how many
quasiparticles tunnel, then it will be essentially random whether an
even or odd number of quasiparticles tunnel across; half of the
time, a NOT gate will be applied and the backscattering probability
(hence the conductance) will change while the other half of the
time, the backscattering probability is unchanged. If the
constriction is pinched down to such an extreme that the $5/2$ state
is disrupted between the quasiparticles, then when it is restored,
there will be an equal probability for the qubit to be in either
state.

This qubit is topologically protected because its state can only be
affected by a charge $e/4$ quasiparticle braiding with it. If a
charge $e/4$ quasiparticle winds around one of the antidots, it
effects a NOT gate on the qubit. The probability for such an event
can be very small because the density of thermally-excited charge
$e/4$ quasiparticles is exponentially suppressed at low
temperatures, $n_{\rm qp} \sim e^{-\Delta/(2T)}$. The simplest
estimate of the error rate $\Gamma$ (in units of the gap) is then of
activated form:
\begin{equation}
{\Gamma}/{\Delta} \sim \left(T/\Delta\right) \, e^{-\Delta/(2T)}
\end{equation}
The most favorable experimental situation \cite{Xia04} considered in
\cite{DasSarma05} has $\Delta \approx 500$ mK and $T\sim 5$ mK,
producing an astronomically low error rate $\sim 10^{-15}$. This
should be taken as an overly optimistic estimate. A more definitive
answer is surely more complicated since there are multiple gaps
which can be relevant in a disordered system. Furthermore, at very
low temperatures, we would expect quasiparticle transport to be
dominated by variable-range hopping of localized quasiparticles
rather than thermal activation. Indeed, the crossover to this
behavior may already be apparent \cite{Pan99b}, in which case, the
error suppression will be considerably weaker at the lowest
temperatures.
%Nevertheless, this estimate is informative since it
%ties the error rate to quantities determined from measurements of
%the resistance.
Although the error rate, which is determined by the
probability for a quasiparticle to wind around the anti-dot, is not
the same as the longitudinal resistance, which is the probability
for it to go from one edge of the system to the other, the two are
controlled by similar physical processes. A more sophisticated
estimate would require a detailed analysis of the quasiparticle
transport properties which contribute to the error rate. In
addition, this error estimate assumes that all of the trapped
(unintended) quasiparticles are kept very far from the
quasiparticles which we use for our qubit so that they cannot
exchange topological quantum numbers with our qubit via tunneling.
We comment on the issues involved in more detailed error estimates
in section \ref{sec:errors}.

The device envisioned above can be generalized to
one with many anti-dots and, therefore, many
qubits. More complicated gates, such as a CNOT
gate can be applied by braiding quasiparticles.
It is not clear how to braid quasiparticles localized in
the bulk -- perhaps by transferring them from one anti-dot to another
in a kind of ``bucket brigade''. This is an important problem
for any realization of topological quantum computing.
However, as we will discuss in section \ref{part3}, even
if this were solved, there would still be the problem that
braiding alone is not sufficient
for universal quantum computation in the $\nu=5/2$
state (assuming that it is the Moore-Read
Pfaffian state). One must either use some unprotected
operations (just two, in fact) or else use the $\nu=12/5$,
if it turns out to be the $\mathbb{Z}_3$ parafermion non-Abelian state.

\subsubsection{Physical Systems and Materials Considerations}
\label{sec:PhysicalSystems}

As seen in the device described in the previous subsection,
topological protection in non-Abelian fractional quantum Hall states
hinges on the energy gap ($\Delta$) separating the many-body
degenerate ground states from the low-lying excited states.  This
excitation gap also leads to the incompressibility of the quantum
Hall state and the quantization of the Hall resistance. Generally
speaking, the larger the size of this excitation gap compared to the
temperature, the more robust the topological protection, since
thermal excitation of stray quasiparticles, which goes as $\exp (-
\Delta/(2T))$, would potentially lead to errors.

%The fact that thermally excited quasiparticle population density
%goes as $e^{-\Delta/k_B T}$ implies that either increasing $\Delta$
%or decreasing $T$ would suppress thermal activation.
It must be emphasized that the relevant $T$ here is the temperature
of the electrons (or more precisely, the quasiparticles) and not
that of the GaAs-AlGaAs lattice surrounding the 2D electron layer.
Although the surrounding bath temperature could be lowered to 1 mK
or below by using adiabatic demagnetization in dilution
refrigerators, the 2D electrons themselves
thermally decouple from the bath at low temperatures and it is very
difficult to cool the 2D electrons below $T \approx$ 20
mK. It will be a great boost to hopes for topological
quantum computation using non-Abelian fractional quantum Hall states
if the electron temperature can be lowered to 1 mK or even below,
and serious efforts are currently underway in several laboratories
with this goal.

Unfortunately, the excitation gaps for the expected non-Abelian
fractional quantum Hall states are typically very small (compared,
for example, with the $\nu=1/3$ fractional
quantum Hall state).
The early measured gap for the 5/2 state was around $\Delta \sim $
25 mK (in 1987) \cite{Willett87}, but steady improvement in materials
quality, as measured by the sample mobility, has considerably
enhanced this gap. In the highest mobility samples currently
(2007) available, $\Delta\approx$ 600 mK \cite{Choi07}.
Indeed, there appears
to be a close connection between the excitation gap $\Delta$
and the mobility (or the sample quality). Although the details of
this connection are not well-understood, it is empirically
well-established that enhancing the 2D mobility
invariably leads to larger measured excitation gaps. In particular,
an empirical relation, $\Delta=\Delta_0-\Gamma$, where $\Delta$ is
the measured activation gap and $\Delta_0$ the ideal excitation gap
with $\Gamma$ being the level broadening arising from impurity and
disorder scattering, has often been discussed in the literature
(see, e.g. \onlinecite{Du93}).
Writing the mobility $\mu=e\tau/m$, with $\tau$ the zero field Drude
scattering time, we can write (an approximation of) the level
broadening as $\Gamma=\hbar/(2\tau)$, indicating $\Gamma \sim
\mu^{-1}$ in this simple picture, and therefore increasing the mobility
should steadily enhance the operational excitation gap, as is found
experimentally.
% There is considerable empirical support for
%the disorder-dependence of the activation gap through a
%$\Gamma=\Gamma_0-\mu$ type relationship.
It has recently been pointed out \cite{Morf02} that by
reducing $\Gamma$, an FQH gap of 2-3 K may be achievable in the 5/2
FQH state. Much less is currently known about the 12/5
state, but recent numerics \cite{Rezayi06} suggest that the
maximal gap in typical samples will be quite a bit lower than for
5/2.

It is also possible to consider designing samples that would
inherently have particularly large gaps.
First of all, the interaction energy (which sets the
overall scale of the gap) is roughly of the $1/r$ Coulomb form, so it
scales as the inverse of the interparticle spacing. Doubling the
density should therefore increase the gaps by roughly 40\%. Although
there are efforts underway to increase the density of
samples \cite{Willett07},
there are practical limitations to how high a density one can obtain
since, if one tries to over-fill a quantum well with electrons, the
electrons will no longer remain strictly two dimensional (i.e., they
will start filling higher subbands, or they will not remain in the
well at all). Secondly, as discussed in section \ref{sec:Non-AbelianQHE}
above, since the non-Abelian states appear generally to be very
sensitive to the precise parameters of the Hamiltonian, another
possible route to increased excitation gap would be to design the
precise form of the inter-electron interaction (which can be
modified by well width, screening layers, and particularly
spin-orbit coupling \cite{Manfra07}) so that  the Hamiltonian is at a
point in the phase diagram with maximal gap. With all approaches for
re-designing samples, however, it is crucial to keep the disorder
level low, which is an exceedingly difficult challenge.

Note that a large excitation gap (and correspondingly low
temperature) suppresses thermally excited quasiparticles but does
not preclude stray localized quasiparticles which could be present
even at $T=0$. As long as their positions are known and fixed, and
as long as they are few enough in number to be sufficiently well
separated, these quasiparticles would not present a problem, as one
could avoid moving other quasiparticles near their positions and one
could then tailor algorithms to account for their presence.  If the
density of stray localized quasiparticles is sufficiently high,
however, this would no longer be possible.   Fortunately, these
stray particles can be minimized in the same way as one of the above
discussed solutions to keeping the energy gap large -- improve the
mobility of the 2D electron sample on which the measurements (i.e.
the computation operations) are being carried out. Improvement in
the mobility leads to both the enhancement of the excitation gap and
the suppression of unwanted quasiparticle localization by disorder.

We should emphasize, however, how extremely high quality the current
samples already are.  Current ``good" sample mobilities are in the
range of $10 - 30 \times 10^6 \, {\rm cm}^2/({\rm
Volt}$-$\rm{sec})$.  To give the reader an idea of how impressive
this is, we note that under such conditions, at low temperatures,
the mean free path for an electron may be a macroscopic
length of a tenth of a millimeter or more. (Compare this to, say, copper at
room temperature, which has a mean free path of tens of
nanometers or less).

Nonetheless, further MBE technique and design improvement may be
needed to push low-temperature 2D electron mobilities to $100 \times
10^6 \, {\rm cm}^2/({\rm Volt}$-$\rm{sec})$ or above for topological
quantum computation to be feasible.  At lower temperatures, $T<100$
mK, the phonon scattering is very strongly suppressed
\cite{Stormer90,Kawamura92},
and therefore, there is essentially no intrinsic limit to how high the
2D electron mobility can be since the extrinsic scattering
associated with impurities and disorder can, in principle, be
eliminated through materials improvement.  In fact, steady materials
improvement in modulation-doped 2D GaAs-AlGaAs heterostructures
grown by the MBE technique has enhanced the 2D electron mobility from
$10^4 \, {\rm cm}^2/({\rm Volt}$-$\rm{sec})$ in the early 1980's to
$30 \times 10^6 \, {\rm cm}^2/({\rm Volt}$-$\rm{sec})$ in 2004, a
three orders of magnitude improvement in materials quality in
roughly twenty years.  Indeed, the vitality of the entire field of
quantum Hall physics is a result of these amazing advances. Another
factor of 2-3 improvement in the mobility seems
possible (L. Pfeiffer, private communication), and will likely be needed for the
successful experimental observation of non-Abelian anyonic
statistics and topological quantum computation.

\subsection{Other Proposed Non-Abelian Systems}
\label{sec:Othersystems}

This review devotes a great deal of attention to
the non-Abelian anyonic properties of certain fractional quantum
Hall states (e.g. $\nu=5/2, 12/5$, etc. states) in two-dimensional semiconductor structures, mainly because theoretical and experimental studies of such (possibly) non-Abelian fractional quantized Hall states is a mature subject, dating back to 1986, with many concrete results and ideas, including a recent proposal \cite{DasSarma05} for the construction of qubits and a NOT gate for topological quantum computation
(described above in subsection \ref{sec:FQHE-qc} and, in greater detail
in section \ref{part3}). But there are several other systems which are
potential candidates for topological quantum computation, and we briefly discuss these systems in this subsection. Indeed, the earliest proposals for fault-tolerant
quantum computation with anyons were based on spin systems,
not the quantum Hall effect \cite{Kitaev97}.

First, we emphasize that the most crucial necessary condition for carrying out topological quantum computation is the existence of appropriate `topological matter',
i.e. a physical system in a topological phase. Such a phase of matter has
suitable ground state properties and quasiparticle excitations manifesting non-Abelian statistics. Unfortunately, the necessary and sufficient conditions for the existence of topological ground states are not known even in theoretical models.
We note that the topological symmetry of the ground state is an emergent
symmetry at low energy, which is not present in the microscopic Hamiltonian
of the system. Consequently, given a Hamiltonian, it is very difficult to determine if its ground state is in a topological phase. It is certainly no easier than showing
that any other low-energy emergent phenomenon occurs in a particular
model. Except for rare exactly solvable models
(e.g. \onlinecite{Kitaev06a}, \onlinecite{Levin05a} which we describe in
section \ref{sec:P-T-Invariant}), topological ground states are
inferred on the basis of approximations and inspired guesswork.
On the other hand, if topological states exist at all, they will be robust
(i.e. their topological nature should be fairly insensitive to local perturbations,
e.g. electron-phonon interaction or charge fluctuations between traps). For this reason,
we believe that if it can be shown that some model
Hamiltonian has a topological ground state, then a real material which
is described approximately by that model is likely to have a topological
ground state as well.

One theoretical model which is known to have a non-Abelian
topological ground state is a $p+ip$ wave superconductor (i.e., a
superconductor where the order parameter is of $p_x + i p_y$
symmetry). As we describe in section \ref{sec:pwave}, vortices in a
superconductor of $p+ip$ pairing symmetry exhibit non-Abelian
braiding statistics. This is really just a reincarnation of the
physics of the Pfaffian state (believed to be realized at the
$\nu=5/2$ quantum Hall plateau) in zero magnetic field. Chiral
$p$-wave superconductivity/superfluidity is currently the most
transparent route to non-Abelian anyons. As we discuss below, there
are multiple physical systems which may host such a reincarnation.
The Kitaev honeycomb model (see also section \ref{sec:P-T-Invariant}
and ) \cite{Kitaev06a} is a seemingly different model which gives
rise to the same physics. In it, spins interact anisotropically in
such a way that their Hilbert space can be mapped onto that of a
system of Majorana fermions. In various parameter ranges, the ground
state is in either an Abelian topological phase, or a non-Abelian
one in the same universality class as a $p+ip$ superconductor.

Chiral $p$-wave superconductors, like quantum Hall states, break
parity and time-reversal symmetries, although they do so
spontaneously, rather than as a result of a large magnetic field.
However, it is also possible to have a topological phase which does
not break these symmetries. Soluble theoretical models of spins on a
lattice have been constructed which have $P,T$-invariant topological
ground states. A very simple model of this type with an {\it
Abelian} topological ground state, called the `toric code', was
proposed in \onlinecite{Kitaev97}. Even though it is not sufficient
for topological quantum computation because it is Abelian, it is
instructive to consider this model because non-Abelian models can be
viewed as more complex versions of this model. It describes $s=1/2$
spins on a lattice interacting through the following Hamiltonian
\cite{Kitaev97}:
\begin{equation}
H = -{J_1}{\sum_i}{A_i} -{J_2}{\sum_p}{F_p}
\label{eqn:toric-code}
\end{equation}
This model can be defined on an arbitrary lattice. The spins are
assumed to be on the links of the lattice.
${A_i}~\equiv~{\prod_{\alpha\in {\cal N}(i)}} \sigma_z^\alpha$,
where ${\cal N}(i)$ is the set of spins on links $\alpha$ which
touch the vertex $i$, and ${F_p}~\equiv~{\prod_{\alpha\in p}}
\sigma_x^\alpha$, where $p$ is a plaquette and $\alpha \in p$ are
the spins on the links comprising the plaquette. This model is
exactly soluble because the $A_i$s and $F_p$s all commute with each
other. For any ${J_1},{J_2}>0$, the ground state $|0\rangle$ is
given by ${A_i}|0\rangle={F_p}|0\rangle=|0\rangle$ for all $i,p$.
Quasiparticle excitations are sites $i$ at which
${A_i}|0\rangle=-|0\rangle$ or plaquettes $p$ at which
${F_p}|0\rangle=-|0\rangle$. A pair of excited sites can be created
at $i$ and $i'$ by acting on the ground state with $\prod_{\alpha\in
{\cal C}} \sigma_x^\alpha$, where the product is over the links in a
chain ${\cal C}$ on the lattice connecting $i$ and $i'$. Similarly,
a pair of excited plaquettes can be created by acting on the ground
state with connected $\prod_{\alpha\in \tilde{\cal C}}
\sigma_z^\alpha$ where the product is over the links crossed by a
chain $\tilde{\cal C}$ on the dual lattice connecting the centers of
plaquettes $p$ and $p'$. Both types of excitations are bosons, but
when an excited site is taken around an excited plaquette, the
wavefunction acquires a minus sign. Thus, these two types of bosons
are {\it relative semions}.

The toric code model is not very realistic,
but it is closely related to some more realistic models
such as the quantum dimer model
\cite{Rokhsar88,Chayes89,Klein82,Moessner01,Nayak01a}.
The degrees of freedom in this model are dimers on the links of
a lattice, which represent a spin singlet bond between the
two spins on either end of a link. The quantum dimer model
was proposed as an effective model for frustrated antiferromagnets,
in which the spins do not order, but instead form singlet bonds
which resonate among the links of the lattice -- the resonating
valence bond (RVB) state \cite{Anderson73,Anderson87,Baskaran87,Kivelson87}
which, in modern language, we would describe as a specific
realization of a simple Abelian topological state
\cite{Balents99,Balents00,Senthil00,Senthil01a,Moessner01}.
While the quantum dimer model on the square lattice does not
have a topological phase for any range of parameter values
(the RVB state is only the ground state at a critical point),
the model on a triangular lattice does have a topological phase \cite{Moessner01}.

\onlinecite{Levin05a,Levin05b} constructed a model which is,
in a sense, a non-Abelian generalization of Kitaev's toric code model.
It is an exactly soluble model of spins on the links
(two on each link) of the honeycomb lattice with three-spin interactions
at each vertex and twelve-spin interactions around each plaquette,
which we describe in section \ref{sec:P-T-Invariant}.
This model realizes a non-Abelian phase which supports Fibonacci
anyons, which permits universal topological quantum computation
(and generalizes straightforwardly to other non-Abelian topological
phases). Other models have been constructed \cite{Freedman05a,Fendley05}
which are not exactly soluble but have
only two-body interactions and can be argued to support topological phases
in some parameter regime. However, there is still a considerable
gulf between models which are soluble or quasi-soluble and
models which might be considered realistic for some material.

Models such as the Kitaev and Levin-Wen models are deep within topological
phases; there are no other competing states nearby in their phase
diagram. However, simple models such as the Heisenberg model or extensions
of the Hubbard model are not of this form. The implication is that
such models are not deep within a topological phase,
and topological phases must compete with other phases, such as broken
symmetry phases. In the quantum dimer model \cite{Rokhsar88,Moessner01},
for instance, an Abelian topological phase must compete with
various crystalline phases which occupy most of the phase diagram.
This is presumably one obstacle to finding topological phases in
more realistic models, i.e. models which would give an approximate
description of some concrete physical system.

There are several physical systems -- apart from fractional quantum
Hall states -- which might be promising hunting grounds for
topological phases, including transition metal oxides and ultra-cold
atoms in optical traps. The transition metal oxides have the
advantage that we already know that they give rise to striking
collective phenomena such as high-$T_c$ superconductivity, colossal
magnetoresistance, stripes, and thermoelectricity. Unfortunately,
their physics is very difficult to unravel both theoretically and
experimentally for this very reason: there are often many different
competing phenomena in these materials. This is reflected in the
models which describe transition metal oxides. They tend to have
many closely competing phases, so that different approximate
treatments find rather different phase diagrams. There is a second
advantage to the transition metal oxides, namely that many
sophisticated experimental techniques have been developed to study
them, including transport, thermodynamic measurements,
photoemission, neutron scattering, X-ray scattering, and NMR.
Unfortunately, however, these methods are tailored for detecting
broken-symmetry states or for giving a detailed understanding of
metallic behavior, not for uncovering a topological phase.
Nevertheless, this is such a rich family of materials that it would
be surprising if there weren't a topological phase hiding there.
(Whether we find it is another matter.) There is one
particular material in this family, Sr$_2$RuO$_4$, for which there is
striking evidence that it is a chiral $p$-wave superconductor at low
temperatures, ${T_c}\approx 1.5$ K \cite{Xia06,Kidwingira06}.
Half-quantum vortices in a thin film of such a superconductor would
exhibit non-Abelian braiding statistics (since Sr$_2$RuO$_4$ is not
spin-polarized, one must use half quantum vortices, not ordinary
vortices). However, half quantum vortices are usually not the lowest
energy vortices in a chiral $p$-wave superconductor, and a direct
experimental observation of the half vortices themselves would be a
substantial milestone on the way to topological quantum computation
\cite{DasSarma06a}.

The current status of research is as follows.
Three-dimensional single-crystals and thin films
of Sr$_2$RuO$_4$ have been fabricated and studied.
The nature of the super-conductivity of these samples has been studied by many experimental probes, with the goal of identifying the symmetry of the Cooper-pair.
There are many indications that support the identification of the Sr$_2$RuO$_4$
as a $p_x+ip_y$ super-conductor. First, experiments that probe the spins of the Cooper pair strongly indicate triplet pairing \cite{Mackenzie03}. Such experiments probe the spin susceptibility through measurements of the NMR Knight shift and of neutron scattering. For singlet spin pairing the susceptibility vanishes at zero temperature, since the spins keep a zero polarization state in order to form Cooper pairs. In contrast, the susceptibility remains finite for triplet pairing, and this is indeed the observed behavior. Second, several experiments that probe time reversal symmetry have indicated that it is broken, as expected from a $p\pm ip$ super-conductor. These experiments include muon spin
relaxation \cite{Mackenzie03} and the polar Kerr effect\cite{Xia06}.
In contrast, magnetic imaging experiments designed to probe the edge currents that are associated with a super-conductor that breaks time reversal symmetry did not find the expected signal \cite{Kirtley07}. The absence of this signal may be attributed
to the existence of domains of $p+ip$ interleaved with those of $p-ip$. Altogether, then, Sr$_2$RuO$_4$ is likely to be a
three dimensional $p+ip$ super-conductor,
that may open the way for a realization of a two-dimensional super-conductor that breaks time reversal symmetry.

The other very promising direction to look for topological phases,
ultra-cold atoms in traps, also has several advantages. The Hamiltonian
can often be tuned by, for instance, tuning the lasers which define
an optical lattice or by tuning through a Feshbach resonance. For instance,
there is a specific scheme for realizing the Hubbard model \cite{Jaksch05}
in this way. At present there are relatively few experimental
probes of these systems, as compared with transition metal oxides or
even semiconductor devices.  However, to look on the bright side,
some of the available probes give information that cannot be
measured in electronic systems. Furthermore, new probes for cold
atoms systems are being developed at a remarkable rate.

There are two different schemes for generating topological phases in ultra-cold atomic gases that seem  particularly promising at the current time.  The first is the approach of using fast rotating dilute bose gases \cite{Wilkin98} to make quantum Hall systems of bosons
\cite{Cooper01}.    Here, the rotation plays the role of an  effective magnetic field, and the filling fraction is given by the ratio of the number of bosons to the number of  vortices caused by rotation.  Experimental techniques \cite{Schweikhard04,Bretin04,Abo-Shaeer01}  have been  developed that can give very large rotation rates and filling fractions can be generated which are as low as  $\nu=500$ \cite{Schweikhard04}.  While this is sufficiently low that all of the bosons are in a single landau  level (since there is no Pauli exclusion, nu> 1 can still be a lowest Landau level state), it is still predicted  to be several orders of magnitude too high to see interesting topological states.  Theoretically, the interesting  topological states occur for
$\nu<10$ \cite{Cooper01}.  In particular, evidence is very strong that
$\nu=1$, should  it be achieved, would be the bosonic analogue of the Moore-Read state, and (slightly less strong) $\nu=3/2$ and  $\nu=2$ would be the Read-Rezayi states, if the inter-boson interaction is appropriately adjusted \cite{Rezayi05,Cooper07}.   In order to access this regime, either rotation rates will need to be increased  substantially, or densities will have to be decreased substantially.   While the latter sounds easier, it then  results in all of the interaction scales being correspondingly lower, and hence implies that temperature would  have to be lower also, which again becomes a challenge.   Several other works have proposed using atomic lattice  systems where manipulation of parameters of the Hamiltonian induces effective magnetic fields and should also  result in quantum hall physics\cite{Sorensen05,Mueller04,Popp04}.  

The second route to generating topological phases in cold atoms is the idea of using a gas of ultra-cold fermions  with a p-wave Feschbach resonance, which could form a spin-polarized chiral p-wave superfluid \cite{Gurarie05}.   Preliminary studies of such p-wave systems have been made experimentally \cite{Gaebler07} and unfortunately, it appears that the decay time of the Feshbach bound states may be so short that thermalization is  impossible.   Indeed, recent theoretical work \cite{Levinsen07} suggests that this may be a generic problem and  additional tricks may be necessary if a $p$-wave superfluid is to be produced in this way.
  
We note that  both the $\nu=1$ rotating boson system and the chiral
$p$-wave superfluid would be quite closely
related to the putative non-Abelian
quantum Hall state at $\nu=5/2$ (as is Sr$_2$RuO$_4$). However,
there is an important difference between a $p$-wave superfluid of
cold fermions and the $\nu=5/2$ state.  Two-dimensional
superconductors, as well as superfluids in any dimension, have a
gapless Goldstone mode. Therefore, there is the danger that the
motion of vortices may cause the excitation of low-energy
modes. Superfluids of cold atoms may, however, be good test grounds
for the detection of localized Majorana modes associated with
localized vortices, as those are expected to have a clear signature
in the absorption spectrum of RF radiation \cite{Tewari07a}, in the
form of a discrete absorption peak whose density and weight are
determined by the density of the vortices \cite{Grosfeld07}. One can
also realize, using suitable laser configurations, Kitaev's
honeycomb lattice model (Eq.~\ref{eqn:Kitaev-honeycomb}) with cold
atoms on an optical lattice \cite{Duan03}. It has recently been
shown how to braid anyons in such a model \cite{Zhang06}.

A major difficulty in finding a topological phase in either
a transition metal oxide or an ultra-cold atomic system
is that topological phases are hard to detect directly. If the
phase breaks parity and time-reversal symmetries, either
spontaneously or as a result of an external magnetic field,
then there is usually an experimental handle through
transport, as in the fractional quantum Hall states or
chiral $p$-wave superconductors. If the state does not break
parity and time-reversal, however, there is no `smoking gun'
experiment, short of creating quasiparticles, braiding them,
and measuring the outcome.

Any detailed discussion of the physics of these `alternative' topological
systems is well beyond the scope of the current review.
We refer the readers to the existing recent literature on these systems for details.
In section \ref{part2} (especially \ref{sec:P-T-Invariant}), however,
we discuss some of the soluble models
which support topological phases because many of their mathematical
features elucidate the underlying structure of topological phases.

\section{Topological Phases of Matter and Non-Abelian Anyons}
\label{part2}

Topological quantum computation is predicated on the existence in
nature of topological phases of matter. In this section, we will
discuss the physics of topological phases from several different
perspectives, using a variety of theoretical tools. The reader who
is interested primarily in the application of topological phases to
quantum computation can skim this section briefly and still
understand section IV. However, a reader with a background in
condensed matter physics and quantum field theory may find it
enlightening to read a more detailed account of the theory of
topological phases and the emergence of anyons from such phases,
with explicit derivations of some of the results
mentioned in section II and used in section IV. These readers may
find topological phases interesting in and of themselves, apart from
possible applications.

Topological phases, the states of matter which support anyons,
occur in many-particle physical systems. Therefore, we will
be using field theory techniques to study these states. A canonical,
but by no means unique, example of a field theory for a topological
phase is Chern-Simons theory. We will frequently use this theory
to illustrate the general points which we wish to make about topological phases.
In section \ref{sec:conclusion}, we will make a few comments about
the problem of classifying topological phases, and how this
example, Chern-Simons theory, fits in the general classification.
In subsection \ref{sec:CS-theory}, we give a more precise definition
of a topological phase and connect this definition with the existence of anyons.
We also introduce Chern-Simons theory, which we
will discuss throughout section III as an example of
the general structure which we discuss in subsection \ref{sec:CS-theory}.
In subsection \ref{sec:pwave}, we will discuss a
topological phase which is superficially
rather different but, in fact, will prove to be a special
case of Chern-Simons theory. This phase can
be analyzed in detail using the formalism of BCS theory.
In subsection \ref{sec:Jones}, we further analyze Chern-Simons theory,
giving a more detailed account of its topological properties,
especially the braiding of anyons. We describe Witten's work \cite{Witten89}
connecting Chern-Simons theory with the knot and link invariants
of Jones and Kauffman \cite{Jones85,Kauffman87}. We show how the latter
can be used to derive the properties of anyons in these topological phases.
In section \ref{sec:FQHE}, we describe a complementary approach
by which Chern-Simons theory can be understood: through its
connection to conformal field theory. We explain how this approach
can be particularly fruitful in connection with fractional
quantum Hall states. In \ref{sec:edge}, we discuss the gapless excitations
which must be present at the edge of any chiral topological phase.
Their physics is intimately connected with the topological properties of
the bulk and, at the same time, is directly probed by transport experiments
in quantum Hall devices. In \ref{sec:experiments2}, we apply the
knowledge which we have gained about the properties of
Chern-Simons theory to the interferometry
experiments which we discussed in \ref{sec:interference}.
Finally, in \ref{sec:P-T-Invariant} we discuss a related but
different class of topological phases which can arise in
lattice models and may be relevant to transition metal oxides
or `artificial' solids such as ultra-cold atoms in optical lattices.

\subsection{Topological Phases of Matter}
\label{sec:CS-theory}

In Section \ref{part1} of this paper, we have used `topological phase'
as essentially being synonymous with any system
whose quasiparticle excitations are anyons.
However, a precise definition is the following.
A system is in a topological phase if, at low temperatures and
energies, and long wavelengths, all observable properties (e.g. correlation
functions) are invariant under smooth deformations (diffeomorphisms)
of the spacetime manifold in which the system lives.
Equivalently, all observable properties are independent of the choice of spacetime coordinates, which need not be inertial or rectilinear.
(This is the `passive' sense of a diffeomorphism,
while the first statement uses
the active sense of a transformation.) By ``at low temperatures and
energies, and long wavelengths,'' we mean that diffeomorphism
invariance is only violated by terms which vanish as
$\sim \text{max}\left(e^{-\Delta/T},e^{-|x|/\xi}\right)$
for some non-zero energy gap $\Delta$ and finite correlation
length $\xi$. Thus, topological phases have, in general, an energy
gap separating the ground state(s) from the lowest excited states.
Note that an excitation gap, while necessary, is not sufficient to ensure
that a system is in a topological phase.

The invariance of all correlation functions under diffeomorphisms
means that the only local operator which has non-vanishing
correlation functions is the identity. For instance, under an
arbitrary change of space-time coordinates $x\rightarrow x'=f(x)$,
the correlations of a scalar operator $\phi(x)$ must satisfy
$\langle{0_i}|\phi({x_1}) \phi({x_2}) \ldots
\phi({x_n})|{0_j}\rangle = \langle{0_i}| \phi(x'_1) \phi(x'_2)
\ldots \phi(x'_n)|{0_j}\rangle$, which implies that
$\langle{0_i}|\phi({x_1}) \phi({x_2}) \ldots
\phi({x_n})|{0_j}\rangle =0$ unless $\phi(x)\equiv c$ for some
constant $c$. Here, $|{0_i}\rangle, |{0_j}\rangle$ are ground states
of the system (which may or may not be different). This property is
important because any local perturbation, such as the environment,
couples to a local operator. Hence, these local perturbations are
proportional to the identity. Consequently, they cannot have
non-trivial matrix elements between different ground states. The
only way in which they can affect the system is by exciting the
system to high-energies, at which diffeomorphism invariance is
violated. At low-temperatures, the probability for this is
exponentially suppressed.

The preceding definition of a topological phase
may be stated more compactly by simply saying that
a system is in a topological phase if its low-energy
effective field theory is a topological quantum field theory (TQFT),
i.e. a field theory whose correlation functions are invariant
under diffeomorphisms.
Remarkably, topological invariance does not imply trivial
low-energy physics.

\subsubsection{Chern-Simons Theory}
\label{sec:SC-theory}

Consider the simplest example of a TQFT, Abelian Chern-Simons
theory, which is relevant to the Laughlin states at filling
fractions of the form $\nu=1/k$, with $k$ an odd integer. Although
there are many ways to understand the Laughlin states, it is useful
for us to take the viewpoint of a low-energy effective theory.
Since quantum Hall systems are gapped, we should be able to describe
the system by a field theory with very few degrees of freedom.  To
this end, we consider the action
\begin{eqnarray}
\label{eqn:Abel-C-S-action}
S_{CS} =  \frac{k}{4\pi} \int {d^2}{\bf r} \,
dt \,\,  {\epsilon^{\mu\nu\rho}}{a_\mu} {\partial_\nu}{a_\rho}
\end{eqnarray}
where $k$ is an integer and $\epsilon$ is the antisymmetric tensor.
Here, $a$ is a U(1) gauge field and indices
$\mu, \nu, \rho$ take the values $0$ (for time-direction), $1$,$2$
(space-directions). This action represents the low-energy
degrees of freedom of the system, which are purely topological.

The Chern-Simons gauge field  $a$ in (\ref{eqn:Abel-C-S-action}) is
an emergent degree of freedom which encodes the low-energy physics
of a quantum Hall system. Although in this particular case, it is
simply-related to the electronic charge density, we will also be
considering systems in which emergent Chern-Simons gauge fields
cannot be related in a simple way to the underlying electronic
degrees of freedom.

In the presence of an external electromagnetic field
and quasiparticles, the action takes the form:
\begin{equation}
\label{eqn:Abel-C-S+field-action}
S = S_{CS} - \int {d^2}{\bf r} \,
dt \,\, \left(   \frac{1}{2\pi}{\epsilon^{\mu\nu\rho}}{A_\mu} {\partial_\nu}{a_\rho}
+ {j^{\rm qp}_\mu}{a_\mu}\right)
\end{equation}
where ${j^{\rm qp}_\mu}$ is the quasiparticle current,
$j^{\rm qp}_0=\rho^{\rm qp}$ is the quasiparticle density,
${\bf j}^{\rm qp}=(j^{\rm qp}_1,j^{\rm qp}_2)$ is the quasiparticle spatial current,
and $A_\mu$ is the external electromagnetic field.
We will assume that the quasiparticles are not dynamical,
but instead move along some fixed classically-prescribed trajectories
which determine ${j^{\rm qp}_\mu}$.
The electrical current is:
\begin{equation}
\label{eq:fluxattach}
j_\mu =
\partial {\cal L}/ \partial A_\mu =
\frac{1}{2 \pi} {\epsilon^{\mu\nu\rho}}
{\partial_\nu}{a_\rho}
\end{equation}
Since the action is quadratic, it is
completely solvable, and one can integrate out
the field $a_\mu$ to obtain the
response of the current to the external electromagnetic field.
The result of such a calculation is
precisely the quantized Hall conductivity $\sigma_{xx} = 0$ and
$\sigma_{xy} = \frac{1}{k} \, e^2 / h$.

The equation of motion obtained by varying $a_0$ is
the Chern-Simons constraint:
\begin{equation}
\label{eqn:CS-constraint}
\frac{k}{2\pi} \nabla \times {\bf a} = j^{\rm qp}_0 + \frac{1}{2\pi}B
\end{equation}
According to this equation, each quasiparticle has
Chern-Simons flux $2\pi/k$ attached to it (the magnetic field
is assumed fixed). Consequently, it has electrical charge $1/k$,
according to (\ref{eq:fluxattach}).
As a result of the Chern-Simons flux, another quasiparticle
moving in this Chern-Simons field picks up an Aharonov-Bohm phase.
The action associated with taking one quasiparticle
around another is, according to Eq.~\ref{eqn:Abel-C-S+field-action},
of the form
\begin{equation}
\frac{1}{2} k \int \,  d{\bf r} \, dt \,{\bf j} \cdot {\bf a} =
 k Q \int_{\cal C} d{\bf r} \cdot {\bf a}
\end{equation}
where $Q$ is the charge of the quasiparticle and the final integral
is just the Chern-Simons flux enclosed in the path. (The factor of
$1/2$ on the left-hand side is due to the action of the Chern-Simons
term itself which, according to the constraint
(\ref{eqn:CS-constraint}) is $-1/2$ times the Aharonov-Bohm phase.
This is cancelled by a factor of two coming from the fact that each
particle sees the other's flux.) Thus the contribution to a path
integral $e^{i S_{CS}}$ just gives an Aharonov-Bohm phase associated
with moving a charge around the Chern-Simons flux attached to the
other charges. The phases generated in this way give the
quasiparticles of this Chern-Simons theory $\theta=\pi/k$ Abelian
braiding statistics.\footnote{ The Chern-Simons effective action for
a hierarchical state is equivalent to the action for the composite
fermion state at the same filling fraction
\cite{Read90,Blok90,Wen92a}. It is a simple generalization of Eq.
\ref{eqn:Abel-C-S-action} which contains several internal gauge
fields $a_\mu^n$ (with $n=1,2,...$), corresponding (in essence) to
the action for the different species of particles (either the
different levels of the hierarchy, or the different composite
fermion Landau levels).}

Therefore, an Abelian Chern-Simons
term implements Abelian anyonic statistics. In fact, it does nothing else.
An Abelian gauge field in $2+1$ dimensions has only one transverse
component; the other two components can be eliminated by fixing
the gauge. This degree of freedom is fixed by the Chern-Simons constraint
(\ref{eqn:CS-constraint}). Therefore, a Chern-Simons gauge field has
no local degrees of freedom and no dynamics.

We now turn to non-Abelian Chern-Simons theory.
This TQFT describes non-Abelian anyons. It is analogous
to the Abelian Chern-Simons described above, but different methods
are needed for its solution, as we describe in this section.
The action can be written on an arbitrary manifold
${\cal M}$ in the form
\begin{eqnarray}
\label{eqn:non-Abel-C-S-action}
S_{CS}[a] &=&
\frac{k}{4\pi}\int_{\cal M} {\rm tr}\left(a\wedge da + \frac{2}{3}
a\wedge a\wedge a\right) \\ &=&  \frac{k}{4\pi} \int_{\cal M}
{\epsilon^{\mu\nu\rho}}\left({a_\mu^{\underline a}}
{\partial_\nu}{a_\rho^{\underline a}} + \frac{2}{3}\,{f_{{\underline
a}\,{\underline b}\,{\underline c}}} {a_\mu^{\underline
a}}{a_\nu^{\underline b}}{a_\rho^{\underline c}}\right) \nonumber
\end{eqnarray}
In this expression, the gauge field now takes values
in the Lie algebra of the group $G$.
$f_{{\underline a}\,{\underline b}\,{\underline c}}$ are the
structure constants of the Lie algebra which are simply
$\epsilon_{{\underline a}\,{\underline b}\,{\underline c}}$ for the
case of SU(2). For the case of SU(2),
we thus have a gauge field $a_\mu^{\underline a}$,
where the underlined indices run from 1 to 3. A
matter field transforming in the spin-$j$ representation of
the SU(2) gauge group will couple to the combination
$a_\mu^{\underline a} x_{\underline a}$, where $x_{\underline a}$
are the three generator matrices of su(2) in the spin-$j$ representation.
For gauge group $G$ and
coupling constant $k$ (called the `level'), we will denote such a theory by $G_k$.
In this paper, we will be primarily concerned with SU(2)$_k$
Chern-Simons theory.

To see that Chern-Simons theory is a TQFT, first note that
the Chern-Simons action (\ref{eqn:non-Abel-C-S-action}) is invariant
under all diffeomorphisms of ${\cal M}$ to itself, $f:{\cal M}\rightarrow{\cal M}$.
The differential form notation in (\ref{eqn:non-Abel-C-S-action})
makes this manifest, but it can be checked in coordinate form
for $x^\mu\rightarrow {f^\mu}(x)$. Diffeomorphism invariance
stems from the absence of the metric tensor
in the Chern-Simons action. Written out in component form,
as in (\ref{eqn:non-Abel-C-S-action}), indices are, instead, contracted
with $\epsilon^{\mu\nu\lambda}$.

Before analyzing the physics of this action (\ref{eqn:non-Abel-C-S-action}),
we will make two observations. First, as a result of the presence
of $\epsilon^{\mu\nu\lambda}$, the action changes
sign under parity or time-reversal transformations. In this
paper, we will concentrate, for the most part, on topological phases
which are chiral, i.e. which break parity and time-reversal symmetries.
These are the phases which can appear in the fractional
quantum Hall effect, where the large magnetic field breaks
$P$, $T$. However, we shall also discuss non-chiral topological
phases in section \ref{sec:P-T-Invariant}, especially in connection
which topological phases emerging from lattice models.

Secondly, the Chern-Simons action is not quite fully invariant under
gauge transformations ${a_\mu} \rightarrow g{a_\mu}g^{-1} +
g{\partial_\mu}g^{-1}$, where $g:{\cal M}\rightarrow G$
is any function on the manifold
taking values in the group $G$. On a
closed manifold, it is only invariant under ``small'' gauge
tranformations. Suppose that the manifold ${\cal M}$ is the
3-sphere, $S^3$. Then, gauge transformations are maps
${S^3}\rightarrow G$, which can be classified topologically
according to it homotopy ${\pi_3}(G)$. For any simple compact group $G$,
${\pi_3}(G)=\mathbb{Z}$, so gauge transformations can be classified
according to their ``winding number''. Under a gauge transformation
with winding $m$,
\begin{equation}
S_{CS}[a] \rightarrow S_{CS}[a] + 2\pi km
\end{equation}
\cite{Deser82}. While the action is invariant under ``small'' gauge
transformations, which are continuously connected to the identity
and have $m=0$, it is not invariant under ``large'' gauge
transformations ($m\neq 0$). However, it is sufficient for
$\exp(iS)$ to be gauge invariant, which will be the case so long as
we require that the level $k$ be an integer. The requirement that
the level $k$ be an integer is an example of the highly rigid
structure of TQFTs. A small perturbation of the microscopic
Hamiltonian cannot continuously change the value of $k$ in the
effective low energy theory; only a perturbation which is large
enough to change $k$ by an integer can do this.

The failure of gauge invariance under large gauge tranformations is also
reflected in the properties of Chern-Simons theory
on a surface with boundary, where the
Chern-Simons action is gauge invariant only up to a surface term.
Consequently, there must be gapless degrees of freedom at
the edge of the system whose dynamics is dictated by
the requirement of gauge invariance of the combined bulk and
edge \cite{Wen92b}, as we discuss in section \ref{sec:edge}.

To unravel the physics of Chern-Simons theory, it
is useful to specialize to the case in which the
spacetime manifold ${\cal M}$ can be decomposed
into a product of a spatial surface
and time, ${\cal M} = \Sigma\times\mathbb{R}$. On such
a manifold, Chern-Simons theory is a theory of the
ground states of a topologically-ordered
system on $\Sigma$. There are no excited states
in Chern-Simons theory because the Hamiltonian vanishes.
This is seen most simply in $a_0 = 0$ gauge, where
the momentum canonically conjugate to $a_1$ is $-\frac{k}{4\pi}\,a_2$,
and the momentum canonically conjugate to $a_2$ is $\frac{k}{4\pi}\,a_1$
so that
\begin{equation}
{\cal H} = \frac{k}{4\pi}{\rm tr}\left({a_2} {\partial_0} {a_1} - {a_1}{\partial_0}{a_2}\right)
- {\cal L} = 0
\end{equation}
Note that this is a special feature of an action with
a Chern-Simons term alone. If the action had both a
Chern-Simons and a Yang-Mills term, then the Hamiltonian would
not vanish, and the theory would have both ground states
and excited states with a finite gap. Since the Yang-Mills
term is subleading compared to the Chern-Simons term
(i.e. irrelevant in a renormalization group (RG) sense), we can forget
about it at energies smaller than the gap and consider the
Chern-Simons term alone.

Therefore, when Chern-Simons theory is viewed as an effective
field theory, it can only be valid at energies much smaller than
the energy gap. As a result, it is unclear, at the moment,
whether Chern-Simons theory has anything to say about the
properties of quasiparticles -- which are excitations above the gap -- or,
indeed, whether those properties
are part of the universal low-energy physics of the system
(i.e. are controlled by the infrared RG fixed point). Nevertheless,
as we will see momentarily, it does and they are.

Although the Hamiltonian vanishes, the theory is
still not trivial because one must solve
the constraint which follows by varying $a_0$.
For the sake of concreteness, we will specialize to the case
$G=$SU(2). Then the
constraint reads:
\begin{equation}
\epsilon_{ij} {\partial_i} {a_j^{\underline a}} +
f^{{\underline a}\,{\underline b}\,{\underline c}}
{a_1^{\underline b}}{a_2^{\underline c}} = 0
\label{eqn:CS-constraint-again}
\end{equation}
where $i,j=1,2$. The left-hand side of this equation is the field strength
of the gauge field $a_i^{\underline a}$, where ${\underline a}=1,2,3$
is an su(2) index. Since the field strength
must vanish, we can always perform a gauge transformation
so that $a_i^{\underline a}=0$ locally. Therefore this theory
has no local degrees of freedom. However, for some field
configurations satisfying the constraint, there may be a global
topological obstruction which prevents us from making the gauge
field zero everywhere. Clearly, this can only happen if $\Sigma$
is topologically non-trivial.

The simplest non-trivial manifold is the annulus, which
is topologically equivalent to the sphere with two punctures. Following
\onlinecite{Elitzur89} (see also \cite{Wen98} for a similar construction
on the torus), let us take coordinates $(r,\phi)$ on the annulus,
with ${r_1}<r<{r_2}$,
and let $t$ be time. Then we can write ${a_\mu} = g {\partial_\mu} g^{-1}$, where
\begin{equation}
\label{eqn:gauge-field-annulus}
g(r,\phi,t) = e^{i\omega(r,\phi,t)}\,e^{i\frac{\phi}{k} \lambda(t)}
\end{equation}
where $\omega(r,\phi,t)$ and $\lambda(t)$ take values in the Lie
algebra su(2) and $\omega(r,\phi,t)$ is a single-valued function of
$\phi$. The functions $\omega$ and $\lambda$ are the dynamical
variables of Chern-Simons theory on the annulus. Substituting
(\ref{eqn:gauge-field-annulus}) into the Chern-Simons action, we see
that it now takes the form:
\begin{equation}
S = \frac{1}{2\pi} \int dt \,
\text{tr}\left(\lambda {\partial_t}\Omega\right)
\end{equation}
where $\Omega(r,t)=\int_0^{2\pi} d\phi\, (\omega({r_1},\phi,t)-\omega({r_2},\phi,t))$.
Therefore, $\Omega$ is canonically conjugate to $\lambda$.
By a gauge transformation, we can always rotate $\lambda$
and $\Omega$ so that they are along the $3$ direction in su(2),
i.e. $\lambda= {\lambda_3}{T^3}$, $\Omega={\Omega_3}{T^3}$.
Since it is defined through the exponential in (\ref{eqn:gauge-field-annulus}),
$\Omega_3$ takes values in $[0,2\pi]$. Therefore, its canonical conjugate,
$\lambda_3$, is quantized to be an integer. From the definition
of $\lambda$ in (\ref{eqn:gauge-field-annulus}), we see that
${\lambda_3}\equiv {\lambda_3} +2k$. However, by a gauge transformation
given by a rotation around the $1$-axis, we can transform
$\lambda\rightarrow -\lambda$. Hence, the independent allowed
values of $\lambda$ are $0,1,\ldots,k$.

On the two-punctured sphere,
if one puncture is of type $a$, the other puncture must be of
type $\bar{a}$. (If the topological charge at one puncture is measured
along a loop around the puncture -- e.g. by a Wilson loop, see
subsection \ref{sec:Jones} -- then the loop can be deformed so that
it goes around the other puncture, but in the opposite direction.
Therefore, the two punctures necessarily have conjugate
topological charges.) For SU(2), $a=\bar{a}$, so both punctures
have the same topological charge. Therefore, the restriction to
only $k+1$ different possible allowed boundary conditions $\lambda$
for the two-punctured sphere implies that there are
$k+1$ different quasiparticle types in SU(2)$_k$
Chern-Simons theory. As we will describe in later subsections,
these allowed quasiparticle types can be identified with the
$j=0,\frac{1}{2},\ldots,\frac{k}{2}$ representations of the
SU(2)$_2$ Kac-Moody algebra.

\subsubsection{TQFTs and Quasiparticle Properties}
\label{sec:TQFTs}

We will continue with our analysis of Chern-Simons theory
in sections \ref{sec:Jones} and \ref{sec:FQHE}. Here, we will
make some more general observations abut TQFTs and the
topological properties of quasiparticles.
We turn to the $n$-punctured sphere,
$\Sigma={S^2}\backslash{P_1}\cup{P_2}\cup\ldots\cup{P_n}$,
i.e. the sphere ${S^2}$ with the points ${P_1}, {P_2} \ldots {P_n}$
deleted, which is equivalent to $n-1$ quasiparticles in the plane (the $n^\text{th}$
puncture becomes the point at $\infty$). This will allow us
to study the topological properties of quasiparticle excitations
purely from ground state properties.
To see how braiding emerges in this approach,
it is useful to note that diffeomorphisms
should have a unitary representation on the ground state
Hilbert space (i.e. they should commute with the Hamiltonian).
Diffeomorphisms which can be smoothly deformed to the identity
should have a trivial action on the Hilbert space of the theory since there
are no local degrees of freedom. However, `large' diffeomorphisms could have
a non-trivial unitary representation on the theory's Hilbert space.
If we take the quotient of the diffeomorphism group by the set of diffeomorphisms which can be smoothly deformed to the identity, then we obtain
the {\it mapping class group}. On the $n$-punctured sphere,
the braid group ${\cal B}_{n-1}$ is a subgroup of the mapping class
group.\footnote{The mapping class group is non-trivial solely as a result
of the punctures. In particular, any diffeomorphism which moves one or more
of the punctures around other punctures cannot be deformed to the
identity; conversely, if two diffeomorphisms move the same punctures
along trajectories which can be deformed into each other, then the
diffeomorphisms themselves can also be deformed into each other.
These classes of diffeomorphisms correspond to the braid group
which is, in fact, a normal subgroup.
If we take the quotient of the mapping class group
by the Dehn twists of $n-1$ of the punctures -- all except the
point at infinity --  we would be left with the braid group ${\cal B}_{n-1}$.}
Therefore, if we study Chern-Simons theory on the
$n$-punctured sphere as we did for the $2$-punctured sphere
above, and determine how the mapping class group acts,
we can learn all of the desired information about quasiparticle braiding.
We do this by two different methods in subsections \ref{sec:Jones}
and \ref{sec:FQHE}.

One extra transformation in the mapping class group,
compared to the braid group, is a $2\pi$ rotation of a
puncture/particle relative to the rest of the system
(a Dehn twist). If we consider particles with a finite extent,
rather than point particles, then we must consider the possibility
of such rotations. For instance, if the particles are small dipoles,
then we can represent their world lines as ribbons. A Dehn twist
then corresponds to a twist of the ribbon. Thickening a world line into
a ribbon is called a {\it framing}. A given world line has multiple
choices of framing, corresponding to how many times the
ribbon twists. A framing is actually essential in Chern-Simons theory
because flux is attached to charge through the constraint
(\ref{eqn:CS-constraint}) or (\ref{eqn:CS-constraint-again}). By putting
the flux and charge at opposite edges of the ribbon, which is a
short-distance regularization of the theory, we can associate a well-defined
phase to a particle trajectory. Otherwise, we wouldn't know how
many times the charge went around the flux.

Any transformation acting on a single particle can only result in
a phase; the corresponding phase is called the twist parameter
${\Theta_a}$. Often, one writes ${\Theta_a}\equiv e^{2\pi i h_a}$,
where $h_a$ is called the {\it spin} of the
particle.\footnote{If $a$ is its own anti-particle,
so that two $a$s can fuse to ${\bf 1}$, then
${R^{aa}_1} = \pm {\Theta_a^*}$, where the minus sign is acquired
for some particle types $a$ which are not quite their own antiparticles
but only up to some transformation which squares to $-1$. This
is analogous to the fact that the fundamental
representation of SU(2) is not real but is pseudoreal. Consequently,
a spin-$1/2$ particle $\psi_\mu$ and antiparticle $\psi^{\mu\dagger}$
can form a singlet, $\psi^{\mu\dagger}\psi_\mu$, but
two spin-$1/2$ particles can as well, ${\psi_\mu}{\psi_\nu}
i({\sigma_y})^{\mu\nu}$,
where $\sigma_y$ is the antisymmetric Pauli matrix.
When some quantities are computed, an extra factor of ${({i\sigma_y})^2}=-1$
results. This $\pm$ sign is called the Froebenius-Schur indicator.
(See, for instance, \onlinecite{Bantay97}.)}
(One must, however, be careful to
distinguish this from the actual spin of the particle, which determines
its transformation properties under the three-dimensional
rotation group and must be half-integral.) However, $h_a$ is
well-defined even if the system is not rotationally-invariant,
so it is usually called the {\it topological spin} of the
particle. For Abelian anyons, it is just the statistics parameter,
$\theta=2\pi i h_a$.

The ground state properties on arbitrary surfaces,
including the $n$-punctured sphere and the torus,
can be built up from more primitive vector spaces in the following way.
An arbitrary closed surface can
be divided into a collection of $3$-punctured spheres which
are glued together at their boundaries. This is called
a `pants decomposition' because of the topological equivalence
of a $3$-punctured sphere to a pair of pants. Therefore, the
$3$-punctured sphere plays a fundamental role in the description
of a topological phase. Its Hilbert space is denoted by $V_{ab}^c$,
if $a$, $b$, and $c$ are the particle types at the three punctures.
If the $a$ and $b$ punctures are fused, a two-punctured sphere will
result. From the above analysis, it has a one-dimensional vector space
if both punctures have topological charge $c$ and a zero-dimensional
vector space otherwise. The dimension of the Hilbert space
of the $3$-punctured sphere is given by
the fusion multiplicity $N_{ab}^c=\text{dim}\left(V_{ab}^c\right)$
which appears in the fusion rule,
${\phi_a}\times{\phi_b} = {\sum_c}{N_{ab}^c}{\phi_c}$.
The Hilbert space on a surface obtained by gluing
together $3$-punctured spheres is
obtained by tensoring together the $V$'s and summing over the
particle types at the punctures where gluing occurs. For instance, the Hilbert space on
the $4$-punctured sphere is given by the direct sum
$V^e_{abd}={\oplus_c} V_{ab}^c V_{cd}^e$;
the Hilbert space on the torus is $V_{T^2} = \oplus_{a} V_{1a}^a V_{a1}^a$.
(If one of the particle types is the vacuum, then the corresponding puncture
can simply be removed; the $3$-punctured
sphere is then actually only $2$-punctured. Gluing two of them together
end to end gives a torus. This is one way of seeing that the
degeneracy on the torus is the number of particle types.)

The Hilbert space of the $n$-punctured sphere with topological
charge $a$ at each puncture can be constructed by sewing together a
chain of $(n-2)$ $3$-punctured spheres. The resulting Hilbert space
is: $V^1_{a\ldots a}={\oplus_{b_i}} V_{aa}^{b_1}
V_{a{b_1}}^{b_2}\,\ldots \,V_{ab_{N-3}}^{a}$.
A simple graphical notation for a set of basis states of
this Hilbert space is given by a {\it fusion chain}
(similar to the fusion tree discussed in appendix A):
\begin{center} 
\begin{picture}(250,30) 
\put(10,3){$a$} 
\put(30,20){$a$} 
\put(60,20){$a$} 
\put(90,20){$a$} 
\put(120,20){$a$} 
\put(40,3){$b_1$}
\put(70,3){$b_2$}
\put(100,3){$b_3$}
\put(10,0){\line(1,0){125}} 
\put(30,0){\line(0,1){15}} 
\put(60,0){\line(0,1){15}} 
\put(90,0){\line(0,1){15}} 
\put(120,0){\line(0,1){15}} 
\put(125,3){$b_4$} 
\put(145,3){$\dots$} 
\put(165,3){$b_{n-4}$}
\put(195,3){$b_{n-3}$} 
\put(165,0){\line(1,0){65}} 
\put(190,0){\line(0,1){15}} 
\put(220,0){\line(0,1){15}} 
\put(178,20){$a$} 
\put(212,20){$a$} 
\put(227,3){$a$} 
\end{picture} 
\end{center} 
The first two $a$s on the far left fuse to $b_1$. The next $a$ fuses with $b_1$
to give $b_2$. The next $a$ fuses with $b_2$ to give $b_3$, and so on.
The different basis vectors in this Hilbert space correspond to
the different possible allowed $b_i$s.
The dimension of this Hilbert space is
$N_{aa}^{b_1} N_{a{b_1}}^{b_2}\,\ldots\,N_{ab_{N-3}}^{a}
=\left({N_a}\right)_{a}^{b_1}
\left({N_a}\right)_{{b_1}}^{b_2}\,\ldots\,
\left({N_a}\right)_{b_{N-3}}^{a}$. On the right-hand-side of this
equation, we have suggested that the fusion multiplicity $N_{ab}^c$
can be viewed as a matrix $\left({N_a}\right)_{b}^{c}$ associated
with quasiparticle species $a$. Let us denote the largest eigenvalue
of the matrix ${N_a}$ by $d_a$. Then the Hilbert space of $M$
quasiparticles of type $a$ has dimension $\sim d_a^{M-2}$ for large
$M$. For this reason, $d_a$ is called the {\it quantum dimension} of
an $a$ quasiparticle. It is the asymptotic degeneracy per particle
of a collection of $a$ quasiparticle. For Abelian particles,
${d_a}=1$ since the multi-particle Hilbert space is one-dimensional
(for fixed particle positions). Non-Abelian particles have
${d_a}>1$. Note that $d_a$ is not, in general, an integer, which is
symptomatic of the non-locality of the Hilbert space: it is {\it
not} the tensor product of $d_a$-dimensional Hilbert spaces
associated locally with each particle.

This non-locality is responsible for the stability of this
degenerate ground state Hilbert space. Not only the Yang-Mills term,
but all possible gauge-invariant terms which we can add to the
action (\ref{eqn:non-Abel-C-S-action}) are irrelevant. This means
that adding such a term to the action might split the $\sim
d_a^{M-2}$-dimensional space of degenerate states in a finite-size
system, but the splitting must vanish as the system size and the
particle separations go to infinity. In fact, we can make an even
stronger statement than that. All ground state matrix elements of
gauge-invariant local operators such as the field strength squared,
$F_{\mu\nu}^{\underline a} F^{\mu\nu{\underline a}}$, vanish
identically because of the Chern-Simons constraint. Therefore, the
degeneracy is not lifted at all in perturbation theory. It can only
be lifted by non-perturbative effects (e.g. instantons/quantum
tunneling), which could cause a splitting $\sim e^{-gL}$ where $g$
is inversely proportional to the coefficient of the Yang-Mills term.
Therefore, the multi-quasiparticle states are degenerate to within
exponential accuracy. At finite-temperatures, one must also consider
transitions to excited states, but the contributions of these will
be $\sim e^{-\Delta/T}$.   Furthermore if we were to add a time
dependent (source) term to the action, these properties will remain
preserved so long as the frequency of this term remains small
compared with the gap.

Aside from the $n$-punctured spheres, the torus is the
most important manifold for considering topological phases.
Although not directly relevant to experiments, the torus is
very important for numerical simulations since periodic boundary
conditions are often the simplest choice. As noted above, the ground state
degeneracy on the torus is equal to the number of quasiparticle
species. Suppose one can numerically solve a Hamiltonian on the
torus. If it has a gap between its ground state(s) and the lowest energy excited
states, then its ground state degeneracy is an
important topological property of the state -- namely the number of
of quasiparticle species.
A simple physical understanding of this degeneracy can be obtained
in the following way. Suppose that we have a system of
electrons in a topological phase. If we consider the system on
the torus, then the electrons must have periodic boundary conditions
around either generator of the torus (i.e. around either handle),
but the quasiparticles need not.
In the Abelian $\nu=1/m$ fractional quantum Hall state,
for instance, it is possible for a quasiparticle to pick up a phase
$e^{2\pi in/m}$ in going around the meridian of the torus,
where $n$ can take any of the values $n=0,1,\ldots,m-1$;
electrons would still have periodic boundary
conditions since they are made up of $m$ quasiparticles.
Indeed, all $m$ of these possibilities occur, so the ground state
is $m$-fold degenerate.

Let us make this a little more precise. We introduce
operators $T_1$ and $T_2$ which create a
quasiparticle-quasihole pair, take the quasiparticle around the
meridian or longitude, respectively, of the torus and annihilate them again.
Then $T_1$ and $T_2$ must satisfy:
\begin{equation}
\label{eqn:T-ops}
T_2^{-1} T_1^{-1} T^{}_{2}T^{}_{1}= e^{2\pi i/m}
\end{equation}
because $T_1^{-1}T^{}_{1} $ amounts to a contractible quasiparticle-quasihole
loop, as does $T_2^{-1}T^{}_{2} $; by alternating these
processes, we cause these loops to be linked.
The quasiparticle trajectories in spacetime (which can be visualized
as a thickened torus) are equivalent
to a simple link between two circles (the Hopf link):
the first quasiparticle-quasihole pair is pulled apart along
the meridian ($T^{}_{1}$); but before they can be brought back together
($T_1^{-1}$), the second pair is pulled apart along the longitude ($T^{}_{2}$).
After the first pair is brought back together and annihilated ($T_1^{-1}$),
the second one is, too ($T_2^{-1}$).
In other words, the phase on the right-hand-side of Eq.~\ref{eqn:T-ops} is simply the
phase obtained when one quasiparticle winds around
another. This algebra can be represented on a vector space
of minimum dimension $m$. Let us call the states in
this vector space $|n\rangle$, $n=0,1,\ldots,m-1$.
Then
\begin{eqnarray}
{T_1}|n\rangle &=& e^{2\pi in/m} |n\rangle\cr {T_2}|n\rangle &=&
|(n+1)\text{ mod }m\rangle
\end{eqnarray}
These $m$ states correspond to $n=0,1,\ldots,m-1$ quanta of flux
threaded through the torus. If we were to cut along a meridian and
open the torus into an annulus, then these states would have flux
$n$ threaded through the hole in the annulus and charge $n/m$ at the
inner boundary of the annulus (and a compensating charge at the
outer boundary). We can instead switch to a basis in which ${T_2}$
is diagonal by a discrete Fourier transform. If we write $|{\tilde
n}\rangle = \frac{1}{\sqrt{m}} \sum_{n=0}^{m-1} e^{2\pi in{\tilde
n}/m} |n\rangle$, then $|{\tilde n}\rangle$ is an eigenstgate of
$T_2$ with eigenvalue $e^{2\pi i{\tilde n}/m}$. In this basis, $T_1$
is an off-diagonal operator which changes the boundary conditions of
quasiparticles around the longitude of the torus. In non-Abelian
states, a more complicated version of the same thing occurs, as we
discuss for the case of Ising anyons at the end of section
\ref{sec:pwave}. The different boundary conditions around the
meridian correspond to the different possible quasiparticle types
which could thread the torus (or, equivalently, could be present at
the inner boundary of the annulus if the torus were cut open along a
meridian). One can switch to a basis in which the boundary
conditions around the longitude are fixed. The desired basis change
is analogous to the discrete Fourier transform given above and is
given by the `$S$-matrix' or `modular $S$-matrix' of the theory.
Switching the longitude and meridian is one of the generators of the
mapping class group of the torus; the $S$-matrix expresses how it
acts on the ground state Hilbert space. The elements of the
$S$-matrix are closely related to quasiparticle braiding. By
following a similar construction to the one with $T_1$, $T_2$ above,
one can see that $S_{ab}$ is equal to the amplitude for creating
$a\bar{a}$ and $b\bar{b}$ pairs, braiding $a$ and $b$, and
annihilating again in pairs. This is why, in an Abelian state, the
elements of the $S$-matrix are all phases (up to an overall
normalization which ensures unitarity), e.g. $S_{nn'} =
\frac{1}{\sqrt{m}}e^{2\pi i n n'/m}$ in the example above. In a
non-Abelian state, the different entries in the matrix can have
different magnitudes, so the basis change is a little more
complicated than a Fourier transform. Entries can even vanish in the
non-Abelian case since, after $a$ and $b$ have been braided, $a$ and
$\bar{a}$ may no longer fuse to ${\bf 1}$.

In the case of Ising anyons on the torus (SU(2)${}_2$), there are
three ground states. One basis is $|{{\bf 1}_m}\rangle$,
$|{\sigma_m}\rangle$, $|{\psi_m}\rangle$, corresponding to the
different allowed topological charges which would be measured at the
inner boundary of the resulting annulus if the torus were cut open
along its meridian. An equally good basis is given by eigenstates of
topological charge around the longitude: $|{{\bf 1}_l}\rangle$,
$|{\sigma_l}\rangle$, $|{\psi_l}\rangle$. As we will see in at the
end of the next section, the basis change between them is given by
\begin{eqnarray}
\label{eqn:Ising-S-matrix}
S =
\left(
 \begin{array}{ccc}
  \frac{1}{2} &  \frac{1}{\sqrt{2}}  &  \frac{1}{2} \\
  \frac{1}{\sqrt{2}} & 0 & - \frac{1}{\sqrt{2}} \\
  \frac{1}{2} &  -\frac{1}{\sqrt{2}}  &  \frac{1}{2}  \\
 \end{array}
 \right)
 \end{eqnarray}
The $S$-matrix not only contains information about braiding,
but also about fusion, according to
Verlinde's formula \cite{Verlinde88} (for a proof, see \onlinecite{Moore88,Moore89}):
\begin{equation}
N_{ab}^c = {\sum_x} \frac{S_{ax}S_{bx}S_{\bar{c}x}}{S_{{\bf 1}x}}
\end{equation}
Consequently, the quantum dimension of a particle of species $a$ is:
\begin{equation}
{d_a} = \frac{S_{{\bf 1}a}}{S_{{\bf 11}}}
\end{equation}

The mathematical structure encapsulating these braiding
and fusion rules is a {\it modular tensor category}
\cite{Walker91,Turaev94,Kassel95,Bakalov01,Kitaev06a}. A category
is composed of objects and morphisms, which are maps
between the objects which preserve their defining structure.
The idea is that one can learn more about the objects by
understanding the morphisms between them.
In our case, the objects are particles with labels (which
specify their species) as well as fixed configurations
of several particles.
The morphisms are particle trajectories, which map a set
of labeled partices at some initial time to a set of
labeled particles at some final time.
A {\it tensor category} has a tensor product structure for multiplying
objects; here, this is simply the fact that one can take two
well-separated (and historically well-separated) collections
of particles and consider their union to be a new `tensor-product'
collection. Since we consider particles in two dimensions,
the trajectories are essentially the elements of the braid group,
but they include the additional possibility of twisting. (Allowing twists
in the strands of a braid yields a {\it braided ribbon category}.)
We will further allow the trajectories to include the fusion
of two particles (so that we now have a {\it fusion category}).
Morphisms can, therefore, be defined by specifying $\Theta_a$,
$V_{ab}^c$, $R$, and $F$.

Why is it necessary to invoke category theory simply to specify the
topological properties of non-Abelian anyons? Could the braid group
not be the highest level of abstraction that we need? The answer is
that for a fixed number of particles $n$, the braid group ${\cal
B}_n$ completely specifies their topological properties (perhaps
with the addition of twists $\Theta_a$ to account for the finite
size of the particles). However, we need representations of ${\cal
B}_n$ for all values of $n$ which are compatible with each other and
with fusion (of which pair creation and annihilation is simply the
special case of fusion to the vacuum). So we really need a more
complex -- and much more tightly constrained -- structure. This is
provided by the concept of a modular tensor category. The $F$ and
$R$ matrices play particularly important roles. The $F$ matrix can
essentially be viewed an associativity relation for fusion: we could
first fuse $i$ with $j$, and then fuse the result with $k$; or we
could fuse $i$ with the result of fusing $j$ with $k$. The
consistency of this property leads to a constraint on the
$F$-matrices called the pentagon equation. (An explicit example of
the pentagon equation is worked out in Section \ref{sec:fibonacci}.)
Consistency between $F$ and $R$ leads to a constraint called the
hexagon equation. Modularity is the condition that the $S$-matrix be
invertible. These self-consistency conditions are sufficiently
strong that a solution to them completely defines a topological
phase.\footnote{Modulo details regarding the central charge $c$ at
the edge. $e^{2\pi i c/8}$ can be obtained from the topological
spins, but not $c$ itself.}

An equivalent alternative to studying punctured surfaces is to add
non-dynamical charges which are coupled to the Chern-Simons gauge
field. Then the right-hand-side of the constraint
(\ref{eqn:CS-constraint-again}) is modified and a non-trivial gauge
field configuration is again obtained which is essentially
equivalent to that obtained around a puncture. In the following
subsections, we will discuss the Hilbert spaces of SU(2)$_k$
Chern-Simons theory, either on the $n$-punctured sphere or in the
presence of non-dynamical sources. These discussions will enable us
to compute the braiding and fusion matrices. The non-trivial
quasiparticle of SU(2)$_1$ is actually Abelian so we do not discuss
this `trivial' case. The next case, SU(2)$_2$, is non-Abelian and
may be relevant to the $\nu=5/2$ fractional quantum Hall state. It
can be understood in several different equivalent ways, which
express its underlying free Majorana fermion structure. Quantum
computation with Majorana fermions is described in Section
\ref{sec:5/2-qubits}. In the next section, we explain this structure
from the perspective of a superconductor with $p+ip$ pairing
symmetry. Although this description is very elegant, it cannot be
generalized to higher $k$. Therefore, in the two sections after
that, we describe two different approaches to solving SU(2)$_k$
Chern-Simons theory for general $k$. We recapitulate the case of
SU(2)$_2$ in these other languages and also describe the case of
SU(2)$_3$. The latter has quasiparticles in its spectrum which are
Fibonacci anyons, a particularly beautiful non-Abelian anyonic
structure which allows for universal topological quantum
computation. It may also underlie the observed $\nu=12/5$ fractonal
quantum Hall state. More details of the Fibonacci theory are given
in Sections \ref{sec:fibonacci}.

\subsection{Superconductors with $p+ip$ pairing symmetry}
\label{sec:pwave}

In this section, we will discuss the topological properties of a
superconductor with $p+ip$ pairing symmetry following the method
introduced by Read and Green \cite{Read00}. This is the most
elementary way in which a non-Abelian topological state can emerge
as the ground state of a many-body system. This non-Abelian
topological state has several possible realizations in various two
dimensional systems: $p+ip$ superconductors, such as Sr$_2$RuO$_4$
(although the non-Abelian quasiparticles are half-quantum vortices
in this case \cite{DasSarma06a}); $p+ip$ superfluids of cold atoms
in optical traps \cite{Gurarie05,Tewari07a}, and the A-phase
(especially the $A_1$ phase\cite{Leggett75,Volovik03}) of $^3$He
films; and the Moore-Read Pfaffian quantum Hall state
\cite{Moore91}. The last of these is a quantum Hall incarnation of this
state: electrons at filling fraction $\nu=1/2$ are equivalent to
fermions in zero field interacting with an Abelian Chern-Simons
gauge field. When the fermions pair and condense in a $p+ip$
superconducting state, the Pfaffian quantum Hall state forms
\cite{Greiter92}. Such a state can occur at
$\frac{5}{2}=2+\frac{1}{2}$ if the lowest Landau level (of both
spins) is filled and inert, and the first excited Landau level is
half-filled.

Ordinarily, one makes a distinction between the fermionic
quasiparticles (or Bogoliubov-De Gennes quasiparticles) of a
superconductor and vortices in a superconductor. This is because,
in terms of electron variables, the former are relatively simple while
the latter are rather complicated. Furthermore, the energy and length
scales associated with the two are very different in the weak-coupling
limit. However, fermionic quasiparticles and vortices are really just different
types of quasiparticle excitations in a superconductor -- i.e. different types of
localized disturbances above the ground state.
Therefore, we will often refer to them both
as simply quasiparticles and use the terms Bogoliubov-de Gennes
or fermionic when referring to the former. In a $p+ip$ superconductor,
the quasiparticles which exhibit non-Abelian statistics are
flux $hc/2e$ vortices.

\subsubsection{Vortices and Fermion Zero Modes}

Let us suppose that we have a system of fully spin-polarized
electrons in a superconducting state of $p_x+ip_y$ pairing symmetry.
The mean field Hamiltonian for such a superconductor is,
\begin{eqnarray}
&H&=\int \!\! \mathrm{d}{\bf r}\,\psi ^{\dagger }({\bf r})h_{0}\psi
({\bf r}) \\ &+& \!\! {\frac{1}{2}} \int \!\! \mathrm{d}{\bf
r}\,\mathrm{d}{\bf r^{\prime }}\left\{ D^{\ast }({\bf r},{\bf r^
{\prime }})\psi ({\bf r^{\prime }})\psi ({\bf r})+D({\bf r},{\bf
r^{\prime }})\psi ^ {\dagger }({\bf r})\psi ^{\dagger }({\bf
r^{\prime }} )\right\} \nonumber \label{bcs-hamiltonian}
\end{eqnarray}
with single-particle term ${h_0} = -\frac{1}{2m}{\nabla^2}-\mu$
and complex $p$-wave pairing function
\begin{equation}
D({\bf r},{\bf r}^{\prime })=\Delta\!\left( {\frac{{\bf r}+{\bf r^
{\prime
}}}{2}}\right) (i\partial _{x^{\prime }}-\partial _{y^{\prime }})\delta
({\bf r}-{\bf
r^{\prime }}).
\end{equation}

The dynamics of $\Delta$ is governed by a Landau-Ginzburg-type
Hamiltonian and will be briefly discussed later. The quadratic
Hamiltonian (\ref{bcs-hamiltonian}) may be diagonalized by solving
the corresponding Bogoliubov-de Gennes equations (BdG) equations,
\begin{eqnarray}
& & E\left(
\begin{array}{c}
u({\bf r}) \\
v({\bf r})
\end{array}
\right) =  \\ & &  \left(
\begin{array}{cc}
-\mu ({\bf r}) & \frac{i}{2}\left\{ \Delta ({\bf r}),\partial _{x}
+i\partial
_{y}\right\} \\
\frac{i}{2}\left\{ \Delta ^{\ast }({\bf r}),\partial _{x}-i\partial
_{y}\right\} & \mu ({\bf r})
\end{array}
\right) \left(
\begin{array}{c}
u({\bf r}) \\
v({\bf r})
\end{array}
\right),\label{BdG} \nonumber
\end{eqnarray}
The Hamiltonian then takes the form:
\begin{equation}
H=E_0+\sum_E E\,\Gamma_E^\dagger\Gamma_E
\label{GammaH}
\end{equation}
where $\Gamma_E^\dagger\equiv\int dr \left [ u_E({\bf r})\psi({\bf
r}) +v_E({\bf r})\psi^\dagger({\bf r})\right ]$ is the creation
operator formed by the positive energy solutions of the
Bogoliubov-de Gennes equations and $E_0$ is
the ground state energy. For the ground state of the
Hamiltonian (\ref{bcs-hamiltonian}) to
be degenerate in the presence of several vortices (which are the most interesting
quasiparticles in this theory) it is essential that the BdG equations have solutions
with eigenvalue zero in this situation.

Before searching for zero eigenvalues of (\ref{BdG}) in the presence
of vortices, however, we focus on a uniform superconductor, where
$\Delta$ is a constant.
Read and Green \cite{Read00} retain only the potential part of
$h_{0} $, which for a uniform superconductor is a constant $-\mu$.
With this simplification, a BdG
eigenstate with momentum $k$ has energy
\begin{equation}
\label{eq:BCS-spectrum}
E_k=\sqrt{\mu^2+ {\Delta^2} |k|^2}
\end{equation}
The ground state of
(\ref{bcs-hamiltonian}) is the celebrated BCS wave function, written
here in an un-normalized form,
\begin{equation}
|{\rm g.s.}\rangle=\prod_{\bf k}\left (1+\frac{v_{\bf k}}{u_{\bf k}}c_{\bf k}^
\dagger c_{\bf -k}^\dagger\right)|{\rm vac}\rangle
=e^{\sum_{\bf k}\frac{v_{\bf k}}{u_{\bf k}}c_{\bf k}^\dagger c_{\bf -k}^\dagger}|
{\rm vac}\rangle
\label{bcswf}
\end{equation}
where
\begin{equation}
\left(
\begin{array}{c}
|u_{\bf k}|^2 \\
|v_{\bf k}|^2
\end{array}
\right) =\frac{1}{2}\left (1\mp\frac{\mu}{\sqrt{\mu^2+|\Delta k|
^2}}\right )
\label{uvsquared}
\end{equation}
are the BCS coherence factors.
%, and ${v_k\over u_k}  = \frac{\sqrt{\mu^2+(\Delta k)^2}+\mu}{\Delta(k_x+ik_y)}$.
The wave function (\ref{bcswf}) describes a coherent state of an
undetermined number of Cooper pairs, each in an internal state of
angular momentum $\ell = -1$. Its projection onto a fixed even
number of particles $N$ is carried out by expanding the exponent in
(\ref{bcswf}) to the $(N/2)^\text{th}$ order. When written in first
quantized language, this wave function describes a properly
anti-symmetrized wave function of $N/2$ Cooper-pairs, each in an
internal state
\begin{equation}
g({\bf r})=\sum_{\bf
k}\frac{v_{\bf k}} {u_{\bf k}}e^{i{\bf k}{\bf r}}
\end{equation}
In first quantized form
the multiparticle BCS wavefunction is then of the form of the Pfaffian
of an antisymmetric matrix whose $i-j$ element is $g({{\bf r}_i} - {{\bf r}_j})$,
an antisymmetrized product of pair wavefunctions
\begin{eqnarray}
\label{eq:PfaffianBCS}
  \Psi_{BCS} &=& {\rm{Pf}}\left[g({\bf r}_i - {\bf r}_j)\right] \\
    &=& {\cal A} \left[ g({\bf r}_1 - {\bf r}_2) g({\bf r}_3 - {\bf r}_4) \ldots g({\bf r}_{N-1} -
  {\bf  r}_N) \right] \nonumber
\end{eqnarray}
with ${\cal A}$ being an antisymmetrization operator.

The function $g({\bf r})$ depends crucially on the sign of $\mu$,
since the small $k$ behavior of $v_{\bf k}/u_{\bf k}$ depends on that sign. When
$\mu  > 0$, we have $g(r)=1/(x+iy) $ in the long distance limit \cite{Read00}.  If we
assume this form holds for all distances, the Pfaffian wave function
obtained is identical to the Moore-Read form discussed below in
connection with the Ising model and the $\nu=5/2$ quantum Hall state
in section \ref{sec:FQHE} (see Eqs. \ref{eqn:MR-Pfaffian1,eq:pfaf2}).
The slow decay of $g(r)$
implies a weak Cooper pairing. (But it does not imply that the state
is gapless. One can verify that electron Green functions
all decay exponentially for any non-zero $\mu$.) When $\mu < 0$ the function $g(r)$
decays much more rapidly with $r$, generically in an exponential
way, such that the Cooper pairs are strongly bound. Furthermore,
there is a topological distinction between the $\mu>0$ and $\mu<0$
phases.
%This distinction is revealed when $(u_k,v_k)$ is viewed as a
%spinor,  from the Pfaffian state by the dependence of the spinor $
%(v_k,u_k)$ on ${\bf k}$.
The distinction, which is discussed in detail in \cite{Read00},
implies that, despite the fact that both states are superconducting,
the $\mu>0$ and $\mu<0$ states must be separated by a phase
transition. (In the analogous quantum Hall state, both states are
characterized by the same Hall conductivity but are separated by a
phase transition, and are distinguished by their thermal Hall
conductivities\cite{Read00}) Indeed, from (\ref{eq:BCS-spectrum}) we
see that the gap vanishes for a uniform $p+ip$ superconductor with
$\mu=0$. The low-energy BdG eigenstates at this second-order phase
transition point form a Dirac cone.

For every solution $(u,v)$ of the BdG equations with
energy $E$, there is a solution $(v^*,u^*)$ of energy $-E$.
A solution with $u=v^*$ therefore has energy zero. We will
soon be considering situations in which there are multiple
zero energy solutions $({u_i},{u_i^*})$, $i=1,2,\ldots$.
If we denote the corresponding operators by $\gamma_i$
(see eq. \ref{zeroenergy} below), then they satisfy:
\begin{equation}
\gamma_i^\dagger=\gamma_i
\label{majorana}
\end{equation}
Eq. (\ref{majorana}) is the definition of a Majorana fermion
operator.

Let us now consider the BdG equations in the presence
of vortices when the bulk of the superconductor is in the
$\mu>0$ phase. As usual, a vortex is
characterized by a point of vanishing $\Delta$, and a
$2\pi$-winding of the phase of $ \Delta$ around that point.
In principle we should, then, solve the BdG equations in
the presence of such a non-uniform $\Delta$. However, we can, instead,
solve them in the presence of a non-uniform $\mu$, which
is much simpler. All that we really need is to make the core
of the superconductor topologically distinct from the bulk -- i.e.
a puncture in the superconductivity.
Making $\mu<0$ in the core is just as good as taking $\Delta$
to zero, as far as topological properties are concerned.
Therefore, we associate the
core of the vortex with a region of $\mu<0$, whereas the bulk is at
$\mu>0$. Thus, there is a $\mu=0$ line encircling the vortex core.
This line is an internal edge of the system.
We will consider the dynamics of edge excitations in more
detail in section \ref{sec:edge}, but here we will be content to
show that a zero energy mode is among them.

The simplest situation to consider is that of azymuthal symmetry,
with the polar coordinates denoted by $r$ and $\theta$. Imagine the
vortex core to be at the origin,
so that $\Delta(r,\theta)=|\Delta(r)|e^{i\theta+i\Omega}$.
Here $\Omega$ is the phase of the
order parameter along the $\theta=0$ line, a phase which will play an
important role later in our discussion. Assume that the $\mu=0$ line
is the circle $r=r_0$, and write
\begin{equation}
\mu(r) = \Delta \,h(r) ,
\end{equation}
with $h(r)$ large and positive for large r,
and $h(r)<0$ for $r<{r_0}$; therefore, the electron
density will vanish for $r\ll{r_0}$.
Such a potential defines an edge at $r={r_0}$.
There are low-energy eigenstates of the BdG Hamiltonian
which are spatially localized near $r=0$ and are exponentially
decaying for $r\rightarrow\infty$:
\begin{equation}
\label{eqn:zero-mode-wvfn}
\phi^{edge}_E(r,\theta) = e^{i\ell\theta} e^{- \int_0^r h(r^\prime) dr^\prime}\,
{e^{-i\theta/2}\choose e^{i\theta/2}} ,
\end{equation}
The spinor on the right-hand-side points in a direction in pseudospin space
which is tangent to the $r={r_0}$ circle at $\theta$.
This wavefunction describes a chiral wave propagating around the edge,
with angular momentum $\ell$ and energy $E=\Delta \ell/{r_0}$.
Since the flux is an odd multiple of $hc/2e$, the Bogoliubov
quasiparticle (\ref{eqn:zero-mode-wvfn}) must be anti-periodic
as it goes around the vortex. However,
the spinor on the right-hand-side of (\ref{eqn:zero-mode-wvfn}) is
also anti-periodic. Therefore, the angular momentum $\ell$ must be
an integer, $\ell\in\mathbb{Z}$. Consequently, a flux $hc/2e$ vortex
has an $\ell=0$ solution, with energy $E=0$.
(Conversely, if the flux through the vortex were an even multiple
of $hc/2e$, $\ell$ would be a half-integer, $\ell\in\mathbb{Z}+\frac{1}{2}$,
and there would be no zero-mode.)
The operator corresponding to this zero mode,
which we will call $\gamma$, can be written in the form:
\begin{equation}
\gamma=\frac{1}{\sqrt{2}}\int dr \, \left [F
({\bf
r})\,e^{-\frac{i}{2}\Omega}\psi({\bf r})+\, F^*
({\bf
r})\,e^{\frac{i}{2}\Omega}\psi^\dagger ({\bf r})\right ]
\label{zeroenergy}
\end{equation}
Here, $F({\bf r})=e^{- \int_0^r h(r^\prime) dr^\prime}e^{-i\theta/2}$.
Since each $\gamma$ is an equal superposition of electron and hole,
it is overall a chargeless, neutral fermion operator

When there are several well separated vortices at positions
${\bf R}_i$, the gap function near the $i^\text{th}$ vortex takes the form
$\Delta({\bf r}) = |\Delta({\bf r})| \exp{(i\theta_i+i\Omega_i)}$,
with $\theta_i=\arg{({\bf r}-{\bf R_i})}$ and $\Omega_i=\sum_{j\ne
i}\arg(({\bf R_j}-{\bf R}_i))$. There is then one zero energy
solution per vortex. Each zero energy solution $\gamma_i$ is localized near
the core of its vortex at ${\bf R_i}$, but the  phase $\Omega_i$
that replaces $\Omega$ in (\ref{zeroenergy}) depends on the position
of all vortices. Moreover, the dependence of the Majorana operators
$\gamma_i$ on the positions ${\bf R_i}$ is not single valued.

While for any $E\ne 0$ the operators $\Gamma_E^\dagger,\Gamma_E$ are
conventional fermionic creation and annihilation operators, the $
\gamma_i$'s are not. In particular, for $E\ne 0$ we have $(\Gamma_E^
\dagger)^2=\Gamma_E^2=0$, but the zero energy operators follow (with
a convenient choice of normalization) $\gamma_i^2=1$. The two types
of fermion operator share the property of mutual anti-commutation,
i.e., the $\gamma$'s satisfy $\{\gamma_i,\gamma_j\}=2\delta_{ij}$.

\subsubsection{Topological Properties of $p+ip$ Superconductors}

The existence of the $\gamma_i$'s implies a degeneracy of the ground
state. The counting of the number of degenerate ground states should
be done with care. A pair of conventional fermionic creation and
annihilation operators span a two dimensional Hilbert space, since
their square vanishes. This is not true for a Majorana operator.
Thus, to count the degeneracy of the ground state when $2N_0$
vortices are present, we construct ``conventional'' complex (Dirac) fermionic
creation and annihilation operators,
\begin{eqnarray}
\psi_i= (\gamma_i +i\gamma_{N_0+i})/2\\
\psi_i^\dagger=(\gamma_i -i\gamma_{N_0+i})/2
\label{majoranatocomplex}
\end{eqnarray}
These operators satisfy $\psi_i^2=\left(\psi_i^\dagger
\right)^2=0$ and thus span
a two-dimensional subspace of degenerate ground states
associated with these operators. Over all, then, the system has $2^{N_0}$
degenerate ground states. If the fermion number is fixed to be even
or odd, then the degeneracy is $2^{{N_0}-1}$. Therefore, the quantum
dimension of a vortex is $d_{\rm vort}=\sqrt{2}$ or, in the
notation introduced in Sec. \ref{sec:Non-Abelian_quantum_statistics}
for Ising anyons, ${d_\sigma}=\sqrt{2}$.

For any two vortices $i$ and $j$, we can associate a two state system.
If we work in the basis of $i{\gamma_i}{\gamma_j}$ eigenstates,
then $i{\gamma_i}{\gamma_j}$ acts as $\sigma_z$
with eigenvalues $\pm 1$, while $\gamma_i$ and $\gamma_j$
act as $\sigma_x$ and $\sigma_y$. (However, it is important
to keep in mind that Majorana fermions $\gamma_k$, $\gamma_l$
anti-commute with $\gamma_i$, $\gamma_j$, {\it unlike} operators
associated with different spins, which commute.)
The two eigenvalues $i{\gamma_i}{\gamma_j}=\mp 1$ are the
two fusion channels of two fermions.
If we form the Dirac fermion $\psi=({\gamma_i}+i{\gamma_j})/2$,
then the two $i{\gamma_i}{\gamma_j}$ eigenstates
have ${\psi^\dagger}\psi=0,1$. Therefore, we will call these
fusion channels ${\bf 1}$ and $\psi$. (One is then tempted to
refer to the state for which $\psi^\dagger \psi=1$ as a ``filled
fermion", and to the $\psi^\dagger \psi=0$
state as an empty fermion. Note however that the eigenvalue of
$\psi^\dagger \psi$ has no bearing on the
occupation of single-particle states.)

Of course, the pairing of vortices to form Dirac fermions
is arbitrary. A given pairing defines a basis, but one
can transform to a basis associated with another pairing.
Consider four vortices with corresponding
zero modes $\gamma_1$, $\gamma_2$, $\gamma_3$, $\gamma_4$.
The $F$-matrix transforms states from the basis in which
$i{\gamma_1}{\gamma_2}$  and $i{\gamma_3}{\gamma_4}$
are diagonal to the basis in which $i{\gamma_1}{\gamma_4}$
and $i{\gamma_2}{\gamma_3}$ are diagonal. Since $i{\gamma_1}{\gamma_4}$
acts as $\sigma_x$ on an $i{\gamma_1}{\gamma_2}$ eigenstate,
the $F$-matrix is just the basis change from
the $\sigma_z$ basis to the $\sigma_x$ basis:
\begin{eqnarray}
\label{eqn:Ising-F-matrix}
\left[ F^{\sigma\sigma\sigma}_{\sigma} \right]= \frac{1}{\sqrt{2}}
\left(
 \begin{array}{cc}
  1 &  1 \\
  1 & -1 \\
 \end{array}
 \right)
 \end{eqnarray}
We will refer to this type of non-Abelian anyons by the name
`Ising anyons'; they are the model introduced in Section
\ref{sec:Non-Abelian_quantum_statistics}. The reason for the
name will be explained in Section \ref{sec:edge}.

In a compact geometry, there must be an even number of vortices
(since a vortex carries half a flux quantum,
and the number of flux quanta penetrating a compact surface
must be integer). In a non-compact geometry, if the number of
vortices is odd, the edge has a zero energy state of its own,
as we show in Section \ref{sec:edge}.

Now, let us examine what happens to the Majorana operators and to
the ground states as vortices move. The positions of the vortices
are parameters in the Hamiltonian (\ref{bcs-hamiltonian}). When
they vary adiabatically in time, the operators $ \gamma_i$ vary
adiabatically in time. In principle, there are two sources for this
variation - the explicit dependence of $\gamma_i$ on the positions
and the Berry phase associated with the motion. The choice of phases
taken at (\ref{zeroenergy}) is such that the Berry phase vanishes,
and the entire time dependence is explicit. The non-single-valuedness
of the phases in (\ref{zeroenergy}) implies then
that a change of $2\pi$ in $\Omega$, which takes place when one
vortex encircles another, does not leave the state unchanged.

As vortices adiabatically traverse trajectories that start
and end in the same set of positions \cite{Ivanov01,
Stern04}, there is a unitary transformation $U$
within the subspace of ground states that takes the initial state
$|\psi(t=0) \rangle$ to the final one $|\psi(t=T)\rangle$,
\begin{equation}
    |\psi(t=T)\rangle = U |\psi(t=0)\rangle.
\end{equation}
Correspondingly, the time evolution of the operators $\gamma_i$
is
\begin{equation}
    \gamma_i(t=T)= U \gamma_i(t=0) U^\dagger.
\end{equation}
By reading the time evolution of $\gamma_i$ from their explicit form
(\ref{zeroenergy}) we can determine $U$ up to a phase. Indeed, one
expects this Abelian phase to depend not only on the topology of the
trajectory but also on its geometry, especially in the analogous
quantum Hall case, where there is an Aharonov-Bohm
phase accumulated as a result of the charge carried
by the quasiparticle.

When vortex $i$ encircles vortex $i+1$, the unitary transformation
is simple: both $\gamma_i$ and $\gamma_{i+1}$
are multiplied by $-1$,
with all other operators unchanged. This is a
consequence of the fact that when the order parameter changes
by a phase factor $2\pi$, fermionic operators change by a phase $\pi$.
Exchange trajectories, in which some of the vortices trade places,
are more complicated, since the phase changes of
$\Omega_{k}$ associated with a particular trajectory do not
only depend on the winding numbers, but also on the details of the
trajectory and on the precise definition of the cut of the function
$\arg ({\bf r})$ where its value jumps by $2\pi$.

The simplest example is the interchange of two vortices. Inevitably,
one of the vortices crosses the branch cut line of the other vortex.
We can place the branch cuts so that a counterclockwise exchange of vortices $1$ and $2$ transforms $c_{1}\rightarrow c_{2}$ and
$c_{2}\rightarrow -c_{1}$ while a clockwise exchange
transforms $c_{1}\rightarrow -c_{2}$ and
$c_{2}\rightarrow c_{1}$ \cite{Ivanov01}.

This may be summarized by writing the representation
matrices for the braid group generators \cite{Nayak96c,Ivanov01}:
\begin{equation}
\rho({\sigma_i}) = e^{i\theta}\,e^{-\frac{\pi}{4}{\gamma_i}\gamma_{i+1}}
\label{eqn:Ising-B_n-rep}
\end{equation}
where $\theta$ is the Abelian part of the transformation.
The two eigenvalues $i{\gamma_i}\gamma_{i+1}=\mp 1$
are the two fusion channels ${\bf 1}$ and $\psi$ of a pair of vortices.
From (\ref{eqn:Ising-B_n-rep}), we see that the $R$-matrices
satisfy $R^{\sigma\sigma}_{\psi} = i\,R^{\sigma\sigma}_{\bf 1}$
(i.e., the phase of taking two $\sigma$ particles around each other
differ by $i$ depending on whether they fuse to $\psi$ or ${\bf 1}$).
It is difficult to obtain the Abelian part
of the phase using the methods of this section, but we will
derive it by other methods in Sections \ref{sec:Jones}
and \ref{sec:FQHE}. The non-Abelian part of (\ref{eqn:Ising-B_n-rep}),
i.e. the second factor on the right-hand-side, is the
same as a $\pi/2$ rotation in the spinor representation
of SO(2n) (see \onlinecite{Nayak96c} for details). The fact that
braiding only enacts $\pi/2$ rotations is the reason why
this type of non-Abelian anyon does not enable universal
topological quantum computation, as we discuss further
in section \ref{part3}.

According to (\ref{eqn:Ising-B_n-rep}), if a system starts in a ground state
$\left| {\rm gs}_{\alpha }\right\rangle$ and vortex $j$ winds around vortex $j+1$, the system's final state is $\gamma_{j}\gamma_{j+1}\left| {\rm gs}_{\alpha }\right\rangle$.
Writing this out in terms of the original electron operators, we have
\begin{equation}
\left( c_{j}e^{\frac{i}{2}\Omega
_{j}}+c_{j}^{\dagger}e^{-\frac{i}{2} \Omega _{j}}\right) \left(
c_{j+1}e^{\frac{i}{2}\Omega _{j+1}}+c_{j+1}^{\dagger}e^{-\frac{
i}{2}\Omega _{j+1}}\right) \left| {\rm gs}_{\alpha }\right\rangle,
\label{unit-trans}
\end{equation}
where $c_{j}^{(\dagger)}$ annihilates a particle in the state
$F(r-R_j)$ and $c_{j+1}^{(\dagger)}$
creates a particle in the state
($F(r-R_{j+1})$) localized very close to the cores of the $j$th and
$(j+1)$th vortex, respectively. Eq.\ (\ref{unit-trans}) seemingly
implies that the motion of the $j$th vortex around the $(j+1)$th
vortex affects the occupations of states very close to the cores of
the two vortices. This is in contrast, however, to the derivation
leading to Eq.\ (\ref {unit-trans}), which explicitly assumes that
vortices are kept far enough from one another so that tunneling
between vortex cores may be disregarded.

This seeming contradiction is analyzed in detail in
\onlinecite{Stern04}, where it is shown that the unitary
transformation (\ref{unit-trans}) does not affect the occupation of
the core states of the $j,j+1$ vortices, because all ground states
are composed of superpositions in which the core states have a
probability of one-half to be occupied and one-half to be empty. The
unitary transformation within the ground state subspace does not
change that probability. Rather, they affect phases in the
superpositions. Using this point of view it is then possible to show
that two ingredients are essential for the non-Abelian statistics of
the vortices. The first is the {\it quantum entanglement} of the
occupation of states near the cores of distant vortices. The second
ingredient is familiar from (Abelian) fractional statistics: the
{\it geometric phase} accumulated by a vortex traversing a closed loop.

Therefore, we conclude that, for $p-$wave
superconductors, the existence of zero-energy intra-vortex modes
leads, first, to a multitude of ground states, and, second, to a
particle-hole symmetric occupation of the vortex cores in all ground
states. When represented in occupation-number basis, a ground state
is a superposition which has equal probability for the vortex core
being empty or occupied by one fermion.
When a vortex traverses a trajectory that encircles another vortex,
the phase it accumulates depends again on the number of fluid
particles it encircles. Since a fluid particle is, in this case, a
Cooper pair, the occupation of a vortex core by a fermion, half a
pair, leads to an accumulation of a phase of $\pi$ relative to the
case when the core is empty. And since the ground state is a
superposition with equal weights for the two possibilities, the
relative phase of $\pi$ introduced by the encircling might in this
case transform the system from one ground state to another.

Now consider the ground state degeneracy of
a $p+ip$ superconductor on the torus. Let us define,
following \onlinecite{Oshikawa07} (see also \onlinecite{Chung07}),
the operators $A_1$, $A_2$ which create a pair of
Bogoliubov-de Gennes quasiparticles, take one around
the meridian or longitude of the torus, respectively,
and annihilate them again. We then define $B_1$, $B_2$ as operators
which create a vortex-antivortex pair, take the vortex
around the meridian or longitude of the torus, respectively,
and annihilate them. $B_1$ increases the flux through the hole
encircled by the longitude of the torus by one half of a flux quantum
while $B_2$ does the same for the other hole.
These operators satisfy the commutation relations $\left[{A_1}, {A_2}\right]=0$
and ${A_1}{B_2}=-{B_2}{A_1}$, ${A_2}{B_1}=-{B_1}{A_2}$.
We can construct a multiplet of ground states as follows.
Since ${A_1}$ and ${A_2}$ commute and square to $1$,
we can label states by their $A_1$ and $A_2$ eigenvalues $\pm 1$.
Let $|1,1\rangle$ be the state with both eigenvalues equal to $1$, i.e.
${A_1}|1,1\rangle={A_2}|1,1\rangle=|1,1\rangle$.
Then ${B_1}|1,1\rangle = |1,-1\rangle$ and ${B_2}|1,1\rangle = |-1,1\rangle$.
Suppose we now try to apply ${B_2}$ to ${B_1}|1,1\rangle= |1,-1\rangle$.
This will create a vortex-antivortex pair; the Majorana zero modes,
$\gamma_a$, $\gamma_b$ associated with the vortex and anti-vortex
will be in the state $|0\rangle$ defined by
$\left({\gamma_a}+i{\gamma_b}\right)|0\rangle = 0$.
When the vortex is taken around the longitude of the torus,
its Majorana mode will be multiplied by $-1$: ${\gamma_a}\rightarrow -{\gamma_a}$.
Now, the vortex-antivortex pair will no longer be in the state
$|0\rangle$, but will instead be in the state $|1\rangle$
defined by $\left({\gamma_a}-i{\gamma_b}\right)|1\rangle = 0$.
Consequently, the vortex-antivortex pair can no longer
annihilate to the vacuum. When they fuse, a fermion is
left over. Therefore, ${B_2}{B_1}|1,1\rangle$ does not give
a new ground state (and, by a similar argument,
neither does ${B_1}{B_2}|1,1\rangle$). Consequently,
a $p+ip$ superconductor has `only' three ground states on the torus.
A basis in which $B_1$ is diagonal is given by:
$(|1,1\rangle \pm |1,-1\rangle)/\sqrt{2}$, with eigenvalue $\pm 1$,
and $|-1,1\rangle$, with eigenvalue $0$ (since there is zero amplitude
for ${B_1}|-1,1\rangle$ to be in the ground state subspace).
They can be identified with the states $|{1_m}\rangle$, $|{\psi_m}\rangle$,
and $|{\sigma_m}\rangle$ in Ising anyon language.
Meanwhile, ${B_2}$ is diagonal in the basis $(|1,1\rangle \pm |-1,1\rangle)/\sqrt{2}$,
$|1,-1\rangle$. By changing from one basis to the other, we find
the $S$-matrix given in the previous subsection follows.

The essential feature of chiral $p$-wave
superconductors is that they have Majorana fermion
excitations which have zero energy modes at vortices
(and gapless excitations at the edge of the system, see section \ref{sec:edge}).
The Majorana character is a result of the superconductivity, which
mixes particle and hole states; the zero modes and gapless edge excitations
result from the chirality. Majorana fermions arise in a completely different
way in the Kitaev honeycomb lattice model \cite{Kitaev06a}:
\begin{equation}
\label{eqn:Kitaev-honeycomb}
H = -{J_x}\sum_{x-\text{links}} {\sigma^x_j}{\sigma^x_j}
-{J_y}\sum_{y-\text{links}} {\sigma^y_j}{\sigma^y_j}
-{J_z}\sum_{z-\text{links}} {\sigma^x_j}{\sigma^z_j}
\end{equation}
where the $z$-links are the vertical links on the honeycomb lattice,
and the $x$ and $y$ links are at angles $\pm\pi/3$ from the vertical.
The spins can be represented by Majorana fermions $b^x$, $b^y$,
$b^z$, and $c$, according to ${\sigma^x_j}=i{b_j^x}{c_j}$, ${\sigma^x_j}=i{b_j^y}{c_j}$,
${\sigma^x_j}=i{b_j^z}{c_j}$ so long as the constraint
${b_j^x}{b_j^y}{b_j^z}{c_j}=1$ is satisfied. Then, the Hamiltonian
is quartic in Majorana fermion operators, but the operators
${b_j^x}{b_k^x}$, ${b_j^y}{b_k^y}$, ${b_j^z}{b_k^z}$ commute
with the Hamiltonian. Therefore, we can take their eigenvalues
as parameters $u_{jk}={b_j^\alpha}{b_k^\alpha}$, with $\alpha=x,y,$ or $z$
appropriate to the $jk$ link. These parameters can be varied to minimize the
Hamiltonian, which just describes Majorana fermions hopping on the
honeycomb lattice:
\begin{equation}
H = \frac{i}{4}\sum_{jk} t_{jk} {c_j}{c_k}
\end{equation}
where $t_{jk}=2 {J_\alpha}u_{jk}$ for nearest neighbor $j$, $k$ and zero
otherwise. For different values of the $J_\alpha$s, the $t_{jk}$'s take
different values. The topological properties of the corresponding $c_j$ bands
are encapsulated by their Chern number \cite{Kitaev06a}.
For a certain range of $J_\alpha$s, a $P, T$-violating perturbation
gives the Majorana fermions a gap in such a way as to support
zero modes on vortex-like excitations (plaquettes on which one of the
$u_{jk}$s is reversed in sign). These excitations are identical in topological
character to the vortices of a $p+ip$ superconductor discussed above.

\subsection{Chern-Simons Effective Field Theories, the Jones Polynomial,
and Non-Abelian Topological Phases}
\label{sec:Jones}

\subsubsection{Chern-Simons Theory and Link Invariants}

In the previous subsection, we have seen an
extremely simple and transparent formulation
of the quasiparticle braiding properties of a particular
non-Abelian topological state which, as we will see
later in this section, is equivalent to SU(2)$_2$ Chern-Simons theory.
It describes the multi-quasiparticle Hilbert space
and the action of braiding operations in terms of free fermions.
Most non-Abelian topological states are not so simple,
however.  In particular, SU(2)$_k$ Chern-Simons theory
for $k>2$ does not have a free fermion or boson
description.\footnote{It is an open question whether there is
an alternative description of an SU(2)$_k$ topological
phase with $k>2$ in terms of fermions or bosons which is
similar to the $p+ip$ chiral superconductor
formulation of SU(2)$_2$.}
Therefore, in the next two subsections, we discuss
these field theories using more general methods.

Even though its Hamiltonian vanishes and it has no local degrees of
freedom, solving Chern-Simons theory is still a non-trivial matter.
The reason is that it is difficult in a non-Abelian gauge theory to
disentangle the physical topological degrees of freedom from the
unphysical local gauge degrees of freedom. There are essentially two
approaches. Each has its advantages, and we will describe them both.
One is to work entirely with gauge-invariant quantities and derive
rules governing them; this is the route which we pursue in this
subsection. The second is to pick a gauge and simply calculate
within this gauge, which we do in the next subsection
(\ref{sec:FQHE}).

Consider SU(2)$_k$ non-Abelian Chern-Simons theory:
\begin{eqnarray}
\label{CS-action-again}
S_{CS}[a] =
\frac{k}{4\pi}\int_{\cal M} {\rm tr}\left(a\wedge da + \frac{2}{3}
a\wedge a\wedge a\right)
\end{eqnarray}
We modify the action by the addition of sources, $j^{\mu {\underline a}}$,
according to ${\cal L} \rightarrow {\cal L} + {\rm tr}\left( j\cdot a\right)$.
We take the sources to be a set of particles on prescribed classical
trajectories. The $i^\text{th}$ particle carries the spin $j_i$ representation
of SU(2). As we saw in subsection \ref{sec:CS-theory}, there are only $k+1$ allowed representations; later in this subsection, we will see that
if we give a particle a higher spin representation
than $j=k/2$, then the amplitude will vanish identically.
Therefore, $j_i$ must be in allowed set of $k+1$ possibilities:
$0,\frac{1}{2},\ldots,\frac{k}{2}$.
The functional integral in the presence of these sources
can be written in terms of Wilson loops,
${W_{{\gamma_i},{j_i}}}[a]$, which are defined
as follows. The holonomy $U_{\gamma,j}[a]$ is an $SU(2)$ matrix
associated with a curve $\gamma$. It is
defined as the path-ordered exponential integral of the
gauge field along the path $\gamma$:
\begin{multline}
  U_{\gamma,j}[a]
  \equiv {\cal P}{e^{i{\oint_\gamma} {{\bf a}^{\underline c}T^{\underline c}
  \cdot d{\bf l}}}}\\
  =  {\sum_{n=0}^\infty} {i^n}{\int_0^{2\pi}}\!\!{ds_1}
  {\int_0^{s_1}}\!\!{ds_2}\ldots
  {\int_0^{s_{n-1}}}\!\!{ds_n}\Bigl[
  \dot{\bf \gamma}({s_1})\cdot{{\bf a}^{{\underline a}_1}}
  \left({\bf \gamma}({s_1})\right)
  {T^{{\underline a}_1}}
  \ldots\,\\ \dot{\bf \gamma}({s_n})\cdot{{\bf a}^{{\underline a}_n}}
  \left({\bf \gamma}({s_n})\right){T^{{\underline a}_n}}\Bigr]
\end{multline}
where ${\cal P}$ is the path-ordering symbol. The
Lie algebra generators $T^{\underline a}$ are taken in
the spin $j$ representation. $\vec{\gamma}(s)$, $s\in[0,2\pi]$
is a parametrization of $\gamma$; the holonomy is clearly
independent of the parametrization. The Wilson
loop is the trace of the holonomy:
\begin{equation}
{W_{\gamma,j}}[a] = {\rm tr}\left(U_{\gamma,j}[a]\right)
\label{eqn:Wilson-loop-def}
\end{equation}
Let us consider the simplest case, in which the source
is a quasiparticle-quasihole
pair of type $j$ which is created out of the ground state,
propagated for a period of time, and then annihilated,
returning the system to the ground state. The amplitude
for such a process is given by:
\begin{equation}
\left\langle 0 | 0 \right\rangle_{\gamma,j}
 = \int {\cal D}a\,e^{iS_{CS}[a]}\,{W_{\gamma,j}}[a]
\end{equation}
Here, $\gamma$ is the spacetime loop formed by the trajectory
of the quasiparticle-quasihole pair. The Wilson loop was introduced
as an order parameter for confinement in a gauge theory
because this amplitude roughly measures the
force between the quasiparticle and the quasihole. If they were to interact
with a confining force $V(r)\sim r$, then the logarithm of
this amplitude would be proportional to the the area of the loop;
if they were to have a short-ranged interaction, it would be proportional to the
perimeter of the loop. However, Chern-Simons theory is independent
of a metric, so the amplitude cannot depend on any length
scales. It must simply be a constant. For $j=1/2$, we will call this constant
$d$. As the notation implies, it is, in fact,
the quantum dimension of a $j=1/2$ particle.
As we will see below, $d$ can be determined in terms of the level $k$, and
the quantum dimensions of higher spin particles can be expressed
in terms of $d$.

\begin{figure}[tb!]
\includegraphics[width=3.5in]{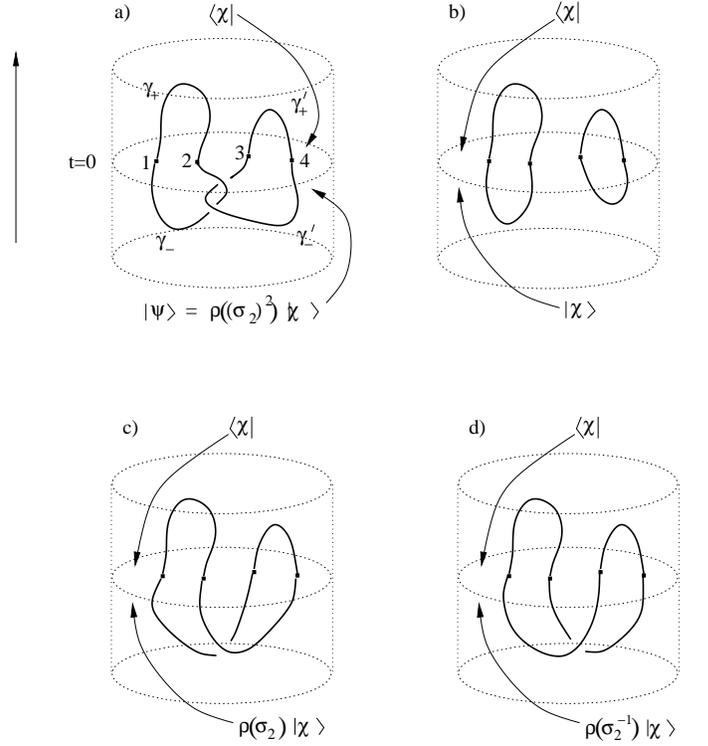}
\caption{The functional integrals which give
(a) $\langle\chi|\rho\!\left({\sigma_2^2}\right)|\chi\rangle$
(b) $\langle\chi|\chi\rangle$,
(c) $\langle\chi|\rho\!\left({\sigma_2}\right)|\chi\rangle$,
(d) $\langle\chi|\rho\!\left({\sigma_2^{-1}}\right)|\chi\rangle$. }
\label{fig:B-calc}
\end{figure}

We can also consider the amplitude
for two pairs of quasiparticles to be created out of
the ground state, propagated for some time, and then annihilated,
returning the system to the ground state:
\begin{equation}
\left\langle 0 | 0 \right\rangle_{{\gamma_1},{j_1};{\gamma_2},{j_2}}
= \int {\cal D}a\,e^{iS_{CS}[a]}\,{W_{{\gamma},{j}}}[a]\,
{W_{{\gamma'},{j'}}}[a]
\end{equation}
This amplitude can take different values depending on how
$\gamma$ and $\gamma'$ are linked as in Fig. \ref{fig:B-calc}a vs \ref{fig:B-calc}b.
If the curves are unlinked the integral must give $d^2$, but when they are linked the value can be nontrivial.
In a similar way, we can formulate the amplitudes for an arbitrary number of sources.

It is useful to think about the history in figure \ref{fig:B-calc}a
as a two step process:
from $t=-\infty$ to $t=0$ and from $t=0$ to $t=\infty$. (The two pairs
are created at some time $t<0$ and annihilated at some time
$t>0$.) At $t=0^-$, the system is in a four-quasiparticle state.
(Quasiparticles and quasiholes are topologically equivalent
if $G=$SU(2), so we will use `quasiparticle' to refer to both.)
Let us call this state $\psi$:
\begin{multline}
\label{eqn:integral-state}
\psi[A] =
\int_{a({\bf x},0)=A({\bf x})}\!\!\!\!
{\cal D}a({\bf x},t)\: W_{{\gamma^{}_-},j}[a]\: W_{{\gamma'_-},j'}[a]\,\times\\
e^{{\int_{-\infty}^0}
dt\int{d^2}x\:{\cal L}_{\rm CS}}
\end{multline}
where ${\gamma_-}$ and ${\gamma'_-}$ are the arcs
given by $\gamma(t)$ and $\gamma'(t)$ for
$t<0$. $A({\bf x})$ is the value of the gauge field on the $t=0$
spatial slice; the wavefunctional $\psi[A]$ assigns an
amplitude to every spatial gauge field configuration. For $G$=SU(2)
and $k>1$, there are actually two different four-quasiparticle states:
if particles $1$ and $2$ fuse to the identity field $j=0$, then particles $3$ and $4$
must as well; if particles $1$ and $2$ fuse to $j=1$, then particles $3$ and $4$
must as well. These are the only possibilities. (For $k=1$, fusion
to $j=1$ is not possible.) Which one the system is in depends on how
the trajectories of the four quasiparticles are intertwined.
Although quasiparticles $1$ and $2$ were created as a pair
from the vacuum, quasiparticle $2$ braided with quasiparticle $3$,
so $1$ and $2$ may no longer fuse to the vacuum.
In just a moment, we will see
an example of a different four-quasiparticle state.

We now interpret the $t=0$ to $t=\infty$ history as the conjugate
of a $t=-\infty$ to $t=0$ history. In other words, it gives us
a four quasiparticle bra rather than a four quasiparticle ket:
\begin{multline}
{\chi^*}[A] =
\int_{a({\bf x},0)=A({\bf x})} \!\!\!\!
{\cal D}a({\bf x},t)\: W_{{\gamma_+},j}[a]\: W_{{\gamma'_+},j'}[a]\,\times\\
e^{{\int_0^{\infty}}
dt\int{d^2}x\:{\cal L}_{\rm CS}}
\end{multline}
In the state $|\chi\rangle$, quasiparticles $1$ and $2$ fuse to
form the trivial quasiparticle, as do quasiparticles $3$ and $4$.
Then we can interpret the functional integral from $t=-\infty$ to $t=\infty$ as
the matrix element between the bra and the ket:
\begin{equation}
\langle \chi | \psi \rangle
= \int {\cal D}a\,e^{iS_{CS}[a]}\,{W_{{\gamma_1},{j_1}}}[a]\,
{W_{{\gamma_2},{j_2}}}[a]
\end{equation}

Now, observe that $|\psi\rangle$ is obtained from $|\chi\rangle$
by taking quasiparticle $2$ around quasiparticle $3$,
i.e. by exchanging quasiparticles $2$ and $3$ twice,
$|\psi\rangle = \rho\!\left({\sigma_2^2}\right) |\chi\rangle$. Hence,
\begin{equation}
\label{eqn:braiding-matrix-integral}
\langle \chi | \rho\!\left({\sigma_2^2}\right) |\chi\rangle
= \int {\cal D}a\,e^{iS_{CS}[a]}\,{W_{{\gamma_1},{j_1}}}[a]\,
{W_{{\gamma_2},{j_2}}}[a]
\end{equation}
In this way, we can compute the entries of the
braiding matrices $\rho\!\left({\sigma_i}\right)$ by computing functional integrals
such as the one on the right-hand-side of
(\ref{eqn:braiding-matrix-integral}).
Note that we should normalize the state $|\chi\rangle$
by computing the figure \ref{fig:B-calc}b, which gives
its matrix element with itself.

Consider, now, the state $\rho\!\left({\sigma_2}\right)|\chi\rangle$, in which particles
$2$ and $3$ are exchanged just once. It is depicted in figure
\ref{fig:B-calc}c. Similarly, the state $\rho\!\left({\sigma_2^{-1}}\right)|\psi\rangle$
is depicted in figure \ref{fig:B-calc}d.
From the figure, we see that
\begin{eqnarray}
\langle\chi|\rho\!\left({\sigma_2}\right)|\chi\rangle &=& d\\
\langle\chi|\rho\!\left({\sigma_2^{-1}}\right)|\chi\rangle &=& d
\end{eqnarray}
since both histories contain just a single unknotted loop.
Meanwhile,
\begin{equation}
\langle\chi|\chi\rangle = d^2
\end{equation}

Since the four-quasiparticle Hilbert
space is two-dimensional, $\rho\!\left({\sigma_2}\right)$ has
two eigenvalues, $\lambda_1$,
$\lambda_2$, so that
\begin{equation}
\rho\!\left({\sigma}\right)-\left({\lambda_1}+{\lambda_2}\right)
+ {\lambda_1}{\lambda_2}\rho\!\left({\sigma^{-1}}\right) = 0
\end{equation}
Taking the expectation value in the state $|\chi\rangle$,
we find:
\begin{equation}
d-\left({\lambda_1}+{\lambda_2}\right){d^2}
+ {\lambda_1}{\lambda_2}d = 0
\end{equation}
so that
\begin{equation}
d=\frac{1+{\lambda_1}{\lambda_2}}{{\lambda_1}+{\lambda_2}}
\end{equation}
Since the braiding matrix is unitary, $\lambda_1$ and $\lambda_2$
are phases. The overall phase is unimportant for quantum
computation, so we really need only a single number. In fact, this
number can be obtained from self-consistency conditions
\cite{Freedman04a}. However, the details of the computation of
$\lambda_1$, $\lambda_2$ within is technical and requires a careful
discussion of framing; the result is \cite{Witten89} that
${\lambda_1}=-e^{-3\pi i/2(k+2)}$, ${\lambda_2}=e^{\pi i/2(k+2)}$.
These eigenvalues are simply
${R^{\frac{1}{2},\frac{1}{2}}_0}~=~{\lambda_1}$,
${R^{\frac{1}{2},\frac{1}{2}}_1}~=~{\lambda_2}$. Consequently,
\begin{equation}
d = 2\,\cos\left(\frac{\pi}{k+2}\right)
\label{eqn:d-value}
\end{equation}
and
\begin{equation}
\label{eqn:Jones-skein}
q^{-1/2}\rho({\sigma_i}) - q^{1/2} {\rho\!\left(\sigma_{i}^{-1}\right)} = q-q^{-1}
\end{equation}
where $q=-e^{\pi i/(k+2)}$ (see Fig. \ref{figure3}). Since this operator
equation applies regardless of the state to which
it is applied, we can apply it locally to any given
part of a knot diagram to relate the amplitude
to the amplitude for topologically simpler processes,
as we will see below \cite{Kauffman01}.
This is an example of a {\it skein relation}; in this case,
it is the skein relation which defines the Jones polynomial.
In arriving at this skein relation, we are retracing
the connection between Wilson loops in Chern-Simons
theory and knot invariants which was made in the remarkable
paper \cite{Witten89}.
In this paper, Witten showed that correlation functions of
Wilson loop operators in SU(2)$_k$ Chern-Simons theory
are equal to corresponding evaluations of the Jones polynomial,
which is a topological invariant of knot theory \cite{Jones85}:
\begin{equation}
\label{eqn:Jones-CS}
\int {\cal D}a\,W_{{\gamma_1},\frac{1}{2}}[a]
\ldots W_{{\gamma_n},\frac{1}{2}}[a]\,e^{iS_{CS}[a]} =
V_{L}(q)
\end{equation}
$V_{L}(q)$ is the Jones polynomial
associated with the link $L={\gamma_1}\cup\ldots\cup{\gamma_n}$,
evaluated at $q=-e^{\pi i/(k+2)}$ using the skein relation
(\ref{eqn:Jones-skein}). Note that we assume here that
all of the quasiparticles transform under the $j=\frac{1}{2}$
representation of $SU(2)$. The other quasiparticle
types can be obtained through the fusion of several $j=1/2$
quasiparticles, as we will discuss below in Section \ref{sec:Kauffman-bracket}.

\subsubsection{Combinatorial Evaluation of Link Invariants
and Quasiparticle Properties}
\label{sec:Kauffman-bracket}

The Jones polynomial \cite{Jones85}
$V_{L}(q)$ is a formal
Laurent series in a variable $q$ which is associated to a
link $L={\gamma_1}\cup\ldots\cup{\gamma_n}$.
It can be computed recursively using (\ref{eqn:Jones-skein}).
We will illustrate how this is done by showing
how to use a skein relation to compute
a related quantity called
the Kauffman bracket $K_{L}(q)$ \cite{Kauffman87},
which differs from the Jones polynomial by a normalization:
\begin{equation}
V_{L}(q) = \frac{1}{d}\,(-q^{3/2})^{w(L)}
 \,K_{L}(q)
\end{equation}
where $w(L)$ is the writhe of the link.
(The Jones polynomial is defined for an oriented link. Given an orientation,
each crossing can be assigned a sign $\pm 1$; the writhe
is the sum over all crossings of these signs.)
The link $L$ embedded
in three-dimensional space (or, rather, three-dimensional
space-time in our case) is projected onto the plane. This can be done faithfully
if we are careful to mark overcrossings and undercrossings. Such a projection
is not unique, but the same Kauffman bracket is obtained for all possible $2D$ projections of a knot (we will see an example of this below).
An unknotted loop $\bigcirc$ is given
the value $K_\bigcirc(q)~=~d~\equiv~-q-q^{-1}~=~2\cos\pi/(k+2)$.
For notational simplicity,
when we draw a knot, we actually mean the Kauffman bracket associated
to this knot. Hence, we write
\begin{equation}
  \pspicture[shift=-0.4](1.0,1.0)
  \pscircle(0.5,0.5){0.333333333333333}
  \endpspicture
   =\: d
\end{equation}
The disjoint union of $n$ unknotted loops is assigned the value $d^n$.

The Kauffman bracket for any given knot can be obtained recursively by
repeated application of the following skein relation which relates it
with the Kauffman brackets for two knots
both of which have one fewer crossing according to the rule:
\begin{equation}
  \label{eq:recursion}
  \pspicture[shift=-0.4](1.0,1.0)
   \psbezier[linewidth=1.0pt](0.333333333333333,0)(0.333333333333333,0.5)(0.666666666666667,0.5)
  (0.666666666666667,1.0)
  \psbezier[linewidth=1.0pt](0.666666666666667,0)(0.666666666666667,0.333333333333333)
  (0.6,0.4)(0.6,0.45)
  \psbezier[linewidth=1.0pt]
  (0.4,0.55)(0.4,0.6)(0.333333333333333,0.666666666666667)(0.333333333333333,1.0)
  \endpspicture
  \:=\: q^{1/2}{\hskip-0.2cm}
  \pspicture[shift=-0.4](1.0,1.0)
   \psbezier[linewidth=1.0pt](0.333333333333333,0)
  (0.333333333333333,0.5)(0.333333333333333,0.5)(0.333333333333333,1.0)
  \psbezier[linewidth=1.0pt](0.666666666666667,0)(0.666666666666667,0.5)
  (0.666666666666667,0.5)(0.666666666666667,1.0)
  \endpspicture
  \; + \; q^{-1/2}{\hskip-0.2cm}
  \pspicture[shift=-0.4](1.0,1.0)
  \psbezier[linewidth=1.0pt](0.333333333333333,0)
  (0.333333333333333,0.333333333333333)
  (0.666666666666667,0.333333333333333)(0.666666666666667,0)
  \psbezier[linewidth=1.0pt](0.666666666666667,1.0)
  (0.666666666666667,0.666666666666667)
  (0.333333333333333,0.666666666666667)(0.333333333333333,1.0)
  \endpspicture
\end{equation}
With this rule, we can eliminate all crossings.
At this point, we are left with a linear combination of the Kauffman brackets
for various disjoint unions of unknotted loops. Adding up these
contributions of the form $d^m$ with their appropriate coefficients
coming from the recursion relation (\ref{eq:recursion}), we obtain the
Kauffman bracket for the knot with which we started.

\begin{figure}[tbh]
\centerline{\includegraphics[width=3.5in]{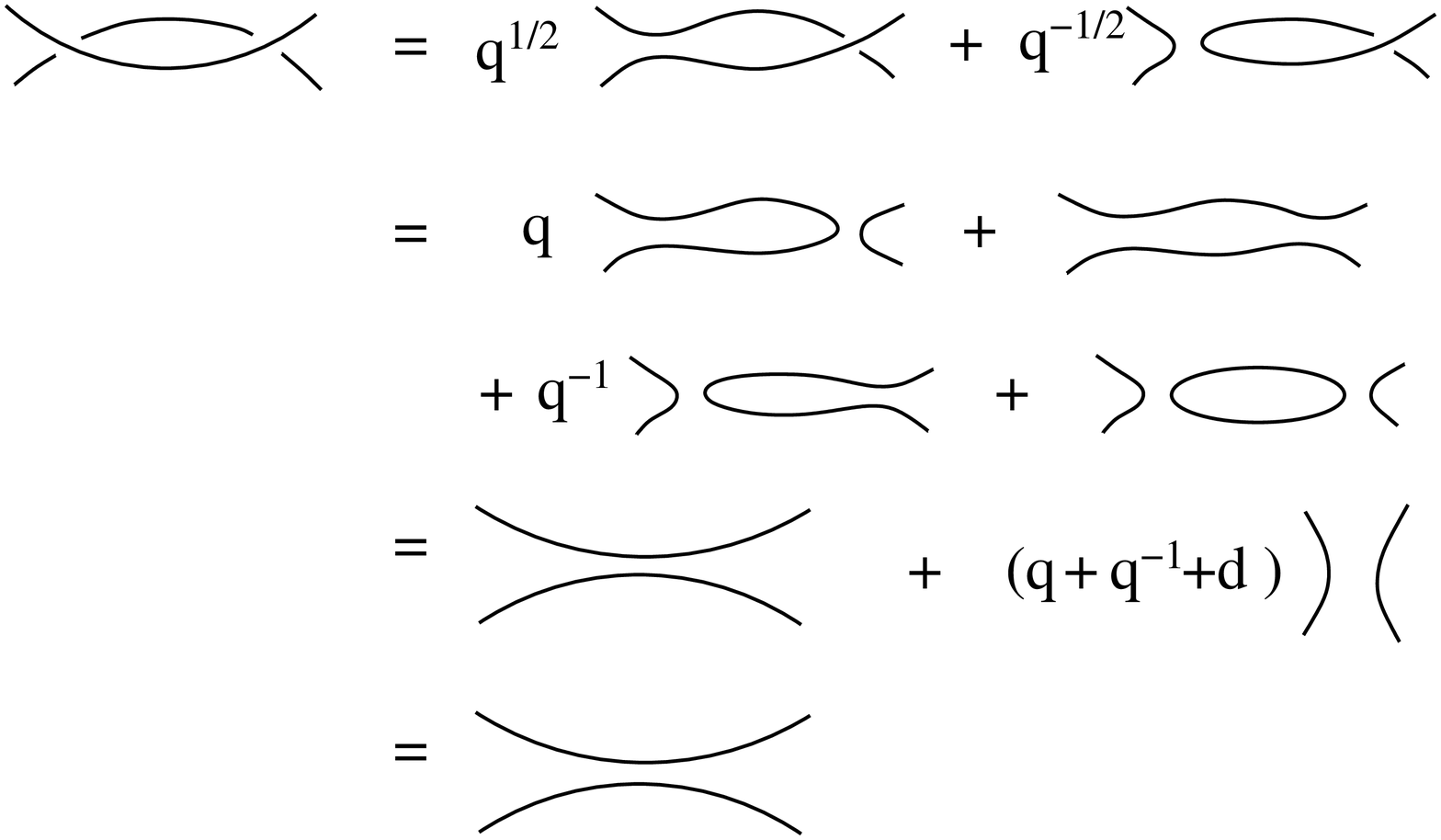}}
\caption{The Kauffman bracket is invariant under continuous motions
of the arcs and, therefore, independent of the particular projection
of a link to the plane.}
\label{fig:Kauffman-reid}
\end{figure}

Let us see how this works with a simple example.
First, consider the two arcs which cross twice in figure \ref{fig:Kauffman-reid}.
We will assume that these arcs continue in some
arbitrary way and form closed loops. By applying the Kauffman
bracket recursion relation in figure \ref{fig:Kauffman-reid},
we see that these arcs can be replaced
by two arcs which do not cross.
In Section \ref{sec:interference}, we will use these methods to
evaluate some matrix elements relevant to interference experiments.

Now, let us consider the two fusion channels of a pair of
quasiparticles in some more detail. When the two quasiparticles
fuse to the trivial particle, as $1$ and $2$ did above, we can
depict such a state, which we will call $|0\rangle$,  as
$\frac{1}{\sqrt{d}}$ times the state yielded by
the functional integral (\ref{eqn:integral-state}) with
a Wilson line which looks like $\bigcup$
because two quasiparticles which are created as a pair out
of the ground state must necessarily fuse to spin $0$
if they do not braid with any other particles.
(The factor $1/\sqrt{d}$ normalizes the state.)
Hence, we can project any two quasiparticles onto the
$j=0$ state by evolving them with a history which looks like:
\begin{equation}
\label{eqn:P_0}
 {\Pi_0}
  \:=\:
  \frac{1}{d}{\hskip-0.2cm}
  \pspicture[shift=-0.4](1.0,1.0)
  \psbezier[linewidth=1.0pt](0.333333333333333,0)
  (0.333333333333333,0.333333333333333)
  (0.666666666666667,0.333333333333333)(0.666666666666667,0)
  \psbezier[linewidth=1.0pt](0.666666666666667,1.0)
  (0.666666666666667,0.666666666666667)
  (0.333333333333333,0.666666666666667)(0.333333333333333,1.0)
  \endpspicture
\end{equation}
On the right-hand-side of this equation, we mean a functional
integral between two times $t_1$ and $t_2$. The functional
integral has two Wilson lines in the manner indicated pictorially.
On the left-hand-side, we have suggested that evolving a state
through this history can be viewed as acting on it with the projection
operator ${\Pi_0}=|0\rangle\langle 0|$.

However, the two quasiparticles could instead be in the state $|1\rangle$,
in which they fuse to form the $j=1$ particle.
Since these states must be orthogonal, $\langle 0|1\rangle=0$,
we must get identically zero if we follow the history (\ref{eqn:P_0})
with a history which defines a projection operator $\Pi_1$
onto the $j=1$ state:
\begin{equation}
  \label{eq:JW1}
{\Pi_1}
  \:=\:
  \pspicture[shift=-0.4](1.0,1.0)
  \psbezier[linewidth=1.0pt](0.333333333333333,0)
  (0.333333333333333,0.5)(0.333333333333333,0.5)(0.333333333333333,1.0)
  \psbezier[linewidth=1.0pt](0.666666666666667,0)(0.666666666666667,0.5)
  (0.666666666666667,0.5)(0.666666666666667,1.0)
  \endpspicture
  \: - \; \frac{1}{d}{\hskip-0.2cm}
  \pspicture[shift=-0.4](1.0,1.0)
  \psbezier[linewidth=1.0pt](0.333333333333333,0)
  (0.333333333333333,0.333333333333333)
  (0.666666666666667,0.333333333333333)(0.666666666666667,0)
  \psbezier[linewidth=1.0pt](0.666666666666667,1.0)
  (0.666666666666667,0.666666666666667)
  (0.333333333333333,0.666666666666667)(0.333333333333333,1.0)
  \endpspicture
\end{equation}
It is easy to see that if this operator acts on a state
which is given by a functional integral which looks like $\bigcup$,
the result is zero.

The projection operators $\Pi_0$ , $\Pi_1$, which are called
{\it Jones-Wenzl projection operators}, project a pair
of a quasiparticles onto the two natural basis states of their
qubit. In other words, we do not need to introduce new types
of lines in order to compute the expectation values of
Wilson loops carrying $j=0$ or $j=1$. We can denote them
with pairs of lines projected onto either of these states.
Recall that a $j=1/2$ loop had amplitude $d$, which was the
quantum dimension of a $j=1/2$ particle. Using the projection
operator (\ref{eq:JW1}), we see that a $j=1$ loop has amplitude
${d^2}-1$ (by connecting the top of the line segments to the bottom
and evaluating the Kauffman bracket).
One can continue in this way to construct projection operators
which project $m$ lines onto $j=m/2$. This projection operator
must be orthogonal to the $j=0,1,3/2,2,\ldots,(m-1)/2$ projection operators
acting on subsets of the $m$ lines, and this condition is sufficient
to construct all of the Jones-Wenzl projection operators recursively. Similarly,
the quantum dimensions can be computed through a recursion
relation. At level $k$, we find that quasiparticles
with $j>k/2$ have quantum dimensions which vanish identically
(e.g. for $k=1$, $d=1$ so the quantum dimension of a $j=1$
particle is ${d^2}-1=0$).
Consequently, these quasiparticle types do not occur. Only
$j=0,\frac{1}{2},\ldots,\frac{k}{2}$ occur.

\begin{figure}[tb]
\centerline{\includegraphics[width=3.5in]{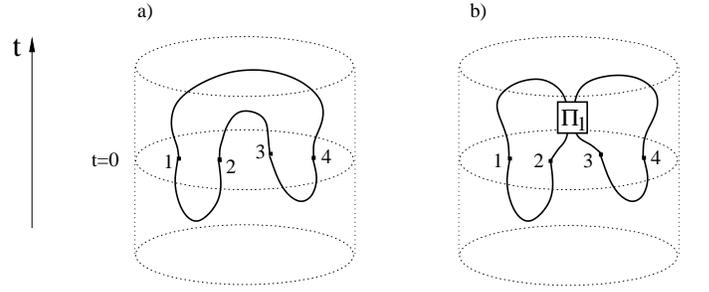}}
\caption{The elements of the $F$-matrix can be obtained by
computing matrix elements between kets in which $1$ and $2$
have a definite fusion channel and bras in which $1$
and $4$ have a definite fusion channel.}
\label{fig:F-matrix}
\end{figure}

The entries in the $F$-matrix can be obtained by graphically
computing the matrix element between a state in which, for instance,
$1$ and $2$ fuse to the vacuum and $3$ and $4$ fuse to the vacuum
and a state in which $1$ and $4$ fuse to the vacuum and $2$ and $3$
fuse to the vacuum, which is depicted in Figure \ref{fig:F-matrix}a.
(The matrix element in this figure must be normalized by the
norms of the top and bottom states to obtain the $F$-matrix elements.)
To compute the matrix element between a state in
which $1$ and $2$ fuse to the vacuum and $3$ and $4$ fuse to the vacuum
and a state in which $1$ and $4$ fuse to $j=1$ and $2$ and $3$
fuse to $j=1$, we must compute the diagram in Figure \ref{fig:F-matrix}b.
For $k=2$, we find the same $F$-matrix as was found for Ising anyons
in Section \ref{sec:pwave}.

\begin{figure}[tbh]
\centerline{\includegraphics[width=3.5in]{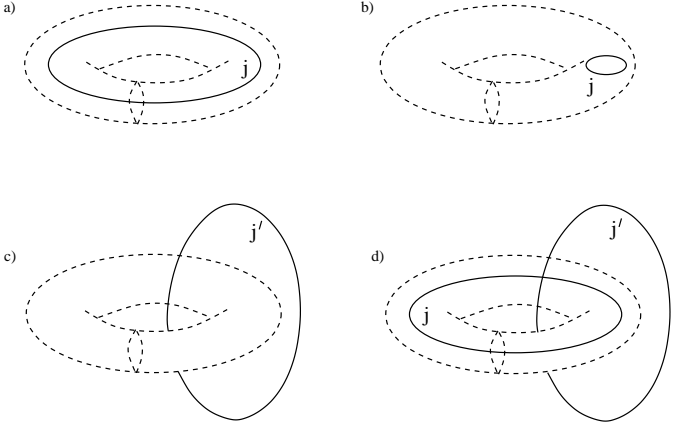}}
\caption{Different degenerate ground states on the torus
are given by performing the functional integral with longitudinal Wilson
loops (a) carrying spin $j=0,\frac{1}{2},\ldots,\frac{k}{2}$. Meridinal
Wilson loops are contractible (b); they do not give new ground states.
The corresponding bras are have Wilson lines in the exterior solid torus (c).
$S$-matrix elements are given by evaluating the history obtained by
combining a bra and ket with their linked Wilson lines.}
\label{fig:torus-states}
\end{figure}

Let us now briefly consider the ground state properties of the
SU(2)$_k$ theory on the torus. As above, we integrate the
Chern-Simons Lagrangian over a $3$-manifold ${\cal M}$ with boundary
$\Sigma$, i.e. ${\cal M} = \Sigma\times(-\infty,0]$ in order to
obtain a $t=0$ state. The boundary $\Sigma$ is the spatial slice at
$t=0$. For the torus, $\Sigma={T^2}$, we take ${\cal M}$ to be the
solid torus, ${\cal M}={S^1}\times D^2$, where $D^2$ is the disk. By
foliating the solid torus, we obtain earlier spatial slices. If
there are no quasiparticles, then there are no Wilson lines
terminating at $\Sigma$. However, the functional integral can have
Wilson loops in the body of the solid torus as in Figure
\ref{fig:torus-states}a. These correspond to processes in the past,
$t<0$, in which a quasiparticle-quashole pair was created, taken
around the meridian of the torus and annihilated. The Wilson loop
can be in any of the $k+1$ allowed representations
$j=0,\frac{1}{2},\ldots,\frac{k}{2}$; in this way, we obtain $k+1$
ground state kets on the torus (we will see momentarily that they
are all linearly independent). Wilson loops around the meridian are
contractible (Figure \ref{fig:torus-states}b), so they can be simply
evaluated by taking their Kauffman bracket; they multiply the state
by $d_j$. Evidently, these Wilson loop operators are diagonal in
this basis. Bras can be obtained by integrating the Chern-Simons
Lagrangian over the $3$-manifold ${\cal
M}'=\Sigma\times[0,\infty)={S^3} \backslash {S^1}\times D^2$, i.e.
the exterior of the torus. Wilson loops in the exterior torus are
now contractible if they are parallel to a longitude but non-trivial
if they are around the meridian, as in in Figure
\ref{fig:torus-states}c. Again, we obtain $k+1$ ground state bras in
this way. The matrix elements between these bras and kets
(appropriately normalized such that the matrix product of a bra with
its conjugate ket is unity) are the entries in the $S$-matrix, which
is precisely the basis change between the longitudinal and meridinal
bases. A matrix element can be computed by evaluating the
corresponding picture. The $ab$ entry in the $S$-matrix is given by
evaluating the Kauffman bracket of the picture in Figure
\ref{fig:torus-states}d (and dividing by the normalization of the
states). This figure makes the relationship between the $S$-matrix
and braiding clear.

Finally, we comment on the difference between SU(2)$_2$ and Ising
anyons, which we have previously described as differing only
slightly from each other (See also the end of section \ref{sec:edge}
below). The effective field theory for Ising anyons contains an
additional U(1) Chern-Simons gauge field, in addition to an
SU(2)$_2$ gauge field \cite{Fradkin98,Fradkin01}. The consequences
of this difference are that ${\Theta_\sigma}=e^{-\pi i/8}$ while
$\Theta_{1/2}=e^{-3\pi i/8}$; ${R^{\sigma\sigma}_1}=e^{-\pi i/8}$
while ${R^{\frac{1}{2},\frac{1}{2}}_0}=-e^{-3\pi i/8}$;
${R^{\sigma\sigma}_\psi}=e^{3\pi i/8}$ while
${R^{\frac{1}{2},\frac{1}{2}}_1}=e^{\pi i/8}$;
$\left[F^{\sigma\sigma\sigma}_{\sigma}\right]_{ab}=
-\left[F^{\frac{1}{2},\frac{1}{2},\frac{1}{2}}_{\frac{1}{2}}\right]_{ab}$.
The rest of the $F$-matrices are the same, as are the fusion
multiplicities $N_{ab}^c$ and the $S$-matrix. In other words, the
basic structure of the non-Abelian statistics is the same in the two
theories, but there are some minor differences in the U(1) phases
which result from braiding. Both theories have threefold ground
state degeneracy on the torus; the Moore-Read Pfaffian state has
ground state degeneracy $6$ because of an extra U(1) factor
corresponding to the electrical charge degrees of freedom.

Of course, in the $k=2$ case we have already obtained all of these results by
the method of the previous subsection.
However, this approach has two advantages: (1) once
Witten's result (\ref{eqn:Jones-CS}) and
Kauffman's recursion relation (\ref{eq:recursion})
are accepted, braiding matrix elements can be obtained
by straightforward high school algebra;
(2) the method applies to all levels $k$, unlike free Majorana
fermion methods which apply only to the $k=2$ case.
There is an added bonus, which is that this formalism is
closely related to the techniques used to analyze lattice
models of topological phases, which we discuss in a later subsection.

\subsection{Chern-Simons Theory, Conformal Field Theory,
and Fractional Quantum Hall States}
\label{sec:FQHE}

\subsubsection{The Relation between Chern-Simons Theory
and Conformal Field Theory}

Now, we consider Chern-Simons theory in a particular gauge, namely
holomorphic gauge (to be defined below). The ground state
wavefunction(s) of Chern-Simons theory can be obtained by performing
the functional integral from the distant past, $t=-\infty$, to time
$t=0$ as in the previous subsection:
\begin{multline}
\psi[A({\bf x})] =
\int_{a({\bf x},0)=A({\bf x})}
{\cal D}a({\bf x},t)\:e^{{\int_{-\infty}^0}
dt{\int_\Sigma}{d^2}x\:{\cal L}_{\rm CS}}
\end{multline}
For the sake of concreteness, let us consider the torus, $\Sigma=T^2$,
for which the spacetime manifold is ${\cal M}=(-\infty,0]\times{T^2}=
{S^1}\times D^2$.
We assume for simplicity that there are no Wilson loops
(either contained within the solid torus or
terminating at the boundary). If $x$ and $y$ are coordinates
on the torus (the fields will be subject to periodicity requirements), we write
$z=x+iy$. We can then change to coordinates $z$, ${\bar z}$,
and, as usual, treat $a_z^{\underline a}$ and $a_{\bar z}^{\underline a}$
as independent variables. Then, we take the holomorphic
gauge, $a_{\bar z}^{\underline a}=0$.
The field ${a_{\bar z}^{\underline a}}$ only appears in the action linearly,
so the functional integral over ${a_{\bar z}^{\underline a}}$ may be performed,
yielding a $\delta$-function:
\begin{eqnarray}
\label{eqn:C-S-delta-action}
\int Da\: {e^{\frac{k}{4\pi}\int_{{D^2}\times S^1}
\epsilon^{\mu\nu\lambda} \left({a_\mu^a}
{\partial_\nu}{a_\lambda^a} + \frac{2}{3}{f_{abc}}{a_\mu^a}
{a_\nu^b}{a_\lambda^c}\right)}} =\cr
\int D{a _i}\,\delta(f_{ij}^{\underline a})\,
{e^{\frac{k}{4\pi}\int_{{D^2}\times S^1}
\epsilon^{ij} {a_i^a}{\partial_{\bar z}}{a_j^a} }}
\end{eqnarray}
where $i,j=t,z$. Here $f_{ij}^{\underline a}=
{partial_i} a_j^{\underline a} - {partial_j} a_i^{\underline a}
+ i\epsilon_{{\underline a}{\underline b}{\underline c}}
a_i^{\underline a} a_j^{\underline b}$ are the
spatial components of the field strength.
There are no other cubic terms in the action
once $a_{\bar z}$ has been eliminated (as is
the case in any such gauge in which one of the
components of the gauge field vanishes).
The constraint imposed by
the $\delta$-function can be solved by taking
\begin{equation}
{a_i^a}  = {\partial_i} U\,{U^{-1}}
\end{equation}
where $U$ is a single-valued function taking values in the
Lie group. Substituting this into the right-hand-side
of (\ref{eqn:C-S-delta-action}),
we find that the action which appears in the exponent
in the functional integral takes the form
\begin{eqnarray}
S &=&  {\frac{k}{4\pi}\int_{{D^2}\times S^1}  \epsilon^{ij}
\text{tr}\left( {\partial_i} U\,{U^{-1}}
{\partial_{\bar z}}\left({\partial_j} U\,{U^{-1}}\right)\right)}\cr
&=& \frac{k}{4\pi}\int_{{D^2}\times S^1}  \epsilon^{ij}
\biggl[\text{tr}\left( {\partial_i} U\,{U^{-1}}
{\partial_{\bar z}} {\partial_j} U\,{U^{-1}}\right)\: +
\cr & & {\hskip 2.7 cm}
\text{tr}\left( {\partial_i} U\,{U^{-1}}
{\partial_j} U\,{\partial_{\bar z}}{U^{-1}}\right) \biggr]
\cr
&=& \frac{k}{4\pi}\int_{{D^2}\times S^1}  \epsilon^{ij}
\biggl[{\partial_j}
\text{tr} \left( {\partial_i} \,{U^{-1}} {\partial_{\bar z}} U \right)
\: + \cr & & {\hskip 2.7 cm}
\text{tr}\left( {\partial_i} U\,{U^{-1}}
{\partial_j} U\,{\partial_{\bar z}} {U^{-1}}\right) \biggr]
 \cr
&=&  \frac{k}{4\pi}\int_{T^2}
\text{tr}\left( {\partial_z} {U^{-1}}
{\partial_{\bar z}} U\right)
\: +\cr & & {\hskip -0.7 cm}
\frac{k}{12\pi}\int_{{D^2}\times S^1}  \epsilon^{\mu\nu\lambda}
\text{tr}\left( {\partial_\mu} U\,{U^{-1}}
{\partial_\nu} U\,{U^{-1}}\,{\partial_\lambda}U\,{U^{-1}}\right)
\label{eqn:WZW-derivation}
\end{eqnarray}
The Jacobian which comes from the $\delta$-function
$\delta(f_{ij}^{\underline a})$ is cancelled by that
associated with the change of integration
variable from $Da$ to $DU$. In the final line, the first term
has been integrated by parts while the second term,
although it appears to be an integral over the
$3D$ manifold, only depends on the boundary values of $U$
\cite{Wess71,Witten83}. This
is the Wess-Zumino-Witten (WZW) action.
What we learn from (\ref{eqn:WZW-derivation}) then, is that,
in a particular gauge, the ground state wavefunction of
$2+1$-D Chern-Simons theory can be viewed as the partition
function of a $2+0$-dimensional WZW model.

For positive integer $k$, the WZW model is a $2D$ conformal
field theory which, in the SU(2) case, has Virasoro
central charge $c={\bar c}= \frac{3k}{k+2}$. (For a brief review of
some of the basics of conformal field theory, see appendix \ref{section:CFT}
and references therein.) However, in computing properties
of the Chern-Simons theory from which we have derived it,
we will couple only to $a_z = {\partial_z}U \cdot U^{-1}$;
i.e. only to the holomorphic or right-moving sector of
the theory. Thus, it is the chiral WZW model which
controls the ground state wavefunction(s) of Chern-Simons
theory.

If we were to follow the same strategy to calculate the
Chern-Simons ground state
wavefunction with Wilson lines or punctures present,
then we would end up with a correlation function of operators
in the chiral WZW model transforming under the corresponding
representations of SU(2). (Strictly speaking, it is not a
correlation function, but a {\it conformal block}, which
is a chiral building block for a correlation function. While
correlation functions are single-valued, conformal blocks
have the non-trivial monodromy properties which we need, as
is discussed in appendix \ref{section:CFT}.)
Therefore, following \cite{Witten89,Elitzur89}, we have mapped
the problem of computing the ground state wavefunction (in $2+0$-dimensions)
of Chern-Simons theory, which is a topological theory with a gap,
to the problem of computing a correlation function in the chiral
WZW model (in $1+1$-dimensions), which is a critical theory.
This is a bit peculiar since one theory is gapped while the other is gapless.
However, the gapless degrees of freedom of the WZW model
for the $t=0$ spatial slice are pure gauge degrees of freedom
for the corresponding Chern-Simons theory. (In the very similar
situation of a surface $\Sigma$ with boundary, however,
the corresponding conformally-invariant $1+1$-D theory
describes the actual dynamical excitations of the edge of
the system, as we discuss in section \ref{sec:edge}.)
Only the topological properties of the chiral
WZW conformal blocks are physically meaningful for us.

More complicated topological states with multiple
Chern-Simons fields and, possibly, Higgs fields \cite{Fradkin98,Fradkin99b,Fradkin01}
correspond in a similar way to other chiral rational conformal field
theories which are obtained by tensoring or
cosetting WZW models. (RCFTs are those CFTs
which have a finite number of primary fields -- see appendix \ref{section:CFT}
for the definition of a primary field -- under some
extended chiral algebra which envelopes the Virasoro
algebra; a Kac-Moody algebra in the WZW case; and,
possibly, other symmetry generators.) Consequently, it is possible
to use the powerful algebraic techniques of rational
conformal field theory to compute the ground state
wavefunctions of a large class of topological
states of matter. The quasiparticles of the topological state
correspond to the primary fields of the chiral RCFT. (It is a matter
of convenience whether one computes correlation functions
with a primary field or one of its descendants since
their topological properties are the same. This is a freedom which
can be exploited, as we describe below.)

The conformal blocks of an RCFT have one property
which is particularly useful for us, namely they are
holomorphic functions of the coordinates. This
makes them excellent candidate wavefunctions for
quantum Hall states. We identify primary fields
with the quasiparticles of the quantum Hall state,
and compute the corresponding conformal block.
However, there is one important issue
which must be resolved: a quantum Hall wavefunction is normally
viewed as a wavefunction for electrons (the quasiparticle
positions, by contrast, are usually viewed merely as
some collective coordinates specifying a given excited state).
Where are the electrons in our RCFT? Electrons have
trivial braiding properties. When one electron is taken
around another, the wavefunction is unchanged, except
for a phase change which is an odd integral multiple of $2\pi$.
More importantly, when any quasiparticle is taken around an electron,
the wavefunction is unchanged apart from a phase change
which is an integral multiple of $2\pi$. Therefore,
the electron must be a descendant of the identity.
In other words, the RCFT must contain a fermionic
operator by which we can extend the chiral algebra.
This new symmetry generator is essentially the electron
creation operator  -- which is, therefore, a descendant of the identity
under its own action. Not all RCFTs have such an operator
in their spectrum, so this is a strong constraint on RCFTs
which can describe quantum Hall states.
If we are interested, instead, in a quantum Hall state of bosons,
as could occur with ultra-cold bosonic atoms in a rotating optical
trap \cite{Cooper01}, then the RCFT must contain a bosonic
field by which we can extend the chiral algebra.

An RCFT correlation function of $N_e$ electron operators therefore
corresponds to the Chern-Simons ground state wavefunction with $N_e$
topologically-trivial Wilson lines. From a purely topological
perspective, such a wavefunction is just as good as a wavefunction
with no Wilson lines, so the Wilson lines would seem superfluous.
However, if the descendant field which represents the electron
operator is chosen cleverly, then the wavefunction with $N_e$ Wilson
lines may be a `good' trial wavefunction for electrons in the
quantum Hall regime.  Indeed, in some cases, one finds that these
trial wavefunctions are the exact quantum Hall ground states of
simple model Hamiltonians \cite{Moore91,Blok92,Wen94,Read99,Greiter91,Ardonne99,Simon07b}. In
the study of the quantum Hall effect, however, a wavefunction is
`good' if it is energetically favorable for a realistic
Hamiltonian, which is beyond the scope of the underlying
Chern-Simons theory, which itself only knows about braiding
properties. It is unexpected good luck that the trial wavefunctions
obtained from Chern-Simons theory are often found to be `good' from
this energetic perspective, which is a reflection of how highly
constrained quantum Hall wavefunctions are, and how central these
braiding properties are to their physics. We emphasize, however,
that a wavefunction obtained in this way will {\it not} be the exact
ground state wavefunction for electrons with Coulomb interactions.
In some cases it might not even have particularly high overlap with
the ground state wavefunction, or have good energetics. The one
thing which it does capture is the topological structure of a
particular universality class.

\subsubsection{Quantum Hall Wavefunctions from Conformal Field Theory}

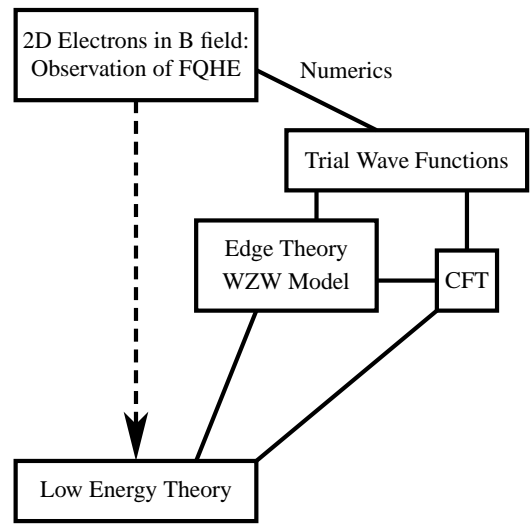
\begin{figure}
 \ifx\JPicScale\undefined\def\JPicScale{.8}\fi
\psset{unit=\JPicScale mm}
\psset{linewidth=0.7,dotsep=1,hatchwidth=0.3,hatchsep=1.5,shadowsize=1,dimen=middle}
\psset{dotsize=0.7 2.5,dotscale=1 1,fillcolor=black}
\psset{arrowsize=2 3,arrowlength=2,arrowinset=0.25,tbarsize=0.7
5,bracketlength=0.15,rbracketlength=0.15}
\begin{pspicture}(0,0)(95,85)
\rput(30,80){2D Electrons in B field:} \rput(30,75){Observation of
FQHE} \pspolygon[](10,85)(50,85)(50,70)(10,70)
\psline[linestyle=dashed,arrows=->](30,70)(30,10) \rput(30,5){Low
Energy Theory} \pspolygon[](10,10)(50,10)(50,0)(10,0)
\rput(75,60){Trial Wave Functions}
\pspolygon[](55,65)(95,65)(95,55)(55,55) \psline(50,75)(70,65)
\rput(65,75){Numerics} \rput(85,40){CFT}
\pspolygon[](80,45)(90,45)(90,35)(80,35) \psline(85,55)(85,45)
\psline(80,40)(70,40) \rput(55,45){Edge Theory} \rput(55,40){WZW
Model} \pspolygon[](40,50)(70,50)(70,35)(40,35)
\psline(80,35)(50,10) \psline(50,35)(40,10) \psline(60,55)(60,50)
\end{pspicture}
\caption{How one arrives at a low-energy theory of the quantum Hall
effect.  At the top, one begins with the experimental observation of
the quantized Hall effect. At the bottom, we know
the low energy theory should be of Chern-Simons form.
One would like to be able to ``integrate out" high energy degrees of
freedom directly to obtain the low energy theory, as shown by the
dotted line, but must instead take a more circuitous route,
as described in the text.} \label{fig:lineofthought}
\end{figure}

Ideally, the logic which would lead us to a particular RCFT
would be as follows, as displayed in Fig.
\ref{fig:lineofthought}. One begins with the experimental
observation of the quantized Hall effect at some filling fraction
$\nu$ (shown at the top).  We certainly know that the Hamiltonian
for the system is simply that of 2D electrons in a magnetic field,
and at the bottom, we know the form of the low energy theory should
be of Chern-Simons form.  One would like to be able to ``integrate
out" high energy degrees of freedom directly to obtain the low-energy
theory.  Given the low-energy Chern-Simons effective field theory,
one can pass to the associated RCFT, as described above.
With the RCFT in hand, one can construct wavefunctions,
as we will describe below.
Indeed, such a procedure has been explicitly
achieved for Abelian quantum Hall
states \cite{Zhang89,Lopez91}. In some special non-Abelian
cases, progress in this direction has been made \cite{Wen91a,Wen99}.

For most non-Abelian
theories, however, the situation is not so simple. The RCFT
is usually obtained through inspired guesswork
\cite{Moore91,Blok92,Read99,Ardonne99,Cappelli01,Simon07a}.
One may try to justify it {\it ex post facto} by
solving for the properties of quasiholes
of a system with some unrealistic (e.g. involving $3$-body or higher
interactions) but soluble Hamiltonian.
The degeneracy can be established by counting \cite{Nayak96c,Read96,Read06}.
The braiding matrices can be obtained by numerically
computing the Berry integrals for the given wavefunctions
\cite{Tserkovnyak03} or by using their connection to
conformal field theory to deduce them
\cite{Moore91,Nayak96c,Gurarie97,Slingerland01}. One can then deduce
the Chern-Simons effective field theory of the state
either from the quasiparticle properties or from
the associated conformal field theory with which both
it and the wavefunctions are connected.

We now show how such wavefunctions can be constructed
through some examples. In appendix \ref{section:CFT}, we review some of the
rudiments of conformal field theory.

 \setcounter{mysubsection}{0}
\vspace*{-5pt} \mysubsection{Wavefunctions from CFTs}
Our goal is to construct a LLL FQH wavefunction $\Psi(z_1,  \ldots, z_N)$
which describes an electron fluid in a circular droplet centered at
the origin.  $\Psi$ must be a homogeneous antisymmetric
analytic function of the $z_i$s, independent of the ${\bar z}_i$s
apart from the Gaussian factor, which we will frequently
ignore (see Sec. \ref{sec:quantumHallreview}).
If we consider the FQHE of bosons,
we would need $\Psi$ to instead be symmetric.
The filling fraction $\nu$ of a FQH wavefunction $\Psi$ is
given by $\nu = N/N_\Phi$ where $N$ is the number of electrons and
$N_\Phi$ is the number of flux quanta penetrating the
droplet \cite{Prange90}. In the LLL, $N_\Phi$ is given by the highest
power of $z$ occurring in $\Psi$.
\mylabel{Gaussian}

We will also frequently need the fact that in an
incompressible state of filling fraction $\nu$, multiplying a
wavefuncton by a factor ${\prod_i} ({z_i}-w)^m$ pushes a charge $\nu m$
away from the point $w$.  This can be understood \cite{Laughlin83}
as insertion of $m$ flux quanta at the point $w$, which, via
Faraday's law creates an azimuthal electric field,
which, then, via the Hall conductivity transfers charge $\nu m$
away from the point $w$.

Our strategy will be to choose a particular chiral RCFT, pick an
``electron"  field $\psi_e$ in this theory (which, by the reasoning
given above, must be a fermionic generator of the extended chiral
algebra of the theory), and write a ground state trial wavefunction
$\Psi_{\rm gs}$ for $N$ electrons as
\begin{equation}\label{eq:gswf}
    \Psi_{\rm gs} = \langle \, \psi_e(z_1) \ldots \psi_e(z_N) \, \rangle
\end{equation}
The field $\psi_e$ must be fermionic since the quantum Hall wavefunction
on the left-hand-side must be suitable for electrons. Not all RCFTs
have such a field in their spectrum, so this requirement
constrains our choice. This requirement also ensures
that we will obtain a wavefunction which has no branch cuts;
in particular, there will only be one conformal block on the
right-hand-side of (\ref{eq:gswf}).
We must do a little more work in choosing $\psi_e$
so that there are no poles either on the right-hand-side of (\ref{eq:gswf}).
As discussed above, the correlation function on the right-hand-side
of (\ref{eq:gswf}) is a ground state wavefunction of Chern-Simons
theory with $N_e$ trivial topological charges at fixed positions
${z_1},{z_2},\ldots,{z_{N_e}}$.

Of course, there isn't a unique choice of RCFT,
even at a given filling fraction. Therefore, there are different
fractional quantum Hall states which can be constructed
in this way. Which fractional quantum Hall state is actually observed at a
particular $\nu$ is determined by comparing the energies of
the various possible competing ground states. Having a good wavefunction
is, by itself, no guarantee that this wavefunction actually describes the
physical system. Only a calculation of its energy gives real
evidence that it is better than other possible states.

The reason for introducing this complex machinery simply
to construct a wavefunction becomes
clearer when we consider quasihole wavefunctions, which
are Chern-Simons ground state wavefunctions with
$N_e$ trivial topological charges and $N_{qh}$ non-trivial topological charges.
In general, there are many possible quasihole operators,
corresponding to the different primary fields of the theory,
so we must really consider $N_{qh1}, N_{qh2}, \ldots N_{qhm}$
numbers of quasiholes if there are $m$ primary fields.
Each different primary field corresponds to a different
topologically-distinct type of ``defect'' in the ground state.
(As in the case of electrons, we are free to choose a descendant
field in place of the corresponding primary field since the two have identical
topological properties although the wavefunction generated by a descendant will be different from that generated by its primary.) Let us suppose that we focus attention
on a particular type of quasiparticle which, in most cases,
will be the quasiparticle of minimal electrical charge.
Then we can write a wavefunction
with quasiholes at positions $w_1, \ldots, w_M$ as
\begin{equation}\label{eq:qhwf}
    \Psi(w_1
\! \ldots w_M\!) \! = \!\langle \psi_{qh}(w_1) \ldots
\psi_{qh}(w_M) \,\,\,  \psi_e(z_1) \ldots \psi_e(z_N)  \rangle
\end{equation}
where $\psi_{qh}$ is the corresponding primary field.
Since $\psi_{qh}$ is a primary field and $\psi_{e}$ is
a descendant of the identity, we are guaranteed that
$\psi_{qh}$ and $\psi_{e}$ are local with respect to each
other, i.e. taking one field around can only produce a phase
which is a multiple of $2\pi$. Consequently, the wavefunction
$\Psi$ remains analytic in the electron coordinates $z_i$
even after the fields $ \psi_{qh}(w_1) \ldots \psi_{qh}(w_M)$
have been inserted into the correlation function.

One important feature of the conformal block on the right-hand-side
of (\ref{eq:qhwf}) is that $\psi_{qh}(w_a)$ and $\psi_{e}(z_i)$
are on roughly the same footing -- they are both fields
in some conformal field theory (or, equivalently, they are both
fixed sources coupled to the Chern-Simons gauge field).
However, when intepreted as a wavefunction on the
left-hand-side of (\ref{eq:qhwf}), the electron coordinates $z_i$ are the variables
for which the wavefunction gives a probability amplitude while
the quasihole coordinates $w_a$ are merely some parameters
in this wavefunction. If we wished to normalize the wavefunction
differently, we could multiply by an arbitrary function of the $w_a$s.
However, the particular normalization which is given by the
right-hand-side of (\ref{eq:qhwf}) is particularly convenient, as we will see momentarily.
Note that since the quasihole positions $w_j$ are merely
parameters in the wavefunction, the wavefunction need not be analytic in these
coordinates.

\mysubsection{Quasiparticle Braiding}
The branch cuts in quasihole positions $w_a$ are symptoms
of the fact that there may be a vector space of conformal blocks
corresponding to the right-hand-side of (\ref{eq:qhwf}).
In such a case, even when the quasihole positions are
fixed, there are several possible linearly independent wavefunctions.
These multiple degenerate states are necessary for non-Abelian
statistics, and they will generically mix when
the quasiholes are dragged around each other.

However, there is still a logical gap in the above reasoning.
The wavefunctions produced by an RCFT have the correct
braiding properties for the corresponding Chern-Simons
ground state wavefunction built into them through their
explicit monodromy properties. As a result of the
branch cuts in the conformal blocks as a function of the $w_a$s,
when one quasihole is taken around another,
the wavefunction $\Psi^\alpha$ transforms into
$M^{\alpha\beta}\Psi^\beta$, where the index
$\alpha=1,2,\ldots, g$ runs over the $g$ different
degenerate $n$-quasihole states.
However, when viewed as quantum
Hall wavefunctions, their quasiparticle braiding properties
are a combination of their explicit monodromy and
the Berry matrix which is obtained from:
\begin{equation}
\label{eq:Berry-matrix}
e^{i\gamma_{\alpha\beta}} =
{\cal P} \exp\left(\oint d{\vec w} \Bigl\langle \Psi^\alpha \Bigl|
\nabla_{\vec w}
\Bigr| \Psi^\beta \Bigr\rangle \right)
\end{equation}
where $\Psi^\alpha$, $\alpha=1,2,\ldots, g$ are the $g$ different
degenerate $n$-quasihole states and {\cal P} is the path ordering
symbol. In this equation, the
$z_i$s are integrated over in order to compute the inner product,
but the $w_a$s are held fixed, except for the one which
is taken around some loop.

Strictly speaking, the effect of braiding is to transform
a state according to $\Psi^\alpha\rightarrow e^{i\gamma_{\alpha\beta}}
M^{\beta\gamma}\Psi^\gamma$. By changing the normalization of the wavefunction, we can alter $e^{i\gamma_{\alpha\beta}}$ and
$M^{\beta\gamma}$. Only the product of the
two matrices on the right-hand-side of this equation is gauge invariant and
physically meaningful. When we presume that the braiding
properties of this wavefunction are given by those of the
corresponding CFT and Chern-Simons theory, we take it
to be equal to $M^{\beta\gamma}$ and ignore  $e^{i\gamma_{\alpha\beta}}$.
This can only be correct if $\gamma_{\alpha\beta}$ vanishes
up to a geometric phase proportional
to the area for a wavefunction given by a CFT conformal block.
In the case of the Laughlin states, it can be verified that
this is indeed correct by repeating the
the Arovas, Schrieffer, Wilczek calculation \cite{Arovas84}
with the Laughlin state normalized according to the
quasihole position dependence given by the corresponding
CFT (see below) \cite{Blok92}. This calculation rests
upon the plasma analogy originally
introduced by Laughlin in his seminal work \cite{Laughlin83}.
For other, more complex states, it is more difficult to
compute the Berry matrix. A version of a plasma analogy
for the MR Pfaffian state was constructed in \onlinecite{Gurarie97};
one could thereby verify the vanishing of the Berry matrix
for a two-quasihole state and, with some further assumptions,
for four and higher multiquasihole states. A direct evaluation
of the integral in (\ref{eq:Berry-matrix}) by the Monte-Carlo
method \cite{Tserkovnyak03} established that it vanishes for MR Pfaffian quasiholes.
The effect of Landau level mixing on statistics has also
been studied \cite{Simon07d}.
Although there has not been a complete proof that the CFT-Chern-Simons
braiding rules are identical to those of the wavefunction,
when it is interpreted as an electron wavefunction (i.e. there
has not been a complete proof that (\ref{eq:Berry-matrix}) vanishes
when the wavefunction is a CFT conformal block), there is
compelling evidence for the MR Pfaffian state, and it is almost
certainly true for many other states as well.
We will, therefore, take it as a given that we can simply
read off the braiding properties
of the wavefunctions which we construct below.

\mysubsection{The Laughlin State}
We now consider wavefunctions generated by perhaps
the simplest CFT, the chiral boson. We suppose that
the chiral boson has compactification radius $\sqrt{m}$,
so that $\phi\equiv \phi + 2\pi\sqrt{m}$. The U(1) Kac-Moody
algebra and enveloping Virasoro algebra can be extended
by the symmetry generator $e^{i\phi\sqrt{m}}$. Since the dimension
of this operator is $m/2$, it is fermionic for $m$ odd and bosonic for even $m$.
The primary fields of this extended chiral algebra are of the form
$e^{in\phi/\sqrt{m}}$, with $n=0,1,\ldots,m-1$.
They are all of the fields which are not descendants and are
local with respect to $e^{i\phi\sqrt{m}}$
(and to the Kac-Moody and Virasoro generators), as may be
seen from the operator product expansion (OPE)
(see Appendix \ref{section:CFT}):
\begin{equation}
e^{i\phi(z)\sqrt{m}}\, e^{in\phi(0)/\sqrt{m}}
\sim {z^n}\,e^{i(n+m)\phi(0)/\sqrt{m}} + \ldots
\end{equation}
When $z$ is taken around the origin, the right-hand-side is unchanged.
It is convenient to normalize the $U(1)$ current as $j=\frac{1}{\sqrt{m}}\partial\phi$;
then the primary field $e^{in\phi/\sqrt{m}}$ has charge $n/m$.
We take ${\psi_e}=e^{i\phi\sqrt{m}}$ as our electron field (which has
charge 1) and consider the resulting ground state wavefunction
according to Eq.~\ref{eq:gswf}.
Using Eq.~\ref{eq:vertex1} we find
\begin{equation}
\label{eq:Laughlin1}
    \Psi_{\rm gs} = \langle \psi_e(z_1) \ldots \psi_e(z_N) \rangle =
    {\mbox{$\prod_{i<j}$}}\,\,\, (z_i - z_j)^{m}
\end{equation}
It is now clear why we have chosen this CFT: to
have $\Psi_{\rm gs}$ given by correlators of a vertex operator
of the form $e^{i\alpha\phi}$ analytic (no branch cuts or poles) we must have
$\alpha^2=m$ a nonnegative integer, and $m$ must be odd to obtain an
antisymmetric wavefunction (or even for symmetric). We recognize
$\Psi_{\rm gs}$  as the $\nu=1/m$ Laughlin wavefunction. The astute
reader will notice that the correlator in Eq.~\ref{eq:Laughlin1}
actually violates the neutrality condition discussed in Appendix \ref{section:CFT}
and so it should actually have zero value. One fix for this problem is to
insert into the correlator (by hand) a neutralizing vertex operator
at infinity $e^{-iN \phi(z=\infty)\sqrt{m}}$ which then makes
Eq.~\ref{eq:Laughlin1} valid (up to a contant factor). Another
approach is to insert an operator that smears the neutralizing
background over the entire system \cite{Moore91}.  This approach also
conveniently results in the neglected Gaussian factors
reappearing! We will ignore
these neutralizing factors for simplicity. From now on, we will
drop the Gaussian factors from quantum Hall wavefunctions, with the
understanding that they result from including a smeared neutralizing background.

The quasihole operator must be a primary field; the primary field
of minimum charge is $e^{i\phi/\sqrt{m}}$.
Using Eq.~\ref{eq:vertex1}, Eq.~\ref{eq:qhwf} yields
\begin{equation} \label{eq:Laughlinqh} \Psi(w_1,
\ldots, w_M \!) \!=\!
 \mathop{\Pi}^{M}_{i<j}
 (w_i - w_j)^{1/m}  \mathop{\Pi}_{i=1}^N  \mathop{\Pi}_{j=1}^M  (z_i - w_j)
 \Psi_{\rm gs}
 %\mathop{\Pi}_{i<j}^N (z_i -
 % z_j)^{m+1} ~~
\end{equation}
As mentioned above, the factor $\mbox{$\prod_j$}\,  ( z_j - w)$
``pushes" charge away from the position $w$ leaving a hole of
charge precisely $Q=+e/m$. The first term on the right of
Eq.~\ref{eq:Laughlinqh} results from the fusion of quasihole
operators with each other, and explicitly shows the fractional
statistics of the quasiholes. Adiabatically taking two quasiholes
around each other results in a fractional phase of $2 \pi/m$.
As promised above, this statistical term appears
automatically in the wavefunction given by this CFT!

\mysubsection{Moore-Read Pfaffian State}

In the Ising CFT (see Appendix \ref{section:CFT}), we
might try to use $\psi_e(z) = \pfepsilon(z)$ as the electron field
\cite{Moore91}.
The $\pfepsilon$ fields can fuse together in pairs to give the
identity (since $\pfepsilon \times \pfepsilon = \bf 1$) so long as
there are an even number of fields. However, when we take two
$\pfepsilon$ fields close to each other, the OPE tells us that
\begin{equation}
\label{eq:epseps}
 \lim_{z_i \rightarrow z_j}   \pfepsilon(z_i) \pfepsilon(z_j) \sim {\bf 1}/(z_i-z_j)
\end{equation}
which diverges as ${z_i}\rightarrow{z_j}$ and is
therefore unacceptable as a wavefunction.
To remedy this problem, we tensor the Ising CFT with the chiral boson
CFT. There is now an operator $\psi\, e^{i\phi\sqrt{m}}$ by
which we can extend the chiral algebra. (If $m$ is even, this
symmetry generator is fermionic; if $m$ is odd, it is bosonic.)
As before, we will take this symmetry generator to be our electron field.
The corresponding primary fields are of the form $e^{in\phi/\sqrt{m}}$,
$\sigma\,e^{i(2n+1)\phi/2\sqrt{m}}$, and $\psi\,e^{in\phi/\sqrt{m}}$,
where $n=0,1,\ldots,m-1$. Again, these are determined by the requirement
of locality with respect to the generators of the chiral algebra, i.e.
that they are single-valued when taken around a symmetry generator,
in particular the elecron field $\psi\, e^{i\phi\sqrt{m}}$. For instance,
\begin{multline}
\psi(z)\, e^{i\phi(z)\sqrt{m}}\cdot\sigma(0)\,e^{i(2n+1)\phi(0)/2\sqrt{m}} \sim\\
 z^{-1/2}\sigma(0)\,z^{n+1/2}e^{i{(2(n+m)+1)}\phi(0)/\sqrt{m}} + \ldots\\
= z^{n}\sigma(0)\,e^{i{(2(n+m)+1)}\phi(0)/\sqrt{m}}
\end{multline}
and similarly for the other primary fields.

Using our new symmetry generator as the electron field,
we obtain the ground state wavefunction according
to Eq.~\ref{eq:gswf}:
\begin{multline}
\label{eq:pfaf2}
    \Psi_{\rm gs} = \langle \pfepsilon(z_1) \ldots \pfepsilon(z_N) \rangle
    \,\,\,
   \mbox{$\prod_{i<j}$} (z_i - z_j)^{m}\\
   = \text{Pf}\!\left(\frac{1}{z_i - z_j}\right)\, \mbox{$\prod_{i<j}$}(z_i - z_j)^{m}
\end{multline}
(See, e.g. \onlinecite{Senechal97} for the calculation of this correlation
function.)
Again, $m$ odd gives an antisymmetric
wavefunction and $m$ even gives a symmetric wavefunction.  For $m=2$
(and even $N$), Eq.~\ref{eq:pfaf2} gives precisely the Moore-Read
Pfaffian wavefunction (Eq.~\ref{eq:PfaffianBCS} with $g = 1/z$ and
two Jastrow factors attached).

To determine the filling fraction of our newly constructed
wavefunction, we need only look at the exponent of the Jastrow
factor in Eq.~\ref{eq:pfaf2}. Recall that
the filling fraction is determined by the highest power of any $z$
(See \ref{sec:FQHE}.1 above).  There are $m(N-1)$
factors of $z_1$ in the Jastrow factor. The Pfaffian
has a factor of $z_1$ in the denominator, so the
highest power of $z_1$ is $m(N-1)-1$.
However, in the thermodynamic limit, the number of factors
scales as $m N$. Thus the filling fraction is $\nu= 1/m$.

We now consider quasihole operators.  As in
the Laughlin case we might consider the primary fields
$\psi_{qh} = e^{in\phi/\sqrt{m}}$.
Similar arguments as in the Laughlin case show that the $n=1$
case generates precisely the Laughlin
quasihole of charge $Q=+e/m$. However we have other options for
our quasihole which have smaller electrical charge.
The primary field $\sigma\,e^{i\phi/2\sqrt{m}}$ has charge
$Q=+e/2m$. We then obtain the wavefunction according
to Eq.~\ref{eq:qhwf}
\begin{eqnarray} \nonumber \!\!\!\!\!
\Psi(\!w_1, \ldots , w_M\!) \! =\! \langle \sigma(w_1) \ldots
\sigma(w_m)
  \pfepsilon(z_1) \ldots \pfepsilon(z_N)  \rangle \times \\
 \!\!\!\!\!\! \mathop{\Pi}^{M}_{i<j}
 (w_i - w_j)^{1/2m}  \mathop{\Pi}_{i=1}^N  \mathop{\Pi}_{j=1}^M  (z_i - w_j)^{1/2}
 \mathop{\Pi}_{i<j}^N (z_i -
  z_j)^{m} ~~ \label{eq:pfafcor}
\end{eqnarray}

Using the fusion rules of the $\sigma$ fields (See Eq.
\ref{eq:isingfusion1},  as well as Fig.~\ref{fig:bratteli2}  and
Table \ref{tab:conformaldata} in Appendix \ref{section:CFT}), we see
that it is impossible to obtain $\bf 1$ from an odd number of
$\sigma$ fields. We conclude that quasiholes $\psi_{qh}$ can only
occur in pairs. Let us then consider the simplest case of two
quasiholes.  If there is an even number of electrons, the
$\pfepsilon$ fields fuse in pairs to form $\bf 1$, and the remaining
two quasiholes must fuse to form $\bf 1$ also. As discussed in
Eq.~\ref{eq:sigmasigma} the OPE of the two $\sigma$ fields will then
have a factor of $(w_1 - w_2)^{-1/8}$. In addition, the fusion of
the two vertex operators $e^{i\phi/2\sqrt{m}}$ results in the first term
in the second line of Eq.~\ref{eq:pfafcor}, $(w_1 -
w_2)^{1/(4m)}$. Thus the phase accumulated by taking the two
quasiholes around each other is $-2 \pi/8 + 2 \pi/4m$.

On the other hand, with an odd number of electrons in the system,
the $\pfepsilon$'s fuse in pairs, but leave one unpaired
$\pfepsilon$. The two $\sigma$'s must then fuse to form a
$\pfepsilon$ which can then fuse with the unpaired $\pfepsilon$ to
give the identity. (See Eq.~\ref{eq:sigmasigma}). In this case, the
OPE of the two $\sigma$ fields will give a factor of $(w_1 -
w_2)^{3/8}$. Thus the phase accumulated by taking the two quasiholes
around each other is $6 \pi/8 + 2 \pi/4m$.

In the language of section \ref{sec:pwave} above, when there is an
even number of electrons in the system, all of these are paired and
the fermion orbital shared by the quasiholes is unoccupied. When an
odd electron is added, it `occupies' this orbital, although the fermion orbital is neutral and the electron is charged (we can think of the electrons'
charge as being screened by the superfluid).

When there are many quasiholes, they may fuse together in many
different ways. Thus, even when the quasihole positions are fixed
there are many degenerate ground states, each corresponding to a
different conformal block (see appendix \ref{section:CFT}).
This degeneracy is precisely what is required for non-Abelian statistics.
Braiding the quasiholes around each other produces a rotation
within this degenerate space.

Fusing $2m$ $\sigma$ fields results in $2^{m-1}$ conformal blocks,
as may be seen by examining the Bratteli diagram of
Fig.~\ref{fig:bratteli2} in appendix \ref{section:CFT}. When two
quasiholes come together, they may either fuse to form $\bf 1$ or
$\bf \pfepsilon$. As above, if they come together to form $\bf 1$
then taking the two quasiholes around each other gives a phase of
$-2 \pi/8 + 2 \pi/4m$.  On the other hand, if they fuse to form
$\pfepsilon$ then taking them around each other gives a phase of $2
\pi 3/8 + 2 \pi/4m$.

These conclusions can be illustrated explicitly in the
cases of two and four quasiholes.
For the case of two quasiholes, the correlation
function (\ref{eq:pfafcor}) can be evaluated to give
\cite{Moore91,Nayak96c} (for an even number
of electrons):
\begin{multline}
\Psi({w_1},{w_2}) ~=~ 
\prod_{j<k} (z_j - z_k)^2\,\times\\\,
 {\rm Pf}\!\left(
  \frac{\left({z_j}-{w_1}\right)\left({z_k}-{w_2}\right)+{z_j}\leftrightarrow{z_k}}{z_j - z_k}\right)~.
\end{multline}
where $w_{12}={w_1}-{w_2}$. For simplicity, we specialize
to the case $m=2$; in general, there would be a prefactor
$\left(w_{12}\right)^{\frac{1}{4m}-\frac{1}{8}}$.
When the two quasiholes at $w_1$ and $w_2$ are brought together at
the point $w$, a single flux quantum Laughlin quasiparticle results,
since two $\sigma$s can only fuse to the identity in this case,
as expected from the above arguments:
\begin{multline}
\Psi_{\rm qh} (w) ~=~ \prod_{j<k} (z_j - z_k)^2
{\prod_i} \left({z_i}-w\right)\:
 {\rm Pf}\!\left(\frac{1}{z_j - z_k}\right)~.
\end{multline}

The situation becomes more interesting when we consider states
with 4 quasiholes. The ground
state is 2-fold degenerate (see appendix
\ref{section:CFT}). If there is an even number
of electrons (which fuse to form the identity), we are then concerned
with the $\langle \sigma \sigma \sigma \sigma \rangle$ correlator.
As discussed in appendix \ref{section:CFT}, two
orthogonal conformal blocks can be specified by whether 1 and 2 fuse
to form either $\bf 1$ or $\pfepsilon$.
The corresponding wavefunctions obtained by evaluating
these conformal blocks are \cite{Nayak96c}:
\begin{multline}
\label{eqn:fourqh}
{\Psi^{(1,\psi)}} =  \frac{\left({w_{13}}{w_{24}}\right)^{\frac{1}{4}}}
{(1 \pm \sqrt{x})^{1/2}}\,
\left( {\Psi_{(13)(24)}} \,\,\pm\,\,\sqrt{x}\,\,
{\Psi_{(14)(23)}}\right)
\end{multline}
where $x~=~w_{14}w_{23}/w_{13}w_{24}$.
(Note that we have taken a slightly different
anharmonic ratio $x$ than in \onlinecite{Nayak96c}
in order to make (\ref{eqn:fourqh}) more compact than
Eqs. (7.17), (7.18) of  \onlinecite{Nayak96c}.)
In this expression,
\begin{multline}
{\Psi_{(13)(24)}}= \prod_{j<k} (z_j - z_k)^2
\,\times\\
{\rm  Pf}\!\left( \frac{(z_j - {w_1}) (z_j - {w_3} )(z_k
-{w_2} ) (z_k - {w_4} ) + (j \leftrightarrow k )}
{z_j - z_k}\right)
\label{qhwf}
\end{multline}
and
\begin{multline}
{\Psi_{(14)(23)}}= \prod_{j<k} (z_j - z_k)^2
\,\times\\
{\rm  Pf}\!\left( \frac{(z_j - {w_1}) (z_j - {w_4} )(z_k
-{w_2} ) (z_k - {w_3} ) + (j \leftrightarrow k )}{z_j - z_k}\right)
\label{qhwf-2}
\end{multline}

Suppose, now, that the system is in the state $\Psi^{(1)}$.
Braiding 1 around 2 or 3 around 4 simply gives a phase
(which is $R^{\sigma\sigma}_{1}$ multiplied by a contribution
from the Abelian part of the theory).
However, if we take $w_2$ around $w_3$, then
after the braiding, the system will be in the state $\Psi^{(\psi)}$
as a result of the branch cuts in (\ref{eqn:fourqh}). Now,
1 and 2 will instead fuse together to form
$\pfepsilon$, as expected from the general argument
in Eq.~\ref{eq:braid1}.
Thus, the braiding yields a rotation in the degenerate
space.  The resulting prediction for the behavior under braiding for
the Moore-Read Pfaffian state is in agreement with the results
obtained in sections \ref{sec:pwave} and \ref{sec:Jones}
above.

\mysubsection{$\mathbb{Z}_3$ Read-Rezayi State (Briefly)} We can
follow a completely analogous procedure with a CFT which
is the tensor product of the $\mathbb{Z}_3$ parafermion CFT
with a chiral boson. As before, the electron operator
is a product of a chiral vertex operator from the bosonic
theory with an operator from the parafermion theory.
The simplest choice is $\psi_e = \psi_1 e^{i\alpha\phi}$. We
would like this field to be fermionic so that it can
be an electron creation operator by which we can
extend the chiral algebra (i.e., so that the electron wavefunction
has no branch cuts or singularities).
(See appendix \ref{section:CFT} for the notation for
parafermion fields.) The fusion rules for $\psi_1$
in the $\mathbb{Z}_3$ parafermion CFT are:
$\psi_1 \times \psi_1 \sim
\psi_2$ but $\psi_1 \times \psi_1 \times \psi_1 \sim {\bf 1}$ so
that the correlator in Eq.~\ref{eq:gswf} is only nonzero if $N$ is
divisible by 3. From the OPE, we obtain $\psi_1(z_1) \psi_1(z_2)
\sim (z_1 - z_2)^{-2/3} \psi_2$ so in order to have the
wavefunction analytic, we must choose $\alpha = \sqrt{m+2/3}$ with
$m \geq 0$ an integer ($m$ odd results in an antisymmetric
wavefunction and even results in symmetric).
The filling fraction in the thermodynamic limit is determined entirely by the
vertex operator $e^{i\alpha\phi}$, resulting in
$\nu= 1/\alpha^2= 1/(m + 2/3)$.

The ground state wavefunction for $N=3n$ electrons takes the form:
\begin{multline}
\Psi_{\rm gs}({z_1},\ldots,z_{3n}) = \prod_{i<j}\left({z_i}-{z_j}\right)^{m} \,\,\times
\\ {\cal S}\left\{ \prod_{0\leq r<s<n}
\chi_{r,s}(z_{3r+1},\ldots,z_{3r+k},z_{3k+1},\ldots z_{3s+3})  \right\}
\end{multline}
where $M$ must be odd for electrons, ${\cal S}$ means the symmetrization
over all permutations, and
\begin{multline}
\chi_{r,s} = (z_{3r+1}-z_{3s+1})(z_{3r+1}-z_{3s+2})\,\times\\
(z_{3r+2}-z_{3s+2})(z_{3r+2}-z_{3s+3})\,\ldots\,\times\\
(z_{3r+3}-z_{3s+3})(z_{3r+3}-z_{3s+1})
\end{multline}

With the electron operator in hand, we can determine
the primary fields of the theory. The primary field of
minimum electrical charge is $\psi_{qh} = \sigma_1 e^{i\phi/3\alpha}$
To see that this field is local with respect to $\psi_e$
(i.e., there should be no branch cuts for the electron coordinates $z_i$),
observe that $\sigma_1(w) \psi_1(z) \sim (z - w)^{-1/3} \pfepsilon$
and $e^{i\phi/3\alpha}(w) e^{i\alpha\phi}(z) \sim (z - w)^{1/3}$.
Constructing the full wavefunction (as in Eq.~\ref{eq:qhwf} and analogous to
Eq.~\ref{eq:pfafcor}) the fusion of of $e^{i\phi/3\alpha}$ (from $\psi_{qh}$)
with $e^{i\alpha\phi}$ (from $\psi_e$) again generates a factor of
$\prod_{i} (z_i - w)^{1/3}$.   We conclude that the
elementary quasihole has charge $Q = +e \nu/3$.

\begin{figure}[tbph]
\setlength{\unitlength}{1mm}
\begin{picture}(60,25)(10,0)
\put(-5,14){$\bf (a)$}\put(0,0){\vector(1,1){4}}
\put(-2,-2){$\bf{1}$} \put(5,4.5){$\sigma_1$}
\put(6.5,6.5){\vector(1,1){4}} \put(6.5,4){\vector(1,-1){4}}
\put(11,11.5){$\sigma_2$} \put(11,-2){$\psi_1$}
\put(13.5,0){\vector(1,1){4}} \put(13.5,11){\vector(1,-1){4}}
 \put(13.5,13.5){\vector(1,1){4}}
 \put(18.5,4.5){$\epsilon$}
 \put(18.5,18.25){$\bf 1$}
\put(20,6.5){\vector(1,1){4}} \put(20,4){\vector(1,-1){4}}
\put(20,17.5){\vector(1,-1){4}} \put(24.25,11.5){$\sigma_1$}
\put(24.5,-2){$\psi_2$} \put(27,0){\vector(1,1){4}}
\put(27,11){\vector(1,-1){4}} \put(27,13.5){\vector(1,1){4}}
\put(32,4.5){$\sigma_2$} \put(31,18.25){$\psi_1$}
\put(33.5,6.5){\vector(1,1){4}} \put(33.5,4){\vector(1,-1){4}}
\put(33.5,17.5){\vector(1,-1){4}} \put(38.25,11.5){$\epsilon$}
\put(38.25,-2){${\bf 1}$} \put(40.5,0){\vector(1,1){4}}
\put(40.5,11){\vector(1,-1){4}} \put(40.5,13.5){\vector(1,1){4}}
\put(44.5,4.75){$\sigma_1$} \put(44.5,18.25){$\psi_2$}
\put(47,6.5){\vector(1,1){4}} \put(47,4){\vector(1,-1){4}}
\put(47,17.5){\vector(1,-1){4}} \put(55,8){$\ldots$}
\end{picture}
\vspace*{10pt}
\begin{picture}(60,15)(0,0)
\put(-15,6){$\bf (b)$}

 \put(0.5,-2){$\bf{1}$}
 \put(2.0,-1){\vector(1,0){4}}
 \put(6.5,-2){$\fib$}
 \put(13.5,4.5){$\fib$}
 \put(8.5,0){\vector(1,1){4}}
 \put(8.5,-1){\vector(1,0){4}}
\put(13.25,-2){$\bf{1}$} \put(15.5,-1){\vector(1,0){4}}
\put(15.5,4){\vector(1,-1){4}} \put(20.5,-2){$\fib$}
\put(16,5){\vector(1,0){4}} \put(20.5,4.0){$\bf{1}$}
\put(22.5,5){\vector(1,0){4}} \put(22,0){\vector(1,1){4}}
\put(22.5,-1){\vector(1,0){4}} \put(27.25,-2){$\bf{1}$}
\put(29,-1){\vector(1,0){4}} \put(29.5,4){\vector(1,-1){4}}
\put(28,4.5){$\fib$} \put(30.5,5){\vector(1,0){4}}
\put(34,-2){$\fib$} \put(34.5,4.0){$\bf{1}$} \put(40,2.5){$\ldots$}
\end{picture}
\caption{
 {\bf (a)} Bratteli diagram for fusion of multiple $\sigma_1$ fields in the $\mathbb{Z}_3$ Parafermion CFT.
 {\bf (b)} Bratteli diagram for Fibonacci anyons.}
 \label{fig:bratteli1}\vspace*{-10pt}
\end{figure}
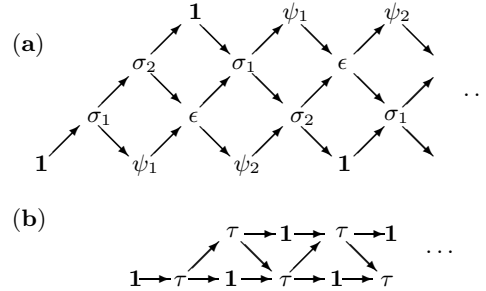

The general braiding behavior for the $\mathbb{Z}_3$ parafermions
has been worked out in \onlinecite{Slingerland01}.  It is trivial,
however, to work out the dimension of the degenerate space by
examining the Bratteli diagram Fig.~\ref{fig:bratteli1}a  (See the
appendix for explanation of this diagram). For example, if the
number of electrons is a multiple of 3 then they fuse together to
form the identity. Then, for example, with 6 quasiholes one has 5
paths of length 5 ending at $\bf 1$ (hence a 5 dimensional
degenerate space). However, if, for example, the number of electrons
is 1 mod 3, then the electrons fuse in threes to form $\bf 1$ but
there is one $\psi_1$ left over. Thus, the quasiholes must fuse
together to form $\psi_2$ which can fuse with the leftover $\psi_1$
to form $\bf 1$. In this case, for example, with 4 quasiholes there
is a 2 dimensional space. It is easy to see that (if the number of
electrons is divisible by 3) the number of blocks with $n$
quasiparticles is given by the $n-1^{st}$ Fibonacci number, notated
$\mbox{Fib(n-1)}$ defined by $\mbox{Fib}(1) = \mbox{Fib}(2)=1$ and
$\mbox{Fib}(n) =\mbox{Fib}(n-1)+\mbox{Fib}(n-2)$ for $n > 2$.

\begin{table}[tbph]

\begin{minipage}{3.25in}
\begin{tabular}{||l|l|c|c|c||}
\hline
State & CFT & $ \nu$ & $\psi_e$ & $\psi_{qh}$   \\
\hline \hline Laughlin & Boson & $\frac{1}{m}$  & $e^{i\phi\sqrt{m}}$
& $e^{i\phi/\sqrt{m}}$ \\
\hline Moore-Read  & Ising & $\frac{1}{m + 1}$ & $\pfepsilon
e^{i\phi\sqrt{m+1}}$ &  $\sigma e^{i\phi/(2 \sqrt{m+1})}$ \\
 \hline  ${\mathbb{Z}}_3$ RR  & $\mathbb{Z}_3$ Paraf. &
$\frac{1}{m + 2/3}$ &  $\psi_1 e^{i\phi\sqrt{m + 2/3}}$ & $\sigma_1
e^{i\phi/(3 \sqrt{m + 2/3})}$ \\ \hline
\end{tabular}
\end{minipage}
\vspace*{5pt}
 \caption{Summary of CFT-wavefunction correspondences discussed
 here.  In all cases $m \geq 0$.  Odd (even) $m$ represents a Fermi (Bose)
 wavefunction.
\label{tab:summary}\vspace*{-10pt} }
\end{table}

\subsection{Edge Excitations}
\label{sec:edge}

When a system in a chiral topological phase has a boundary
(as it must in any experiment), there must be gapless
excitations at the boundary \cite{Halperin81,Wen92b}.
To see this, consider the Chern-Simons
action on a manifold ${\cal M}$ with boundary $\partial{\cal M}$
\cite{Witten89,Elitzur89},
Eq. \ref{eqn:non-Abel-C-S-action}.
The change in the action under a gauge transformation,
${a_\mu} \rightarrow g{a_\mu}g^{-1} +
g{\partial_\mu}g^{-1}$, is:
\begin{equation}
\label{eqn:CS+boundary-variation}
S_{CS}[a] \rightarrow S_{CS}[a] + \frac{k}{4\pi} \int_{\partial {\cal M}}
\text{tr}(g^{-1}dg \wedge a)
\end{equation}
In order for the action to be invariant, we fix
the boundary condition so that the second term on the r.h.s.
vanishes. For instance,
we could take boundary condition
$\left(a_0^{\underline a}\right)_{\scriptscriptstyle |\partial{\cal M}} = 0$,
where $x_0$, $x_1$ are coordinates on the boundary of ${\cal M}$
and $x_2$ is the coordinate perpendicular to the
boundary of ${\cal M}$.
Then the action is invariant under all transformations
which respect this boundary condition, i.e. which satisfy
${\partial_0}g = 0$ on the boundary. We separate these into
gauge and global symmetries.
Functions $g:{\cal M}\rightarrow G$ satisfying
$g^{}_{\scriptscriptstyle |\partial{\cal M}}=1$ are the
gauge symmetries of the theory. (They necessarily satisfy
${\partial_0}g = 0$ since $x_0$ is a coordinate along the boundary.) Meanwhile,
functions $f:{\cal M}\rightarrow G$
which are independent of $x_0$ are really global symmetries
of the theory. The representations of this global symmetry
form the spectrum of edge excitations of the theory.
(The distinction between gauge and global transformations
is that a gauge transformation can leave the $t=0$ state unchanged
while changing the state of the system at a later time $t$. Since it
is, therefore, not possible for a given initial condition to
uniquely define the state of the system at a later time, all
physically-observable quantities must be invariant under
the gauge transformation. By contrast, a global symmetry, even if it
acts differently at different spatial points, cannot leave
the $t=0$ state unchanged while changing the state of the
system at a later time $t$. A global symmetry does not
prevent the dynamics from uniquely defining the state of the
system at a later time for a given initial condition. Therefore,
physically-observable quantities need not be invariant under
global transformations. Instead, the spectrum of the theory
can be divided into representations of the symmetry.)

With this boundary condition, the natural gauge choice for
the bulk is $a_0^{\underline a}=0$. We can then transform
the Chern-Simons functional integral into the chiral WZW functional
integral following the steps in Eqs.
\ref{eqn:C-S-delta-action}-\ref{eqn:WZW-derivation}
\cite{Elitzur89}:
\begin{multline}
S = \frac{k}{4\pi}\int_{\partial{\cal M}}
\text{tr}\left( {\partial_0} {U^{-1}}
{\partial_1} U\right)
\: +\\
\frac{k}{12\pi}\int_{{\cal M}}  \epsilon^{\mu\nu\lambda}
\text{tr}\left( {\partial_\mu} U\,{U^{-1}}
{\partial_\nu} U\,{U^{-1}}\,{\partial_\lambda}U\,{U^{-1}}\right)
\label{eqn:chiral-WZW}
\end{multline}
Note the off-diagonal form of the quadratic term (analogous
to the $z-\bar{z}$ form in Eq. \ref{eqn:WZW-derivation}), which follows
from our choice of boundary condition. This boundary condition
is not unique, however. The topological order of the bulk state
does not determine the boundary condition. It is determined by
the physical properties of the edge. Consider, for instance, the
alternative boundary condition
$\left(a_0^{\underline a}+
v a_1^{\underline a}\right)_{\scriptscriptstyle |\partial{\cal M}}=0$
for some constant $v$ with dimensions of velocity.
With this boundary condition, the quadratic term in the Lagrangian
will now be $\text{tr}\left( \left({\partial_0}+v{\partial_1}\right){U^{-1}}
{\partial_1} U\right)$ and the edge theory is the chiral
WZW model with non-zero velocity.

It is beyond the scope of this paper to discuss the chiral WZW model
in any detail (for more details, see
\onlinecite{Knizhnik84,Gepner86,Gepner87} ). However, there are a
few key properties which we will list now. The chiral WZW model is a
conformal field theory. Therefore, although there is a gap to all
excitations in the bulk, there are gapless excitations at the edge
of the system. The spectrum of the WZW model is organized into
representations of the Virasoro algebra and is further organized
into representations of the $G_k$ Kac-Moody algebra. For the sake of
concreteness, let us consider the case of SU(2)$_k$. The SU(2)$_k$
WZW model contains primary fields $\phi_j$, transforming in the
$j=0,1/2,1,\ldots,k/2$ representations. These correspond precisely
to the allowed quasiparticle species: when the total topological
charge of all of the quasiparticles in the bulk is $j$, the edge
must be in the sector created by acting with the spin $j$ primary
field on the vacuum.

The $G_k$ case is a generalization of
the U(1)$_m$ case, where $g=e^{i\phi}$ and the WZW model
reduces to a free chiral bosonic theory:
\begin{equation}
\label{eqn:chiral-boson-edge}
S = \frac{m}{4\pi}\int {d^2}x\, \left({\partial_t}+v{\partial_x}\right)\phi
\,{\partial_x}\phi
\end{equation}
(In Sec. \ref{sec:CS-theory}, we used $k$ for the coefficient
of an Abelian Chern-Simons term; here, we use $m$ to
avoid confusion with the corresponding coupling of the
SU(2) Chern-Simons term in situations in which both gauge
fields are present.)
The primary fields are $e^{in\phi}$, with $n=0,1,...,m-1$.
(The field $e^{im\phi}$ is either fermionic or bosonic
for $m$ odd or even, respecitvely, so it is not a primary field, but is, rather,
included as a generator of an extended algebra.)
A quantum Hall state will always have such a term
in its edge effective field theory; the U(1) is the symmetry responsible
for charge conservation and the gapless chiral excitations
(\ref{eqn:chiral-boson-edge}) carry the quantized Hall current.

Therefore, we see that chiral topological phases, such as fractional
quantum Hall states, must have gapless chiral edge excitations.
Furthermore, the conformal field theory which models the low-energy
properties of the edge is {\it the same} conformal field theory
which generates ground state wavefunctions of the corresponding
Chern-Simons action. This is clear from the fact that the two
derivations (Eqs. \ref{eqn:WZW-derivation} and \ref{eqn:chiral-WZW})
are virtually identical. The underlying reason is that Chern-Simons
theory is a topological field theory. When it is solved on a
manifold with boundary, it is unimportant whether the manifold is a
fixed-time spatial slice or the world-sheet of the edge of the
system. In either case, Chern-Simons theory reduces to the same
conformal field theory (which is an example of `holography'). One
important difference, however, is that, in the latter case, a
physical boundary condition is imposed and there are real gapless
degrees of freedom. (In the former case, the CFT associated with a
wavefunction for a fixed-time spatial slice may have apparent
gapless degrees of freedom which are an artifact of a gauge choice,
as discussed in Section \ref{sec:FQHE}.)

The WZW models do not, in general,
have free field representations. One well-known exception
is the equivalence between the SU(N)$_1$ $\times$ U(1)$_N$
chiral WZW model and $N$ free chiral Dirac fermions.
A somewhat less well-known exception is the
SU(2)$_2$ chiral WZW model, which
has a representation in terms of 3 free chiral Majorana fermions.
Before discussing this representation, we first
consider the edge excitations of a $p+ip$ superconductor,
which supports Ising anyons which, in turn,
differ from SU(2)$_2$ only by a U(1) factor.

Let us solve the Bogoliubov-de Gennes Hamiltonian (\ref{BdG})
with a spatially-varying chemical potential, just as
we did in Section \ref{sec:pwave}. However, instead of
a circular vortex, we consider an edge at $y=0$:
\begin{equation}
\mu(y) = \Delta \,h(y) ,
\end{equation}
with $h(y)$ large and positive for large $y$,
and $h(y)<0$ for $y<0$; therefore, the electron
density will vanish for $y$ large and positive.
Such a potential defines an edge at $y=0$.
There are low-energy eigenstates of the BdG Hamiltonian
which are spatially localized near $y=0$:
\begin{equation}
\label{eqn:zero-mode-wvfn2} \phi^{edge}_E({\bf x}) \approx  e^{ik
x} e^{- \int_0^y h(y^\prime) dy^\prime} \phi_{0} ,
\end{equation}
with $\phi_{0} = {1\choose 1}$ an eigenstate of $\sigma^x$.  This
wavefunction describes a chiral wave propagating in the $x-$direction
localized on the edge, with wave vector $k = E/\Delta$.
A more complete solution of the superconducting Hamiltonian in
this situation would involve
self-consistently solving the BdG equations, so that
both the density and the gap $\Delta(y)$ would vanish
for large positive $y$. The velocity of the chiral edge mode
would then depend on how sharply $h(y)$ varies. However,
the solutions given above with fixed constant $\Delta$ are sufficient
to show the existence of the edge mode.

If we define an edge fermion operator $\psi({\bf x})$:
$$
\psi({\bf x}) = e^{-\int_0^y h(y^\prime) dy^\prime}
\sum_{k >0} [ {\psi}_k e^{ikx} \phi_0 + {\psi}_{-k}
e^{-ikx} \phi_0 ].
$$
The fermion operators, $\psi_{k}$,  satisfy $\psi_{-k} = \psi_k^\dagger$, so
$\psi(x) = \sum_k \psi_k e^{ikx}$ is a real
Majorana field, $\psi(x) = \psi^\dagger(x)$.
The edge Hamitonian is:
\begin{equation}
\label{eqn:BdG-edge-Ham}
\hat{\cal H}_{edge}
= \sum_{k>0} v_n k \,\psi_k^\dagger \psi_k
= \int dx \,\psi(x) (-iv_n \partial_x ) \psi(x)  ,
\end{equation}
where the edge velocity $v = \Delta$. The Lagrangian density takes the form:
\begin{equation}
{\cal L}_\text{fermion} =  i {\psi}(x) (\partial_t + v_n \partial_x) {\psi}(x)
\end{equation}

The $2D$ Ising model can be mapped onto the problem
of (non-chiral) Majorana fermions on a lattice. At the critical
point, the Majorana fermions become massless. Therefore,
the edge excitations are the right-moving chiral part of the
critical Ising model. (This is why the vortices of a $p+ip$ superconductor
are call Ising anyons.) However, the edge excitations have
non-trivial topological structure for the same reason that
correlation functions of the spin field are non-trivial in the Ising model:
while the fermions are free, the Ising spin field is non-local in
terms of the fermions, so its correlations are non-trivial.
The Ising spin field $\sigma(z)$ inserts a branch cut running from
$z={v_n}x+it$ to infinity for the fermion $\psi$. This is precisely what
happens when a flux $hc/2e$ vortex is created in a $p+ip$ superconductor.

The primary fields of the free Majorana fermion are ${\bf 1}$, $\sigma$,
and $\psi$ with respective scaling dimensions $0$, $1/16$, and $1/2$,
as discussed in Section \ref{sec:FQHE}. When there is an odd number
of flux $hc/2e$ vortices in the bulk, the edge is in the $\sigma(0)|0\rangle$
sector. When there is an even number, the edge is in either the
$|0\rangle$ or $\psi(0)|0\rangle$ sectors, depending on whether there
is an even or odd number of fermions in the system. So long as quasiparticles
don't go from the edge to the bulk or vice versa, however, the system
remains in one of these sectors and all excitations are simply free fermion
excitations built on top of the ground state in the relevant sector.

However, when a quasiparticle tunnels from the edge to the bulk (or through
the bulk), the edge goes from one sector to another -- i.e. it is acted
on by a primary field. Hence, in the presence of a constriction at which
vortices of fermions can tunnel from one edge to another, the
edge Lagrangian of a $p+ip$ superconductor is \cite{Fendley07a}:
\begin{multline}
\label{eqn:p+ip-action}
S = \int d\tau\,dx\, \left({{\cal L}_\text{fermion}}({\psi_a})
+ {{\cal L}_\text{fermion}}({\psi_b})\right) \\
+ \int d\tau\, \lambda_\psi\, i{\psi_a}{\psi_b}
+ \int d\tau\, \lambda_\sigma {\sigma_a}{\sigma_b}
\end{multline}
where $a,b$ denote the two edges. (We have dropped all irrelevant
terms, e.g descendant fields.) In other words, although the edge
theory is a free theory in the absence of coupling to the bulk or
to another edge through the bulk, it is perturbed by primary fields when
quasiparticles can tunnel to or from the edge through the bulk.
The topological structure of the bulk constrains the edge through
the spectrum of primary fields.

As in the discussion of Section \ref{sec:FQHE}, the edge of the
Moore-Read Pfaffian quantum Hall state is a chiral Majorana fermion
together with a free chiral boson $\phi$ for the charge sector of the theory.
As in the case of a $p+ip$ superconductor,
the primary fields of this theory determine how the edge is
perturbed by the tunneling of quasiparticles between two edges
through the bulk \cite{Fendley06,Fendley07a}:
\begin{multline}
\label{eqn:five-halves-action}
S = \int d\tau\,\left[\int dx\, \left(
{\cal L}_\text{edge}({\psi_a},{\phi_a})
+ {\cal L}_\text{edge}({\psi_b},{\phi_b})\right)\right.\\
+ \lambda_{1/2} \, \cos(({\phi_a}(0)-{\phi_b}(0))/\sqrt{2})
+ \lambda_{\psi,0}\, i{\psi_a}{\psi_b}\\
+ \lambda_{1/4} \,{\sigma_a}(0) {\sigma_b}(0)\,
\cos(({\phi_a}(0)-{\phi_b}(0))/2\sqrt{2})\Big]
\end{multline}
The most relevant coupling is $\lambda_{1/4}$, so
the tunneling of charge $e/4$ quasiparticles dominates
the transport of charge from one edge to the other at
the point contact. (The tunneling of charge $e/2$ quasiparticles
makes a subleading contribution while the tunneling of neutral
fermions contributes only to thermal transport.)
At low enough temperatures, this relevant tunneling process
causes the point contact to be pinched off \cite{Fendley06,Fendley07a},
but at temperatures that are not too low, we can treat the tunneling
of $e/4$ quasiparticles perturbatively and neglect other the
other tunneling operators. Of course, the structure of the edge
may be more complex than the minimal structure dictated by
the bulk which we have analyzed here. This depends on the details of the confining
potential defining the system boundary, but at low enough temperatures,
the picture described here should still apply.
Interesting information about the non-Abelian character of the
Moore-Read Pfaffian state can be obtained from
the temperature dependence of the tunneling conductance
\cite{Fendley06,Fendley07a} and from current noise \cite{Bena06}.

Finally, we return to SU(2)$_2$.
The SU(2)$_2$ WZW model is a triplet of chiral Majorana fermions,
$\psi_1$, $\psi_2$, $\psi_3$ -- i.e. three identical copies of
the chiral Ising model. This triplet is the spin-$1$ primary
field (with scaling dimension $1/2$). The spin-$1/2$ primary field
is roughly $\sim {\sigma_1}{\sigma_2}{\sigma_3}$ with dimension $3/16$
(a more precise expression involves the sum of products such as
${\sigma_1}{\sigma_2}{\mu_3}$, where $\mu$ is the Ising disorder
operator dual to $\sigma$). This is one of the primary
differences between the Ising model and SU(2)$_2$:
$\sigma$ is a dimension $1/16$ field, while the spin-$1/2$ primary field
of SU(2)$_2$ has dimension $3/16$.
Another way to understand the difference between the two models
is that the SU(2)$_2$ WZW model has two extra Majorana fermions.
The pair of Majorana fermions can equally well be viewed as
a Dirac fermion or, through bosonization, as a free chiral boson,
which has U(1) symmetry. Thus, the Ising model is often written
as SU(2)$_2$/U(1) to signify that the the U(1) chiral boson has been
removed. (This notion can be made precise with the notion of
a {\it coset} conformal field theory \cite{Senechal97}
or by adding a U(1) gauge field
to the 2D action and coupling it to a U(1) subgroup of the SU(2)
WZW field $g$ \cite{Gawedzki88,Karabali89}.
The gauge field has no Maxwell term, so it serves only
to eliminate some of the degrees of freedom, namely the U(1) piece.)
As we discussed in subsection \ref{sec:Jones}, these differences
are also manifested in the bulk, where they lead to some differences
in the Abelian phases which result from braiding but do not change the
basic non-Abelian structure of the state.

On the other hand, the edge of the Moore-Read Pfaffian quantum Hall state
is a chiral Majorana fermion together with a free chiral boson $\phi$
which carries the charged degrees of freedom. So we restore
the chiral boson which we eliminated in passing from SU(2)$_2$ to
the Ising model, with one important difference. The compactification
radius $R$ (i.e., the theory is invariant under
$\phi\rightarrow \phi + 2 \pi R$)
of the charged boson need not be the same as that
of the boson which was removed by cosetting.
For the special case of bosons at $\nu=1$, the boson is, in fact, at
the right radius. Therefore, the charge boson can be fermionized
so that there is a triplet of Majorana fermions. In this case, the edge theory is the
SU(2)$_2$ WZW model \cite{Fradkin98}. In the
case of electrons at $\nu=2+1/2$, the chiral boson is not at this
radius, so the edge theory is U(1)${_2}\times$Ising, which is not
quite the SU(2)$_2$ WZW model.

\subsection{Interferometry with Anyons}

\label{sec:experiments2}

In Section \ref{part1} of this review we described an interference
experiment that is designed to demonstrate the non-Abelian
statistics of quasiparticles in the $\nu=5/2$ state. We start this
section by returning to this experiment, and using it as an exercise for
the application of the calculational methods reviewed above. We then
generalize our analysis to arbitrary SU(2)$_k$ non-Abelian states
and also describe other experiments that share the same goal.

In the experiment that we described in Section \ref{part1},
a Fabry-Perot interference device is made of a Hall bar
perturbed by two constrictions (see Fig. \ref{fig:Fabry-Perot}). The
back-scattered current is measured as a function of the area of
the cell enclosed by the two constrictions and of the magnetic
field. We assume that the system is at $\nu=5/2$ and consider interference
experiments which can determine if the electrons are in the Moore-Read Pfaffian
quantum Hall state.

Generally speaking, the amplitude for back--scattering is a sum
over trajectories that wind the cell $\ell$ times, with
$\ell=0,1,2...$ an integer. The partial wave that winds the cell
$\ell$ times, winds the $n$ quasiparticles localized inside the
cell $\ell$ times. From
the analysis in Section \ref{sec:pwave}, if the electrons are in the
Pfaffian state, the unitary transformation that the tunneling quasiparticle
applies on the wave function of the zero energy modes is
\begin{equation}
\left ({{\hat U}_n}\right )^\ell=
\left [e^{i\alpha_n}\gamma_a^n\prod_{i=1}^n\gamma_i\right]^\ell
\label{unitarytrans}
\end{equation}
where the $\gamma_i$'s are the Majorana modes of the localized bulk
quasiparticles, $\gamma_a$ is the Majorana mode of the
quasiparticle that flows around the cell, and $\alpha_n$ is an
Abelian phase that will be calculated below.

The difference between the even and odd values of $n$, that we
described in Section \ref{part1} of the review, is evident from Eq.
(\ref{unitarytrans}) when we we look at the lowest order,
$\ell=1$. For even $n$, ${\hat U}_n$ is independent of $\gamma_a$.
Thus, each tunneling quasiparticle applies the same unitary
transformation on the ground state. The flowing current then {\it
measures} the operator ${\hat U}_n$ (more precisely, it measures
the interference term, which is an hermitian operator. From that
term the value of ${\hat U}_n$ may be extracted). In contrast,
when $n$ is odd the operator ${\hat U}_n$ depends on $\gamma_a$. Thus, a
different unitary operation is applied by every incoming
quasiparticle. Moreover, the different unitary operators do not
commute, and share no eigenvectors. Thus, their expectation values
average to zero, and no interference is to be observed. This
analysis holds in fact for all odd values of $\ell$.

The phase $\alpha_n$ is composed of two parts. First, the
quasiparticle accumulates an Aharonov-Bohm phase of $2\pi
e^*\Phi/hc$, where $e^*=e/4$ is the quasiparticle charge for
$\nu=5/2$ and $\Phi$ is the flux enclosed.
And second, the tunneling quasiparticle accumulates a
phase as a consequence of its interaction with the $n$ localized
quasiparticles. When a charge $e/4$ object goes around $n$ flux
tubes of half a flux quantum each, the phase it accumulates is
$n\pi/4$.

Altogether, then, the unitary transformation (\ref{unitarytrans}) has
two eigenvalues. For even $n$, they are  $\left(\pm
i\right)^{nl/2}$. For odd $n$, they are $\left(\pm i\right
)^{(n-1)l/2}$. The back--scattered current then assumes the following
form \cite{Stern06},
\begin{equation}
I_{bs}=\sum_{m=0}^\infty I_m \cos^2{mn\frac{\pi}{2}} \cos{m(\phi+\frac{{\tilde n}\pi}{4}+\frac{\pi\alpha}{2})}
\label{generalresistance}
\end{equation}
where ${\tilde n}=n$ for $n$ even, and ${\tilde n}=n+1$ for $n$ odd.
The $m^\text{th}$ term of this sum is the contribution from a process that loops around $m$ times, which vanishes if $n$ and $m$
are both odd.

We can restate this analysis using the CFT description of the
Moore-Read Pfaffian state. Charge $e/4$ quasiparticles are
associated with the operator $\sigma \, e^{i\phi/\sqrt{8}}$ operators.
The fusion of $n$ such quasiparticles is then to
\begin{equation}
e^{in\phi/\sqrt{8}}\times\left \{
\begin{array}{c}
1\\
\psi\\
\sigma
\end{array}
\right.
\label{fusionbulk}
\end{equation}
where either of the first two is possible for even $n$, and the last is
the outcome of the fusion for odd $n$. In order to determine the
effect of braiding an incoming quasiparticle around the $n$ bulk ones,
we consider the possible fusion channels of one quasiparticle
with (\ref{fusionbulk}). The fusion of the bosonic factors (i.e. the electrical
charge) is:
\begin{equation}
e^{in\phi(z_1)/\sqrt{8}}\times e^{i\phi(z_2)/\sqrt{8}}\rightarrow e^{i(n+1)\phi(z_1)/\sqrt{8}}(z_1-z_2)^{-n/8}
\label{bosonicfusion}
\end{equation}
Thus, when the incoming quasiparticle, at coordinate $z_2$,
encircles the bulk $\ell$ times, it accumulates a phase of
$2\pi\times(n/8)\times\ell=n\ell\pi/4$ purely as a result of the U(1)
part of the theory. Now consider the neutral sector. The fusion of the $\sigma$
operator depends on the state of the bulk. When the bulk is has total topological
charge ${\bf 1}$, the fusion is trivial, and does not involve any
accumulation of phases. When the bulk has total topological charge $\psi$,
the fusion is:
\begin{equation}
\sigma(z_2)\times\psi(z_1)\rightarrow \sigma(z_1)\times (z_1-z_2)^{-1/2}
\label{fusionpsi}
\end{equation}
and an extra phase of $\pi \ell$ is accumulated when the incoming
quasiparticle winds the bulk quasiparticles $\ell$ times. When the
bulk has total topological charge $\sigma$, i.e. when $n$ is odd, the non-Abelian
fusion rule applies (see Eq. \ref{eq:sigmasigma}), and
\begin{equation}
\sigma(z_1)\times\sigma(z_2)\rightarrow
(z_1-z_2)^{-1/8}\left[1+(z_1-z_2)^{1/2}\psi(z_1)\right]
\label{non-Abelianfusion}
\end{equation}
Since the probability for the two fusion outcomes is equal
\footnote{This follows from $N_{\sigma\sigma}^{\bf 1} =
N_{\sigma\sigma}^{\psi}=1$.}, for
any odd $\ell$ we get two interference patterns that are mutually
shifted by $\pi$, and hence mutually cancel one another, while for
even $\ell$ we get an extra phase of $\ell\pi/4$. Altogether, this
reproduces the expression (\ref{generalresistance}).

Now let us consider the same calculation using the relation
between Chern-Simons theory and the Jones polynomial.
For simplicity, we will just compute the current due to a single
backscattering and neglect multiple tunneling processes, which
can be computed in a similar way. The elementary quasiparticles
have $j=1/2$. These are the quasiparticles which will tunnel at the point
contacts, either encircling the bulk quasiparticles or not. (Other
quasiparticles will give a sub-leading contribution to the current
because their tunneling amplitudes are smaller and less relevant in
the RG sense.) First, consider the case in
which there is a single $j=1/2$ quasiparticle in the bulk.
The back-scattered current is of the form:
\begin{equation}
I_{bs} = {I_0} + {I_1}\mbox{Re}\!\left\{e^{i\phi}
\bigl\langle \chi \bigl| \rho\!\left({\sigma_2^2}\right) \bigr| \chi \bigr\rangle\right\}
\label{eqn:bscurrent-non-Abelian}
\end{equation}
The matrix element on the right-hand-side is given by
the evaluation of the link in Figure \ref{fig:B-calc}a
\cite{Fradkin98,Bonderson06a}
(up to a normalization of the bra and ket; see Sec. \ref{sec:Jones}).
It is the matrix element between a state $|\chi\rangle$ is the state
in which $1$ and $2$ fuse to the trivial particle as do $3$ and $4$
and the state $\rho\!\left({\sigma_2^2}\right) | \chi \rangle$.
The former is the state in which the tunneling quasparticle (qp. $3$)does
not encircle the bulk quasiparticle (qp. $2$); the latter is the state in which
it does. The matrix element between these two states determines the
interference.

\begin{figure}[tb]
\centerline{\includegraphics[width=3.5in]{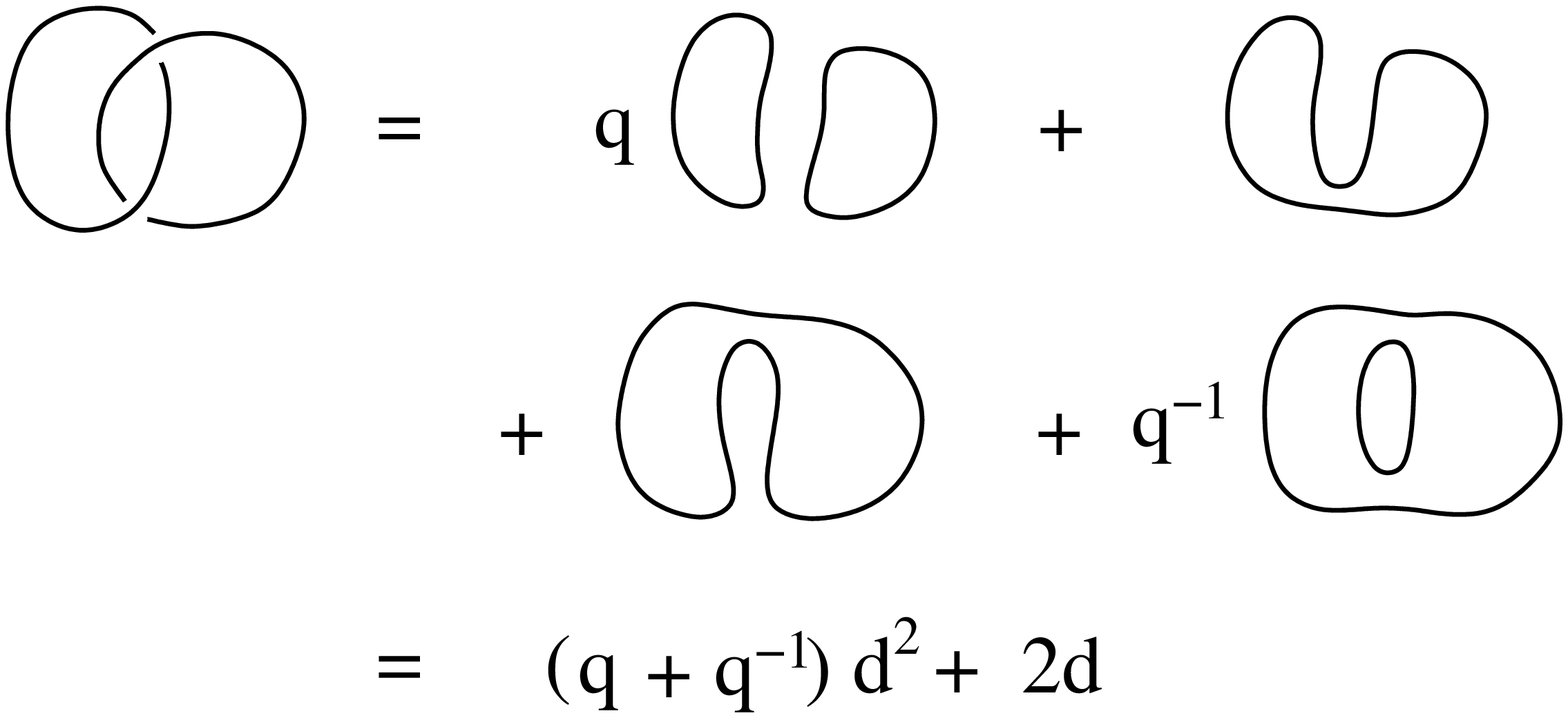}}
\caption{Using the recursion relation (\ref{eq:recursion}),
we can evaluate $\langle \chi | \rho\!\left({\sigma_2^2}\right)
|\chi\rangle$.}
\label{fig:hopf}
\end{figure}

Using the recursion relation (\ref{eq:recursion}) as shown
in Figure \ref{fig:hopf}, we obtain:
\begin{eqnarray}
\langle \chi | \rho\!\left({\sigma_2^2}\right) |\chi\rangle &=&
\left(q+q^{-1}\right) {d^2} + 2d\cr
&=& -{d^3} + 2d
\end{eqnarray}
For $k=2$, $d=\sqrt{2}$, so this vanishes. Consequently, the
interference term in (\ref{eqn:bscurrent-non-Abelian}) also vanishes,
as we found above by other methods.
The case of an arbitrary odd number of quasiparticles in the island
is similar.

\begin{figure}[thb]
\includegraphics[width=3.25in]{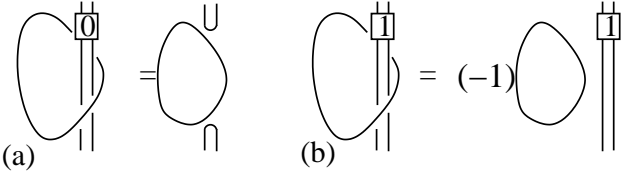}
\caption{We can obtain the result of taking a $j=1/2$ quasiparticle
around a qubit from the two diagrams in this figure.
In (a) the qubit is in the state $0$,
while in (b) it is in state $1$. These figures are similar to the left-hand-side
of Fig. \ref{fig:hopf}, but with the loop on the right replaced by
a loop with (a) $j=0$ or (b) $j=1$.}
\label{fig:interfere-link}
\end{figure}

Now consider the case in which there are an even number of
quasiparticles in the island. For the sake of simplicity, we
consider the case in which there are two quasiparticles in
the bulk, i.e. a qubit. The pair can either fuse to
$j=0$ or $j=1$. In the former case, it is clear that
no phase is acquired, see Fig. \ref{fig:interfere-link}a.
In the latter case, the recursion
rule (\ref{eq:recursion}) gives us a $-1$, as depicted
in figure \ref{fig:interfere-link}. This difference allows us to
read out the value of a topologically-protected qubit \cite{DasSarma05}.

What happens if the qubit is in a superposition of $j=0$ and $j=1$?
The interference measurement causes the tunneling quasiparticles
to become entangled with the bulk quasiparticle
\cite{Overbosch01,Freedman06,Bonderson07}. When the integrated current
is large enough that many quasiparticles have tunneled and
equilibrated at the current leads, the $j=0$ and $j=1$ possibilities
will have decohered. The measurement will see one of the two possibilities
with corresponding probabilities.

The experiment that we analyzed above for $\nu=5/2$ may be analyzed
also for other non-Abelian states. The computation using knot invariants
can be immediately adapted to other SU(2)$_k$ states by
simply replacing $d=\sqrt{2}$ with $d=2\,\cos\pi/(k+2)$.
We should calculate the value of the Hopf link as in figures \ref{fig:hopf}
and \ref{fig:interfere-link}, with one of the loops corresponding to the tunneling
quasiparticle and the other loop corresponding to the
total topological charge of the bulk quasiparticles.
The result can be written in the more general form \cite{Bonderson06b}:
\begin{equation}
I_{bs}(a) = {I_0} + {I_1} \left|M_{ab}\right| \, \cos(\beta+\theta_{ab})
\label{eqn:bscurrent-general}
\end{equation}
where $M_{ab}$ is defined in terms of the $S$-matrix:
\begin{equation}
M_{ab} = \frac{S_{ab}S_{\bf 11}}{S_{{\bf 1}a}S_{{\bf 1}b}}
\end{equation}
and $M_{ab}=\left|M_{ab}\right|\,e^{i\theta_{ab}}$.
The expression (\ref{eqn:bscurrent-general}) gives the
current to due to $a$ quasiparticles if the quasiparticles
in the bulk fuse to $b$. If the contribution of $j=1/2$ quasiparticles
dominates, as in the $\nu=5/2$ case, then we should set $a=\frac{1}{2}$
in this expression. For the level $k=3$ case, taking $a=\frac{1}{2}$,
$\left|M_{ab}\right|=1$ for $b=0,\frac{3}{2}$ while
$\left|M_{ab}\right|=\phi^{-2}$ for $b=\frac{1}{2}, 1$,
where $\phi$ is the golden mean, $\phi=(1+\sqrt{5})/2$.
(In $\mathbb{Z}_3$ parafermion language, $b=0,\frac{3}{2}$
correspond to the fields ${\bf 1},\psi^{}_{1,2}$ while $b=\frac{1}{2},1$
correspond to the fields $\sigma^{}_{1,2}, \varepsilon$.)

Finally, we can analyze the operation of an interferometer
using the edge theory (\ref{eqn:five-halves-action}). The preceding discussion esentially
assumed that the current is carried by non-interacting anyonic
quasiparticles. However, the edge is gapless and, in general,
does not even have well-defined quasiparticles. Therefore,
a computation using the edge theory is more complete.
The expected results are recovered since they are determined
by the topological structure of the state, which is shared by both the
bulk and the edge. However, the edge theory also enables one
to determine the temperature and voltage dependences of $I_0$, $I_1$, ... in
(\ref{generalresistance}), (\ref{eqn:bscurrent-non-Abelian})
\cite{Ardonne07b,Fidkowski07,Bishara07}.
As is discussed in these papers, at finite temperature,
interference will not be visible if the two
point contacts are further apart than
the thermal length scale $L_\phi$, where
$L^{-1}_\phi = {k_B}T \left(\frac{1/8}{v_c}+\frac{1/8}{v_n}\right)$,
if the charged and neutral mode velocities are ${v_c}$, ${v_n}$.
Another important feature is that
the interference term (when it is non-vanishing) is oscillatory
in the source-drain voltage while the $I_0$ term has a simple power law
dependence.

The assumption that the edge and the bulk are well separated is
crucial to that above calculations of interference, but in practice this
may not be the case. When there is bulk-edge tunneling one might
imagine that a quasiparticle moving along the edge may tunnel into
the bulk for a moment and thereby evade encircling some of the
localized quasiparticles thus smearing out any interference pattern.
The first theoretical steps to analysing this situation have been taken in
\cite{Overbosch07,Rosenow07b} where tunnling to a single impurity is
considered. Surprisingly it is found that the interfernece pattern is
full strength both in the strong tunneling limit as well as in the weak tunneling limit. 

While the experiment we described for the
$\nu=5/2$ state does not require a precise determination of $n$, as
it is only its parity that determines the amplitude of the
interference pattern, it does require that the number $n$ does not
fluctuate within the duration of the experiment. Generally,
fluctuations in $n$ would be suppressed by low temperature, large
charging energy and diminished tunnel coupling between the bulk and
the edge. However, when their suppression is not strong enough, and
$n$ fluctuates over a range much larger than $1$ within the time of
the measurement, two signatures of the non-Abelian statistics of the
quasiparticles would still survive, at least as long as the
characteristic time scale of these fluctuations is much longer than
the time between back--scattering events. First, any change in $n$
would translate to a change in the back--scattered current, or the
two-terminal conductance of the device. Hence, fluctuations in $n$
would introduce current noise of the telegraph type, with a unique
frequency dependence \cite{Grosfeld06b}.
Second, fluctuations in $n$ would suppress all
terms in Eq. (\ref{generalresistance}) other than those where $m=4k$
with $k$ an integer. Thus, the back--scattered current will have a
periodicity of one flux quatum $\Phi_0$, and the visibility of the
flux oscillations, for weak back--scattering, would be
$\frac{I_4}{I_0}\propto{I_0^3}$.

A similar relation holds also for another type of interference
experiment, in which the interferometer is of the Mach-Zehnder type,
rather than the Fabry-Perot type. (A Mach-Zehnder interferometer has
already been constructed in the integer quantum Hall regime
\cite{Ji03}). If we are to describe the Mach-Zehnder interferometer
in a language close to that we used for the Fabry-Perot one, we
would note the following important differences: first, no multiple
back--scattering events are allowed. And second, since the area
enclosed by the interfering partial waves now encompasses the inner
edge, the quantum state of the encircled area {\it changes with each
tunneling quasiparticle}. Thus, it is not surprising that the
outcome of an interference experiment in a Mach-Zehnder geometry
will be very close to that of a Fabry-Perot experiment with strong
fluctuations in $n$.  The telegraph noise in the Fabry-Perot
case\cite{Grosfeld06b} becomes shot noise in the Mach-Zehnder case.
Remarkably\cite{Feldman07} the effective charge extracted from that
noise carries a signature of the non-Abelian statistics: as the flux
is varied, the charge changes from $e/4$ to about $3e$.

Other than interference experiments, there are several proposals for
experiments that probe certain aspects of the physics of non-Abelian
states. The degeneracy of the ground state in the presence of
vortices may be probed\cite{Grosfeld06a} by the consequences of its
removal: when the filling factor is $\nu=5/2+\epsilon$ with
$\epsilon\ll 1$, quasiparticles are introduced  into the bulk of the
system,  with a density proportional to $\epsilon$. For a clean
enough sample, and a low enough density, the quasiparticles form a
lattice. In that lattice, the Majorana zero modes of the different
quasiparticles couple by tunneling, and the degeneracy of the ground
states is removed. The subspace of multiply-degenerate ground states
is then replaced by a band of excitations. The neutrality of the
Majorana modes is removed too, and the excitations carry a charge
that is proportional to their energy. This charge makes these modes
weakly coupled to an externally applied electric field, and provides
a unique mechanism for a dissipation of energy, with a
characteristic dependence on the wave vector and frequency of the
electric field. Since the tunnel coupling between neighboring
quasiparticles depends exponentially on their separation, this
mechanism will be exponentially sensitive to the distance of the
filling factor from 5/2 \cite{Grosfeld06a}.

\subsection{Lattice Models with $P,T$-Invariant
Topological Phases}
\label{sec:P-T-Invariant}

Our discussion of topological phases has revolved around fractional
quantum Hall states because these are the only ones known to occur
in nature (although two dimensional $^{3}$He-A
\cite{Leggett75,Volovik03} and Sr$_2$RuO$_4$ may join this list
\cite{Xia06,Kidwingira06}). However, there is nothing inherent in
the definition of a topological phase which consigns it to the
regime of high magnetic fields and low temperatures. Indeed, highly
idealized models of frustrated magnets also show such phases, as we
have discussed in section \ref{sec:Othersystems}. Of course, it is
an open question whether these models have anything to do with any
real electronic materials or their analogs with cold atoms in
optical lattices, i.e. whether the idealized models can be
adiabatically connected to more realistic models. In this section,
we do not attempt to answer this question but focus, instead, on
understanding how these models of topological phases can be solved.
As we will see, their solubility lies in their incorporation of the
basic topological structure of the corresponding phases.

One way in which a topological phase can emerge from
some microscopic model of interacting electrons, spins, or
cold atoms is if the low-lying degrees of freedom of the microscopic
model can be mapped to the degrees of freedom of the topological
phase in question. As we have seen in section \ref{sec:Jones},
these degrees of freedom are Wilson loops (\ref{eqn:Wilson-loop-def}).
Loops are the natural degrees of freedom in a topological
phase because the topological charge of a particle or collection
of particles can only be determined, in general, by taking a test particle
around the particle or collection in question.
Therefore, the most direct way in which a system can settle
into a topological phase is if the microscopic degrees of freedom
organize themselves so that the low-energy degrees of freedom
are loops or, as we will see below, string nets
(in which we allow vertices into which three lines
can run). As we will describe more fully below, the Hilbert space
of a non-chiral topological phase can be described very roughly
as a `Fock space for loops' \cite{Freedman04a}. Wilson loop operators
are essentially creation/annihilation operators for loops. The Hilbert
space is spanned by basis states which can be built up by acting
with Wilson loop operators on the state with no loops, i.e.
$|{\gamma_1}\cup\ldots\cup{\gamma_n}\rangle =
W[{\gamma_n}]\ldots W[{\gamma_1}]|\emptyset\rangle$
is (vaguely) analgous to $|{k_1},\ldots,{k_2}\rangle\equiv
{a^\dagger_{k_n}}\ldots{a^\dagger_{k_1}}|0\rangle$.
An important difference is that the states in the topological
theory must satisfy some extra constraints in order to
correctly represent the algebra of the operators $W[{\gamma}]$.
If we write an arbitrary state $|\Psi\rangle$ in the basis given above,
$\Psi[{\gamma_1}\cup\ldots\cup{\gamma_n}]
=\langle\Psi | {\gamma_1}\cup\ldots\cup{\gamma_n}\rangle$,
then the ground state(s) of the theory are linearly independent
$\Psi[{\gamma_1}\cup\ldots\cup{\gamma_n}]$
satisfying some constraints.

In fact, we have already seen an example of this in section
\ref{sec:Othersystems}: Kitaev's toric code model
(\ref{eqn:toric-code}). We now represent the solution in a way which
makes the emergence of loops clear. We color every link of the
lattice on which the spin points up. Then, the first term in
(\ref{eqn:toric-code}) requires that there be an even number of
colored links emerging from each site on the lattice. In other
words, the colored links form loops which never terminate. On the
square lattice, loops can cross, but they cannot cross on the
honeycomb lattice; for this reason, we will often find it more
convenient to work on the honeycomb lattice. The second term in the
Hamiltonian requires that the ground state satisfy three further
properties: the amplitude for two configurations is the same if one
configuration can be transformed into another simply by (1)
deforming some loop without cutting it, (2) removing a loop which
runs around a single plaquette of the lattice, or (3) cutting open
two loops which approach each other within a lattice spacing and
rejoining them into a single loop (or vice-versa), which is called
{\it surgery}. A vertex at which the first term in the Hamiltonian
is not satisfied is an excitation, as is a plaquette at which the
second term is not satisfied. The first type of excitation acquires
a $-1$ when it is taken around the second.

The toric code is associated with the low-energy physics of the
deconfined phase of $\mathbb{Z}_2$ gauge theory
\cite{Fradkin79,Kogut79};
see also \onlinecite{Senthil00} for an application to strongly-correlated
electron systems). This low-energy physics can be described by
an Abelian BF-theory \cite{Hansson04}:
\begin{eqnarray}
\label{eqn:Abelian-BF}
{\cal S} &=&
\frac{1}{\pi}\int {e_\mu}\epsilon^{\mu\nu\lambda}{\partial_\nu}{a_\lambda}\cr
&=& {\cal S}_{CS}\left(a+\mbox{$\frac{1}{2}$}e\right) -
{\cal S}_{CS}\left(a-\mbox{$\frac{1}{2}$}e\right)
\end{eqnarray}
${e_\mu}$ is usually denoted $b_\mu$ and $\epsilon^{\mu\nu\lambda}
{\partial_\nu}{a_\lambda}=\frac{1}{2}\epsilon^{\mu\nu\lambda}f_{\nu\lambda}$,
hence the name. Note that this theory is non-chiral. Under a combined
parity and time-reversal transformation, $e_\mu$ must change sign,
and the action is invariant. This is important since it enables the
fluctuating loops described above to represent the Wilson loops of the
gauge field $a_\mu$. In a chiral theory, it is not clear how to do this
since $a_1$ and $a_2$ do not commute with each other.
They cannot both be diagonalized; we must arbitrarily
choose one direction in which Wilson loops are diagonal operators.
It is not clear how this will emerge from some microscopic model,
where we would expect that loops would not have a preferred direction,
as we saw above in the toric code.
Therefore, we focus on non-chiral phases, in particular,
the SU(2)$_k$ analog of (\ref{eqn:Abelian-BF}) \cite{Cattaneo95}:
\begin{eqnarray}
\label{eqn:non-Abelian-BF}
{\cal S} &=& {\cal S}^{k}_{CS}(a+e) - {\cal S}^{k}_{CS}(a-e)\cr
&=& \frac{k}{4\pi}\int \text{tr}\left(e\wedge f + \frac{1}{3}e\wedge e\wedge e\right)
\end{eqnarray}
We will call this theory {\it doubled} SU(2)$_k$ Chern-Simons theory
\cite{Freedman04a}.

We would like a microscopic lattice model whose low-energy Hilbert space
is composed of wavefunctions
$\Psi[{\gamma_1}\cup\ldots\cup{\gamma_n}]$
which assign a complex amplitude
to a given configuration of loops. The model must differ from the
toric code in the constraints which it imposes on these wavefunctions.
The corresponding constraints for (\ref{eqn:non-Abelian-BF})
are essentially the rules for Wilson loops which we discussed
in subsection \ref{sec:Jones} \cite{Freedman04a}. For instance,
ground state wavefunctions should {\it not} give the same
the amplitude for two configurations if one configuration
can be transformed into another simply by removing a loop which runs
around a single plaquette of the lattice. Instead, the amplitude
for the former configuration should be larger by a factor of $d=2\cos\pi/(k+2)$,
which is the value of a single unknotted Wilson loop.
Meanwhile, the appropriate
surgery relation is not the joining of two nearby loops into a single one,
but instead is the condition that when $k+1$ lines come close together,
the amplitudes for configurations in which they are cut open and rejoined
in different ways satisfy some linear relation. This relation is essentially
the requirement that the $j=(k+1)/2$ Jones-Wenzl projector should
vanish within any loop configuration, as we might expect since a
Wilson loop carrying the corresponding SU(2) representation should vanish.

The basic operators in the theory are Wilson loops, $W[\gamma]$, of the
gauge field $a_\mu^{\underline a}$ in (\ref{eqn:non-Abelian-BF})
in the fundamental ($j=1/2$) representation of SU(2).
A Wilson loop in a higher $j$ representation can be constructed
by simply taking $2j$ copies of a $j=1/2$ Wilson loop
and using the appropriate Jones-Wenzl projector to eliminate
the other representations which result in the fusion of $2j$
copies of $j=1/2$. If the wavefunction satisfies the constraint
mentioned above, then it will vanish identically if acted on
by a $j>k/2$ Wilson loop.

These conditions are of a topological nature, so they are
most natural in the continuum. In constructing a lattice model
from which they emerge, we have a certain amount of freedom in deciding
how these conditions are realized at the lattice scale. Depending
on our choice of short-distance regularization, the model may be more
of less easily solved. In some cases, an inconvenient choice of
short-distance regularization may actually drive the system out
of the desired topological phase. Loops on the lattice
prove not to be the most convenient regularization of loops in the continuum,
essentially because when $d$ is large, the lattice fills up with loops
which then have no freedom to fluctuate \cite{Freedman04a}.
Instead, trivalent graphs on the lattice prove to be
a better way of proceeding
(and, in the case of SU(3)$_k$ and other gauge groups,
trivalent graphs are essential \cite{Turaev92,Kuperberg96}).
The most convenient lattice is the honeycomb lattice, since each
vertex is trivalent. A trivalent graph is simply a subset of the links
of the honeycomb lattice such that no vertex has only a single
link from the subset emanating from it. Zero, two, or three links
can emanate from a vertex, corresponding to vertices which are not
visited by the trivalent graph, vertices through which a curve passes,
and vertices at which three curves meet. We will penalize energetically
vertices from which a single colored link emanates. The ground state
will not contain such vertices, which will be quasiparticle excitations.
Therefore, the ground state $\Psi[\Gamma]$ assigns a
complex amplitude to a trivalent graph $\Gamma$.

\begin{figure}[thb]
\includegraphics[width=3.25in]{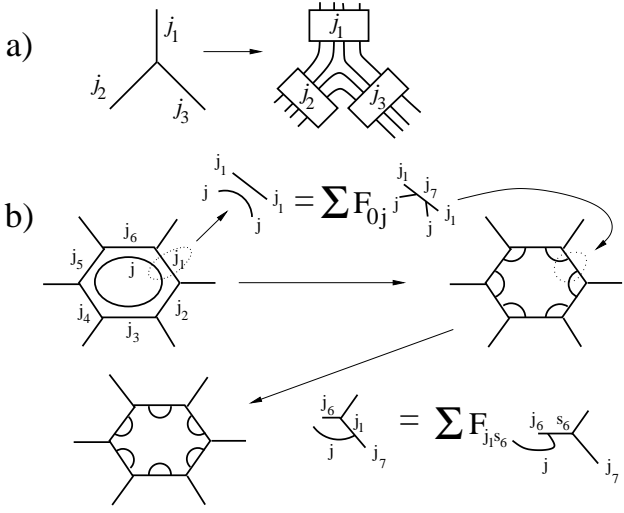}
\caption{(a) $j/2$ parallel lines projected onto representation $j$ are represented
by the label $j$ on a link. (b) The plaquette terms add a rep.-$j$ loop.
This can be transformed back into a trivalent graph on the
lattice using the $F$-matrix as shown.}
\label{fig:lattice-models}
\end{figure}

Such a structure arises in a manner analogous to the loop
structure of the toric code: if we had spins on the links of
the honeycomb lattice, then an appropriate choice of interaction
at each vertex will require that colored links (on which the spin points up)
form a trivalent graph. We note that links can be given a further labeling,
although we will not discuss this more complicated situation in any detail.
Each colored link can be assigned a $j$ in the set $\frac{1}{2},1,\ldots,\frac{k}{2}$.
Uncolored links are assigned $j=0$. Rather than spin-$1/2$ spins on each link,
we should take spin-$k/2$ on each link, with ${S_z}=-k/2$ corresponding to
$j=0$, ${S_z}=-k/2+1$ corresponding to $j=1/2$, etc. (or perhaps, we
may want to consider models with rather different microscopic degrees
of freedom). In this case, we would further require
that the links around each vertex should satisfy the branching rules
of SU(2)$_k$: $\left|{j_1}-{j_2}\right|\geq{j_3}
\leq\text{min}\left({j_1}+{j_2},\frac{k}{2}-{j_1}-{j_2}\right)$.
The case which we have described in the previous paragraph,
without the additional $j$ label could be applied to the level $k=1$
case, with colored links carrying $j=1/2$ or to level $k=3$,
with colored links carrying $j=1$, as we will discuss further
below. A trivalent graph represents a loop configuration in the manner depicted
in Figure \ref{fig:lattice-models}a. One nice feature is that
the Jones-Wenzl projections are enforced on every link from the start, so
no corresponding surgery constraint is needed.

If we would like a lattice model to be in the doubled
SU(2)$_3$ universality class, which has quasiparticle
excitations which are Fibonacci anyons, then its Hamiltonian
should impose the following: all low-energy states should
have vanishing amplitude on configurations which are are
not trivalent graphs, as defined above; and the amplitude
for a configuration with a contractible loop should be
larger than the amplitude for a configuration without this
loop by a factor of $d=2\cos\frac{\pi}{5}=\phi=(1+\sqrt{5})/2$
for a closed, contractible loop. These conditions can be
imposed by terms in the Hamiltonian which are more complicated
versions of the vertex and plaquette terms of (\ref{eqn:toric-code}).
It is furthermore necessary for the ground state wavefunction(s)
to assign the same amplitude to any two trivalent graphs which
can be continuously deformed into each other. However, as mentioned
above, surgery is not necessary. The Hamiltonian takes the form
\cite{Levin05a} (see also \onlinecite{Turaev92}):
\begin{equation}
H = -{J_1}\sum {A_i} -{J_2}\sum_{p}{\sum_{j=0}^{k/2}}{F^{(j)}_p}
\label{eqn:Levin-Wen}
\end{equation}
Here and below, we specialize to $k=3$.
The degrees of freedom on each link are $s=1/2$ spins;
${s_z}=+\frac{1}{2}$ is interpreted as a $j=1$ colored link,
while ${s_z}=-\frac{1}{2}$ is interpreted as a $j=0$ uncolored link.
The vertex terms impose the triangle inequality,
$\left|{j_1}-{j_2}\right|\geq{j_3}
\leq\text{min}\left({j_1}+{j_2},\frac{3}{2}-{j_1}-{j_2}\right)$, on the three
$j$'s on the links neighboring each vertex. For Fibonacci
anyons (see Sec. \ref{sec:fibonacci}), which can only have $j=0,1$, this means that
if links with $j=1$ are colored, then the colored links must
form a trivalent graph, i.e. no vertex can have only a single up-spin
adjacent to it. (There is no further requirement, unlike
in the general case, in which there are additional labels on the
trivalent graph.)

The plaquette terms in the Hamiltonian are complicated in
form but their action can be understood in the following simple
way: we imagine adding to a plaquette a loop $\gamma$ carrying
representation $j$ and require that the amplitude
for the new configuration $\Psi[\Gamma\cup\gamma]$ be larger
than the amplitude for the old configuration by a factor of $d_j$.
For Fibonacci anyons, the only non-trivial representation is $j=1$;
we require that the wavefunction change by a factor of $d=\phi$
when such a loop is added.
If the plaquette is empty, then `adding a loop' is simple.
We simply have a new trivalent graph with one extra loop.
If the plaquette is not empty, however, then we need to
specify how to `add' the additional loop to the occupied links.
We draw the new loop in the interior of the plaquette so that
it runs alongside the links of the plaquette, some of which are occupied.
Then, we use the $F$-matrix, as depicted
in Figure \ref{fig:lattice-models}b, to recouple the links of the plaquette
\cite{Levin05a} (see also \onlinecite{Turaev92}).
This transforms the plaquette so that it is now in a superposition
of states with different $j$'s, as depicted in Figure \ref{fig:lattice-models}b;
the coefficients in the superposition are sums of products of
elements of the $F$-matrix. The plaquette term commutes with
the vertex terms since adding a loop to a plaquette cannot violate
the triangle inequality (see Figure \ref{fig:lattice-models}a).
Clearly vertex terms commute with each other,
as do distant plaquette terms. Plaquette terms on adjacent plaquettes
also commute because they just add loops to the link which they share.
(This is related to the pentagon identity, which expresses
the associativity of fusion.)
Therefore, the model is exactly soluble since all terms
can be simultaneously diagonalized. Vertices with a single
adjacent colored (ie. monovalent vertices) are non-Abelian anyonic excitations
carrying $j=1$ under the SU(2) gauge group of $a_\mu^{\underline a}$ in
(\ref{eqn:non-Abelian-BF}). A state at which the plaquette term in
(\ref{eqn:Levin-Wen}) is not satisfied is a non-Abelian anyonic
excitation carrying $j=1$ under the SU(2) gauge group of
$e_\mu^{\underline a}$ (or, equivalently, $a_\mu^{\underline a}$ flux).

One interesting feature of the ground state wavefunction
$\Psi[\Gamma]$ of (\ref{eqn:Levin-Wen}), and of related models
with loop representations \cite{Freedman04a,Fendley05,Fidkowski06}
is their relation to the Boltzmann weights of statistical mechanical
models. For instance, the norm squared of ground state of of (\ref{eqn:Levin-Wen}),
satisfies ${|\Psi[\Gamma]|^2}=e^{-\beta H}$, where $\beta H$ is the
Hamiltonian of the $q=\phi+2$ state Potts model. More precisely, it
is the low-temperature expansion of the $q=\phi+2$ state Potts model
extrapolated to infinite temperature $\beta=0$. The square of the ground state
of the toric code (\ref{eqn:toric-code}) is the low-temperature expansion
of the Boltzmann weight of the $q=2$ state Potts model
extrapolated to infinite temperature $\beta=0$.
On the other hand, the squares of the ground states
$|\Psi[{\gamma_1}\cup\ldots\cup{\gamma_n}]|^2$
of loop models \cite{Freedman04a}, are equal to the
partition functions of O(n) loop gas models
of statistical mechanics, with $n=d^2$. These relations
allow one to use known results from statistical mechanics
to compute equal-time ground state correlation functions
in a topological ground state, although the interesting
ones are usually of operators which are non-local in the
original quantum-mechanical degrees of freedom of the model.

It is also worth noting that a quasi-one-dimensional analog
has been studied in detail \cite{Feiguin07a,Bonesteel07}.
It is gapless for a single chain and
has an interesting phase diagram for ladders.

Finally, we note that the model of Levin and Wen is,
admittedly, artificial-looking. However, a model
in the same universality class might emerge from simpler models
\cite{Fidkowski06}. Since (\ref{eqn:Levin-Wen}) has a gap,
it will be stable against small perturbations. In the case of the toric
code, it is known that even fairly large perturbations do not
destabilize the state \cite{Trebst07}.

This brings to a close our survey of the physics of topological
phases. In section \ref{part3}, we will consider their application
to quantum computing.

\section{Quantum Computing with Anyons}
\label{part3}

\subsection{$\nu=5/2$ Qubits and Gates}
\label{sec:5/2-qubits}

A topological quantum computer is constructed using
a system in a non-Abelian topological phase. A computation
is performed by creating quasiparticles, braiding them, and measuring their
final state. In section \ref{sec:FQHE-qc}, we saw how
a qubit could be constructed with the $\nu=5/2$ state
and a NOT gate applied.
In this section, we discuss some ideas about
how a quantum computer could be built by extending these ideas.

The basic feature of the Ising TQFT and its close relative,
SU(2)$_2$, which we exploit for storing quantum
information is the existence of two fusion channels
for a pair of $\sigma$ quasiparticles, $\sigma\times\sigma\sim {\bf 1}+ \psi$.
When the fusion outcome is ${\bf 1}$, we say that the qubit is
in the state $|0\rangle$; when it is $\psi$, the state $|1\rangle$.
When there are $2n$ quasiparticles, there is a $2^{n-1}$-dimensional
space of states. (This is how many states there are with total topological charge
${\bf 1}$; there is an equal number with total topological charge $\psi$.)
We would like to use this $2^{n-1}$-dimensional space to store
quantum information; the most straightforward way to do so
is to view it as $n-1$ qubits.

\begin{figure}[tbh]
\includegraphics[width=3.5in]{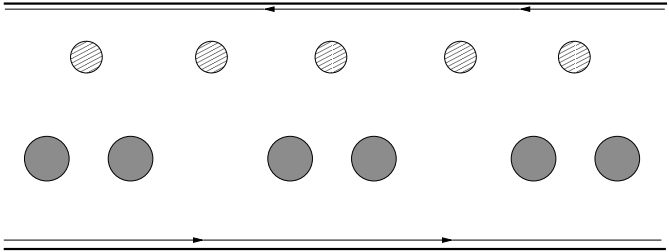}
\caption{A system with $n$ quasihole pairs (held at pairs of anti-dots,
depicted as shaded circles) supports $n$ qubits. Additional antidots
(hatched) can be used to move the quasiparticles.}
\label{fig:multi-qhs}
\end{figure}

Generalizing the construction of section \ref{sec:FQHE-qc}
to many pairs of anti-dots, we can envision \cite{Freedman06}
an $(n-1)$-qubit system which is a Hall bar with
$2n$ antidots at which quasiholes are pinned,
as in Figure \ref{fig:multi-qhs}.

The NOT gate discussed in section \ref{sec:FQHE-qc}
did not require us to move the quasiparticles comprising
the qubit, only additional quasiparticles which we
brought in from the edge. However, to implement other
gates, we will need to move the quasiparticles on the
anti-dots. In this figure, we have also depicted additional
anti-dots which can be used to move quasiparticles
from one anti-dot to another (e.g. as a `bucket brigade'),
see, for instance, \onlinecite{Simon00}.
If we exchange two quasiparticles from the same qubit,
then we apply the phase gate
$U=e^{\pi i/8}\,\text{diag}(R^{\sigma\sigma}_1,R^{\sigma\sigma}_\psi)$
(the phase in front of the matrix comes from the U(1) part of the theory).
However, if the two quasiparticles
are from different qubits, then we apply the transformation
\begin{eqnarray}
\label{eqn:2-qubit-couple}
U = \frac{1}{\sqrt{2}}\left(
 \begin{array}{cccc}
  1 & 0 & 0 &  -i \\
  0 & 1 &  -i & 0 \\
  0 & -i & 1 &  0   \\
   -i & 0 & 0 & 1
 \end{array}
\right).
\end{eqnarray}
to the two-qubit Hilbert space.

By coupling two qubits in this way,
a CNOT gate can be constructed. Let us suppose that
we have 4 quasiparticles. Then, the first pair can fuse to either ${\bf 1}$ or $\psi$,
as can the second pair. Naively, this is $4$ states but, in fact, it is really
two states with total topological charge ${\bf 1}$ and two states with
total topological charge $\psi$. These two subspaces cannot mix
by braiding the four quasiparticles. However, by braiding our qubits
with additional quasiparticles, we can mix these four states. (In our
single qubit NOT gate, we did this by using quasiparticles from the edge.)
Therefore, following \onlinecite{Georgiev06b}, we consider a system
with 6 quasiparticles. Quasiparticles $1$ and $2$ will be qubit $1$;
when they fuse to $1$ or $\psi$, qubit 1 is in state $|0\rangle$ or $|1\rangle$.
Quasiparticles $5$ and $6$ will be qubit 2; when they fuse to ${\bf 1}$ or $\psi$,
qubit 2 is in state $|0\rangle$ or $|1\rangle$. Quasiparticles 3 and 4
soak up the extra $\psi$, if necessary to maintain total topological charge ${\bf 1}$
for the entire six-quasiparticle system. In the four states $|0,0\rangle$,
$|1,0\rangle$, $|0,1\rangle$, and $|1,1\rangle$, the quasiparticle
pairs fuse to ${\bf 1}, {\bf 1}, {\bf 1}$, to $\psi,\psi,{\bf 1}$, to
${\bf 1},\psi,\psi$, and to $\psi,{\bf 1},\psi$,
respectively.

In this basis, $\rho({\sigma_1})$, $\rho({\sigma_3})$, $\rho({\sigma_5})$
are diagonal, while $\rho({\sigma_2})$ and $\rho({\sigma_4})$
are off-diagonal (e.g. $\rho({\sigma_2})$ is (\ref{eqn:2-qubit-couple})
rewritten in the two qubit/six quasiparticle basis).
By direct calculation (e.g. by using
$\rho({\sigma_i})=e^{\frac{\pi}{4}{\gamma_i} \gamma_{i+1}}$),
it can be shown \cite{Georgiev06b} that:
\begin{eqnarray}
\rho(\sigma_3^{-1} \sigma_{4} \sigma_{3} \sigma_{1} \sigma_{5}
\sigma_{4} \sigma_3^{-1}) =
\left(
 \begin{array}{cccc}
  1 & 0 & 0 &  0 \\
  0 & 1 &  0 & 0 \\
  0 & 0 & 0 &  1   \\
   0 & 0 & 1 & 0
 \end{array}
\right).
\end{eqnarray}
which is simply a controlled NOT operation.

One can presumably continue in this way, with
one extra pair of quasiparticles, which is used to soak
up an extra $\psi$ if necessary. However, this is not a particularly
convenient way of proceeding since various gates will be
different for different numbers of particles: the CNOT gate above
exploited the extra quasiparticle pair which is shared equally
between the two qubits acted on by the gate, but this will not
work in the same way for more than two qubits. Instead, it
would be easier to encode each qubit in four quasiparticles.
If each quartet of quasiparticles has total topological charge ${\bf 1}$,
then it can be in either of two states since a given pair within a
quartet can fuse to either ${\bf 1}$ or $\psi$. In other words, each
quasiparticle pair comes with its own spare pair of quasiparticles
to soak up its $\psi$ if necessary.

Unfortunately, the SU(2)$_2$ phase of matter is not capable of universal
quantum computation, i.e. the transformations generated by braiding operations are not sufficient to implement all possible unitary
transformations \cite{Freedman02a,Freedman02b}.
The reason for this shortcoming is that in this
theory, braiding of two particles has the effect of
a 90 degree rotation \cite{Nayak96c} in
the multi-quasiparticle Hilbert space.
Composing such 90 degree rotations will
clearly not allow one to construct arbitrary unitary operations (the
set of 90 degree rotations form a finite closed set).

However, we do not need to supplement braiding with much in order
to obtain a universal gate set. All that is needed is a
single-qubit $\pi/8$ phase gate and a two-qubit measurement.
One way to implement these extra gates is to
use some non-topological operations \cite{Bravyi06}.
First, consider the single-qubit phase gate. Suppose quasiparticles
$1,2,3,4$ comprise the qubit. The states $|0\rangle$ and $|1\rangle$
correspond to $1$ and $2$ fusing to ${\bf 1}$ or $\psi$ ($3$ and $4$
must fuse to the same as $1$ and $2$, since the total topological charge is
required to be ${\bf 1}$). If we bring quasiparticles $1$ and $2$
 close together then their splitting will become
appreciable. We expect it to depend on the separation $r$
as $\Delta\!E(r) \sim e^{- r\Delta/c}$,
where $r$ is the distance
between the quasiparticles and $c$ is some constant with dimensions of
velocity. If we wait a time $T_p$ before pulling the quasiparticles apart again,
then we will apply the phase gate \cite{Freedman06}
${U_P} = \text{diag}(1,e^{i\Delta\! E(r)\,T_p} )$.
If the time $T$ and distance $r$ are chosen so that
$\Delta\! E(r)\,T_p = \pi/4$, then up to an overall
phase, we would apply the phase gate:
\begin{eqnarray}
\label{eqn:piover8phase-gate}
U_{\pi/8} = \left(
 \begin{array}{cc}
  e^{-\pi i /8} & 0   \\
  0  & e^{\pi i/ 8}
   \end{array}
\right)
\end{eqnarray}
We note that, in principle, by measuring the energy when
the two quasiparticles are brought together, the state of the
qubit can be measured.

%\begin{figure}[tbh]
%\includegraphics[width=3.25in]{unprotected.eps}
%\caption{By bringing together quasiparticles 1 and 2, which
%form a qubit, we can apply a single qubit $\pi/8$ phase gate.}
%\label{fig:unprotected}
%\end{figure}

The other gate which we need for universal quantum computation
is the non-destructive measurement of the total topological charge
of any four quasiparticles. This can be done with an interference
measurement. Suppose we
have two qubits which are associated with
quasiparticles $1,2,3,4$ and quasiparticles $5,6,7,8$ and we
measure the total topological charge of $3,4,5,6$.
The interference measurement is of the type described in
subsection \ref{sec:interference}: edge currents tunnel
across the bulk at two points on either side of the set of four quasiparticles.
Depending on whether the four quasiparticles have total topological
charge ${\bf 1}$ or $\psi$, the two possible trajectories interfere with
a phase $\pm 1$. We can thereby measure the total parity
of two qubits. (For more details, see \onlinecite{Freedman06}.)

Neither of these gates can be applied exactly, which means
that we are surrendering some of the protection which we
have worked so hard to obtain and need some software error
correction. However, it is not necessary
for the $\pi/8$ phase gate or the two qubit measurement to
be extremely accurate in order for error correction to work. The former needs
to be accurate to within $14\%$ and the latter to within $38\%$
\cite{Bravyi06}. Thus, the requisite quantum error correction protocols
are not particularly stringent.

An alternate solution, at least in principle, involves
changing the topology of the manifold on which
the quasiparticles live \cite{Bravyi01}. This can be realized
in a device by performing interference measurements
in the presence of moving quasiparticles \cite{Freedman06}.

However, a more elegant approach is to work with a non-Abelian
topological state which supports universal topological quantum computation
through quasiparticle braiding alone. In the next subsection,
we give an example of such a state and how quantum computation
can be performed with it. In subsection \ref{sec:universal-tqc},
we sketch the proof that a large class of such states is universal.

\subsection{Fibonacci Anyons: a Simple Example
which is Universal for Quantum Computation}
\label{sec:fibonacci}
 \setcounter{mysubsection}{0}

One of the simplest models of non-Abelian statistics is known as the
Fibonacci anyon model, or ``Golden
theory" \cite{Freedman02a,Preskill04,Bonesteel05,Hormozi07}. In this
model, there are only two fields, the identity ($\bf 1$) as well as
single nontrivial field usually called $\fib$ which represents the
non-Abelian quasiparticle.  (Note there is no field representing the
underlying electron in this simplified theory).  There is a single
nontrivial fusion rule in this model
\begin{equation}
\label{eq:fibonaccifusion} \fib \times \fib = {\bf 1} + \fib
\end{equation}
which results in the Bratteli diagram given in
Fig.~\ref{fig:bratteli1}b. This model is particularly simple in that
any cluster of quasiparticles can fuse only to $\bf 1$ or $\fib$.

The $j=0$ and $j=1$ quasiparticles in SU(2)$_3$ satisfy the fusion
rules of Fibonacci anyons. Therefore, if we simply omit the $j=1/2$
and $j=3/2$ quasiparticles from SU(2)$_3$, we will have FIbonacci
anyons. This is perfectly consistent since half-integral $j$ will
never arise from the fusions of integral $j$s; the model with only
integer spins can be called SO(3)$_2$ or, sometimes, `the even part
of SU(2)$_3$'. As a result of the connection to SU(2)$_3$, sometimes
$\bf 1$ is called q-spin ``0" and $\fib$ is called q-spin ``1'' (see
\cite{Hormozi07}). $\mathbb{Z}_3$ parafermions are equivalent to a
coset theory SU(2)$_3$/U(1). This can be realized with an SU(2)$_3$
WZW model in which the U(1) subgroup is coupled to a gauge field
\cite{Gawedzki88,Karabali89}. Consequently, $\mathbb{Z}_3$
parafermions have essentially the same fusion rules as SU(2)$_3$;
there are some phase differences between the two theories which show
up in the $R$ and $F$-matrices. In the $\mathbb{Z}_3$ parafermion
theory, the field $\epsilon$ which results from fusing $\sigma_1$
with $\psi_1$ satisfies the Fibonacci fusion rule Eq.
\ref{eq:fibonaccifusion}, i.e., $\epsilon \times \epsilon = {\bf 1}
+ \epsilon$.

As with the $\mathbb{Z}_3$ parafermion model described above, the
dimension of the Hilbert space with $n$ quasiparticles (i.e., the
number of paths through the Bratteli diagram \ref{fig:bratteli1}b
terminating at $\bf 1$) is given by the Fibonacci number
$\mbox{Fib}(n-1)$, hence the name Fibonacci anyons.  And similarly
the number terminating at $\fib$ is $\mbox{Fib}(n)$. Therefore, the
quantum dimension of the $\fib$ particle is the golden mean,
${d_\fib}=\phi\equiv (1+\sqrt{5})/2$ (from which the theory receives
the name ``golden" theory). The Fibonacci model is the simplest
known non-Abelian model that is capable of universal quantum
computation \cite{Freedman02a}. (In the next section, the proof will
be described for SU(2)$_3$, but the Fibonacci theory, which is its
even part, is also universal.) It is thus useful to study this model
in some detail.  Many of the principles that are described here will
generalize to other non-Abelian models. We note that a detailed
discussion of computing with the Fibonacci model is also given in
\onlinecite{Hormozi07}.

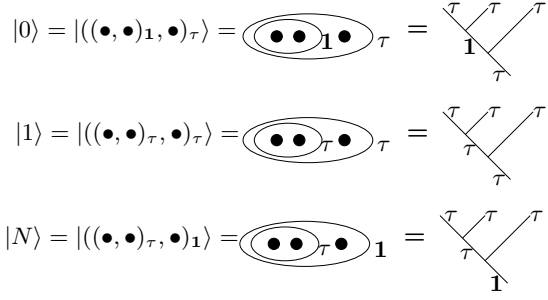
\begin{figure}
\ifx\JPicScale\undefined\def\JPicScale{.3}\fi
\psset{unit=\JPicScale} \psset{unit=\JPicScale mm}
%PSTricks content-type (pstricks.sty package needed)
%Add \usepackage{pstricks} in the preamble of your LaTeX file
%You can rescale the whole picture (to 80% for instance) by using the command \def\JPicScale{0.8}
\psset{unit=.3 mm}
\psset{linewidth=0.3,dotsep=1,hatchwidth=0.3,hatchsep=1.5,shadowsize=1,dimen=middle}
\psset{dotsize=0.7 2.5,dotscale=1 1,fillcolor=black}
\psset{arrowsize=1 2,arrowlength=1,arrowinset=0.25,tbarsize=0.7
5,bracketlength=0.15,rbracketlength=0.15} \hspace*{45pt}
\begin{pspicture}(0,0)(70,77.5)
\rput{0}(27.5,67.5){\psellipse[fillcolor=black,fillstyle=solid](0,0)(2.5,-2.5)}
\rput{0}(37.5,67.5){\psellipse[fillcolor=black,fillstyle=solid](0,0)(2.5,-2.5)}
\rput{0}(57.5,67.5){\psellipse[fillcolor=black,fillstyle=solid](0,0)(2.5,-2.5)}
\rput{0}(32.5,67.5){\psellipse[](0,0)(15,-7.5)} \rput(50,65){$\bf
1$} \rput{0}(41.25,67.5){\psellipse[](0,0)(28.75,-10)}
\rput(75,65){$\fib$}
\rput(-40,70){$|0\rangle=|((\bullet,\bullet)_{\bf 1}, \bullet)_\tau
\rangle=$} \rput(90,70){\boldmath$=$}
\end{pspicture}
\hspace*{5pt} \psset{unit=\JPicScale} \psset{unit=\JPicScale mm}
\psset{linewidth=0.3,dotsep=1,hatchwidth=0.3,hatchsep=1.5,shadowsize=1,dimen=middle}
\psset{dotsize=0.7 2.5,dotscale=1 1,fillcolor=black}
\psset{arrowsize=1 2,arrowlength=1,arrowinset=0.25,tbarsize=0.7
5,bracketlength=0.15,rbracketlength=0.15}
\begin{pspicture}
(0,0)(60,80) \psline(20,80)(50,50) \psline(30,70)(40,80)
\psline(40,60)(60,80) \rput(25,80){$\fib$} \rput(43,80){$\fib$}
\rput(63,80){$\fib$} \rput(32,63){$\bf 1$} \rput(45,50){$\fib$}
\end{pspicture}
\\
\vspace*{-30pt}
 \psset{unit=\JPicScale} \psset{unit=\JPicScale mm}
%PSTricks content-type (pstricks.sty package needed)
%Add \usepackage{pstricks} in the preamble of your LaTeX file
%You can rescale the whole picture (to 80% for instance) by using the command \def\JPicScale{0.8}
\psset{unit=.3 mm}
\psset{linewidth=0.3,dotsep=1,hatchwidth=0.3,hatchsep=1.5,shadowsize=1,dimen=middle}
\psset{dotsize=0.7 2.5,dotscale=1 1,fillcolor=black}
\psset{arrowsize=1 2,arrowlength=1,arrowinset=0.25,tbarsize=0.7
5,bracketlength=0.15,rbracketlength=0.15} \hspace*{45pt}
\begin{pspicture}(0,0)(70,77.5)
\rput{0}(27.5,67.5){\psellipse[fillcolor=black,fillstyle=solid](0,0)(2.5,-2.5)}
\rput{0}(37.5,67.5){\psellipse[fillcolor=black,fillstyle=solid](0,0)(2.5,-2.5)}
\rput{0}(57.5,67.5){\psellipse[fillcolor=black,fillstyle=solid](0,0)(2.5,-2.5)}
\rput{0}(32.5,67.5){\psellipse[](0,0)(15,-7.5)} \rput(50,65){$\fib$}
\rput{0}(41.25,67.5){\psellipse[](0,0)(28.75,-10)}
\rput(75,65){$\fib$} \rput(-40,70){
$|1\rangle=|((\bullet,\bullet)_{\fib}, \bullet)_\fib \rangle=$}
\rput(90,70){\boldmath$=$}
\end{pspicture}
\hspace*{5pt} \psset{unit=\JPicScale} \psset{unit=\JPicScale mm}
\psset{linewidth=0.3,dotsep=1,hatchwidth=0.3,hatchsep=1.5,shadowsize=1,dimen=middle}
\psset{dotsize=0.7 2.5,dotscale=1 1,fillcolor=black}
\psset{arrowsize=1 2,arrowlength=1,arrowinset=0.25,tbarsize=0.7
5,bracketlength=0.15,rbracketlength=0.15}
\begin{pspicture}
(0,0)(60,80) \psline(20,80)(50,50) \psline(30,70)(40,80)
\psline(40,60)(60,80) \rput(25,80){$\fib$} \rput(43,80){$\fib$}
\rput(63,80){$\fib$} \rput(32,63){$\fib$} \rput(45,50){$\fib$}
\end{pspicture}
\\
\vspace*{-30pt} \psset{unit=\JPicScale} \psset{unit=\JPicScale mm}
%PSTricks content-type (pstricks.sty package needed)
%Add \usepackage{pstricks} in the preamble of your LaTeX file
%You can rescale the whole picture (to 80% for instance) by using the command \def\JPicScale{0.8}
\psset{unit=.3 mm}
\psset{linewidth=0.3,dotsep=1,hatchwidth=0.3,hatchsep=1.5,shadowsize=1,dimen=middle}
\psset{dotsize=0.7 2.5,dotscale=1 1,fillcolor=black}
\psset{arrowsize=1 2,arrowlength=1,arrowinset=0.25,tbarsize=0.7
5,bracketlength=0.15,rbracketlength=0.15} \hspace*{45pt}
\begin{pspicture}(0,0)(70,77.5)
\rput{0}(27.5,67.5){\psellipse[fillcolor=black,fillstyle=solid](0,0)(2.5,-2.5)}
\rput{0}(37.5,67.5){\psellipse[fillcolor=black,fillstyle=solid](0,0)(2.5,-2.5)}
\rput{0}(57.5,67.5){\psellipse[fillcolor=black,fillstyle=solid](0,0)(2.5,-2.5)}
\rput{0}(32.5,67.5){\psellipse[](0,0)(15,-7.5)} \rput(50,65){$\fib$}
\rput{0}(41.25,67.5){\psellipse[](0,0)(28.75,-10)} \rput(75,65){$\bf
1$} \rput(-40,70){$|N\rangle=|((\bullet,\bullet)_{\fib},
\bullet)_{\bf 1} \rangle=$}  \rput(90,70){\boldmath$=$}
\end{pspicture}
\hspace*{5pt}
\psset{unit=\JPicScale} \psset{unit=\JPicScale mm}
\psset{linewidth=0.3,dotsep=1,hatchwidth=0.3,hatchsep=1.5,shadowsize=1,dimen=middle}
\psset{dotsize=0.7 2.5,dotscale=1 1,fillcolor=black}
\psset{arrowsize=1 2,arrowlength=1,arrowinset=0.25,tbarsize=0.7
5,bracketlength=0.15,rbracketlength=0.15}
\begin{pspicture}
(0,0)(60,80) \psline(20,80)(50,50) \psline(30,70)(40,80)
\psline(40,60)(60,80) \rput(25,80){$\fib$} \rput(43,80){$\fib$}
\rput(63,80){$\fib$} \rput(32,63){$\fib$} \rput(45,50){$\bf 1$}
\end{pspicture}
\vspace*{-40pt} \caption{The three possible states of three
Fibonacci particles, shown in several common notations.  The
``quantum number" of an individual particle is $\fib$.    In the
parenthesis and ellipse notation (middle), each particle is shown as
a black dot, and each pair of parenthesis or ellipse around a group
of particles is labeled at the lower right with the total quantum
number associated with the fusion of that group.   Analogously in
the fusion tree notation (right) we group particles as described by
the branching of the tree, and each line is labeled with the quantum
number corresponding to the fusion of all the particles in the
branches above it. For example on the top line the two particles on
the left fuse to form $\bf 1$ which then fuses with the remaining
particle on the right to form $\fib$.  As discussed below in section
\ref{sec:fibonacci}.c, three Fibonacci particles
can be used to represent a qubit.  The
three possible states are labeled  (far left) as the logical
$|0\rangle$, $|1\rangle$ and $|N\rangle$ (noncomputational) of the
qubit.} \label{fig:fusion1}
\end{figure}

\mysubsection{Structure of the Hilbert Space} \mylabel{Structure}
An important feature of non-Abelian systems is the detailed structure
of the Hilbert space.  A given state in the space will be described
by a ``fusion path", or ``fusion tree"  (See appendix
\ref{section:CFT}).  For example, using the fusion rule
(\ref{eq:fibonaccifusion}), or examining the Bratteli diagram we see
that when two $\fib$ particles are present, they may fuse into two
possible orthogonal degenerate states  -- one in which they fuse to
form $\bf 1$ and one in which they fuse to form $\fib$. A convenient
notation \cite{Bonesteel05} for these two states is $|(\bullet,
\bullet)_{\bf 1}\rangle$ and $|(\bullet, \bullet)_{\fib}\rangle$.
Here, each $\bullet$ represents a particle.   From the fusion rule,
when a third is added to two particles already in the $\bf 1$ state
(i.e., in $|(\bullet, \bullet)_{\bf 1}\rangle$) it must fuse to form
$\fib$. We denote the resulting state as $|((\bullet, \bullet)_{\bf
1}, \bullet)_\fib\rangle \equiv |0\rangle$. But if the third is
added to two in the $\fib$ state, it may fuse to form either $\fib$
or $\bf 1$, giving the two states $|((\bullet, \bullet)_{\fib},
\bullet)_\fib\rangle \equiv |1\rangle$ and $|((\bullet,
\bullet)_{\fib}, \bullet)_{\bf 1} \rangle \equiv |N\rangle$
respectively.   (The notations $|0\rangle, |1 \rangle$ and
$|N\rangle$ will be discussed further below).  Thus we have a three
dimensional Hilbert space for three particles shown using several
notations in Fig. \ref{fig:fusion1}.

In the previous example, and in Fig.~\ref{fig:fusion1} we have
always chosen to fuse particles together starting at the left and
going to the right.  It is, of course,  also possible to fuse
particles in the opposite order, fusing the two particles on the
right first, and then fusing with the particle furthest on the left
last.  We can correspondingly denote the three resulting states as
$|(\bullet, (\bullet, \bullet)_{\bf 1})_{\fib}\rangle$, $|(\bullet,
(\bullet, \bullet)_{\fib})_{\fib}\rangle$, and $|(\bullet, (\bullet,
\bullet)_{\fib})_{\bf 1}\rangle$.   The space of states that is spanned
by fusion of non-Abelian particles is independent of the fusion
order. However, different fusion orders results in a different basis
set for that space.  This change of basis is precisely that
given by the $F$-matrix. For
Fibonacci anyons it is easy to see that
\begin{equation}
\label{eq:onestate}
 |(\bullet, (\bullet,
\bullet)_{\fib})_{\bf 1}\rangle = |((\bullet, \bullet)_{\fib},
\bullet)_{\bf 1} \rangle
\end{equation}
since in either fusion order there is only a single state that has
total topological charge $\bf 1$ (the overall quantum number of a
group of particles is independent of the basis).   However, the
other two states of the three particle space transform nontrivially
under change of fusion order.   As described in appendix
\ref{section:CFT}, we can write a change of basis using
the $F$-matrix as
\begin{equation}
\label{eq:Fdef}
 |(\bullet, (\bullet, \bullet)_i)_k\rangle =
\mbox{$\sum_j$} \,\, [F^{\fib \fib \fib}_{k}]_{ij} \,\,\,
|((\bullet, \bullet)_{j}, \bullet)_{k} \rangle
\end{equation}
where $i,j,k$ take the values of the fields $\bf 1$ or $\fib$. (This
is just a rewriting of a special case of Fig. \ref{fig:fusiontree}).
Clearly from Eq. \ref{eq:onestate}, $F^{\fib\fib\fib}_{\bf 1}$ is trivially unity.
However, the two-by-two matrix $F^{\fib\fib\fib}_\fib$ is nontrivial
\begin{equation} \label{eq:Fmatrix}
[F^{\tau\tau\tau}_\tau]=\left(\begin{array}{cc} F_{{\bf 1 1}} &
F_{{\bf 1} \fib }
\\
  F_{\fib {\bf 1}} &  F_{\fib \fib}
\end{array}
 \right)
 =
\left(\begin{array}{cc} \phi^{-1} &  \sqrt{\phi^{-1}} \\
\sqrt{\phi^{-1}} & -\phi^{-1}
\end{array}
 \right)
\end{equation}
Using this $F$ matrix, one can translate between bases that describe arbitrary
fusion orders of many particles.

For the Fibonacci theory \cite{Preskill04},
it turns out to be easy to calculate the $F$-matrix using a
consistency condition known as the pentagon equation
\cite{Moore88,Moore89,Fuchs92,Gomez96}.
This condition simply says that one should be able to make changes
of basis for four particles in several possible ways and get the
same result in the end.   As an example, let us consider
\begin{eqnarray} \nonumber
  |(\bullet,(\bullet,(\bullet,\bullet)_{\bf 1})_\fib)_{\bf 1}\rangle &=&
  |((\bullet,\bullet)_{\bf 1},(\bullet,\bullet)_{\bf 1})_{\bf 1}\rangle \\
  &=& |((\bullet,\bullet)_{\bf 1},\bullet)_\fib,\bullet)_{\bf 1}
  \rangle \label{eq:Feq1}
\end{eqnarray}
where both equalities, as in Eq. \ref{eq:onestate} can be deduced
from the fusion rules alone.   For example, in the first equality,
given (on the left hand side) that the overall quantum number is
$\bf 1$ and the rightmost two particles are in a state $\bf 1$, then
(on the right hand side) when we fuse the leftmost two particles
they must fuse to $\bf 1$ such that the overall quantum number
remains $\bf 1$.  On the other hand, we can also use the $F$-matrix
(Eq. \ref{eq:Fdef}) to write
\begin{eqnarray}
& &   |(\bullet,(\bullet,(\bullet,\bullet)_{\bf 1})_\fib)_1\rangle
=~~~~
  ~~~~~~~~~~~~~~~~~~
  \\ \nonumber \rule{0pt}{10pt}
& &   F_{{\bf 1 1}} |(\bullet, ((\bullet, \bullet)_{\bf
1},\bullet)_\fib)_{\bf 1} \rangle +
 F_{{\bf 1 \fib}} |(\bullet, ((\bullet, \bullet)_{\fib},\bullet)_\fib)_{\bf 1}
 \rangle =
\\ \nonumber \rule{0pt}{10pt}
 && F_{{\bf 1 1}} |((\bullet, (\bullet, \bullet)_{\bf 1})_\fib,\bullet)_{\bf 1} \rangle +
  F_{{\bf 1 \fib}} |((\bullet, (\bullet, \bullet)_{\fib})_\fib,\bullet)_{\bf 1}
 \rangle = \\ \nonumber \rule{0pt}{10pt}
& &  \mbox{$\sum_{j}$} \left( F_{{\bf 1 1}} F_{{\bf 1} j} \right. +
\left. F_{{\bf 1 \fib}} F_{{\fib
 j}} \right)   |((\bullet,\bullet)_{j},\bullet)_\fib,\bullet)_{\bf 1} \rangle
\end{eqnarray}
Comparing to Eq. \ref{eq:Feq1}, yields $F_{{\bf 1} \fib}
(F_{\bf 11} + F_{\fib \fib} )  = 0$ and $F_{{\bf 1 1}} F_{{\bf 1 1}}
+ F_{{\bf 1} \fib} F_{\fib {\bf 1} } =1$.  This, and other similar
consistency identities, along with the requirement that $F$ be
unitary, completely fix the Fibonacci $F$-matrix to be precisely
that given in Eq. \ref{eq:Fmatrix} (up to a gauge freedom in the
definition of the phase of the basis states).

\mysubsection{Braiding Fibonacci Anyons}  As discussed in the
introduction, for non-Abelian systems, adiabatically braiding
particles around each other results in a unitary operation on the
degenerate Hilbert space.  Here we attempt to determine which unitary
operation results from which braid.   We start by considering what
happens to two Fibonacci particles when they are braided around each
other. It is known \cite{Fuchs92} that the topological spin $\Theta_\fib$ of a
Fibonacci field $\fib$ is $\Theta_\tau \equiv e^{2\pi i \Delta_\tau} = e^{4\pi i /5}$. (Note that $\Delta_\tau$ is also the dimension of the $\epsilon$
field of the $\mathbb{Z}_3$ theory, see Appendix \ref{section:CFT}.) With
this information, we can use the OPE
(see Appendix \ref{section:CFT})as in section \ref{sec:FQHE}
above, to determine the phase accumulated when two particles wrap
around each other. If the two $\fib$ fields fuse together to form
$\bf 1$, then taking the two fields around each other clockwise
results in a phase $-8 \pi/5 = 2 \pi (-2 \Delta_\fib) $ whereas if
the two fields fuse to form $\fib$, taking the two fields around
each other results in a phase $-4 \pi/5 = 2 \pi (-\Delta_\fib)$.
Note that a Fibonacci theory with the opposite chirality can exist
too (an ``antiholomorphic theory"), in which case one accumulates
the opposite phase. A particularly interesting non-chiral (or
``achiral") theory also exists which is equivalent to a combination
of two chiral Fibonacci theories with opposite
chiralities. In section \ref{sec:P-T-Invariant}, we discussed
lattice spin models \cite{Levin05a} which give rise to
a non-chiral (or ``achiral") theory which is equivalent to a combination
of two chiral Fibonacci theories with opposite
chiralities. We will not discuss these theories further here.

\begin{figure} \ifx\JPicScale\undefined\def\JPicScale{.3}\fi
\psset{unit=.4 mm} \vspace*{30pt} \psset{unit=0.012mm}
\psset{yunit=0.018mm} \psset{xunit=-0.016mm} \hspace*{0pt}
\psset{linewidth=50,dotsep=1,hatchwidth=0.3,hatchsep=1.5,shadowsize=1,dimen=middle}
%\psset{dotsize=0.7 2.5,dotscale=1 1,fillcolor=black}
%\psset{arrowsize=12,arrowlength=1,arrowinset=0.25,tbarsize=0.75,bracketlength=0.15,rbracketlength=0.15}
\begin{pspicture}(0,7600)(400,8400)
\psline[linestyle=dashed,arrows=->](1400,7600)(1400,8800)
\rput[angle=90](1400,7400){\large time}
\psline(0.00,7600.00)(-0.04,8000.00)
\psline(400.00,7600.00)(399.96,8000.04)
\psbezier(-0.04,8000.00)(-0.04,8200.00)(399.96,8200.02)(399.96,8400.00)
\pspolygon[linewidth=0pt,fillcolor=white,fillstyle=solid,linecolor=white](199.96,8033.35)(33.29,8200.00)(199.96,8366.67)(366.63,8200.02)
\psbezier(399.96,8000.04)(399.96,8200.02)(-0.04,8200.00)(-0.04,8400.00)
\psline[arrows=->](-0.04,8400.00)(-0.04,8800.00)
\psline[arrows=->](399.96,8400.00)(399.96,8800.00)
\psline(0.00,7600.00)(-0.04,8000.00)
\psline[arrows=->](-380,7600)(-380,8800)
\end{pspicture}
\hspace*{110pt}  \psset{unit=0.005 mm} \psset{unit=0.012mm}
\psset{yunit=0.018mm} \psset{xunit=-0.016mm}
%\psset{linewidth=0.3,dotsep=1,hatchwidth=0.3,hatchsep=1.5,shadowsize=1,dimen=middle}
%\psset{dotsize=0.7 2.5,dotscale=1 1,fillcolor=black}
%\psset{arrowsize=12,arrowlength=1,arrowinset=0.25,tbarsize=0.75,bracketlength=0.15,rbracketlength=0.15}
\begin{pspicture}
\psline(0.00,7600.00)(-0.04,8000.00)
\psline(400.00,7600.00)(399.96,8000.04)
\psbezier(-0.04,8000.00)(-0.04,8200.00)(399.96,8200.02)(399.96,8400.00)
\pspolygon[linewidth=0pt,fillcolor=white,fillstyle=solid,linecolor=white](199.96,8033.35)(33.29,8200.00)(199.96,8366.67)(366.63,8200.02)
\psbezier(399.96,8000.04)(399.96,8200.02)(-0.04,8200.00)(-0.04,8400.00)
\psline[arrows=->](-0.04,8400.00)(-0.04,8800.00)
\psline[arrows=->](399.96,8400.00)(399.96,8800.00)
\psline(0.00,7600.00)(-0.04,8000.00)
\psline[arrows=->](750,7600)(750,8800) \rput(1900,8150){\large
$\sigma_1$} \rput(-175,8150){\large $\sigma_2$}
\end{pspicture}
\\ \vspace*{10pt} \hspace*{35pt}
\ifx\JPicScale\undefined\def\JPicScale{.3}\fi \psset{unit=.4 mm}
\psset{linewidth=0.3,dotsep=1,hatchwidth=0.3,hatchsep=1.5,shadowsize=1,dimen=middle}
\psset{dotsize=0.7 2.5,dotscale=1 1,fillcolor=black}
\psset{arrowsize=1 2,arrowlength=1,arrowinset=0.25,tbarsize=0.7
5,bracketlength=0.15,rbracketlength=0.15}
\begin{pspicture}(0,0)(92.5,87.5)
\rput{0}(17.5,77.5){\psellipse[fillstyle=solid](0,0)(2.5,-2.5)}
\rput{0}(32.5,77.5){\psellipse[fillstyle=solid](0,0)(2.5,-2.5)}
\rput{0}(47.5,77.5){\psellipse[fillstyle=solid](0,0)(2.5,-2.5)}
\rput{0}(25,77.5){\psellipse[](0,0)(15,-7.5)} \rput(42.5,72.5){$j$}
\rput{0}(32.5,77.5){\psellipse[](0,0)(27.5,-10)} \rput(62.5,70){$k$}
\end{pspicture}
\hspace*{-10pt}
\begin{pspicture}(0,0)(92.5,87.5)
\rput{0}(17.5,77.5){\psellipse[fillstyle=solid](0,0)(2.5,-2.5)}
\rput{0}(32.5,77.5){\psellipse[fillstyle=solid](0,0)(2.5,-2.5)}
\rput{0}(47.5,77.5){\psellipse[fillstyle=solid](0,0)(2.5,-2.5)}
\rput{0}(25,77.5){\psellipse[](0,0)(15,-7.5)} \rput(42.5,72.5){$j$}
\rput{0}(32.5,77.5){\psellipse[](0,0)(27.5,-10)} \rput(62.5,70){$k$}
\end{pspicture}
\\
\vspace*{-30pt}
 \psset{unit=0.008 mm} \hspace*{-150pt}
%\psset{xunit=0.008mm}
\psset{linewidth=50,dotsep=1,hatchwidth=0.3,hatchsep=1.5,shadowsize=1,dimen=middle}
\begin{pspicture}(6000,0)(10000,1000)
{\psellipse[fillstyle=solid](7000,0)(50,50)}
{\psellipse[fillstyle=solid](7000,400)(50,50)}
{\psellipse[fillstyle=solid](7000,800)(50,50)}
\psellipse[linewidth=20](7000,400)(350,900)
\psellipse[linewidth=20](7000,600)(150,400)
\psline(7600.00,0.00)(8500.00,-0.09)
\psline(7600.00,0.00)(8500.00,-0.09)
\psline(7600.00,400.00)(8500.04,399.91)
\psline(7600.00,800.00)(8500.08,799.91)
\psline(8500.08,799.91)(9400.00,799.91)
\psbezier(8500.00,-0.09)(8950.00,-0.09)(8950.02,399.91)(9400.00,399.91)
\pspolygon[linewidth=0pt,fillcolor=white,fillstyle=solid,linecolor=white](8575.02,199.91)(8950.00,33.24)(9325.00,199.91)(8950.02,366.58)
\psbezier(8500.04,399.91)(8950.02,399.91)(8950.00,-0.09)(9400.00,-0.09)
\psline(9400.00,-0.09)(10300.00,-0.09)
\psbezier(9400.00,399.91)(9850.00,399.91)(9850.00,799.91)(10300.00,799.91)
\pspolygon[linewidth=0pt,fillcolor=white,fillstyle=solid,linecolor=white](9475.00,599.91)(9850.00,433.24)(10225.00,599.91)(9850.00,766.58)
\psbezier(9400.00,799.91)(9850.00,799.91)(9850.00,399.91)(10300.00,399.91)
\psline(10300.00,-0.09)(11200.00,-0.09)
\psbezier(10300.00,399.91)(10750.00,399.91)(10750.00,799.91)(11200.00,799.91)
\pspolygon[linewidth=0pt,fillcolor=white,fillstyle=solid,linecolor=white](10375.00,599.91)(10750.00,433.24)(11125.00,599.91)(10750.00,766.57)
\psbezier(10300.00,799.91)(10750.00,799.91)(10750.00,399.91)(11200.00,399.91)
\psline(11200.00,799.91)(12100.00,799.91)
\psbezier(11200.00,399.91)(11650.00,399.91)(11650.00,-0.09)(12100.00,-0.09)
\pspolygon[linewidth=0pt,fillcolor=white,fillstyle=solid,linecolor=white](11275.00,199.91)(11650.00,33.24)(12025.00,199.91)(11650.00,366.57)
\psbezier(11200.00,-0.09)(11650.00,-0.09)(11650.00,399.91)(12100.00,399.91)
\psline(12100.00,799.91)(13000.00,799.91)
\psbezier(12100.00,399.91)(12550.00,399.91)(12550.00,-0.09)(13000.00,-0.09)
\pspolygon[linewidth=0pt,fillcolor=white,fillstyle=solid,linecolor=white](12175.00,199.91)(12550.00,33.24)(12925.00,199.91)(12550.00,366.57)
\psbezier(12100.00,-0.09)(12550.00,-0.09)(12550.00,399.91)(13000.00,399.91)
\psline(13000.00,-0.09)(13900.00,-0.10)
\psbezier(13000.00,399.91)(13450.00,399.91)(13450.00,799.91)(13900.00,799.90)
\pspolygon[linewidth=0pt,fillcolor=white,fillstyle=solid,linecolor=white](13075.00,599.91)(13450.00,433.24)(13825.00,599.90)(13450.00,766.57)
\psbezier(13000.00,799.91)(13450.00,799.91)(13450.00,399.91)(13900.00,399.90)
\psline[arrows=->](13900.00,-0.10)(14800.00,-0.10)
\psline[arrows=->](13900.00,399.90)(14800.00,399.90)
\psline[arrows=->](13900.00,799.90)(14800.00,799.90)
\psline[linestyle=dashed,arrows=->](9300,1550)(13500,1550)
\rput(11500,1800){\large time}
\psline[linestyle=dashed,linewidth=20](9400,1100)(9400,-500)
\psline[linestyle=dashed,linewidth=20](10300,1100)(10300,-500)
\psline[linestyle=dashed,linewidth=20](11200,1100)(11200,-500)
\psline[linestyle=dashed,linewidth=20](12100,1100)(12100,-500)
\psline[linestyle=dashed,linewidth=20](13000,1100)(13000,-500)
\rput(9050,-550){$\sigma_2^{\phantom{{-1}}}$}
\rput(9950,-550){$\sigma_1^{\phantom{{-1}}}$}
\rput(10850,-550){$\sigma_1^{\phantom{{-1}}}$}
\rput(11750,-550){$\sigma_2^{-1}$}
\rput(12650,-550){$\sigma_2^{-1}$}
\rput(13550,-550){$\sigma_1^{\phantom{{-1}}}$}
\end{pspicture}
\vspace*{20pt}
 \caption{{\bf Top:} The two elementary braid operations $\sigma_1$ and  $\sigma_2$ on three particles.
{\bf Bottom:} Using these two braid operations and their inverses,
an arbitrary braid on three strands can be built.  The braid shown
here is written as $\sigma_2 \sigma_1 \sigma_1 \sigma_2^{-1}
\sigma_2^{-1} \sigma_1$. }
  \label{fig:threebraid}
\end{figure}
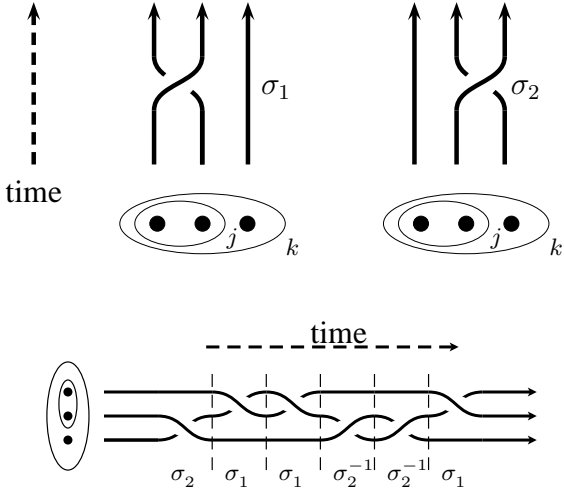

Once we have determined the phase accumulated for a full wrapping of
two particles, we then know that clockwise exchange of two particles
(half of a full wrapping) gives a phase of $\pm 4 \pi/5$ if the
fields fuse to $\bf 1$ or $\pm 2 \pi/5$ if the fields fuse to
$\fib$.  Once again we must resort to consistency conditions to
determine these signs. In this case, we invoke the so-called
``hexagon"-identities \cite{Moore88,Moore89,Fuchs92}
which in essence assure that the
rotation operations are consistent with the $F$-matrix, i.e., that
we can rotate before or after changing bases and we get the same
result. (Indeed, one way of proving that $\Delta_\fib=2/5$  is by
using this consistency condition).  We thus determine that the
$R$-matrix is given by
\begin{eqnarray}
\label{eq:R1}
\hat R \, |(\bullet,\bullet)_{\bf 1} \rangle  &=  e^{- 4 \pi i/5} & |(\bullet,\bullet)_{\bf 1} \rangle \\
\hat R \, |(\bullet,\bullet)_{\fib} \rangle  &=   -e^{- 2 \pi i/5} &
|(\bullet,\bullet)_{\fib} \rangle \label{eq:Rfib}
\end{eqnarray}
i.e., $R_{\tau \tau}^1 = e^{-4 \pi i /5}$ and $R_{\tau \tau}^\tau  = -e^{2 \pi i/5}$.
Using the $R$-matrix, as well as the basis changing $F$-matrix,
we can determine the unitary operation that results from performing
any braid on any number of particles.  As an example, let us
consider three particles.  The braid group is generated by
$\sigma_1$ and $\sigma_2$. (See Fig. \ref{fig:threebraid})
As discussed above, the Hilbert space of three
particles is three-dimensional as shown in Fig. \ref{fig:fusion1}.
We can use Eqs. \ref{eq:R1} and \ref{eq:Rfib} trivially to determine
that the unitary operation corresponding to  the braid $\sigma_1$ is
given by
\begin{equation} \left( \begin{array}{c}
                   |0\rangle \\
                   |1\rangle \\
                   |N\rangle
                 \end{array} \right)
 \rightarrow  \underbrace{\left( \begin{array}{cc|c} e^{- 4 \pi i/5} & 0 & 0 \\ 0 & -e^{-2 \pi i/5} & 0 \\
\hline  0 & 0& -e^{-2 \pi i/5}
\end{array}\right)}_{\mbox{$\rho({\sigma_1})$}}  \left( \begin{array}{c}
                   |0\rangle \\
                   |1\rangle \\
                   |N\rangle
                 \end{array} \right) \label{eq:sigma1}
\end{equation}
where we have used the shorthand notation (See
Fig.~\ref{fig:fusion1}) for the three particle states.
Evaluating the effect of $\sigma_2$ is less
trivial. Here, we must first make a basis change (using $F$) in
order to determine how the two rightmost particles fuse.
Then we can make the rotation using
$\hat R$ and finally undo the basis change. Symbolically, we can
write $ \rho({\sigma_2}) = F^{-1} \hat R F $ where $\hat R$ rotates the
two rightmost particles.   To be more explicit, let us consider what
happens to the state $|0\rangle$. First, we use Eq. \ref{eq:Fdef} to
write $|0 \rangle = F_{\bf 1 1} |(\bullet, (\bullet, \bullet)_{\bf
1})_\fib\rangle + F_{\fib \bf 1} |(\bullet, (\bullet,
\bullet)_{\fib})_\fib\rangle$. Rotating the two right particles then
gives $e^{-4 \pi i/5} F_{\bf 1 1} |(\bullet, (\bullet, \bullet)_{\bf
1})_\fib\rangle - e^{-2 \pi i/5} F_{\fib \bf 1} |(\bullet, (\bullet,
\bullet)_{\fib})_\fib\rangle$, and then we transform back to the
original basis using the inverse of Eq. \ref{eq:Fdef} to yield
$\rho({\sigma_2})|0\rangle=  ( [F^{-1}]_{\bf 11} e^{-4 \pi i/5} F_{\bf 1
1} - [F^{-1}]_{\bf 1 \fib} e^{-2 \pi i/5} F_{\fib \bf 1} ) |0
\rangle + ([F^{-1}]_{\fib \bf 1} e^{-4 \pi i/5} F_{\bf 1 1} -
[F^{-1}]_{\fib  \fib} e^{-2 \pi i/5} F_{\fib \bf 1} ) |1 \rangle =
-e^{- \pi i/5}/\phi \, |0 \rangle -i e^{-i
 \pi/10}/\sqrt{\phi} \, | 1 \rangle$.  Similar results can be derived
 for the other two basis states to give the matrix
\begin{equation} \rho({\sigma_2}) = \left( \!\!\! \begin{array}{cc|c} -e^{- \pi i/5}/\phi & -i e^{-i
 \pi/10}/\sqrt{\phi} & 0
 \\   -i e^{-i \pi/10}/\sqrt{\phi}& -1/\phi & 0 \\
\hline  0 & 0& -e^{-2 \pi i/5}
\end{array} \!\!\! \right) \label{eq:sigma2}
\end{equation}
Since the braid operations $\sigma_1$ and $\sigma_2$ (and their
inverses) generate all possible braids on three strands (See
Fig.~\ref{fig:threebraid}), we can use Eqs. \ref{eq:sigma1} and
\ref{eq:sigma2} to determine the unitary operation resulting from
any braid on three strands, with the unitary operations being built
up from the elementary matrices $\rho({\sigma_1})$ and $\rho({\sigma_2})$
in the same way that the complicated braids are built from the braid
generators $\sigma_1$ and $\sigma_2$. For example, the braid
$\sigma_2 \sigma_1 \sigma_1 \sigma_2^{-1} \sigma_2^{-1} \sigma_1$
shown in Fig. \ref{fig:threebraid} corresponds to the unitary matrix
$\rho({\sigma_1}) \rho(\sigma_2^{-1}) \rho(\sigma_2^{-1}) \rho({\sigma_1})
\rho({\sigma_1}) \rho({\sigma_2})$  (note that the order is reversed since
the operations that occur at earlier times are written to the left
in conventional braid notation, but on the right when multiplying
matrices together).

\mysubsection{Computing with Fibonacci Anyons}
\mylabel{Fibonacci-Computing}  Now that we know
many of the properties of Fibonacci anyons, we would like to show
how to compute with them.  First, we need to construct our qubits.
An obvious choice might be to use two particles for a qubit and
declare the two states $|(\bullet, \bullet)_{\bf 1}\rangle$ and
$|(\bullet, \bullet)_\fib\rangle$ to be the two orthogonal states of
the qubit. While this is a reasonably natural looking qubit, it
turns out not to be convenient for computations. The reason for this
is that we will want to do single qubit operations (simple
rotations) by braiding. However, it is not possible to change the
overall quantum number of a group of particles by braiding within
that group.  Thus, by simply braiding the two particles around each
other, we can never change $|(\bullet, \bullet)_{\bf 1}\rangle$ to
$|(\bullet, \bullet)_\fib\rangle$.  To remedy this problem, it is
convenient to use three quasiparticles to represent a qubit as
suggested by ~\onlinecite{Freedman02a} (many other schemes for
encoding qubits are also possible \cite{Freedman02a,Hormozi07}).
Thus, we represent the two states of the qubit as the $|0\rangle $
and $|1 \rangle$ states shown in Fig. \ref{fig:fusion1}.  The
additional state $|N\rangle$ is a ``noncomputational" state.  In
other words, we arrange so that at the beginning and end of our
computations, there is no amplitude in this state.  Any amplitude
that ends up in this state is known as ``leakage error".   We note,
however, that the braiding matrices $\rho({\sigma_1})$ and
$\rho({\sigma_2})$ are block diagonal and therefore never mix the
noncomputational state $|N\rangle$ with the computational space
$|0\rangle$ and $|1 \rangle$ (This is just another way to say that
the overall quantum number of the three particles must remain
unchanged under any amount of braiding). Therefore, braiding the
three particles gives us a way to do single qubit operations with no
leakage.

In section \ref{sec:universal-tqc}, we will describe a proof that
the set of braids has a ``dense image" in
the set of unitary operations for the Fibonacci theory. This means
that there exists a braid that corresponds to a unitary operation
arbitrarily close to any desired operation. The closer one wants to
approximate the desired unitary operation, the longer the braid
typically needs to be, although only logarithmically so (i.e, the
necessary braid length grows only as the log of the allowed error
distance to the target operation). The problem of actually finding
the braids that correspond to desired unitary operations, while
apparently complicated, turns out to be
straightforward \cite{Bonesteel05,Hormozi07}.  One simple approach is
to implement a brute force search on a (classical) computer to
examine all possible braids (on three strands) up to some certain
length, looking for a braid that happen to generate a unitary
operation very close to some desired result. While this approach
works very well for searching short
braids \cite{Bonesteel05,Hormozi07}, the job of searching all braids
grows exponentially in the length of the braid, making this scheme
unfeasible if one requires high accuracy long braids. Fortunately,
there is an iterative algorithm by Solovay and Kitaev
(see \onlinecite{Nielsen00}) which allows one to put together many short
braids to efficiently construct a long braid arbitrarily close to
any desired target unitary operation. While this algorithm does not
generally find the shortest braid for performing some operation
(within some allowed error), it does find a braid which is only
polylogarithmically long in the allowed error distance to the
desired operation. Furthermore,  the (classical) algorithm for
finding such a braid is only algebraically hard in the length of the
braid.

\begin{figure}
\centerline{\includegraphics[width=3.4in,angle=0]{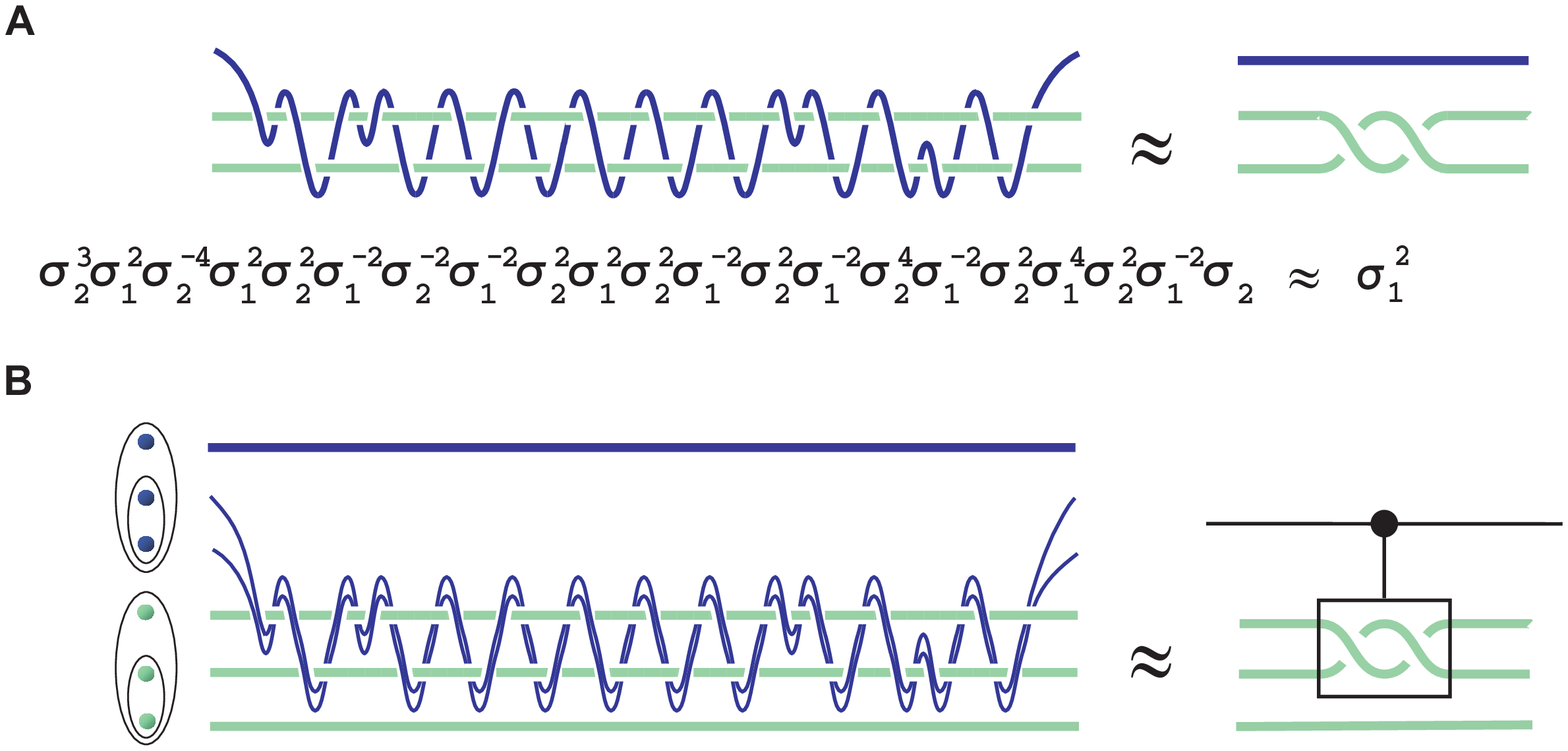}}
\caption{Construction of a two qubit gate from a certain three
particle problem. ime flows from left to right in this picture.
In the top we construct a braid on three strands
moving only the blue particle which has the same effect as
interchanging the two green strands.  Using this same braid
(bottom), then constructs a controlled rotation gate.  If the state
of the upper (control) qubit is $|0\rangle$, i.e., the control pair
is in state $\bf 1$ then the braid has no effect on the Hilbert
space (up to a phase).  if, the upper (control) qubit is in the
state $|1\rangle$ then the braid has the same effect as winding two
of the particles in the lower qubit.  Figure from
\onlinecite{Bonesteel05}} \label{fig:twoqubitgate}
\end{figure}

Having solved the single qubit problem, let us now imagine we have
multiple qubits, each encoded with three particles.  To perform
universal quantum computation, in addition to being able to perform
single qubit operations,  we must also be able to perform two-qubit
entangling gates \cite{Nielsen00,Bremner02}. Such two-qubit gates will
necessarily involve braiding together (physically ``entangling"!)
the particles from two different qubits. The result of
\onlinecite{Freedman02a} generally guarantees that braids exist
corresponding to any desired unitary operation on a two-qubit
Hilbert space. However, finding such braids is now a much more
formidable task.  The full Hilbert space for six Fibonacci particles
(constituting two qubits) is now 13 dimensional, and searching for a
desired result in such a high dimensional space is extremely hard
even for a powerful classical computer. Therefore, the problem needs
to be tackled by divide-and-conquer approaches, building up
two-qubit gates out of simple braids on three
particles \cite{Bonesteel05,Hormozi07}.  A simple example of such
a construction is sketched in Fig.~\ref{fig:twoqubitgate}.  First, in
Fig.~\ref{fig:twoqubitgate}.a, we consider braids on three strands
that moves (``weaves" \cite{Simon06}) only a single particle (the blue
particle in the figure) through two stationary particles (the green
particles). We search for such a braid whose action on the Hilbert
space is equivalent to exchanging the two green particles twice.
Since this is now just a three particle problem, finding such a
braid, to arbitrary accuracy, is computationally tractable.  Next,
for the two qubit problem, we label one qubit the control (blue in
Fig.~\ref{fig:twoqubitgate}.b) and another qubit the target (green).
We take a pair of particles from the control qubit (the control
pair) and weave them as a group through two of the particles in the
target, using the same braid we just found for the three particle
problem. Now, if the quantum number of the control pair is $\bf 1$
(i.e, control qubit is in state $|0\rangle$) then any amount of
braiding of this pair will necessarily give just an Abelian phase
(since moving $\bf 1$ around is like moving nothing around).
However, if the quantum number of the control pair is $\fib$ (i.e,
the control qubit is in state $|1\rangle$) then we can think of this
pair as being equivalent to a single $\fib$ particle, and we will
cause the same nontrivial rotation as in
Fig.\ref{fig:twoqubitgate}.a above (Crucially, this is designed to
not allow any leakage error!). Thus, we have constructed a
``controlled rotation" gate, where the state of the target qubit is
changed only if the control qubit is in state $|1\rangle$, where the
rotation that occurs is equivalent to exchanging two particles of
the target qubit as shown in Fig.~\ref{fig:twoqubitgate}.b.  The
resulting two-qubit controlled gate, along with single qubit
rotations, makes a universal set for quantum
computation \cite{Bremner02}.  More conventional two-qubit gates, such
as the controlled NOT gates (CNOT), have also been designed using
braids \cite{Bonesteel05,Hormozi07}.

\mysubsection{Other theories}
\mylabel{other}
The Fibonacci theory is a particularly interesting theory to study,
not only because of its
simplicity, but also because of its close relationship (see the discussion
at the beginning of section \ref{sec:fibonacci})
with the $\mathbb{Z}_3$ parafermion theory  ---
a theory thought to actually describe \cite{Rezayi06} the observed
quantum Hall state at $\nu=12/5$ \cite{Xia04}.
It is not hard to show that a given
braid will perform the same quantum computation in either
theory \cite{Hormozi07} (up to an irrelevant overall Abelian phase).
Therefore, the Fibonacci theory and the associated braiding may be physically
relevant for fractional quantum Hall topological quantum computation
in high-mobility 2D semiconductor structures.

However, there are many other non-Abelian theories, which are not
related to Fibonacci anyons.  Nonetheless, for arbitrary
non-Abelian theories, many of the themes we have discussed in this
section continue to apply.  In all cases, the Hilbert space can be
understood via fusion rules and an $F$-matrix; rotations of two
particles can be understood as a rotation $\hat R$ operator that
produces a phase dependent on the quantum number of the two
particles;  and one can always encode qubits in the quantum number
of some group of particles.  If we want to be able to do single
qubit operations by braiding particles within a qubit (in a theory
that allows universal quantum computation) we always need to encode
a qubit with at least three particles (sometimes more).  To perform
two-qubit operations we always need to braid particles constituting
one qubit with the particles constituting another qubit.   It is
always the case that for any unitary operation that can be achieved
by braiding $n$ particles around each other with an arbitrary braid
can also be achieved by weaving a single particle around $n-1$
others that remain stationary \cite{Simon06} (Note that we implicitly
used this fact in constructing Fig.~\ref{fig:threebraid}.a).  So long
as the state is among the ones known to have braid group representations
with dense images in the unitary group, as described in
Section \ref{sec:universal-tqc} below, it will be able to support universal quantum computation. Finally, we note that it seems to
always be true that the practical construction of complicated braids
for multi-qubit operations needs to be subdivided into more
manageable smaller problems for the problem to be tractable.

\subsection{Universal Topological Quantum Computation}
\label{sec:universal-tqc}

\begin{figure}[htpb]
\begin{center}
\includegraphics[height=0.8in]{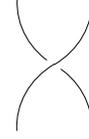}
\caption{The entire gate set needed in a state supporting universal
quantum computation.}
\label{fig:braid-gate}
\end{center}
\end{figure}

As we have seen in subsection \ref{sec:5/2-qubits}, even
if the $\nu=5/2$ state is non-Abelian, it
is not non-Abelian enough to function as a universal
quantum computer simply by braiding anyons.
However, in subsection \ref{sec:fibonacci}, we described
Fibonacci anyons which, we claimed, were capable of supporting
universal topological quantum computation.
In this subsection, we sketch a proof of this claim within
the context of the more general question:
which topological states are universal for quantum computation or, in
starker terms, for which topological
states is the entire gate set required to efficiently
simulate an arbitrary quantum circuit to arbitrary accuracy
simply that depicted in Figure \ref{fig:braid-gate} (see also
\onlinecite{Kauffman04,Kauffman07}).

The discussion in this section is more mathematical
than the rest of the paper and can skipped by less
mathematically-inclined readers.

In other words, the general braid is composed of copies of a single
operation (depicted in Figure \ref{fig:braid-gate}) and its inverse.
(Actually, as we will see, ``positive braids'' will prove to be sufficient,
so there is no necessity to ever use the inverse operation.)
Fibonacci anyons, which we discussed in subsection \ref{sec:fibonacci},
are an example which have this property. In this subsection, we
will see why.

For the sake of concreteness,
let us assume that we use a single species of quasiparticle,
which we will call $\sigma$. When there are $n$
$\sigma$'s at fixed positions $z_1, \dots , z_n$, there
is an exponentially-large ($\sim \left({d_\sigma}\right)^n$-dimensional)
ground state subspace of Hilbert space. Let us call this
vector space $V_n$. Braiding the $\sigma$'s produces a representation
$\rho_n$ characteristic of the topological phase in question,
$\rho_n: {\cal B}_n \rightarrow$ U($V_n$) from the braid group on $n$ strands
into the unitary transformations of $V_n$. We do not care
about the overall phase of the wavefunction, since only the projective
reduction in PU($V_n$) has physical significance.
(PU($V_n$) is the set of unitary transformations on $V_n$ with
two transformations identified if they differ only by a phase.)
We would like to be able to enact an arbitrary unitary transformation,
so $\rho({\cal B}_n)$ should be dense in PU(${V_n}$), i.e. dense up
to phase. By `dense' in PU(${V_n}$), we mean that the intersection of all closed sets
containing $\rho({\cal B}_n)$ should simply be PU(${V_n}$).
Equivalently,
it means that an arbitrary unitary transformation can be approximated, up to a phase,
by a transformation in $\rho({\cal B}_n)$ to within any desired accuracy.
This is the condition which our topological phase must satisfy.

For a modestly large number ($\geq 7$) of $\sigma$s,
it was shown \cite{Freedman02a,Freedman02b} that the braid group
representations associated with $SU(2)$ Chern-Simons theory
at level $k \neq 1,2,4$ are dense in $SU(V_{n,k})$ (and hence in
$PU(V_{n,k})$). With only a small number of low-level and small
anyon number exceptions, the same articles show density for almost
all SU(N)$_k$.

These Jones-Witten (JW) representations satisfy a key ``two
eigenvalue property'' (TEVP), discussed below, derived in this
$SU(N)$ setting from the Hecke relations, and corresponding to the
HOMFLY polynomial (see, for instance, \onlinecite{Kauffman01}
and refs. therein). 
The analysis was extended with similar
conclusions in \cite{Larsen05}
to the case where the Lie group $G$ is of type BCD and braid
generators have three eigenvalues, corresponding to the BMW algebra
and the two variable Kauffman polynomial.  For JW-representations of
the exceptional group at level $k$, the number of eigenvalues of
braid generators can be composite integers (such as 4 for $G_2$) and
this has so far blocked attempts to prove density for these
JW-representations.

In order to perform quantum computation with anyons,
there are many details needed to align the topological picture with the usual
quantum-circuit model from computer science. First, qubits must be
located in the state space $V_n$.  Since $V_n$ has no natural tensor
factoring (it can have prime dimension) this alignment \cite{Freedman02a} is
necessarily a bit inefficient\footnote{Actually, current
schemes use approximately half the theoretical
number of qubits.  One finds $\alpha \text{log}_2(\text{dim}{V_n})$
computational qubits in $V_n$, for $\alpha = (\text{log}_2
\tau^3)^{-1} \approx 0.48$, $\phi = \frac{1 + \sqrt{5}}{2}$.}; some
directions in $V_n$ are discarded from the computational space and
so we must always guard against unintended ``leakage'' into the
discarded directions.  A possible research project is how to adapt
computation to ``Fibonacci'' space (see subsection \ref{sec:fibonacci}) rather than
attempting to find binary structure within $V_n$.  A somewhat forced
binary structure was explained in subsection \ref{sec:fibonacci}
in connection with encoding qubits into SU(2)$_ 3$,
as it was done for level $2$ in subsection \ref{sec:5/2-qubits}.
(A puzzle for readers:  Suppose we write integers out as ``Fibonacci
numerals'':  0 cannot follow 0, but 0 or 1 can follow 1.  How do you
do addition and multiplication?) However, we will not dwell on
these issues but instead go directly to the essential
mathematical point: How, in practice, does one tell which braid
group representations are dense and which are not, i.e. which ones
are sufficient for universal topological quantum computation and which ones
need to be augmented by additional non-topological gate operations?

We begin by noting that the fundamental skein relation of Jones'
theory is:

\begin{figure}[htpb]
\begin{center}
\includegraphics[height=1in]{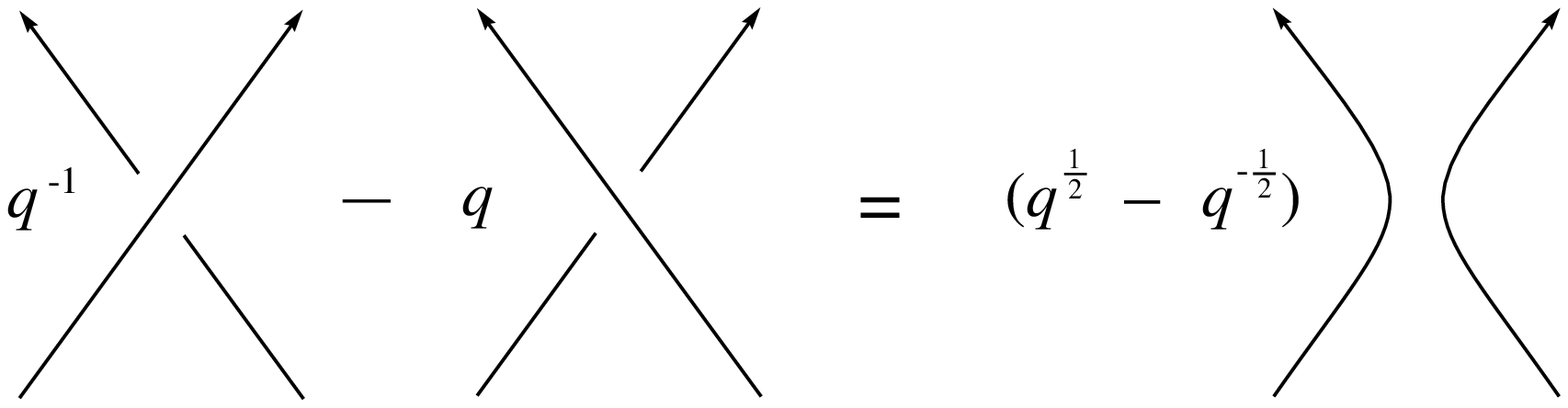}
\caption{Jones skein relation. (See (\ref{eqn:Jones-skein})}
\label{figure3}
\end{center}
\end{figure}

 \noindent (see (\ref{eqn:Jones-skein}) and the associated relation for the
Kauffman bracket (\ref{eq:recursion})) This is a quadratic relation
in each braid generator $\sigma_i$ and by inspection any
representation of $\sigma_i$ will have only two distinct eigenvalues
$q^{\frac{3}{2}}$ and $-q^{\frac{1}{2}}$.  It turns out to be
exceedingly rare to have a representation of a compact Lie group $H$
where $H$ is densely generated by elements $\sigma_i$ with this
eigenvalue restriction. This facilitates the identification of the
compact closure $H = \overline{\text{image}(\rho)}$ among the
various compact subgroups of $U(V_n)$.

\begin{defn}
\label{defn1} Let $G$ be a compact Lie group and $V$ a faithful,
irreducible, unitary representation.  The pair $(G,V)$ has the {\it
two eigenvalue property} (TEVP) if there exists a conjugacy class
$[g]$ of $G$ such that:
\begin{enumerate}
\item $[g]$ generates a dense set in $G$
\item For any $g \in [g]$, $g$ acts on $V$ with exactly two distinct
eigenvalues whose ratio is not $-1$.
\end{enumerate}
\end{defn}

Let $H$ be the closed image of some Jones representation $\rho : B_n
\rightarrow U(V_n)$.  We would like to use figure \ref{figure3} to
assert that the fundamental representation of $U(V_n)$ restricted to
$H$, call it $\theta$, has the TEVP.  All braid generators
$\sigma_i$ are conjugate and, in nontrivial cases, the eigenvalue
ratio is $-q \neq -1$.  However, we do not yet know if the
restriction is irreducible.  This problem is solved by a series of
technical lemmas in \cite{Freedman02a}.  Using TEVP, it is shown first that the
further restriction to the identity component $H_0$ is isotipic and
then irreducible.  This implies that $H_0$ is reductive, so its
derived group $[H_0, H_0]$ is semi-simple and, it is argued, still
satisfies the TEVP.  A final (and harmless) variation on $H$ is to
pass to the universal cover $H' := \widetilde{[H_0, H_0]}$.  The
pulled back representation $\theta'$ still has the TEVP and we are
finally in a situation, namely irreducible representations of
semi-simple Lie groups of bounded dimension, where we can hope to
apply the classification of such representations \cite{McKay81}
to show that our mysterious $H'$
is none other than $SU(V_n)$.  If this is so, then it will follow
that the preceding shenanigans $H \rightarrow H_0 \rightarrow [H_0,
H_0] \rightarrow \widetilde{[H_0, H_0]}$ did nothing (beyond the
first arrow, which may have eliminated some components of $H$ on
which the determinant is a nontrivial root of unity).

In general, milking the answer (to the question of which Jones
representations are projectively dense) out of the classification
requires some tricky combinatorics and rank-level \cite{Freedman02b} duality.
Here we will be content with doing the easiest nontrivial case.
Consider six Fibonacci anyons $\tau$ with total charge $= 1$.  The
associated $V_6 \cong \C^5 \cong 2 \text{\;qubits} \oplus
\text{non-computational\;} \C$ as shown:

\begin{figure}[htpb]
\begin{center}
\includegraphics[height=1in]{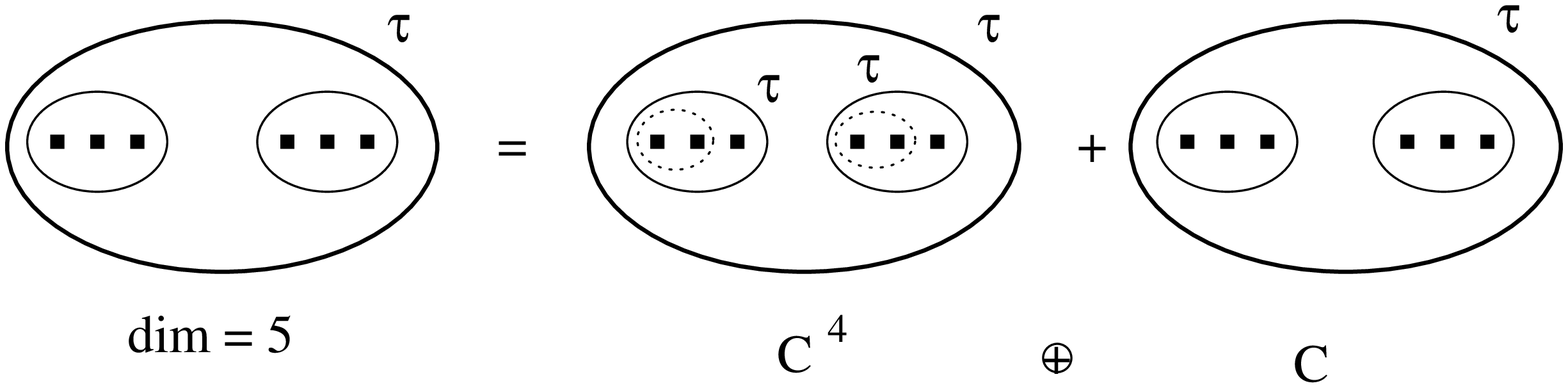}
\caption{The charge on the dotted circle can be 1 or $\tau$
providing the qubit.} \label{figure4}
\end{center}
\end{figure}

In coordinates, $\rho$ takes the braid generators (projectively) to
these operators:
$$\sigma_1 \longmapsto \left[ \begin{array}{rrrrr} -1 \\ & q \\ && -1 \\ &&& q \\ &&&& q \end{array} \right],\;\;\;\;\;\;\; q = e^{-\frac{2\pi i}{5}} $$

$$\sigma_2 \longmapsto \left[ \begin{array}{rrrrr}\frac{q^2}{q+1} & -\frac{q\sqrt{[3]}}{q+1} \\ -\frac{q\sqrt{[3]}}{q+1} & -\frac{1}{q+1} \\
&& \frac{q^2}{q+1} & -\frac{q\sqrt{[3]}}{q+1} \\ &&
-\frac{q\sqrt{[3]}}{q+1} & -\frac{1}{q+1} \\ &&&& q
\end{array} \right]$$

\noindent where $[3] = q + q^{-1} + 1$, and $\sigma_i$, for $i = 3,4,5$, are similar.
See \onlinecite{Funar99} for details.

The closed image of $\rho$ is $H \subset U(5)$, so our irreducible
representation $\theta'$ of $H'$, coming from $U(5)$'s fundamental,
is exactly 5 dimensional (we don't yet know the dimension of $H'$).
From \onlinecite{McKay81}, there are four 5-dimensional irreducible
representations, which we list by rank:
\begin{enumerate}
\item rank = 1:  $(SU(2), 4\pi_1)$
\item rank = 2:  $(Sp(4), \pi_2)$
\item rank = 4:  $(SU(5), \pi_i)$, $i=1,4$
\end{enumerate}

Suppose $x \in SU(2)$ has eigenvalues $\alpha$ and $\beta$ in
$\pi_1$.  Then under $4\pi_1$, it will have $\alpha^i \beta^j$, $i +
j = 4 \; (i,j \geq 0)$ as eigenvalues, which are too many (unless
$\frac{\alpha}{\beta} = -1$).  In case (2), since 5 is odd, every
element has at least one real eigenvalue, with the others coming in
reciprocal pairs.  Again, there is no solution.  Thus, the TEVP
shows we are in case (3), i.e. that $H' \cong SU(5)$.  It follows
from degree theory that $[H_0, H_0] \cong SU(5)$ and from this we
get the desired conclusion: $SU(5) \subset H \subset U(5)$.

We have not yet explained in what sense the topological
implementations of quantum computations are efficient.  Suffice it
to say that there are (nearly) quadratic time algorithms due to
Kitaev and Solvay \cite{Nielsen00} for finding the braids that
approximate a given quantum circuit.  In practice, brute force, load
balanced searches for braids representing fundamental gates, should
yield accuracies on the order of $10^{-5}$ (within the ``error
threshold''). Note that these are systematic, unitary errors
resulting from the fact that we are enacting a unitary
transformation which is a little different from what an algorithm
may ask for. Random errors, due to decoherence, are caused by
uncontrolled physical processes, as we discuss in the next
subsection.

\subsection{Errors}
\label{sec:errors}

As we discussed in section \ref{sec:Topological_Quantum_Computation},
small inaccuracies in the trajectories along which we move our
quasiparticles are not a source of error. The topological class of
the quasiparticles' trajectories (including undesired quasiparicles)
must change in order for an error to
occur. Therefore, to avoid errors, one must keep
careful track of all of the quasiparticles in the system and move
them so that the intended braid is performed.   As mentioned in
the introduction section \ref{sec:Topological_Quantum_Computation},
stray thermally excited quasiparticles could form unintended braids
with the quasiparticles of our system and cause errors in the
computation. Fortunately, as we mentioned in section
\ref{sec:Topological_Quantum_Computation}, there is a large class of
such processes that actually do not result in errors.  We will
discuss the two most important of these.

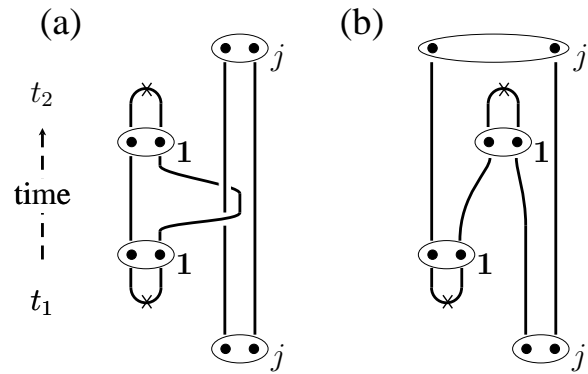
\begin{figure}[htbp]
\vspace*{100pt} \hspace*{-80pt}
\ifx\JPicScale\undefined\def\JPicScale{.3}\fi \psset{unit=.25 mm}
\psset{linewidth=0.3,dotsep=1,hatchwidth=0.3,hatchsep=1.5,shadowsize=1,dimen=middle}
\psset{dotsize=0.7 2.5,dotscale=1 1,fillcolor=black}
\psset{arrowsize=1 2,arrowlength=1,arrowinset=0.25,tbarsize=0.7
5,bracketlength=0.15,rbracketlength=0.15}
\begin{pspicture}(0,0)(140,180)
% this is qubit j
\rput{0}(77.5,27.5){\psellipse[fillstyle=solid](0,0)(2.5,-2.5)}
\rput{0}(92.5,27.5){\psellipse[fillstyle=solid](0,0)(2.5,-2.5)}
%\rput{0}(47.5,77.5){\psellipse[fillstyle=solid](0,0)(2.5,-2.5)}
\rput{0}(85,27.5){\psellipse[](0,0)(15,-7.5)}
\rput(105.5,22.5){\large $j$}
\rput{0}(77.5,187.5){\psellipse[fillstyle=solid](0,0)(2.5,-2.5)}
\rput{0}(92.5,187.5){\psellipse[fillstyle=solid](0,0)(2.5,-2.5)}
%\rput{0}(47.5,77.5){\psellipse[fillstyle=solid](0,0)(2.5,-2.5)}
\rput{0}(85,187.5){\psellipse[](0,0)(15,-7.5)}
\rput(105.5,182.5){\large $j$}
% this is created qubit
 \rput{0}(27.5,77.5){\psellipse[fillstyle=solid](0,0)(2.5,-2.5)}
\rput{0}(42.5,77.5){\psellipse[fillstyle=solid](0,0)(2.5,-2.5)}
\rput{0}(35,77.5){\psellipse[](0,0)(15,-7.5)}
\rput(55.5,72.5){\large $\bf 1$}
 \rput{0}(27.5,137.5){\psellipse[fillstyle=solid](0,0)(2.5,-2.5)}
\rput{0}(42.5,137.5){\psellipse[fillstyle=solid](0,0)(2.5,-2.5)}
\rput{0}(35,137.5){\psellipse[](0,0)(15,-7.5)}
\rput(55.5,132.5){\large $\bf 1$}
% motion of the created qubit
\psset{linewidth=1.5} \psline(27,85)(27,130)
\psline(42.5,85)(42.5,90) \psbezier(42.5,90)(42.5,95)(85,95)(85,100)
\psline(42.5,125)(42.5,130)
\psbezier(42.5,125)(42.5,120)(85,115)(85,110)
\psline(85,100)(85,110)
\pspolygon[linewidth=0pt,fillcolor=white,fillstyle=solid,linecolor=white]
(74,110)(80,110)(80,120)(74,120)
% this is the creation process
\psset{linewidth=1.5} \rput{0}(35,60) {\pscustom{\scale{1 1}
\psarc(0,0){8}{180}{360} }} \psline(43,70.5)(43,60)
\psline(27,70.5)(27,60) \psset{linewidth=.75} \psline(33,56)(38,49)
\psline(38,56)(33,49)
% this is anticreation
\psset{linewidth=1.5} \rput{0}(35,158) {\pscustom{\scale{1 1}
\psarc(0,0){8}{0}{180} }} \psline(43,145.5)(43,158)
\psline(27,145.5)(27,158) \psset{linewidth=.75}
\psline(33,169)(38,162) \psline(38,169)(33,162)
% this is motion of j qubit
\psset{linewidth=1.5} \psline(77,35)(77,93)
\psline(93,35)(93,180)\psline(77,100)(77,180)
% this is the timeline
\psline[linestyle=dashed,arrows=->](-20,75)(-20,145)
\rput[angle=90](-20,110){\large time} \rput(-20,52){\large $t_1$}
\pspolygon[linewidth=0pt,fillcolor=white,fillstyle=solid,linecolor=white]
(-25,100)(-15,100)(-15,118)(-25,118) \rput[angle=90](-20,110){\large
time} \rput(-20,52){\large $t_1$}
 \rput(-20,163){\large $t_2$} \rput[angle=90](-10,200){\Large (a)}
%%%%
% this is the second process
\rput[angle=90](150,200){\Large (b)}
\ifx\JPicScale\undefined\def\JPicScale{.3}\fi \psset{unit=.25 mm}
\psset{linewidth=0.3,dotsep=1,hatchwidth=0.3,hatchsep=1.5,shadowsize=1,dimen=middle}
\psset{dotsize=0.7 2.5,dotscale=1 1,fillcolor=black}
\psset{arrowsize=1 2,arrowlength=1,arrowinset=0.25,tbarsize=0.7
5,bracketlength=0.15,rbracketlength=0.15}
% this is created qubit
 \rput{0}(187.5,77.5){\psellipse[fillstyle=solid](0,0)(2.5,-2.5)}
\rput{0}(202.5,77.5){\psellipse[fillstyle=solid](0,0)(2.5,-2.5)}
\rput{0}(195,77.5){\psellipse[](0,0)(15,-7.5)}
\rput(215.5,72.5){\large $\bf 1$}
\rput{0}(217.5,137.5){\psellipse[fillstyle=solid](0,0)(2.5,-2.5)}
\rput{0}(232.5,137.5){\psellipse[fillstyle=solid](0,0)(2.5,-2.5)}
\rput{0}(225,137.5){\psellipse[](0,0)(15,-7.5)}
\rput(245.5,132.5){\large $\bf 1$}
\rput{0}(237.5,27.5){\psellipse[fillstyle=solid](0,0)(2.5,-2.5)}
\rput{0}(252.5,27.5){\psellipse[fillstyle=solid](0,0)(2.5,-2.5)}
\rput{0}(245,27.5){\psellipse[](0,0)(15,-7.5)}
\rput(265.5,22.5){\large $j$}
% this is the final qubit
\rput{0}(187.5,187.5){\psellipse[fillstyle=solid](0,0)(2.5,-2.5)}
\rput{0}(252.5,187.5){\psellipse[fillstyle=solid](0,0)(2.5,-2.5)}
%\rput{0}(47.5,77.5){\psellipse[fillstyle=solid](0,0)(2.5,-2.5)}
\rput{0}(220,187.5){\psellipse[](0,0)(40,-7.5)}
\rput(266.5,182.5){\large $j$}
% this is anticreation
\psset{linewidth=1.5} \rput{0}(225,158) {\pscustom{\scale{1 1}
\psarc(0,0){8}{0}{180} }} \psline(233,145.5)(233,158)
\psline(217,145.5)(217,158) \psset{linewidth=.75}
\psline(223,169)(228,162) \psline(228,169)(223,162)
% this is creation
\psset{linewidth=1.5} \rput{0}(195,60) {\pscustom{\scale{1 1}
\psarc(0,0){8}{180}{360} }} \psline(203,70.5)(203,60)
\psline(187,70.5)(187,60) \psset{linewidth=.75}
\psline(193,56)(198,49) \psline(198,56)(193,49)
% this is motion of j qubit
\psset{linewidth=1.5} \psline(237,35)(237,93)
\psline(253,35)(253,180) \psline(187,85)(187,180)
% and the curvy part
\psbezier(237,93)(237,110)(232,122)(232,129)
\psbezier(202,85)(202,110)(218,122)(218,129)
\end{pspicture}
 \caption{Two processes involving excited quasiparticle-quasihole pairs
 that do not cause errors in a topological quantum computation.
 (a) In the process shown on the left,
 a quasiparticle-quasihole pair is excited at
 time $t_1$ (marked by an $\times$), one of these particles wraps around a
 quasiparticle of our computational system, and then comes back to
 its partner and re-annihilates at a later time $t_2$.   When the
 pair is created it necessarily has the identity quantum number $\bf 1$
 of the vacuum, and when it annihilates, it also necessarily has
 this vacuum quantum number.  As a result (as discussed in the text) the
 quantum number of the computational system is not changed by this
 process.  (b) In the process shown on the right,  a
 quasiparticle-quasihole pair is excited at
 time $t_1$ (marked by an $\times$), one of these particles annihilates
an existing quasiparticle of our computational system at a later
time $t_2$, and leaves behind its partner to replace the the
annihilated quasiparticle of the computational system.  Again, when
the pair is created, it necessarily has the identity quantum number
$\bf 1$ of the vacuum.  Similarly the annihilating pair has the
quantum number of the vacuum.  As a result, the two particles
remaining in the end have the same quantum numbers as the two
initial quantum numbers of the computational system.}
  \label{fig:noerror}
\end{figure}

Perhaps the simplest such process that does not cause errors is when
a quasiparticle-quasihole pair is thermally (or virtually) excited
from the vacuum, one of the two excited particles wanders around a
single quasiparticle in our system then returns to reannihilate its
partner.  (See Figure~\ref{fig:noerror}.a).   For the sake of
argument, let us imagine that our initial computational system is a
pair of quasiparticles in state $j$.  At some time $t_1$ (marked by
an $\times$ in the figure), we imagine that a
quasiparticle-quasihole pair becomes excited from the vacuum. Since
the pair comes from the vacuum, it necessarily has overall quantum
number $\bf 1$ (i.e., fusing these particles back together gives the
vacuum $\bf 1$). Thus the overall quantum number of all four
particles is $j$.  (In the above notation, we could draw a circle
around all four particles and label it $j$). We then imagine that
one of our newly created quasiparticles wanders around one of the
quasiparticles of our computational system as shown in the figure.
Using $F$ matrices or braiding matrices $\hat \sigma$ we could
compute the full state of the system after this braiding operation.
Importantly, however, the overall quantum number $j$ of all four
particles is preserved.

Now at some later time $t_2$ the two created particles reannihilate
each other and are returned to the vacuum as shown by the second
$\times$ in Figure \ref{fig:noerror}.a.   It is crucial to point out
that in order for two particles to annihilate, they must have the
identity quantum number $\bf 1$ (i.e., they must fuse to $\bf 1$).
The annihilation can therefore be thought of as a measurement of the
quantum number of these two particles. The full state of the system,
then collapses to a state where the annihilating particles have
quantum number $\bf 1$. However, the overall quantum number of all
four particles must remain in the state $j$.  Further, in order for
the overall state of the four particles to be $j$ and the two
annihilating particles to be $\bf 1$ the two other (original)
particles must have quantum number $j$.  Thus, as shown in the
figure, the two original quasiparticles must end up in their
original state $j$ once the created particles are re-annihilated.
Similarly, if the original particles had started in a superposition
of states, that superposition would be preserved after the
annihilation of the two excited particles. (Note that an arbitrary
phase might occur, although this phase is independent of the quantum
number $j$ and therefore is irrelevant in the context of quantum
computations).

Another very important process that does not cause errors is shown
in Figure \ref{fig:noerror}.b.  In this process, one of the members
of a thermally excited quasiparticle-quasihole pair annihilate with
one of the particles in our computational system, leaving behind its
partner as a replacement.  Again, since both the created pair and
the annihilating particles  have the same quantum numbers as the
vacuum, it is easy to see  (using similar arguments as above) that
the final state of the two remaining particles must be the same as
that of the original two particles, thus not causing any errors so
long as the new particle is used as a replacement for the
annihilated quasiparticle.

The fact that the two processes described above do not cause errors
is actually essential to the notion of topological quantum
computation. Since the created quasiparticles need not move very far
in either process, these processes can occur very frequently, and
can even occur virtually since they could have low total action.
Thus it is crucial that these likely processes do not cause errors.
The simplest processes that can actually cause error would require a
thermally (or virtually) created  quasiparticle-quasihole pair to
braid nontrivially with at least two quasiparticles of our
computational system. Since it is assumed that all of the
quasiparticles that are part of our system are kept very far from
each other, the action for a process that wraps a (virtually)
created quasiparticle around two different particles of our system
can be arbitrarily large, and hence these virtual processes can be
suppressed. Similarly, it can be made unlikely that thermally
excited quasiparticles will wrap around two separate particles of
our system before re-annihilating. Indeed, since in two dimensions a
random walk returns to its origin many times, a wandering
quasiparticle may have many chances to re-annihilate before it wraps
around two of the particles of our computational system and causes
errors. Nonetheless, in principle, this process is a serious consideration
and has the potential to cause errors if too many quasiparticle-quasihole
pairs are excited.

The probability for these error-causing processes is naively $\sim
e^{-\Delta/(2T)}$ (thermally-excited quasiparticles) or $\sim
e^{-\Delta L/v}$ (virtual quasiparticles), where $T$ is the
temperature, $\Delta$ is quasiparticle energy gap, $L$ is the
distance between the quasiparticles comprising a qubit, and $v$ is a
characteristic velocity. However, transport in real systems is, in
fact, more complicated. Since there are different types of
quasiparticles, the gap measured from the resistance may not be the
smallest gap in the system. For instance, neutral fermionic
excitations in the Pfaffian state/SU(2)$_2$ may have a small gap,
thereby leading to a splitting between the two states of a qubit if
the two quasiparticles are too close together. Secondly, in the
presence of disorder, the gap will vary throughout the system.
Processes which take advantage of regions with small gaps may
dominate the error rate. Furthermore, in a disordered system,
variable-range hopping, rather than thermally-activated transport is
the most important process. Localized quasiparticles are an
additional complication. If they are truly fixed, then they can be
corrected by software, but if they drift during the course of a
calculation, they are a potential problem. In short, quasiparticle
transport, even ordinary electrical transport, is very complicated
in semiconductor quantum Hall systems. A complete theory does not
exist. Such a theory is essential for an accurate prediction of the
error rate for topological quantum computation in non-Abelian
quantum Hall states in semiconductor devices and is an important
future challenge for solid state theory.

\section{Future Challenges for Theory and Experiment}
\label{sec:conclusion}

Quantum mechanics represented a huge revolution in thought.  It was
such a stretch of the imagination that many great minds and much
experimental information were required to put it into place.  Now,
eighty years later, another collaborative effort is afoot to
revolutionize computation by a particularly
rich use of quantum mechanics.  The preceding information revolution,
which was based on the MOSFET, rested on the 1-electron physics
of semiconductors. The revolution which we advocate will require
the understanding and manipulation of strongly-interacting electron
systems.  Modern condensed matter physics has powerful tools
to analyze such systems: renormalization group (RG), CFT, Bethe Ansatz,
dualities, and numerics.  Even without the quantum computing
connection, many of the most interesting problems in physics lie in
this direction.  Prominent here is the problem of creating,
manipulating, and classifying topological states of matter.

\begin{figure}[htpb]
\begin{center}
\includegraphics[height=2in]{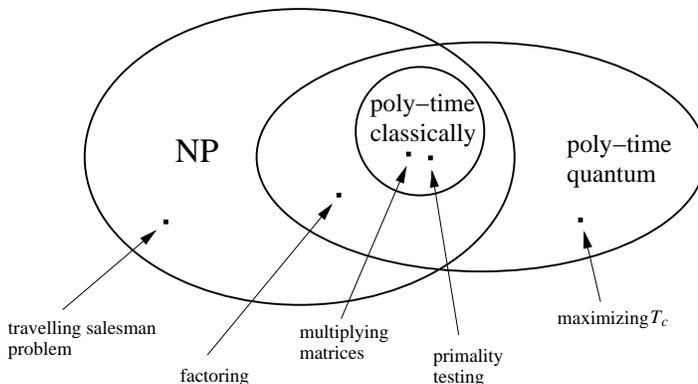}
\caption{A conjectural view of relative computational complexity}
\label{fig:complexity}
\end{center}
\end{figure}

There is a second ``richness'' in the connection between quantum
mechanics and computation.  The kind of computation which will
emerge is altogether new.  While the MOSFET-based silicon revolution facilitated
the same arithmetic as done on the abacus, the quantum computer will
compute in superposition.  We have some knowledge about what this
will allow us to do.  Select mathematical problems (factoring,
finding units in number fields, searching) have efficient solutions in the
quantum model.  Many others may succumb to quantum heuristics (e.g.
adiabatic computation \cite{Farhi00}) but we will not know until we can play with
real quantum computers.  Some physical problems, such as maximizing
$T_c$ within a class of superconductors, should be advanced by
quantum computers, even though, viewed as math problems, they lie
even outside class NP (i.e. they are {\it very} hard).  A
conjectural view of relative computational complexity
is shown in Fig. \ref{fig:complexity}.

But, before we can enter this quantum computing paradise, there are
fundamental issues of physics to be tackled. The first problem is
to find a non-Abelian topological phase in nature. The
same resistance to local perturbation that makes topological phases
astonishing (and, we hope, useful) also makes them somewhat covert.
An optimist might hope that they are abundant and that we are merely
untutored and have trouble noticing them.
At present, our search is guided primarily by a process of elimination:
we have focussed our attention on those systems in which
the alternatives don't occur -- either quantum Hall states for which
there is no presumptive Abelian candidate or frustrated magnets
which don't order into a conventional broken-symmetry state.
What we need to do is
observe some topological property of the system, e.g.
create quasiparticle excitations above the ground state,
braid them, and observe how the state of
the system changes as a result. In order to do this, we need
to be able to (1) create a specified number of quasiparticles at known
positions, (2) move them in a controlled way, and (3) observe their
state. All of these are difficult, but not impossible.

It is instructive to see how these difficulties are manifested
in the case of quantum Hall states and other possible topological
states. The existence of a topological phase in the quantum Hall regime is
signaled by the quantization of the Hall conductance.
This is a special feature of those chiral topological phases in which there is
a conserved current $J_\mu$ (e.g. an electrical charge current or
spin current). Topological invariance and $P,T$-violation
permit a non-vanishing correlation function of the form
\begin{equation}
\left\langle J_\mu (q) J_\nu (-q)\right\rangle = C\,\epsilon_{\mu\nu\lambda} {q_\lambda}
\end{equation}
where $C$ is a topological invariant. If the topological phase does not break $P$
and $T$ or if there is no conserved current in the low-energy effective field theory,
then there will not be such a dramatic signature.
However, even in the quantum Hall context, in which we have a leg up thanks
to the Hall conductance, it is still a subtle matter to determine which topological
phase the system is in.

As we have described, we used theoretical input to focus our attention
on the $\nu=5/2$ and $\nu=12/5$ states. Without such input, the available
phase space is simply too large and the signatures of a topological phase are too subtle.
One benefit of having a particular theoretical model of
a topological phase is that experiments can be done
to verify other (i.e. non-topological) aspects of the model. By corroborating
the model in this way, we can gain indirect evidence about the
nature of the topological phase. In the case of the $\nu=5/2$ state,
the Pfaffian model wavefunction \cite{Moore91,Greiter92} for this state
is fully spin-polarized. Therefore, measuring the spin polarization
at $\nu=5/2$ would confirm this aspect of the model, thereby strengthening
our belief in the the model as a whole -- including its topological features
(see \onlinecite{Tracy07} for such a measurement at $\nu=1/2$).
In the case of Sr$_2$RuO$_4$, the $p+ip$ BCS model
predicts a non-zero Kerr rotation \cite{Xia06}.
This is not a topological invariant, but when it is non-zero and the superconducting
order parameter is known to be a spin-triplet, we can infer a non-zero
spin quantum Hall effect (which is a topological invariant but is
much more difficult to measure). Thus, non-topological
measurements can teach us a great deal when we have a particular
model in mind.

In frustrated magnets, one often cuts down on the complex many-dimensional
parameter space in the following way: one focusses on systems
in which there is no conventional long-range order. Although it is
possible for a system to be in a topological phase and simultaneously
show conventional long-range order (quantum Hall ferromagnets are
an example), the absence of conventional
long-range order is often used as circumstantial evidence that
the ground state is `exotic' \cite{Coldea03,Shimizu03}. 
This is a reasonable place to start, but
in the absence of a theoretical model predicting a specific topological state,
it is unclear whether the ground state is expected to be
topological or merely `exotic' in some other way (see below for a further
discussion of this point).

While theoretical models and indirect probes can help to identify
strong candidates, only the direct measurement of a topological
property can demonstrate that a system is in a topological phase.
If, as in the quantum Hall effect, a system has been shown to be
in a topological phase through the measurement of one property
(e.g. the Hall conductance), then there is still the problem of
identifying which topological phase. This requires the complete
determination of all of its topological properties (in principle, the
quasiparticle species, their topological spins, fusion rules,
$R$- and $F$-matrices). Finding non-trivial
quasiparticles is the first step. In the quantum Hall regime, quasiparticles
carry electrical charge (generally fractional). Through capacitive measurements of
quasiparticle electric charges \cite{Goldman95}
or from shot noise measurements \cite{Picciotto97,Saminadayar97},
one can measure the minimal electric charges and infer
the allowed quasiparticle electric charges. The observation of
charge $e/4$ quasiparticles by either of these methods
would be an important step in characterizing
the $\nu=5/2$ state.
Detecting charged quasiparticles capacitatively
or through noise measurements necessitates
gated samples: anti-dots and/or point contacts. In the case
of delicate states such as $\nu=5/2$,
this is a challenge; we don't want the gates to reduce the
quality of the device and excessively degrade the robustness of the states.
Even if this proves not to be surmountable, it only solves
the problem of measuring charged quasiparticles; it  does not
directly help us with non-trivial neutral quasiparticles (such as those which
we believe exist at $\nu=5/2$).

Again, a particular theoretical model of the state can be extremely
helpful. In the case of the toric code, an excited plaquette or $\mathbb{Z}_2$
vortex (see Secs. \ref{sec:Othersystems} \ref{sec:P-T-Invariant})
is a neutral spinless excitation and, therefore, difficult to probe.
However, when such a phase arises in models of superconductor-Mott insulator
transitions, $\mathbb{Z}_2$ vortices can be isolated by going back and forth
through a direct second-order phase transition between a topological phase and
a superconducting phase \cite{Senthil01a}.
Consider a superconductor in an annular geometry with a single
half-flux quantum vortex through the hole in the annulus. Now suppose that some
parameter can be tuned so that the system undergoes a second-order
phase transition into an insulating state which is a topological phase
of the toric code or $\mathbb{Z}_2$ variety. Then the single vortex ground state
of the superconductor will evolve into a state with a $\mathbb{Z}_2$
vortex in the hole of the annulus. The magnetic flux will escape,
but the $\mathbb{Z}_2$ vortex will remain. (Eventually, it will
either quantum tunnel out of the system or, at finite temperature,
be thermally excited out of the system. It is important to perform the
experiment on shorter time scales.) If the system is then taken back
into the superconducting state, the $\mathbb{Z}_2$ vortex will evolve
back into a superconducting vortex; the flux must be regenerated, although
its direction is arbitrary. Although Senthil and Fisher considered the case
of a $\mathbb{Z}_2$ topological phase, other topological phases with
direct second-order phase transitions into superconducting states
will have a similar signature. On the other hand, in a non-topological phase,
there will be nothing left in the insulating phase after the flux has escaped.
Therefore, when the system is taken back into the superconducting phase,
a vortex will not reappear.
The effect described above is not a feature of the topological
phase alone, but depends on the existence of a second-order quantum
phase transition between this topological state and a superconducting
state. However, in the happy circumstance that such a transition
does exist between two such phases of some material, this experiment
can definitively identify a topologically non-trivial neutral excitation.
In practice, the system is not tuned through a quantum phase transition
but instead through a finite-temperature one; however, so long as the temperature
is much smaller than the energy gap for a $\mathbb{Z}_2$ vortex,
this is an unimportant distinction. This experiment was
performed on an underdoped cuprate superconductor
by \onlinecite{Wynn01}. The result was negative, implying that
there isn't a topological phase in the low-doping part of the phase diagram
of that material, but the experimental technique may still prove to
be a valuable way to test some other candidate material in the future.
It would be interesting and useful to design analogous experiments
which could exploit the possible proximity of topological phases to
other long-range ordered states besides superconductors.

Even if non-trivial quasiparticles have been found, there is still
the problem of determining their braiding properties.
In the quantum Hall case, we have described in Secs.
\ref{sec:interference}, \ref{sec:experiments2}
how this can be done using quasiparticle tunneling and interferometry
experiments. This requires even more intricate gating.
However, even these difficult experiments are the most concrete
that we have, and they work only because these states
are chiral and have gapless edge excitations -- and, therefore, have
non-trivial DC transport properties -- and because charged anyons
contribute directly to these transport properties.
Neutral quasiparticles are an even bigger challenge. Perhaps
they can be probed through thermal transport or even, if they carry spin,
through spin transport.

As we have seen in Sec. \ref{sec:interference},
abelian and non-Abelian interference effects are qualitatively different.
Indeed, the latter may actually be easier to observe in practice.
It is striking that quasiparticle interferometry,
which sounds like an {\it application} of
topological phases, is being studied as a basic probe of the state.
The naive logical order is reversed:  to see if a system is in a
topological phase, we are (ironically) saying ``shape the system
into a simple computer and if it computes as expected, then it must
have been in the suspected phase.''  This is a charming inversion,
but it should not close the door on the subject of probes.
It is, however, important to pause and note that
we now know the operational principles and
methodology for carrying out quasiparticle braiding in a concrete physical
system. It is, therefore, possible that non-Abelian anyons will
be observed in the quantum Hall regime in the near future.
This is truly remarkable. It would not close the book on non-Abelian
anyons, but open a new chapter and encourage us to
look for non-Abelian anyons elsewhere even while trying to
build a quantum computer with a quantum Hall state.

One important feature of non-Abelian anyons is that
they generally have multiple fusion channels.
These different fusion channels can be distinguished
interferometrically, as discussed in Secs. \ref{sec:interference}, \ref{sec:experiments2}.
This is not the only possibility. In ultra-cold neutral atom systems, 
they can be optically detected \cite{Tewari07a,Grosfeld07} in the
case of states with Ising anyons. Perhaps, in a solid, it will be possible
to measure the force between two anyons. Since the two fusion channels
will have different energies when the anyons are close together, there will
be different forces between them depending on how the anyons fuse.
If an atomic force microscope can `grab' an anyon
in order measure this force, perhaps it can also be used to
drag one around and perform a braid.

Thus, we see that new ideas would be extremely helpful in
the search for non-Abelian topological phases.
It may be the case that each physical system, e.g. FQHE, cold atoms,
Sr$_2$RuO$_4$ films, etc\dots, may be suited
to its own types of measurements, such as the ones described above
and in Secs. \ref{sec:interference}, \ref{sec:experiments2},
but general considerations, such as topological entropy
\cite{Kitaev06b,Levin06}, may inform and unify these
investigations. Another difficulty is that, as mentioned
above, we are currently searching for non-Abelian topological
phases in those systems in which there is an absence of alternatives.
It would be far better to have positive {\it a priori} reasons to
look at particular systems.

This state of affairs points to the dire need for 
general principles, perhaps of a mathematical nature, which
will tell us when a system is likely to have a topological phase.
Equivalently, can we define the necessary conditions for the existence of a topological
phase with non-Abelian quasiparticle statistics? 
For contrast, consider the case of magnetism.
Although there is a great deal which we don't know about magnetism,
we do know that we need solids containing ions with partially
filled $d$ or $f$ shells. Depending on the effective Coulomb interaction
within these orbitals and their filling fractions, we understand how
various mechanisms such as exchange and superexchange can lead
to effective spin-spin interactions which, in turn, can lead
to ferromagnetism, antiferromagnetism, spin-density-waves, etc..
We need a comparable understanding of topological phases.
One direction, which we have described in
Sec. \ref{sec:P-T-Invariant}, is to analyze models
in which the interactions encode some combinatorial relations,
such as those associated with string nets or loop gases
\cite{Levin05a,Fidkowski06,Freedman05a,Fendley05,Fendley07c}.
However, we only have a few examples of microscopic interactions
which give rise to these intermediate scale structures. We sorely need
more general guidelines which would enable us to look at
a given Hamiltonian and determine if it is likely to have
a non-Abelian topological phase; a more detailed analysis or experimental
study could then be carried out. This is a particularly important direction
for future research because, although nature has given us the quantum
Hall regime as a promising hunting ground for topological phases,
the energy scales are very low. A topological phase in a transition
metal oxide might have a much larger gap and, therefore,
be much more robust.

An important problem on the mathematical side is a
complete classification of topological phases. In this review,
we have focussed on a few examples of topological phases:
those associated with SU(2)$_k$ Chern-Simons theory,
especially the $k=2,3$ cases.
These are part of a more general class associated with
an arbitrary semi-simple Lie group $G$ at level $k$.
Another class is associated with discrete groups, such as phases
whose effective field theories are lattice gauge theories with
discrete gauge group. New topological phases can be obtained
from both of these by coset constructions and/or tensoring together
different effective field theories. However, a complete classification
is not known. With a complete classification in hand, if we were to
observe a topological phase in nature, we could identify it by
comparing it against the list of topological phases. Since we
have observed relatively few topological phases in nature,
we have not needed a complete classification. If, however,
many more are lurking, waiting to be observed, then a complete
classification could be useful in the way that the closely-related problem
of classifying rational conformal field theories has proved useful
in understanding classical and quantum critical points. 

We refer here, as we have throughout this article, to
topological phases as we have defined them in Sec. \ref{part2}
(and which we briefly recapitulate below).
There are many other possible `exotic' phases which share some
characteristics of topological phases, such as the emergence of gauge
fields in their low-energy theories \cite{Wen04}, but do not
satisfy all of the criteria. These do not appear to be useful
for quantum computation.

Finally, the three-dimensional frontier must be mentioned.  Most
theory (and experiment) pertains to 2D or quasi-2D systems.  In
3+1-dimensions, even the underlying mathematical structure of TQFTs
is quite open.  Little is known beyond finite group gauge theories.
For example, we do not know if quantum information can (in the
thermodynamic limit) be permanently stored at finite temperature in
any 3-dimensional system.  (By \onlinecite{Dennis02}, this is possible in
4+1-dimensions, not possible in 2+1-dimensions, and is an open question in
3+1-dimensions.)  The case of 2+1-dimensions has been the
playground of anyons for 30 years.  Will loop-like ``particles''
in 3+1-dimensions be as
rich a story 30 years from now?

Perhaps it is fitting to end this review with a succinct statement
of the definition of a topological phase: the ground state in the
presence of multiple quasiparticles or in a non-trivial topology has
a stable degeneracy which is immune to weak (but finite) local
perturbations. Note that the existence of an excitation gap is not
needed as a part of this definition although, as should be obvious
by this point, the stability of the ground state degeneracy to local
perturbations almost always necessitates the existence of an
excitation gap. We make three comments about this definition before
concluding: (1) incompressible FQH states satisfy our definition and
they are, so far, the only experimentally-established topological
phases. (2) The existence of a topological phase does not, by
itself, enable topological quantum computation -- one needs
quasiparticles with non-Abelian braiding statistics, and for {\it
universal} topological quantum computation, these quasiparticles'
topological properties must belong to a class which includes
SU(2)$_k$, with $k=3,5,6,7,8,9,\ldots$, as we have discussed
extensively in this article. (3) Possible non-Abelian quantum Hall
states, such as $\nu=5/2$ and $12/5$ are the first among several
possible candidates, including Sr$_2$RuO$_4$, which has recently
been shown to be a chiral $p$-wave superconductor
\cite{Xia06,Kidwingira06}, and $p$-wave paired cold atom
superfluids.

{\it Note added in proof:} A measurement of the charge of a quasi-particle  
in a $\nu=5/2$ fractional quantum Hall state has been recently  
reported by Dolev et al. in arXiv:0802.0930 (to appear in Nature).
In that measurement, current tunnels across a constriction between  
two opposite edge states of a Hall bar, and the quasi-particle charge is extracted from  
the current shot noise. Dolev et al. have found the charge to be consistent with $e/4$, and  
inconsistent with $e/2$.  A quasi-particle charge of $e/4$ is consistent
with paired states at $\nu=5/2$, including both the Moore-Read state,
the anti-Pfaffian state, and also Abelian paired states. Thus, the observation of
charge $e/4$ quasiparticles is necessary but not sufficient to show that
the $\nu=5/2$ state is non-Abelian.

{\it Note added in proof:}

Dolev {\it et al.} (arXiv:0802.0930; Nature, in press) have recently measured the
low-frequency current noise (`shot noise') at a point contact in
the $\nu=5/2$ state. They find the noise to be consistent with
charge-$e/4$ quasiparticles, and inconsistent with $e/2$.
A quasi-particle charge of $e/4$ is consistent with paired states at
$\nu=5/2$, including both the Moore-Read (Pfaffian) state, the anti-Pfaffian state, and
also Abelian paired states.

In another recent experiment, Radu {\it et al.} (arXiv:0803.3530) measured
the dependence on voltage and temperature of the tunneling current
at a point contact in the $\nu=5/2$ state. They find that the current is
well fit by the form $I = T^\alpha F(e^* V/k_B T)$ where $e^* = e/4$,
and the exponent $\alpha$ and scaling function $F(x)$ are at least
consistent with the anti-Pfaffian state, although it is premature to
rule out other states.

In a recent preprint (arXiv:0803.0737), Peterson {\it et al.} have
performed finite-system exact diagonalization studies which find the
correct ground state degeneracy on the torus at $\nu=5/2$ and also observe the
expected degeneracy between Pfaffian and anti-Pfaffian states.
The key new ingredient in their calculation is the inclusion of the effects of the finite-thickness
of the 2D layer which also appears to enhance the overlap between the
non-Abelian states and the exact numerical finite-system wavefunction at
$\nu=5/2$.

The first two papers provide the first direct experimental evidence in
support of the 5/2 state being non-Abelian while the third paper strengthens the
case from numerics.

\acknowledgements

The authors are grateful for support from Microsoft Station Q,
the National Science Foundation under grant DMR-0411800,
the Army Research Office under grant W911NF-04-1-0236, the
Israel Science Foundation, the U.S.-Israel Binational Science
Foundation, and Alcatel-Lucent Bell Labs.

\appendix

\section{Conformal Field Theory (CFT) for Pedestrians}
\label{section:CFT}

We consider chiral CFTs in 2 dimensions.  ``Chiral" means that all
of our fields will be functions of $z = x + i y$ only and not
functions of $\bar z$.  (For a good introduction to CFT
see \cite{Belavin84,Senechal97}).

\setcounter{mysubsection}{0}
\mysubsection{OPE} To describe a CFT we give its ``conformal
data", including a set of primary fields, each with a conformal
dimension $\Delta$, a table of fusion rules of these fields and a
central charge $c$ (which we will not need here, but is
fundamental to defining each CFT). Data for three CFTs are given
in Table \ref{tab:conformaldata}.

The operator product expansion (OPE) describes what happens to two
fields when their positions approach each other.  We write the OPE
for two arbitrary fields $\phi_i$ and $\phi_j$ as
\begin{equation}
\lim_{z \rightarrow w}  \phi_i(z) \phi_j(w) = \mbox{$ \sum_k  C_{i
j}^k$}\,\, (z - w)^{\Delta_k - \Delta_i - \Delta_j} \, \phi_k(w)
\end{equation}
where the structure constants $C_{ij}^k$ are only nonzero as
indicated by the fusion table. (For our purposes, we can assume
that all fields $\phi_k$ are primary fields.  So called
``descendant" fields, which are certain types of ``raising
operators" applied to the primary fields, can also occur on the
right hand side, with the dimension of the descendant being
greater than that of its primary by an integer. Since we will be
concerned only with leading singularities in the OPE, we will
ignore descendants. For all the CFTs that we consider the
coefficient of the primary on the right hand side will not vanish,
although this can happen.) Note that the OPE  works {\em inside} a
correlator.  For example, in the $\mathbb{Z}_3$ parafermion CFT
(see Table \ref{tab:conformaldata}), since  $\sigma_1 \times
\psi_1 = \epsilon$, for arbitrary fields $\phi_i$ we have
\begin{eqnarray}
& & \lim_{z \rightarrow w}  \,  \langle \,  \phi_1(z_1) \ldots
\phi_M(z_M) \, \sigma_1(z) \psi_1(w) \, \rangle   \\  \nonumber  &
\sim& (z - w)^{2/5 - 1/15 - 2/3} \langle \, \phi_1(z_1) \ldots
\phi_M(z_M) \epsilon(w) \,
 \rangle
\end{eqnarray}

In addition to the OPE, there is also an important ``neutrality"
condition: a correlator is zero unless all of the fields can fuse
together to form the identity field $\bf 1$. For example, in the
$\mathbb{Z}_3$ parafermion field theory $\langle \psi_2 \psi_1
\rangle \neq 0$ since $\psi_2 \times \psi_1 = {\bf 1}$, but
$\langle \psi_1 \psi_1 \rangle=0$ since $\psi_1 \times \psi_1 =
\psi_2 \neq {\bf 1}$.

\begin{table}[tbph]
 \begin{minipage}{3.5in}
\begin{minipage}{1.7in}
Chiral Bose Vertex: ($c=1$)

\begin{tabular}{|c||c|} \hline
$ $ & $\Delta$  \\
\hline \hline $e^{i\alpha\phi}$ & $\alpha^2/2$  \\ \hline
\end{tabular}
\hspace*{2pt}
\begin{tabular}{ |c||c||} \hline
$\times$ & $e^{i \alpha \phi}$  \\
\hline \hline $e^{i \beta \phi} $ & $e^{i(\alpha+\beta)\phi}$    \\ \hline
\end{tabular}
\end{minipage}
\begin{minipage}{1.7in}
 Ising CFT:  ($c=1/2$)

\begin{tabular}{|c||c|} \hline
$ $ & $\Delta$  \\
\hline \hline $\pfepsilon$ & $1/2$  \\ \hline
 $\sigma$ & $1/16$
\\ \hline
\end{tabular}
\hspace*{2pt}
\begin{tabular}{ |c||c|c||} \hline
$\times$ & $\pfepsilon$ & $\sigma$  \\
\hline \hline $\pfepsilon $ & {\bf 1} &   \\ \hline
 $\sigma $ & $\sigma$ &  ${\bf 1} + \pfepsilon $ \\ \hline
\end{tabular}
\end{minipage}

\vspace*{5pt}

$\mathbb{Z}_3$ Parafermion CFT:  ($c=4/5$)

\begin{tabular}{|c||c|} \hline
$ $ & $\Delta$  \\
\hline \hline $\psi_{1}$ & $2/3$  \\ \hline $\psi_{2}$ & $2/3$  \\
\hline
 $\sigma_{1}$ & $1/15$ \\
 \hline
 $\sigma_{2}$ & $1/15$ \\
 \hline
 $\epsilon$ & 2/5
\\ \hline
\end{tabular}
\hspace*{5pt}
\begin{tabular}{ |c||c|c|c|c|c||} \hline
$\times$ & $\psi_1 $ & $\psi_2 $ & $\sigma_1$ & $\sigma_ 2 $& $\epsilon$  \\
\hline \hline $\psi_1 $ & $\psi_2$ &  & &    & \\
\hline $\psi_2 $ & ${\bf 1}$ &  $\psi_1$  & & &  \\
\hline $\sigma_1 $ & $\epsilon$ &  $\sigma_2$  &  $\sigma_2 + \psi_1$ & &  \\
\hline $\sigma_2 $ & $\sigma_1$ &  $\epsilon$  &  ${\bf 1} + \epsilon$  & $\sigma_1 + \psi_2$ &   \\
\hline $\epsilon $ & $\sigma_2$ &  $\sigma_1$  &  $\sigma_1 + \psi_2$  & $\sigma_2 + \psi_1$ &  ${\bf 1}+ \epsilon$ \\
\hline
\end{tabular}
%\end{minipage}
%\begin{minipage}{1.5in}
\end{minipage}
\vspace*{5pt}
 \caption{Conformal data for three CFTs.   Given is the list of primary fields in the CFT with their
conformal dimension $\Delta$, as well as the fusion table.
 In addition, every CFT has an identity field ${\bf 1}$ with dimension $\Delta=0$
 which fuses trivially with any field (${\bf 1} \times
 \phi_i = \phi_i$ for any $\phi_i$).
 Note that fusion tables are symmetric so only
 the lower part is given.  In the Ising CFT the field $\pfepsilon$
 is frequently notated as $\pfnotepsilon$.
 This fusion table indicates the nonzero elements of the fusion
 matrix $N_{ab}^c$.  For example in the $\mathbb{Z}_3$ CFT,
 since ${\sigma_1} \times {\sigma_2} =
 {\bf 1} + \epsilon$, $N_{{\sigma_1} {\sigma_2}}^1 =
 N_{{\sigma_1} {\sigma_2}}^\epsilon = 1$ and
 $N_{{\sigma_1} {\sigma_2}}^c = 0$ for all $c$ not equal to ${\bf 1}$ or
 $\epsilon$.
 \vspace*{-10pt}
\label{tab:conformaldata}}
\end{table}

\mysubsection{Conformal Blocks}
\mylabel{Blocks}
Let us look at what
happens when a fusion has more than one possible result. For
example, in the Ising CFT, $\sigma \times \sigma = {\bf 1} + {\bf
\pfepsilon}$. Using the OPE, we have
\begin{equation}
 \lim_{w_1 \rightarrow w_2} \! \sigma(w_1) \sigma(w_2) \! \sim \frac{{\bf 1}}{(w_1 -
  w_2)^{1/8}}
   \! + (w_1 - w_2)^{3/8} \, {\pfepsilon} \label{eq:sigmasigma}
\end{equation}
where we have neglected the constants $C_{ij}^k$.  If we consider
$\langle \sigma \sigma \rangle$, the neutrality condition picks out
only the first term in Eq.~\ref{eq:sigmasigma} where the two
$\sigma$'s fuse to form $\bf 1$. Similarly, $\langle \sigma \sigma
\pfepsilon\rangle$ results in the second term of
Eq.~\ref{eq:sigmasigma} where the two $\sigma$'s fuse to form
$\pfepsilon$ which then fuses with the additional $\pfepsilon$ to
make $\bf 1$.

Fields may also fuse to form the identity in more than one way.
For example, in the correlator $\langle \sigma(w_1) \sigma(w_2)
\sigma(w_3) \sigma(w_4) \rangle$ of the Ising CFT, the identity is
obtained via two possible fusion paths
--- resulting in two different so-called ``conformal blocks".  On
the one hand, one can fuse $\sigma(w_1)$ and $\sigma(w_2)$ to form
$\bf 1$ and similarly fuse $\sigma(w_3)$ and $\sigma(w_4)$ to form
$\bf 1$. Alternately, one can fuse $\sigma(w_1)$ and $\sigma(w_2)$
to form $\pfepsilon$ and fuse $\sigma(w_3)$ and $\sigma(w_4)$ to
form $\pfepsilon$ then fuse the two resulting $\pfepsilon$ fields
together to form $\bf 1$.  The correlator generally gives a linear
combination of the possible resulting  conformal blocks. We should
thus think of such a correlator as living in a vector space rather
than having a single value. (If we instead choose to fuse $1$ with
$3$,  and $2$ with $4$,  we would obtain two blocks which are linear
combinations of the ones found by fusing 1 with 2 and 3 with 4. The
resulting vectors space, however, is independent of the order of
fusion). Crucially, transporting the coordinates $w_i$ around each
other makes a rotation within this vector space.

To be more clear about the notion of conformal blocks, let us look
at the explicit form of the Ising CFT correlator
\begin{eqnarray}
\label{eq:F1fuse}
 & & \lim_{w \rightarrow \infty}   \langle \sigma(0)\sigma(z) \sigma(1) \sigma(w) \rangle
    =
    a_+ \, F_+ +  a_- \, F_- \\
& &   F_{\pm}(z) \sim (w z(1- z))^{-1/8}  \sqrt{1 \pm \sqrt{1
-z}}\label{eq:F2fuse}
\end{eqnarray}
where $a_+$ and $a_-$ are arbitrary coefficients.
(Eqs.~\ref{eq:F1fuse}-\ref{eq:F2fuse} are results of calculations
not given here \cite{Senechal97}).  When $z \rightarrow 0$ we have $F_+
\sim z^{-1/8}$ whereas $F_- \sim z^{3/8}$. Comparing to
Eq.~\ref{eq:sigmasigma} we conclude that $F_+$ is the result of
fusing $\sigma(0) \times \sigma(z) \rightarrow {\bf 1}$ whereas
$F_-$ is the result of fusing $\sigma(0) \times \sigma(z)
\rightarrow \pfepsilon$.   As $z$ is taken in a clockwise circle
around the point $z=1$, the inner square-root changes sign,
switching $F_+$ and $F_-$.  Thus, this ``braiding" (or ``monodromy")
operation transforms
\begin{equation}
\label{eq:braid1}
    \mbox{${a_+}\choose{a_-}$} \rightarrow  e^{2 \pi i/8}    \mbox{${0 ~~ 1}\choose{1 ~~
    0}$}    \mbox{${a_+}\choose{a_-}$}
\end{equation}
Having a multiple valued correlator (I.e., multiple conformal
blocks) is a result of having such branch cuts. Braiding the
coordinates ($w$'s) around each other results in the correlator
changing values within its allowable vector space.

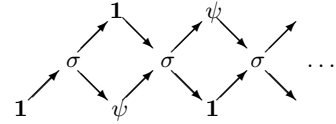
\begin{figure}[tbph]
\setlength{\unitlength}{1mm}
\begin{picture}(60,15)(0,0)
\put(-5,14){}\put(0,0){\vector(1,1){4}}
\put(-2,-2){$\bf{1}$} \put(5,4.5){$\sigma$}
\put(6.5,6.5){\vector(1,1){4}} \put(6.5,4){\vector(1,-1){4}}
\put(10.75,11){$\bf 1$}  \put(11,-2){$\pfepsilon$}
\put(12.5,0){\vector(1,1){4}} \put(12.5,11){\vector(1,-1){4}}
\put(17.5,4.5){$\sigma$} \put(19,6.5){\vector(1,1){4}}
\put(19,4){\vector(1,-1){4}} \put(23.5,11){$\pfepsilon$}
\put(23.5,-2){$\bf 1$} \put(25,0){\vector(1,1){4}}
\put(25,11){\vector(1,-1){4}} \put(29.5,4.5){$\sigma$}
\put(31.5,6.5){\vector(1,1){4}} \put(31.5,4){\vector(1,-1){4}}
\put(37,4.5){$\ldots$}
\end{picture}
\caption{Bratteli diagram for fusion of multiple $\sigma$ fields in
the Ising CFT.} \label{fig:bratteli2}\vspace*{-10pt}
\end{figure}

A useful technique for counting conformal blocks is the ``Bratteli
diagram."  In Fig.~\ref{fig:bratteli2} we give the Bratteli diagram
for the fusion of multiple $\sigma$ fields in the Ising CFT.
Starting with $\bf 1$ at the lower left, at each step moving from
the left to the right, we fuse with one more $\sigma$ field.  At the
first step, the arrow points from $\bf 1$ to $\bf \sigma$ since
${\bf 1} \times \sigma = \sigma$. At the next step $\sigma$ fuses
with $\sigma$ to produce either $\pfepsilon$ or $\bf 1$ and so
forth. Each conformal block is associated with a path through the
diagram. Thus to determine the number of blocks in $\langle \sigma
\sigma \sigma \sigma \rangle$ we count the number of paths of four
steps in the diagram starting at the lower left and ending at $\bf
1$.

\begin{figure}
\vspace*{30pt} \ifx\JPicScale\undefined\def\JPicScale{.5}\fi
\psset{unit=\JPicScale} \psset{unit=\JPicScale mm}
\psset{xunit=-.5mm}
\psset{linewidth=0.3,dotsep=1,hatchwidth=0.3,hatchsep=1.5,shadowsize=1,dimen=middle}
\psset{dotsize=0.7 2.5,dotscale=1 1,fillcolor=black}
\psset{arrowsize=1 2,arrowlength=1,arrowinset=0.25,tbarsize=0.7
5,bracketlength=0.15,rbracketlength=0.15}
\begin{pspicture}
(0,0)(60,80) \psline(20,80)(50,50) \psline(30,70)(40,80)
\psline(40,60)(60,80) \rput(26,80){$\phi_k$} \rput(43,80){$\phi_j$}
\rput(63,80){$\phi_i$} \rput(30,63){$\phi_p$} \rput(42,50){$\phi_m$}
\rput(0,60){\large$=\sum [F^{ijk}_{m}]_{pq}$} \rput(10,52){$q$}
\end{pspicture}
\psset{unit=\JPicScale} \psset{unit=\JPicScale mm}
\psset{linewidth=0.3,dotsep=1,hatchwidth=0.3,hatchsep=1.5,shadowsize=1,dimen=middle}
\psset{dotsize=0.7 2.5,dotscale=1 1,fillcolor=black}
\psset{arrowsize=1 2,arrowlength=1,arrowinset=0.25,tbarsize=0.7
5,bracketlength=0.15,rbracketlength=0.15} \hspace*{15pt}
\begin{pspicture}
(0,0)(60,80) \psline(20,80)(50,50) \psline(30,70)(40,80)
\psline(40,60)(60,80) \rput(25,80){$\phi_i$} \rput(44,80){$\phi_j$}
\rput(64,80){$\phi_k$} \rput(32,63){$\phi_q$} \rput(45,50){$\phi_m$}
\psline[linewidth=1.5pt](15,40)(15,90)
\psline[linewidth=1.5pt](15,90)(20,90)
\psline[linewidth=1.5pt](15,40)(20,40)
\psline[linewidth=1.5pt](70,40)(70,90)
\psline[linewidth=1.5pt](70,40)(65,40)
\psline[linewidth=1.5pt](70,90)(65,90)
\end{pspicture}
\vspace*{-60pt} \caption{The basis states obtained by fusing fields
together depends on the order of fusion (although the space spanned
by these states is independent of the order).  The $F$-matrix
converts between the possible bases.}
 \label{fig:fusiontree}
\end{figure}

\mysubsection{Changing Bases}
\mylabel{Basis}
As mentioned above,
the space spanned by the conformal blocks resulting from the fusion
of fields is independent of the order of fusion (which field is
fused with which field first). However, fusing fields together in
different orders results in a different basis for that space.   A
convenient way to notate fusion of fields is a particular order is
using fusion tree diagrams as shown in Fig. \ref{fig:fusiontree}.
Both diagrams in this figure show the fusion of three initial fields
$\phi_i, \phi_j, \phi_k$.  The diagram on the left shows $\phi_j$
and $\phi_k$ fusing together first to form $\phi_p$ which then fuses
with $\phi_i$ to form $\phi_m$.  One could equally well have chosen
to fuse together $\phi_i$ and $\phi_j$ together first before fusing
the result with $\phi_k$, as shown on the right of Fig.
\ref{fig:fusiontree}.  The mathematical relation between these two
bases is given in the equation shown in Fig. \ref{fig:fusiontree} in
terms of the so-called $F$-matrix (for ``fusion"), which is an
important property of any given CFT or TQFT.  An example
of using the $F$-matrix is given in section \ref{sec:fibonacci}.

\mysubsection{The Chiral Boson}
A particularly important CFT is
obtained from a free Bose field theory in 1+1 dimension by keeping
only the left moving modes \cite{Senechal97}. The free chiral Bose
field $\phi(z)$, which is a sum of left moving creation and
annihilation operators, has a correlator $ \langle \phi(z)
\phi(z') \rangle = -\log(z - z') $.   We then define the normal
ordered ``chiral vertex operator" $:\:e^{i \alpha\phi(z)}\!:$ ,
which is a conformal field. Note that we will typically not write the
normal ordering indicators `: :'. Since $\phi$ is
a free field, Wick's theorem can be used to obtain \cite{Senechal97}
\begin{eqnarray} \nonumber
    \left\langle \, e^{i{\alpha_1}\phi}(z_1) \ldots e^{i{\alpha_N}\phi}(z_N) \, \right\rangle
&=&
    e^{- \sum_{i < j} \alpha_i \alpha_j \langle \phi(z_i)
    \phi(z_j) \rangle} \\ &=& {\mbox{$\prod_{i<j}$}} \,\,\,  (z_i - z_j)^{\alpha_i
    \alpha_j}
    \label{eq:vertex1}
\end{eqnarray}
(Strictly speaking thi identity holds only if the
neutrality condition $\sum_i \alpha_i =
0$ is satisfied, otherwise the correlator vanishes).

%\nocite{*}
%\bibliographystyle{apsrmp-mod}
%\bibliography{rmp-refs}
%\end{document}

\end{document}